\documentclass[numberedappendix,iop,revtex4]{emulateapj}
\usepackage{lscape}
\usepackage{url}

\newcommand{\kms}{\ensuremath{\rm{km}~\rm{s}^{-1}}}
\newcommand{\HII}{\ion{H}{2}}
\newcommand{\CO}{$^{12}$\rm{CO}}
\newcommand{\Msun}{\ensuremath{M_{\odot}}}
\newcommand{\Mlum}{\ensuremath{M_{\rm lum}}}
\newcommand{\Mvir}{\ensuremath{M_{\rm vir}}}
\newcommand{\LCO}{\ensuremath{L_{\rm CO}}}

\begin{document}

\shorttitle{ALMA View of GMCs in NGC 300}
\shortauthors{Faesi, Lada, \& Forbrich}
\slugcomment{Accepted for publication in the Astrophysical Journal}

\title{The ALMA view of GMCs in NGC 300: Physical Properties and Scaling Relations at 10~pc Resolution}

\author{Christopher M. Faesi}

\author{Charles J. Lada}
\affiliation{Harvard-Smithsonian Center for Astrophysics \\ 60 Garden Street, Cambridge, MA 02138}

\author{Jan Forbrich}
\affiliation{University of Hertfordshire, Hatfield, Hertfordshire, United Kingdom}

\email{cfaesi@cfa.harvard.edu}

\begin{abstract}

We have conducted a $^{12}$CO(2-1) survey of several molecular gas complexes in the vicinity of \ion{H}{2} regions within the spiral galaxy NGC 300 using the Atacama Large Millimeter Array. Our observations attain a resolution of 10 pc and 1 {\kms}, sufficient to fully resolve Giant Molecular Clouds (GMCs), and are the highest to date obtained beyond the Local Group. We use the CPROPS algorithm to identify and characterize 250 GMCs across the observed regions. GMCs in NGC 300 appear qualitatively and quantitatively similar to those in the Milky Way disk: they show an identical scaling relationship between size $R$ and linewidth $\Delta V$ ($\Delta V \propto R^{0.48\pm0.05}$), appear to be mostly in virial equilibrium, and are consistent with having a constant surface density of $60~\Msun$ pc$^{-2}$. The GMC mass spectrum is similar to those in the inner disks of spiral galaxies (including the Milky Way). Our results suggest that global galactic properties such as total stellar mass, morphology, and average metallicity may not play a major role in setting GMC properties, at least within the disks of galaxies on the star-forming main sequence. Instead, GMC properties may be more strongly influenced by local environmental factors such as the mid-plane disk pressure. In particular, in the inner disk of NGC 300 we find this pressure to be similar to that in the local Milky Way but markedly lower than that in the disk of M51 where GMCs are characterized by systematically higher surface densities and a higher coefficient for the size-linewidth relation.

\end{abstract}



\section{Introduction}
\label{sec:intro}

Giant Molecular Clouds (GMCs) are the cold, dense, turbulent structures within the interstellar medium (ISM) in which essentially all stars form. GMCs thus represent the initial conditions for star formation, and so understanding their properties and life cycle is the key to understanding the interplay between gas and stars within galaxies. GMCs are most easily traced (particularly in external galaxies) by emission from the low-lying rotational ($J$) states of the CO molecule, which are readily excited via collisions at the 5--20~K temperatures found in clouds, and been shown to be co-extensive with molecular hydrogen (H$_2$) above modest hydrogen gas column densities~\citep[i.e., $N_{\rm H} \gtrsim 1.4 \times 10^{21}$~cm$^{-2}$;][]{1988ApJ...334..771V}. The term ``GMC'' was first coined to describe structures with H$_2$ masses in excess of $10^5~\Msun$~\citep{1980gmcg.work.....S}; here, for simplicity, we will use the terms ``GMC'' and ``cloud'' interchangeably to refer to all molecular gas structures we detect.

Based primarily on Milky Way observations over the last several decades, GMCs have become a well-characterized class of objects. They typically have masses $M$ of a few thousand to a few million $\Msun$, sizes $R$ of a few to about a hundred pc, and highly supersonic bulk velocity dispersions $\sigma_v$ \citep[e.g.,][and references therein]{Heyer:2015ee}. In the inner several kpc of the Milky Way, GMCs are distributed by mass as a power law with slope $\gamma \approx -1.6$ \citep[e.g.,][]{Solomon:1987uq,Rosolowsky:2005gt,Rice:2016ko}. \cite{Larson:1981vv} identified three key empirical GMC scaling relationships that have become canonical diagnostics for the structure and physical conditions of clouds: (1) a power law scaling between size and velocity dispersion (or equivalently, linewidth) such that $R \propto \sigma_v^{0.5}$, suggesting that GMCs reflect the scale-dependent, hierarchical turbulent structure of the ISM; (2) an approximate balance between gravitational potential energy and kinetic energy, implying that clouds are on average gravitationally bound; and, (3) $M \propto R^2$, which implies a constant mass surface density. These relations have subsequently been studied in great detail and verified to hold in Milky Way molecular clouds \citep[e.g.,][]{Solomon:1987uq,2010A&A...519L...7L,Rice:2016ko,2017ApJ...837..154N}.

While Milky Way observations have established the empirical foundation for GMCs, Galactic studies are subject to a number of observational challenges including cloud blending along the line-of-sight and difficulty in measuring distances due to our unique embedded vantage point within the Galaxy's disk. With the advent of modern (sub)millimeter interferometers and large single dish telescopes, it has recently become possible to achieve tens-of-parsec resolution and sufficient surface brightness sensitivity to marginally resolve and characterize GMCs in many nearby galaxies. The distance to all clouds in (low inclination) extragalactic systems is essentially identical, meaning that uncertainties in distance-dependent properties such as size and mass are only affected in a relative sense, and blending is no longer an issue. Over the past two decades high-resolution CO observations have been conducted for a handful of galaxies beyond the Milky Way, including M33 \citep{2003ApJS..149..343E,2003ApJ...599..258R,2007ApJ...661..830R,Gratier:2012km}, M31 \citep{1988ApJ...328..143L,2007ApJ...654..240R,2008ApJ...675..330S}, the Large Magellanic Cloud \citep[LMC;][]{2001PASJ...53L..41F,2008ApJS..178...56F,2010MNRAS.406.2065H,Wong:2011ib}, IC~342 \citep{Hirota:2011dt}, M51 \citep{Schinnerer:2013jy,2014ApJ...784....3C}, NGC~300~\citep{Faesi:2016bl}, M83 \citep{Freeman:2017bb}, various Local Group dwarf galaxies \citep[e.g.,][and references therein]{Schruba:2017ck,2008ApJ...686..948B}, and several other nearby ($D \sim $~3 -- 8 Mpc) spiral galaxies \citep[e.g.,][]{Rebolledo:2012ex,2013ApJ...772..107D}.

GMC properties in other galaxies appear to broadly span a similar range of physical properties as those observed in the Milky Way, within the limited resolution and sensitivity available to these studies. However, some studies have begun to show the importance of environment in setting macroscopic cloud properties \citep[e.g.,][]{2013ApJ...779...46H,2014ApJ...784....3C}. The GMC mass spectrum in M33 is significantly steeper beyond 2~kpc than in the inner galaxy~\citep{Gratier:2012km}, analogous to the radial variation seen in the Milky Way~\citep{Rice:2016ko}. In M51, the mass spectrum slope is steeper in the interarm regions than in the spiral arms \citep{2014ApJ...784....3C}. In some extragalactic studies, general consistency with the Milky Way size-linewidth relation is reported \citep{2003ApJ...599..258R,2007ApJ...654..240R,2008ApJ...686..948B,Hirota:2011dt,Rebolledo:2012ex,Faesi:2016bl}. In other investigations, the authors find a distinct lack of correlation between size and linewidth \citep{Wong:2011ib,Gratier:2012km,2013ApJ...779...46H}. Similar disagreement is present regarding the virial state and surface density of clouds.

Due to the differing spatial and spectral resolution, limited dynamic range, and variety of cloud decomposition schemes employed in extragalactic studies, it is difficult to disentangle true physical differences from methodological effects. In all previous extragalactic studies (with the exception of those toward nearby dwarf galaxies) no prior observations have obtained spatial resolution better than $\sim20$~pc or spectral resolution better than $\sim2~\kms$, and sample sizes have often been limited. It may be that the Larson relations observed in the central Milky Way are universal, but that extragalactic observations have insufficient resolution and/or sensitivity to detect and disentangle molecular gas into its constituent structures. The apparent universality seen locally may also potentially be restricted to certain environments, likely those similar to the inner disk of our galaxy. To test these possibilities, it is necessary to match as nearly as possible the spatial and spectral resolution and sensitivity of Milky Way studies.

In the present study we have conducted the highest resolution to-date investigation into an extragalactic GMC population beyond the Local Group via Atacama Large (sub)Millimeter Array (ALMA) observations of the nearby spiral NGC~300. By targeting a nearby, southern galaxy, we are able to achieve $< 10$~pc and $1~\kms$ resolution at the 1.93~Mpc distance of the galaxy \citep{2004AJ....128.1167G}, thereby enabling direct and salient comparison with the well-characterized Milky Way GMC population. We focus our attention on regions within the galaxy in which we have measured the bulk molecular gas content (via CO(2-1) spectroscopy) and star formation rate (via multiwavelength observations and population synthesis modeling) on 250~pc scales~\citep[][hereafter F14]{Faesi:2014ib}. This investigation follows the pilot study of \cite{Faesi:2016bl} in which we successfully detected the molecular gas substructure within a subset of these regions with the Submillimeter Array (SMA) but only marginally resolved GMC scales ($\sim 40$~pc).

This manuscript is organized as follows. In Section~\ref{sec:obs} we present our ALMA observations and data processing. Section~\ref{sec:clouds} describes our GMC identification and characterization procedure and defines the physical quantities of relevance. In Section~\ref{sec:results} we present a description of our results, including the basic statistics of our catalog of 250 clouds. We compile a number of empirical GMC diagnostics and compare them to the Milky Way and other galaxies in Section~\ref{sec:GMCpop}. Section~\ref{sec:disc} involves the role of global and local galactic environment in setting GMC properties, and discusses the implications of our results in the context of previous GMC studies. We summarize our results in Section~\ref{sec:summary}. Finally, the Appendix discusses algorithmic details, presents zeroth moment maps of all observed regions with the detected clouds overlaid, and includes a table presenting the full cloud catalog.


\section{Observations and Data}
\label{sec:obs}

\begin{figure*}
\includegraphics[width=\linewidth]{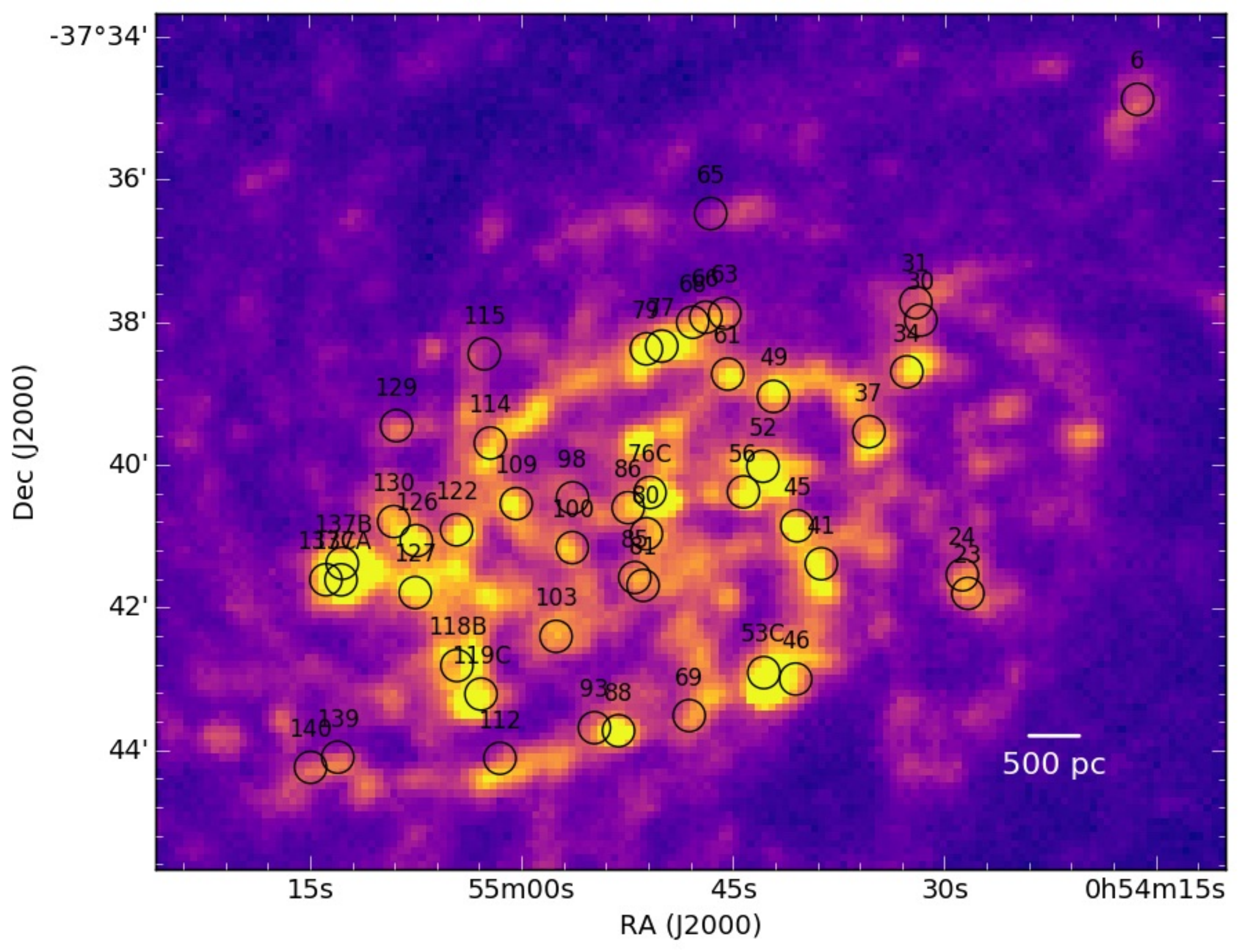}
\caption{\textit{Herschel}/SPIRE 250~$\mu$m map of NGC~300 with the 48 ALMA-observed regions presented in this study overlaid as black circles. The circle sizes are scaled to the ALMA primary beam FWHM of $26.1\arcsec$ (250~pc at the 1.93~Mpc distance of NGC~300). The numbers refer to the region numbers from the \cite{Deharveng:1988wh} catalog, which we also carry forward in our catalog.}
\label{fig:almaobs}
\end{figure*}

We obtained Atacama Large Millimeter Array (ALMA) Cycle 2 observations (project code 2013.1.00586.S; PI: C. Lada) in the CO(2-1) line at 230.538 GHz towards 48 positions in NGC 300 representing all 42 CO detections plus 6 selected CO nondetections from the Atacama Pathfinder EXperiment (APEX) survey of F14. The observed pointing centers were originally drawn from the {\HII} region catalog of \cite{Deharveng:1988wh}, with refinements made to better center on CO peaks for the subsample of 10 regions detected with the SMA \citep{Faesi:2016bl}. We note that our ALMA primary beam FWHM size is much larger than the vast majority of the \cite{Deharveng:1988wh} {\HII} regions, and so our study encompasses both actively star-forming and quiescent portions of the galaxy. Figure~\ref{fig:almaobs} shows the observed regions on a 250~$\mu$m \textit{Herschel}/SPIRE image obtained by our team (PI: J. Forbrich), and Table~\ref{tab:targets} lists the observed regions.

\begin{deluxetable}{l c c r r r}
\centering
\tabletypesize{\scriptsize}
\tablecolumns{6}
\tablewidth{0pt}
\tablecaption{NGC 300 Regions Observed \label{tab:targets}}
\tablehead{
	\colhead{Region} &
	\colhead{R.A.} &
	\colhead{decl.} &
	\colhead{$I_{\rm CO}$\tablenotemark{a}} &
	\colhead{$v_0$\tablenotemark{a}} &
	\colhead{FWHM\tablenotemark{a}}
 \\
 	\colhead{} &
	\colhead{(J2000)} &
	\colhead{(J2000)} &
	\colhead{(K)} &
	\colhead{{\kms}} &
	\colhead{{\kms}}
}
\startdata
dcl6     & 00:54:16.39 & -37:34:52.5 & $<0.014$ & 187.6\tablenotemark{b} & \nodata \\
dcl23   & 00:54:28.36 & -37:41:48.3 & 0.364 & 173.6 & 10.1 \\
dcl24   & 00:54:28.75 & -37:41:32.7 & 0.045 & 179.6 & 15.0 \\
dcl30   & 00:54:31.71 & -37:37:58.4 & 0.255 & 195.7 & 11.5 \\
dcl31   & 00:54:32.07 & -37:37:43.7 & 0.153 & 198.1 & 6.1  \\
dcl34   & 00:54:32.70 & -37:38:42.0 & 0.313 & 191.0 & 7.9  \\
dcl37   & 00:54:35.38 & -37:39:32.4 & 0.431 & 185.0 & 9.4  \\
dcl41   & 00:54:38.75 & -37:41:23.5 & 0.899 & 167.1 & 14.9 \\
dcl45   & 00:54:40.48 & -37:40:51.6 & 0.057 & 174.9 & 15.0 \\
dcl46   & 00:54:40.56 & -37:43:00.5 & 0.322 & 167.2 & 11.4 \\
dcl49   & 00:54:42.15 & -37:39:02.8 & 0.185 & 172.2 & 8.0  \\ 
dcl52   & 00:54:42.89 & -37:40:01.5 & 0.602 & 179.6 & 10.6 \\
dcl53C & 00:54:42.82 & -37:42:55.1 & 0.243 & 160.6 & 15.0 \\
dcl56   & 00:54:44.27 & -37:40:23.1 & 0.128 & 167.5 & 4.5  \\
dcl61   & 00:54:45.39 & -37:38:44.0 & 0.251 & 161.3 & 6.4  \\
dcl63   & 00:54:45.60 & -37:37:53.0 & 0.178 & 152.0 & 5.7  \\
dcl65   & 00:54:46.60 & -37:36:29.0 & 0.188 & 159.6 & 5.3  \\
dcl66   & 00:54:46.94 & -37:37:56.1 & 0.245 & 153.1 & 7.0  \\
dcl68   & 00:54:47.87 & -37:38:00.4 & 0.517 & 155.1 & 9.2  \\
dcl69   & 00:54:48.30 & -37:43:40.1 & 0.412 & 141.2 & 7.4  \\
dcl76C& 00:54:50.89 & -37:40:23.6 & 0.643 & 152.8 & 14.8 \\
dcl77   & 00:54:50.06 & -37:38:20.5 & \nodata & \nodata & \nodata \\
dcl79   & 00:54:51.15 & -37:38:22.8 & 1.140 & 156.6 & 9.0  \\
dcl80   & 00:54:51.15 & -37:40:58.3 & 0.179 & 144.4 & 13.0 \\
dcl81   & 00:54:51.38 & -37:41:41.8 & 0.201 & 138.8 & 6.2  \\
dcl85   & 00:54:51.97 & -37:41:35.1 & 0.284 & 138.4 & 6.7  \\
dcl86   & 00:54:52.44 & -37:40:36.4 & 0.353 & 146.4 & 16.9 \\
dcl88   & 00:54:53.12 & -37:43:44.0 & 0.591 & 126.4 & 4.1  \\
dcl93   & 00:54:54.83 & -37:43:41.5 & 0.319 & 125.6 & 9.3  \\
dcl98   & 00:54:56.38 & -37:40:28.1 & 0.471 & 134.4 & 23.2 \\
dcl100  & 00:54:56.40 & -37:41:10.0 & 0.283 & 123.1 & 9.5  \\
dcl103  & 00:54:57.55 & -37:42:24.7 & 0.413 & 122.5 & 7.6  \\
dcl109  & 00:55:00.37 & -37:40:33.0 & 0.626 & 123.0 & 17.9 \\
dcl112  & 00:55:01.53 & -37:44:07.2 & 0.366 & 112.3 & 23.3 \\
dcl114  & 00:55:02.34 & -37:39:53.8 & 0.971 & 123.7 & 16.5 \\
dcl115  & 00:55:02.63 & -37:38:27.0 & $<0.158$ & 126.4\tablenotemark{b} & \nodata \\
dcl118B & 00:55:04.58 & -37:42:49.2 & 0.267 & 101.1 & 8.5  \\
dcl119C & 00:55:02.87 & -37:43:13.2 & 0.336 & 109.2 & 5.3  \\
dcl122  & 00:55:04.60 & -37:40:55.0 & 0.423 & 106.9 & 8.4  \\
dcl126  & 00:55:07.45 & -37:41:04.1 & 0.370 & 98.9  & 6.1  \\
dcl127  & 00:55:07.53 & -37:41:47.8 & 0.745 & 97.2  & 8.7  \\
dcl129  & 00:55:08.85 & -37:39:27.4 & 0.251 & 115.3 & 4.6  \\
dcl130  & 00:55:09.05 & -37:40:48.0 & 0.726 & 100.9 & 24.1 \\
dcl137A & 00:55:12.79 & -37:41:37.0 & 0.324 & 87.1  & 7.6  \\
dcl137B & 00:55:12.70 & -37:41:23.1 & 0.620 & 95.5  & 10.1 \\
dcl137C & 00:55:13.86 & -37:41:36.9 & 0.735 & 97.3  & 6.5  \\
dcl139  & 00:55:13.03 & -37:44:06.2 & 0.274 & 86.3  & 5.2  \\
dcl140  & 00:55:14.96 & -37:44:14.7 & 0.312 & 84.6  & 7.6
\enddata
\tablenotetext{a}{CO integrated intensity $I_{\rm CO}$, velocity centroid $v_0$, and spectral Gaussian fit FWHM from the APEX single dish observations of \cite{Faesi:2014ib}}
\tablenotetext{b}{$v_0$ for CO nondetections is instead the expected velocity at that position based on the \ion{H}{1} moment 1 map of \cite{1990AJ....100.1468P}.}
\end{deluxetable}

\subsection{ALMA observations}

All targets were observed with the ALMA 12 m array during Cycle 2 on 9, 12, and 15 December 2014. Each of the 48 regions was observed for two  sets of five 6-second scans each night, adding up to a total on-source time of 3.03 minutes per target. Table~\ref{tab:obslog} lists the observing dates and conditions. Observations of a bandpass calibrator (J0334-4008 on 9 December, J2357-5311 on 12 and 15 December) and flux calibrator (Uranus) preceded science observations each night. J0106-4034 was observed as a phase calibrator every $\sim6$ minutes, for a total of ten 30-second observations per night. ALMA was in configuration C34-2/1 for all three observing nights, with 32, 37, and 36 antennas (by night, sequentially) arranged with baselines ranging from 15.0 m to 348.5 m, implying a minimum beam angular dimension of $0.9\arcsec$ and a maximum recoverable scale of $10.7\arcsec$ (at 230.5 GHz). This corresponds to sensitivity to physical scales of 8.4 to 100~pc at the 1.93~Mpc distance of NGC~300 -- ideal for detecting emission from GMCs. The primary beam for the 12 m array was measured to be $26.07\arcsec$ at FWHM.

The ALMA Band 6 correlator was set up to have three spectral windows: two spectral line windows with resolution 122 kHz ($\sim0.16~\kms$), bandwidth 469 MHz, and centered at 230.538~GHz (to cover $^{12}$CO(2-1)) and 219.560~GHz (for C$^{18}$O(2-1)), adjusted for the local standard-of-rest velocity of NGC~300, $144~\kms$\footnote{\url{https://ned.ipac.caltech.edu}} (heliocentric), plus one continuum spectral window with resolution 15.625 MHz ($\sim20~\kms$), bandwidth 1.875 GHz, and centered at $\sim231.5$~GHz. We did not detect significant emission in C$^{18}$O(2-1) in the pipeline-produced images and defer further analysis and discussion of that data to a future paper.

\begin{deluxetable*}{l c c c c c c c}
\centering
\tabletypesize{\scriptsize}
\tablecolumns{8}
\tablewidth{0pt}
\tablecaption{ALMA Observation Log \label{tab:obslog}}
\tablehead{
	\colhead{Execution Block} &
	\colhead{Number of} &
	\colhead{Start Time} &
	\colhead{End Time} &
	\colhead{Median} &
	\colhead{Median} &
	\colhead{Median} &
	\colhead{On-source} \\
	\colhead{} &
	\colhead{Antennas} &
	\colhead{(UT)} &
	\colhead{(UT)} &
	\colhead{elevation} &
	\colhead{$T_{\rm sys}$\tablenotemark{a}} &
	\colhead{PWV} &
	\colhead{time} \\
	\colhead{} &
	\colhead{} &
	\colhead{} &
	\colhead{} &
	\colhead{(degrees)} &
	\colhead{(K)} &
	\colhead{(mm)} &
	\colhead{(minutes)}
}
\startdata
uid://A002/X9630c0/X4d3 &	 32 &		2014 Dec 9 01:45:05 &	2014 Dec 9 03:02:54 &	58 &		67 &	0.62 &	48.5 \\
uid://A002/X96bfab/X585 &	37 &		2014 Dec 13 00:08:11 &	2014 Dec 13 01:24:55 &	71 &		71 &	0.83 &	48.5 \\
uid://A002/X9707f1/X548 &	36 &		2014 Dec 15 00:11:45 &	2014 Dec 15 01:28:09 &	69 &		91 &	2.33 &	48.5
\enddata
\tablenotetext{a}{for spectral window 0, which contains the CO(2-1) line}
\end{deluxetable*}

\subsection{Data processing}

Our data were processed and imaged entirely using the Common Astronomy Software Applications (CASA) package (\url{http://casa.nrao.edu}). Calibration was performed manually by the North American ALMA Science Center team using CASA version 4.3.0. Briefly, the processing followed these steps: a priori flagging of autocorrelation data and data from shadowed antennas, if any; calibration using the Water Vapor Radiometer data, $T_{\rm sys}$, and antenna positions; bandpass calibration to correct for varying frequency response; gain calibration to adjust for phase and amplitude variations with time; and, flux calibration using the Butler-JPL-Horizons 2012 model for Uranus to set the flux scale. We adopt ALMA's fiducial 1\% relative and 5\% absolute flux calibration uncertainty\footnote{\url{https://safe.nrao.edu/wiki/bin/view/ALMA/CalAmp}} and propagate the latter into all calculations based on absolute flux density measurements. 

\subsection{Imaging and deconvolution}
\label{sec:imaging}

While the pipeline-imaged data cubes delivered to us demonstrated clear signal in CO(2-1) across the majority of observed regions, we re-did the entire imaging process in order to tailor the deconvolution algorithm for our particular science goals. For one, we aimed to maximize sensitivity in order to most reliably identify and detect the full extent of GMCs, thus we used natural weighting. Since we expect to easily resolve GMCs at our $\sim10$~pc resolution and thus need sensitivity to extended structure, we use the \texttt{multiscale CLEAN} algorithm~\citep{2008ISTSP...2..793C} implemented in \texttt{CASA}. This approach incorporates basis functions consisting of a series of circular Gaussians in addition to the delta functions utilized by the standard \texttt{CLEAN}. Small scales are weighted higher than larger scales such that the area-normalized weighting is approximately equal. The multiscale approach has been shown to better recover extended emission, reducing the depth of negative emission features and removing low-level flux missed by standard \texttt{CLEAN} algorithms \citep[e.g.,][]{2008AJ....136.2897R}. To increase our image fidelity at all potential scales of emission, we set the \texttt{multiscale} parameter to factors of 1, 2, 5, 10, and 15 times the synthesized beam, where the latter scale corresponds to the maximum recoverable scale for the array configuration. In practice, emission structures were well-fit by the smallest two to three scales in most cases. To balance sensitivity with the need to spectrally resolve GMCs, we smoothed our spectral images to $1~\kms$ velocity resolution. We imaged 200 velocity channels centered on the NGC~300 systemic velocity of $144~\kms$ -- sufficient to easily include all emission detected by APEX while placing all regions on the same velocity grid. The final cubes have $0.15~\arcsec$ pixels and a spatial extent of $48~\arcsec$.

We \texttt{CLEAN}ed each of the 48 regions manually and iteratively. To begin, we drew by hand a ``\texttt{CLEAN} box'' (in this case, a polygon) around all significant and coherent emission in the single velocity channel having the highest peak brightness in the entire data cube. Here we define ``significant'' to mean a factor of at least a few higher by-eye than the noise, and ``coherent'' to mean that the emission extends over at least one synthesized beam in area. We then examined successive contiguous velocity channels within $20~\kms$ of the APEX peak velocity and extended the existing \texttt{CLEAN} box created as described above to cover any additional significant, coherent emission found at those velocities. We then used this extended box as a mask for deconvolution of all image planes. We ran \texttt{CLEAN} for 100 iterations at a time with loop gain set to 0.1 and a threshold of $4\sigma$, where $\sigma$ is the measured RMS noise in emission-free channels of the individual data cube. After each cycle of 100 iterations we examined the residuals and extended the \texttt{CLEAN} box to encompass any additional significant emission that was now apparent, then re-ran the deconvolution for an additional 100 iterations at a time. This process continued until no additional emission peaks were identified and the final residuals were \texttt{CLEAN}ed to 4$\sigma$. Once this procedure was complete, we ran \texttt{CLEAN} again using the final \texttt{CLEAN} box to a threshold of $2\sigma$ in order to recover faint emission to the edges of clouds (see Section~\ref{sec:clouds}).

Since the spatial structure of the emission often varies between even nearby velocity channels, we also tested the results of using individual, channel-by-channel \texttt{CLEAN} boxes instead of a single box for all channels. We compared to the approach described above for three test regions by (a) examining the RMS noise in the final data cubes, and (b) identifying and characterizing clouds in each cube using \texttt{CPROPS}. We found no significant difference in either RMS noise or cloud properties in comparing these approaches, and thus utilized the far less time-intensive method of a single \texttt{CLEAN} box for each data cube.

To further test the level of convergence in the deconvolution procedure, we conducted a series of experiments on one of our data cubes, that of DCL88-126. We choose this region because it has resolved emission distributed in multiple structures throughout the primary beam area and is relatively bright (peak signal-to-noise of $\sim27$). This analysis is motivated by the convergence tests presented in \cite{Pety:2013fw}. First, we measured the total CLEANed flux after each major cycle of 100 iterations in each channel, and tabulated this for 1000 major cycles (using no threshold). We find that in all channels with significant signal, the total CLEANed flux increases rapidly within 300 iterations then plateaus, remaining nearly constant for the remainder of the experiment. It takes 800 iterations to reach the 2$\sigma$ threshold defined above, by which all channels have essentially leveled off in their total CLEANed flux. As the second test, we compare the residuals in channels with signal after the CLEANing reaches the 2$\sigma$ threshold to the images in line-free channels of the dirty image, and find that they are statistically indistinguishable: the residuals look like noise. Finally, we run the CLEAN procedure for 1600 iterations -- a factor of two longer than that needed to reach the 2$\sigma$ threshold -- and subtract the resulting residuals on a channel-by-channel basis from the residuals of the final (800 iteration-CLEANed) cubes. The resultant channel images again look like noise, which we take to be a sign that the deconvolution has converged to a reasonable extent. Note that since the threshold and mask are defined in each region individually, the number of CLEAN iterations varies from a few hundred to several thousand across the sample, but since these definitions are based on local noise levels, the conclusions drawn from our experiments should be representative.

The RMS noise in the final data cubes ranges between 6.6 and 6.8~mJy~beam$^{-1}$, which corresponds to a surface brightness sensitivity of about 130~mK (using the higher value). Assuming the median CO-to-H$_2$ conversion factor and median GMC linewidth from our sample (see Section~\ref{sec:clouds}), this implies a 3$\sigma$ mass surface density sensitivity of $13.2~\Msun$~pc$^{-2}$. This calculation also assumed a mean molecular weight of $\mu = 2.2$ (i.e. a helium mass fraction of 0.1) and a CO(2-1) to CO(1-0) line ratio of 0.7. The synthesized beam (i.e., final angular resolution) is approximately $1.5\arcsec \times 0.9\arcsec$, with a position angle of roughly $90^{\circ}$ (measured E from N). This translates into a physical resolution of 14~pc by 8.4~pc at the 1.93~Mpc distance of NGC~300. From here on we will refer to size scales primarily in pc, noting the dependence of our assumed distance to the galaxy.

\subsection{Continuum data and dust emission upper limits}

We also imaged the continuum, averaging the line-free channels in the two line spectral windows in addition to the full continuum spectral window. This resulted in a total effective bandwidth of approximately 2.6~GHz after excising a narrow peak at 231~GHz that was obvious in the raw visibility data in most sources. No continuum emission was detected in any of the regions studied, and thus we did not attempt to continuum-subtract the spectral line data. The RMS noise level was approximately 0.13-0.15~mJy~beam$^{-1}$, which implies a $3\sigma$ upper limit to the dust mass of $1.6 \times 10^3~\Msun$ at the 1.93~Mpc distance of NGC~300 for unresolved structures (assuming a dust temperature of 20~K and an opacity of 0.19~cm$^2$~g$^{-1}$, the latter based on the \citealp{2001ApJ...548..296W} Milky Way R$_V=3.1$ dust models\footnote{\url{https://www.astro.princeton.edu/$\sim$draine/dust/dustmix.html}}). This is equivalent to an upper limit of $\sim2$ -- $3 \times 10^5~\Msun$ for the gas mass based on a dust-to-gas ratio of $1/160$ (the appropriate metallicity-dependent value from \cite{2009ASPC..414..453D} for the average metallicity of NGC~300). The highest (CO-determined) mass unresolved GMC in our sample is about $5 \times 10^4~\Msun$, thus our nondetections in 1.3~mm continuum are consistent with the CO-inferred gas mass estimates we measure. Based on these calculations, we are within a factor of $\sim5$ of detecting the peaks of cold dust emission in NGC~300 GMCs. Equivalently, assuming an area the size of the synthesized beam, the continuum RMS noise implies a $3\sigma$ upper limit to the dust surface density of $14~\Msun$~pc$^{-2}$.

\subsection{Comparison to APEX single dish observations}

F14 presented single-pointing APEX single dish observations of several regions in NGC~300, including the vast majority of those also observed here. To assess the degree to which our ALMA interferometric observations recover the single dish flux measured by APEX, we have performed a comparison between the integrated ALMA spectrum and APEX single dish spectrum in each region observed at the same position in both campaigns (44 regions total). The procedure follows that of \cite{Faesi:2016bl}, which we briefly summarize again here. First we created a spatial mask for the ALMA data cube which restricts the region of analysis to include only pixels with at least one channel having signal greater than three times the RMS noise, measured individually for each pixel. Next, we multiplied each plane of the data cube by the APEX beam profile, which we estimate as a Gaussian with FWHM 27$\arcsec$. Then, we integrated all unmasked pixels in each channel to create a single ALMA spectrum. Finally, we summed the spectral flux in a $40~\kms$ wide window centered on the \ion{H}{1} velocity at the region's position (taken from F14). The resulting measurement of the CO integrated intensity $I_{\rm ALMA}$ is then divided by $I_{\rm APEX}$, the equivalent quantity from F14, to arrive at the desired ratio $\mathcal{F_{\rm rec}} \equiv I_{\rm ALMA} / I_{\rm APEX}$. We also compute an uncertainty on this ratio by propagating the measured uncertainties in the integrated intensities.

\begin{figure}
\includegraphics[width=\linewidth]{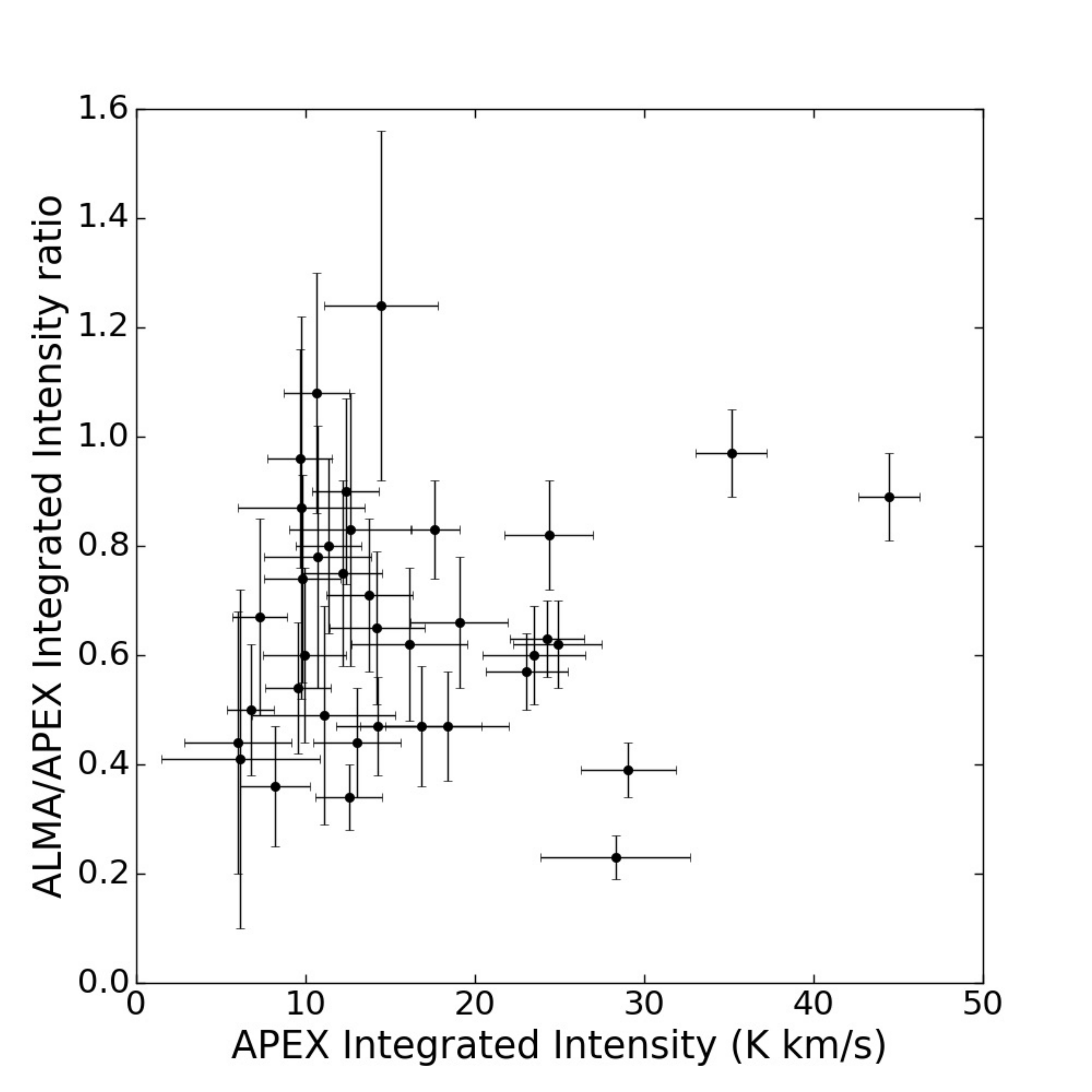}
\caption{Ratio of ALMA interferometric to APEX single dish integrated intensity as a function of APEX intensity for the sources in our sample with matched observations in both data sets. $1\sigma$ error bars are shown. The median ratio is 0.63, and the majority of regions are consistent with full single dish flux recovery to within the $3\sigma$ errors. Any missing flux is likely due to a combination of limited surface brightness sensitivity and some undetected extended emission on large ($> 100$~pc) scales. However, the array configuration used for these observations is sensitive to the full scale of GMCs from 10 to 100~pc, and thus any such extended emission would not likely change our measured cloud properties significantly.}
\label{fig:apexalma}
\end{figure}

Figure~\ref{fig:apexalma} shows the distribution of $\mathcal{F_{\rm rec}}$ for the 37 regions with (1) $>3\sigma$ APEX detections; (2) uncertainties on $\mathcal{F_{\rm rec}}$ less than 0.5; (3) observations at identical positions in the APEX and ALMA campaigns. We find a median $\mathcal{F_{\rm rec}}$ of 0.63 with typical uncertainties of about 0.2, and a range from 0.34 to approximately unity across our sample. 62\% of the regions are consistent to within $3\sigma$ formal uncertainties of recovering all the APEX flux. The range and median are consistent with the fraction of single dish intensity recovered in the ten NGC 300 regions observed by the SMA, as presented in \cite{Faesi:2016bl}. Here, as in that paper, we propose that the majority of the emission missed by the interferometer is due to limited surface brightness sensitivity. Our ALMA observations have a typical sensitivity of $94$~mK when scaled to the APEX channel width of $1.4~\kms$, while the APEX observations achieved 11~mK on average. This means there may be low surface brightness CO emission we are unable to detect in the ALMA data. However note that the procedure we use to characterize GMCs does attempt to account for low brightness emission at the edges of clouds by extrapolating cloud properties to the 0~K contour (see Section~\ref{sec:clouds}). In other words, even if the edges of clouds are not detected, cloud properties should still be robust \citep[see][]{Rosolowsky:2006cb}. There may also be faint and/or very cold clouds in the regions observed that we are entirely unable to detect in CO due to our limited sensitivity, and these clouds may contribute to the flux missed by ALMA. There could also be potential systematic uncertainties unaccounted for such as in the absolute flux scaling, residual baseline subtraction errors in the single dish data, or over-inclusion of noise pixels in the integrated ALMA spectrum. Finally, it may also be that there is emission at spatial frequencies below those sampled by our interferometer configuration, particularly in the few regions that show the lowest values of $\mathcal{F_{\rm rec}}$ with small uncertainties. However, since our maximum recoverable scale corresponds to a physical size comparable to the maximum size of GMCs in the Milky Way \citep[$\sim100$~pc; e.g., ][]{Heyer:2015ee}, any emission from larger scales is not likely part of the GMCs in these regions. It would instead imply the presence of a diffuse component of the molecular ISM, as has been seen in other galaxies including the Milky Way \citep{2016ApJ...818..144R} and M51 \citep{Pety:2013fw}, and speculated to also be present in NGC~300 (F14). Since the present analysis is concerned principally with the GMC population of NGC~300, we leave this consideration to a future study.

\section{Cloud identification and characterization}
\label{sec:clouds}

To decompose the emission in our data cubes into structures and characterize the properties of these structures, we use a modified version of the \texttt{CPROPS} algorithm \citep[][hereafter RL06]{Rosolowsky:2006cb}. We will hereafter refer to such structures as ``GMCs'' or ``clouds.'' The cloud identification scheme follows \cite{Faesi:2016bl}, which are briefly summarized here.

\begin{enumerate}
\item{Identify all pairs of adjacent three-dimensional pixels (``voxels'') having signal-to-noise $>4\sigma$, where $\sigma$ is the local RMS noise computed on a pixel-by-pixel basis for each data cube.}
\item{Extend each such emission peak outwards to include all connected emission to a boundary of $2\sigma$. Each structure at this stage is labeled an ``island,'' which may consist of one or more spatially overlapping clouds.}
\item{Decompose each island into clouds by searching for all local maxima in the island using a 15~pc $\times$ 15~pc $\times$ $2~\kms$ moving cube and then temporarily assigning to each such maximum all emission uniquely associated with it. These collections of voxels are candidate clouds, which must then (a) be larger than one synthesized beam in area; (b) have a peak brightness temperature $>1~K$ ($\sim51$~mJy for the typical synthesized beam sizes in these observations) above the merge level with other candidate clouds, and; (c) have moments that change by more than a user-defined fraction when computed at contour levels just above and just below the merge level with adjacent candidates.The value of this fraction is set by the \texttt{SIGDISCONT} parameter. For our final catalog presented here, we use \texttt{SIGDISCONT}=5, corresponding to a conservative requirement of a change of a factor 5 in the moments to separate clouds. We arrived at this choice through a series of experiments which we discuss in detail in Appendix~\ref{sec:cpropsex}.}
\item{Assign all emission amongst surviving clouds using a modified \texttt{CLUMPFIND} \citep{1994ApJ...428..693W} algorithm. Using this method for assignment instead of the original RL06 algorithm ensures that all detected emission is assigned to one of the detected structures (see below).}
\end{enumerate}

The version of \texttt{CPROPS} we used in this analysis has been modified to accommodate elliptical synthesized beams in the deconvolution (E. Rosolowsky, private communication). In addition, due to the APEX observation scheme of F14, several of the regions we observed with ALMA overlap slightly. We removed duplicate clouds, i.e. those with central pixel within $1\arcsec$ of one another in separate regions, by keeping only the cloud nearest its respective pointing center. Comparing the physical properties within a set of duplicates also allows for an additional measure of the reliability of both the observations and the cloud decomposition. We find an average change of -21\%, -4\%, and -13\% in size, linewidth, and luminous mass, respectively, between sets of duplicates in our sample, in the sense that the cloud labeled as a ``duplicate'' (i.e., the cloud farther from a pointing center) tends to have smaller size, linewidth, and mass. This is perhaps unsurprising, as the absolute noise level increases with distance from the pointing center, making GMC recovery more difficult and likely excluding several pixels that should be in the cloud due to noise fluctuations. However, we note that these changes are smaller than the typical uncertainties in the respective properties, and so we treat this result as an additional validation that the GMCs detected at the same position in multiple pointings are actually real, and therefore a validation on our GMC detection scheme in general.

\subsection{Cloud properties}

Once all significant emission has been partitioned into GMCs, this collection of voxels is used to calculate GMC physical properties, summarized as follows (see RL06 and \citealp{Faesi:2016bl} for details). Note that only the first three properties are fully independent; the others are derived from various physically meaningful combinations of these three.

\subsubsection{Independent properties}

\begin{itemize}
\item{The \textbf{physical size $R$} is taken to be the geometric mean of the intensity-weighted second moments along the major and minor axes of the cloud multiplied by a scaling factor $\eta = 1.91$. This particular numerical value of $\eta$ was originally derived by \cite{Solomon:1987uq} as the empirical scaling factor between the measured (rectangular) angular extent of a GMC on the sky and the average second moment in the latitude and longitude directions, scaled such that the size reflects the radius of a circle with area equal to the rectangular area on the sky. It is also within a factor of 30\% of the theoretically-derived value of $\eta$ obtained for a spherical cloud with radial density profile $\rho \propto R^{-1}$. To follow convention and facilitate comparison with previous work, we adopt the \cite{Solomon:1987uq} value, which has become the default in the literature. Note that in the final calculation of $R$, the effects of finite resolution are mitigated by deconvolving the synthesized beam from the measured major and minor axis lengths, and the effects of finite sensitivity are accounted for by extrapolating the size linearly to a 0~K contour by computing the size at a range of contour levels (see RL06). We then also multiply by an additional factor of 1.2 to account for the presence of CO-dark molecular gas at the cloud edges based on theoretical models of \citealp{2010ApJ...716.1191W} (see the ensuing section for discussion). Note that for GMCs with observed sizes near our angular resolution, $R$ can formally be smaller than the synthesized beam. For clouds with extent smaller than the synthesized beam along any dimension, $R$ is undefined.}
\item{The \textbf{linewidth $\Delta V$} is computed as the intensity-weighted second moment of the velocity, after deconvolution from a Gaussian approximation to the spectral response function. $\Delta V$ includes a factor of $\sqrt{8 \ln{2}}$ to account for the conversion from velocity moment to FWHM, and is also extrapolated to 0~K as discussed above.}
\item{The \textbf{luminous mass {\Mlum}} is calculated by summing the emission over all voxels to get the CO luminosity $L_{\rm CO}$ as follows:
\begin{equation}
L_{\rm CO} = \sum_i T_i \,\delta x \, \delta y \, \delta v \,\, D^2,
\end{equation}
where $D$ is the distance (taken to be 1.93~Mpc for all clouds), $T_i$ is the brightness temperature of at voxel $i$, the $\delta$ terms are the voxel sizes in the two spatial and one velocity dimension, and the sum runs over all voxels in the cloud. We then converted luminosity to mass using the metallicity-dependent conversion factors $\alpha_{\rm CO}$ for each region calculated by F14 (see also Section~\ref{sec:improved} below) as
\begin{equation}
\label{eqn:Mlum}
\Mlum = \alpha_{\rm CO} R^{-1}_{21} L_{\rm CO},
\end{equation}
where we assume the CO(2-1) to CO(1-0) line intensity ratio $R_{21}=0.7$ in this calculation. {\Mlum} is extrapolated to 0~K using a quadratic function, which better recovers low-level extended flux at the edges of model clouds than a linear fit (RL06).}

\subsubsection{Derived properties}
\label{sec:derivedprops}

\item{The \textbf{virial mass {\Mvir}} is computed from the size and linewidth as 
\begin{equation}
\label{eqn:Mvir}
\Mvir = 189 \, (\Delta V)^2 \, R \,\, \Msun,
\end{equation}
where $\Delta V$ is in {\kms} and $R$ in pc. This formulation \citep[e.g.,][RL06]{Solomon:1987uq} assumes a truncated power law density distribution $\rho \propto R^{-\beta}$, taking $\beta = 1$, and that magnetic fields and external pressure are negligible. {\Mvir} is only defined for clouds with finite $R$, i.e. for resolved clouds.}
\item{The \textbf{virial parameter $\alpha_{\rm vir}$} describes the ratio of kinetic to gravitational potential energies, and in the absence of other forces, the level of gravitational boundedness. Clouds with $\alpha_{\rm vir} \lesssim 2$ are typically considered ``bound''; those with larger $\alpha_{\rm vir}$ must have some other constraining force to survive more than a dynamical time; and, those with $\alpha_{\rm vir} << 1$ must have some internal source of support (such as from magnetic fields). We apply the formulation of \cite{1992ApJ...395..140B}, who assume non-magnetized, constant density clouds with no external pressure.}
\begin{equation}
\alpha_{\rm vir} = \frac{5 \sigma_v^2 R}{G \Mlum} = \frac{210 \Delta V^2 R}{\Mlum}
\label{eqn:alphavir}
\end{equation}
\item{The \textbf{mass surface density $\Sigma$} is derived from the luminous mass and size assuming that the latter represents the radial extent if the cloud had a circular cross-section, i.e.
\begin{equation}
\label{eqn:Sigma}
\Sigma = \frac{\Mlum}{\pi R^2}
\end{equation}
}
\item{The \textbf{size-linewidth scaling coefficient}, which we call $C_{R \Delta V}$, relates the cloud linewidth to its size assuming a scaling exponent of 0.5. For a cloud in virial equilibrium ($\Mlum = \Mvir$), this quantity can be related to the mass surface density by combining Equations~\ref{eqn:Mvir} and \ref{eqn:Sigma}~\citep[e.g.,][]{Heyer:2009ii} as
\begin{eqnarray}
\label{eqn:szlwcoeff}
C_{R \Delta V} &=& \frac{\sigma_v}{R^{1/2}} = \left( \frac{\pi G \Sigma}{5}\right)^{1/2} \nonumber \\
&=& 0.052 \left( \frac{\Sigma}{\Msun~{\rm pc}^{-2}}\right)^{1/2}~\kms~\rm{pc}^{-1/2}.
\end{eqnarray}
Note we have converted the linewidth $\Delta V$ to the one-dimensional velocity dispersion $\sigma_v = \Delta V / 2.355$ to facilitate comparison with previous studies.
}
\end{itemize}

Uncertainties in the independent properties are computed using a bootstrap method with 1000 iterations. We calculate the uncertainties in derived properties using standard propagation of errors.

\subsection{Improving physical parameter estimates}
\label{sec:improved}

We made two important adjustments to the procedure used in \cite{Faesi:2016bl} for determining cloud properties. Both attempt to account for the potentially significant quantity of CO-dark molecular gas that has been inferred to exist in the outskirts of GMCs, both observationally~\cite[e.g.,][]{2005Sci...307.1292G,2007ApJ...658.1027L} and theoretically \citep[e.g.,][]{1985ApJ...291..722T,1995ApJS...99..565S}.

The first adjustment is to GMC sizes. While our measurements of cloud size are extrapolated to the scale where CO emission is expected to disappear entirely, this does not account for the moderate column density region where gas is still primarily molecular but CO is unable to survive. The extent of such a region can be significant, particularly at low metallicity and/or integrated column density \citep{2010ApJ...716.1191W}. Since we have no direct method of tracing CO-dark molecular gas on GMC scales, we simply assume the prediction of \cite{2010ApJ...716.1191W} that $R_{\rm{H}_2}/R_{\rm CO} \approx 1.2$ (their Figure~8), i.e. that the size of the molecular region is 20\% larger than the size of the region in which CO exists. Instead of interpolating their solar and half-solar metallicity models to the appropriate metallicity for each region, we assume the fiducial ratio of 1.2 for all regions for simplicity. We thus multiply the inferred sizes of all resolved clouds by 1.2 and use the corrected values for the remainder of our analysis.

The second adjustment comes in calculating the cloud masses. Just as in F16 we derive a unique region-by-region CO-to-H$_2$ conversion factor $\alpha_{\rm CO}$ based on metallicities derived in F14 from the well-characterized metallicity gradient measured by \cite{Bresolin:2009hh}. NGC 300 has a metallicity near 0.8 solar near its center, decreasing radially to about 0.5 solar at 4~kpc (see F14, equation [4]). We do not have individual gas phase metallicities for the observed regions, so we assume that the radial gradient dominates over local region-to-region variations in metallicity. We again take the parameterization of $\alpha_{\rm CO}$ from \cite{Bolatto:2013hl}, who suggest a CO-to-H$_2$ conversion factor based on physical models as follows:
\begin{equation}
\label{eqn:B13}
\alpha_{\rm CO} = \alpha_{\rm CO,MW} \times 0.67 \exp{\left(\frac{0.4}{Z' \, \Sigma^{100}_{\rm GMC}}\right)},
\end{equation}
where $Z'$ is the metallicity in solar units, $\Sigma^{100}_{\rm GMC}$ is the characteristic GMC surface density in units of $100~\Msun$~pc$^{-2}$, and $\alpha_{\rm CO, MW} = 4.35$ is the CO-to-H$_2$ conversion factor for the Milky Way in units of {\Msun}~pc$^{-2}$~(K~{\kms})$^{-1}$. The adjustment we make here is to self-consistently determine a value for $\Sigma^{100}_{\rm GMC}$ using our data instead of assuming a value of unity. To accomplish this, we combine Equations~(\ref{eqn:Sigma}) and (\ref{eqn:B13}) with the definition of {\Mlum}, which yields
\begin{equation}
\exp{\left(\frac{0.4}{Z' \, \Sigma^{100}_{\rm GMC}}\right)} = \left(\frac{\pi R^2}{0.67 \, \alpha_{\rm CO,MW} \,{L_{\rm CO}}}\right) \Sigma.
\label{eqn:prodlog}
\end{equation}
We then make $\Sigma$ dimensionless (i.e. divide it by $100~\Msun$~pc$^{-2}$) so that there are factors of $\Sigma^{100}_{\rm GMC}$ on both sides. The solution to an equation of the form $\exp{(a/x)}=bx$, where $a$ and $b$ are constants, is $x = a/W(ab)$, where $W(z)$ is the product log function. To solve for $\Sigma^{100}_{\rm GMC}$, we plug in to Equation~(\ref{eqn:prodlog}) the median values of $Z'$, $R$, and $L_{\rm CO}$ across the resolved sample of GMCs, then take the real part of the product log, arriving at $\Sigma^{100}_{\rm GMC} = 0.69$, i.e. a GMC surface density of $69~\Msun$~pc$^{-2}$. We use this single value as the input to the CO-to-H$_2$ conversion factor (Equation~[\ref{eqn:B13}]) for the entire sample. We opt not to derive individual $\Sigma^{100}_{\rm GMC}$ values for each cloud because (1) less than half the sample is well-resolved (i.e. has a defined size $R$) and thus we are unable to calculate $\Sigma^{100}_{\rm GMC}$, and (2) $\Sigma^{100}_{\rm GMC}$ reflects a characteristic surface density for a population of clouds \citep{Bolatto:2013hl}, and thus it would be potential overreach to apply Equation~(\ref{eqn:B13}) to individual clouds or regions.


\section{Results}
\label{sec:results}

Among the 48 regions observed, 46 of them (all but DCL88-6 and DCL88-115) have at least one GMC identified and characterized by \texttt{CPROPS}. Using our conservative decomposition scheme, our sample consists of 253 clouds across these 46 regions after removing duplicates (32 GMCs) and clouds with central pixel more than $29\arcsec$ from the pointing center (i.e., more than a factor of 10\% in angular distance outside the primary beam FWHM; 20 GMCs). 153 of these 253 clouds are resolved by the $\sim 14$~pc $\times 8.4$~pc synthesized beam. When the beam size is similar to the observed size, uncertainties are greatly magnified by the deconvolution process, making derived small $R$ unreliable (see Appendix~\ref{sec:sizes}). We thus exclude all clouds with $R<5$~pc (approximately half our spatial resolution) from the majority of the remaining analysis presented here. With these cuts, our final cloud catalog comprises 121 GMCs -- 48\% of the 253 in the sample. Table~\ref{tab:gmcprops} presents the final cloud catalog, which contains columns as follows:

\begin{enumerate}
\item{ID, cloud designation by DCL region and assigned number preceded by `A' to signify this ALMA study;}
\item{R.A. (J2000), cloud right ascension as measured by the intensity-weighted first moment along this direction;}
\item{decl. (J2000), cloud declination measured as above;}
\item{$v_0$, cloud central velocity as measured by the intensity-weighted first moment along the velocity axis;}
\item{$a/b_{\rm dc}$, cloud deconvolved aspect ratio, i.e. the deconvolved major axis diameter divided by the deconvolved minor axis diameter;}
\item{PA$_{\rm dc}$, cloud deconvolved position angle in degrees'}
\item{$R$, cloud size in pc, extrapolated and deconvolved from the spatial beam;}
\item{$\Delta V$, cloud FWHM linewidth in {\kms}, extrapolated and deconvolved from the spectral response;}
\item{{\Mlum}, cloud luminous mass in solar masses, calculated from CO luminosity and a metallicity-dependent X-factor;}
\item{{\Mvir}, cloud virial mass in solar masses;}
\item{$L_{\rm CO}$, CO(2-1) luminosity in K~{\kms}~pc$^2$;}
\item{$T_{\rm max}$, cloud peak CO(2-1) brightness temperature in K;}
\item{S2N, cloud peak signal-to-noise;}
\item{note, including: ``F'' if the cloud is in the final sample (i.e., $R\geq5$~pc), ``R'' if the cloud is resolved (but not in the final sample), ``O'' if the cloud's position is $>10$\% of the primary beam FWHM beyond the area of the primary beam, and if the cloud is a duplicate, the name of the cloud so duplicated.}
\end{enumerate}

\begin{figure}
\includegraphics[width=\linewidth]{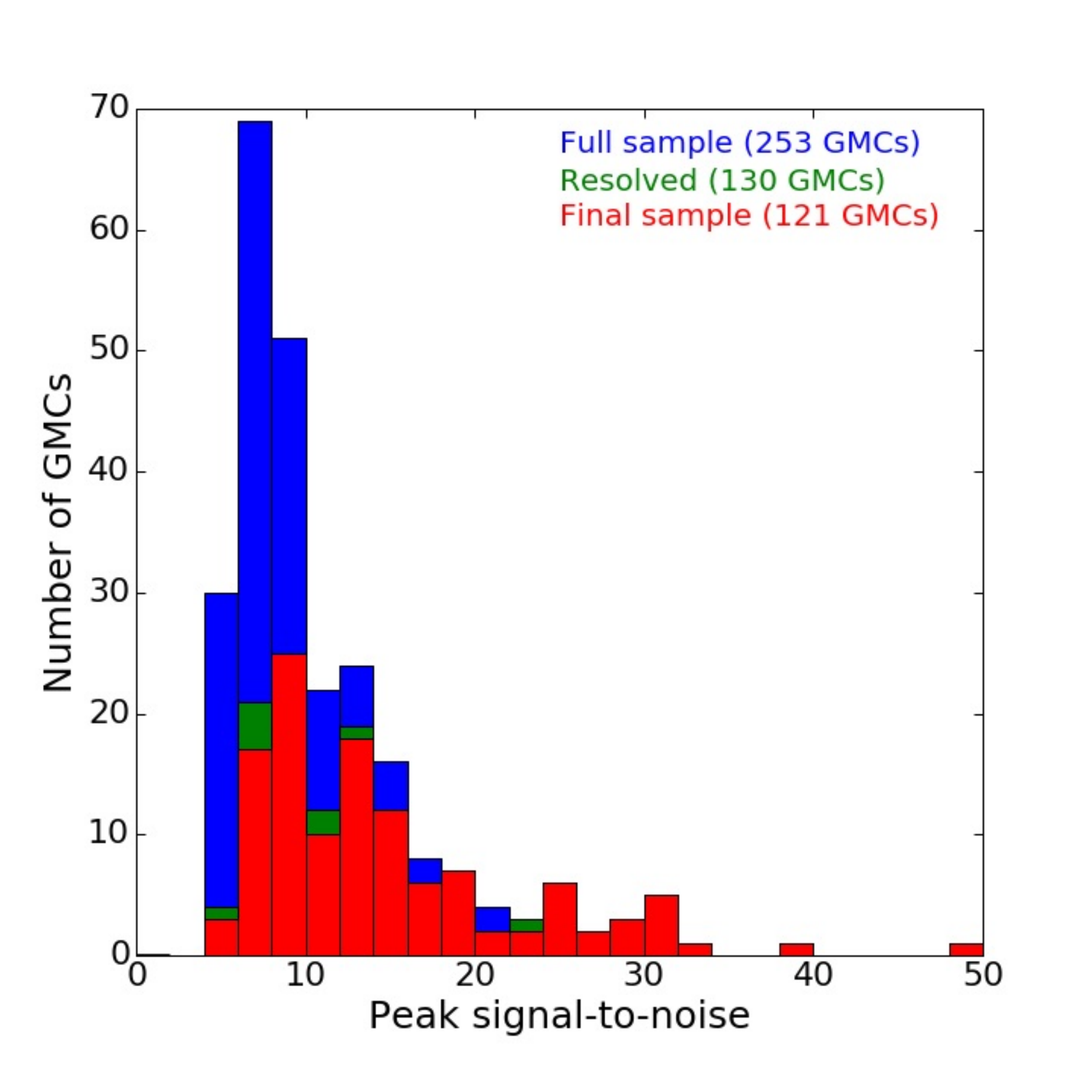}
\caption{Histogram of cloud peak signal-to-noise, color coded by full sample (all unique \texttt{CPROPS}-identified GMCs), resolved subsample (all those with size $R$ larger than the synthesized beam), and the ``final'' sample ($R > 5$~pc).}
\label{fig:s2nhist}
\end{figure}

We present a histogram of the peak signal-to-noise in our GMC sample in Figure~\ref{fig:s2nhist}. Due to the imposed requirement of adjacent $4\sigma$ voxels to define clouds in \texttt{CPROPS}, we recover all GMCs with at least this significance. The median peak signal-to-noise is 8.9 in the full sample, 11.4 for the resolved clouds, and 12.6 for the final sample (see below). RL06 showed through simulations that using \texttt{CPROPS} with data having a peak signal-to-noise of $> 10$ provides an accuracy of $\sim 10\%$ in determining sizes, linewidths, and masses, while lower signal-to-noise can potentially lead to the mis-estimation of cloud properties. Thus our high sensitivity should produce robust estimates of these quantities.

Table~\ref{tab:GMCstats} shows a statistical summary of the GMC properties in our full GMC sample (with $R$, {\Mvir}, and $\Sigma$ for the final sample only). Our study is the first beyond the Local Group to achieve $\sim10$~pc, $1~\kms$ resolution and limiting sensitivity of $M \sim 10^4~\Msun$ -- i.e., to be able to detect and spatially and spectrally resolve analogs of nearby Milky Way clouds such as Taurus, Perseus, and Ophiuchus -- and thus to sample most of the wide dynamic range in cloud sizes, linewidths, and masses observed in the Milky Way.

\begin{deluxetable}{l c c c c}
\centering
\tabletypesize{\scriptsize}
\tablecolumns{5}
\tablewidth{0pt}
\tablecaption{GMC statistics \label{tab:GMCstats}}
\tablehead{
	\colhead{Property} &
	\colhead{Unit} &
	\colhead{Minimum} &
	\colhead{Maximum} &
	\colhead{Median}
}I
\startdata
S2N &			\nodata &			4.5 &				48.1 &			8.9 \\
$T_{\rm max}$ &	K &				0.6 &				8.9 &				1.8 \\
$L_{\rm CO}$ &	K~{\kms}~pc$^2$ &	$1.52\times10^2$ &	$8.85\times10^4$ &	$1.36\times10^3$ \\
$R$ &			pc &				5.0 &				63.4 &			15.0 \\
$\Delta V$ &		{\kms} &			1.1 &				10.6 &			3.5 \\
{\Mlum} &			{\Msun} &			$5.3\times10^3$ &	$1.02\times10^6$ &	$1.67\times10^4$ \\
{\Mvir} &			{\Msun} &			$4.68\times10^3$ &	$1.18\times10^6$ &	$4.46\times10^4$ \\
$\Sigma$ &		{\Msun}~pc$^{-2}$ &	13.8 &			397.0 &			60.6
\enddata
\end{deluxetable}

\begin{figure}
\includegraphics[width=\linewidth]{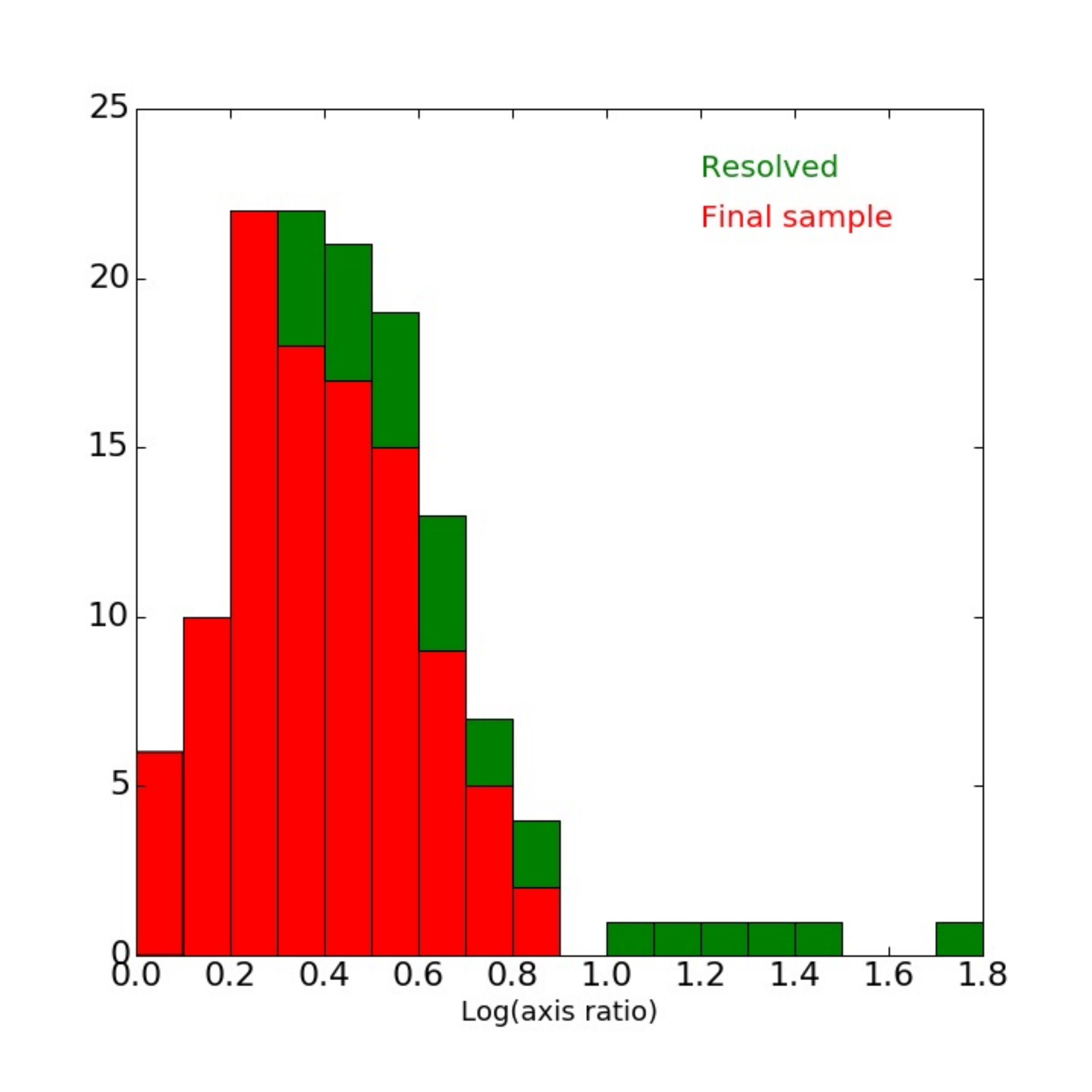}
\label{fig:axisratio}
\caption{Histogram of projected, deconvolved aspect ratios (on a logarithmic scale) in the resolved sample (green), and the final sample, which consists of all GMCs with $R > 5$~pc (red, overlaid). The median axis ratio is 2.6, suggesting that many GMCs are highly elliptical.}
\end{figure}

For resolved clouds, we also measure the projected deconvolved aspect ratio, defined as the length of the deconvolved major axis divided by the length of the deconvolved minor axis. We present a histogram of the resolved GMC aspect ratios in Figure~\ref{fig:axisratio}. Since clouds with $R < 5$~pc are not included in this sample, it is unlikely that our elliptical beam shape is causing the trend toward ellipticity in cloud shape seen in the Figure, as beam effects are only dominant for smaller clouds (see Appendix~\ref{sec:sizes}). However, there are a few GMCs in the final sample with minor axis lengths that are sufficiently small (less than 5 pc), so we do not consider them in the present discussion of aspect ratios. With these clouds removed, GMCs still clearly tend toward being more elliptical than circular, with a median value of 2.4 within the final sample (a value of unity would be purely circular). 75\% of the final sample have aspect ratios less than 3.5, while 90\% have aspect ratios less than 4.6. 

The filamentary structure seen in several regions such as DCL52, DCL80, DCL88, DCL93, and DCL137 is qualitatively reminiscent of that observed in the Milky Way molecular ISM~\citep[e.g.,][]{2009A&A...505..405P,2009ApJ...700.1609M} as well as other galaxies including M51~\citep{2014ApJ...784....3C}, suggesting that molecular filaments are a common feature of spiral galaxies. In particular, the median and range of aspect ratios we find in our sample are quite similar to those observed in the Milky Way infrared dark cloud catalog of \citep{2009A&A...505..405P} as well as in LMC GMCs \citep{Wong:2011ib}.

\begin{figure}
$\begin{array}{cc}
\includegraphics[width=0.5\linewidth]{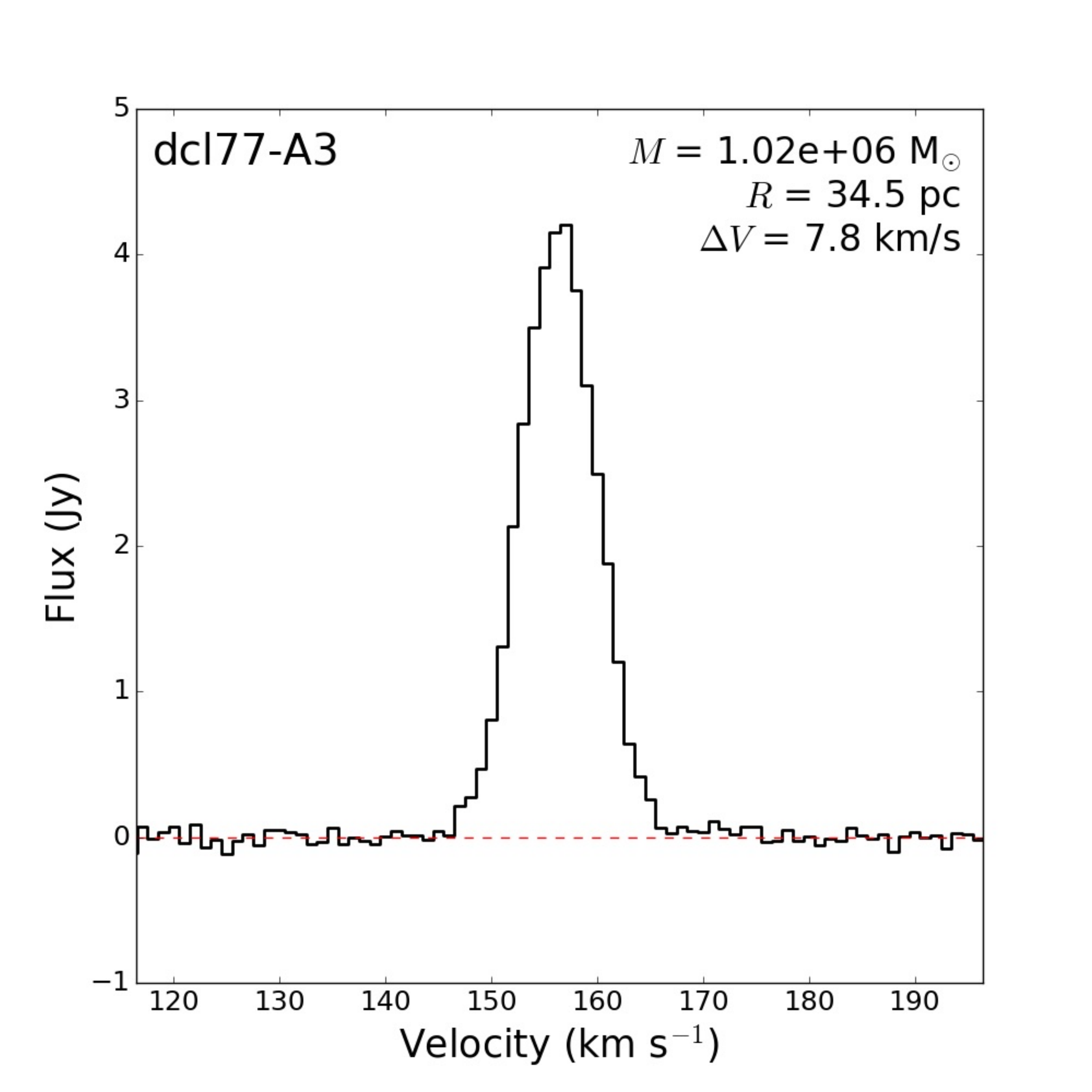} &
\includegraphics[width=0.5\linewidth]{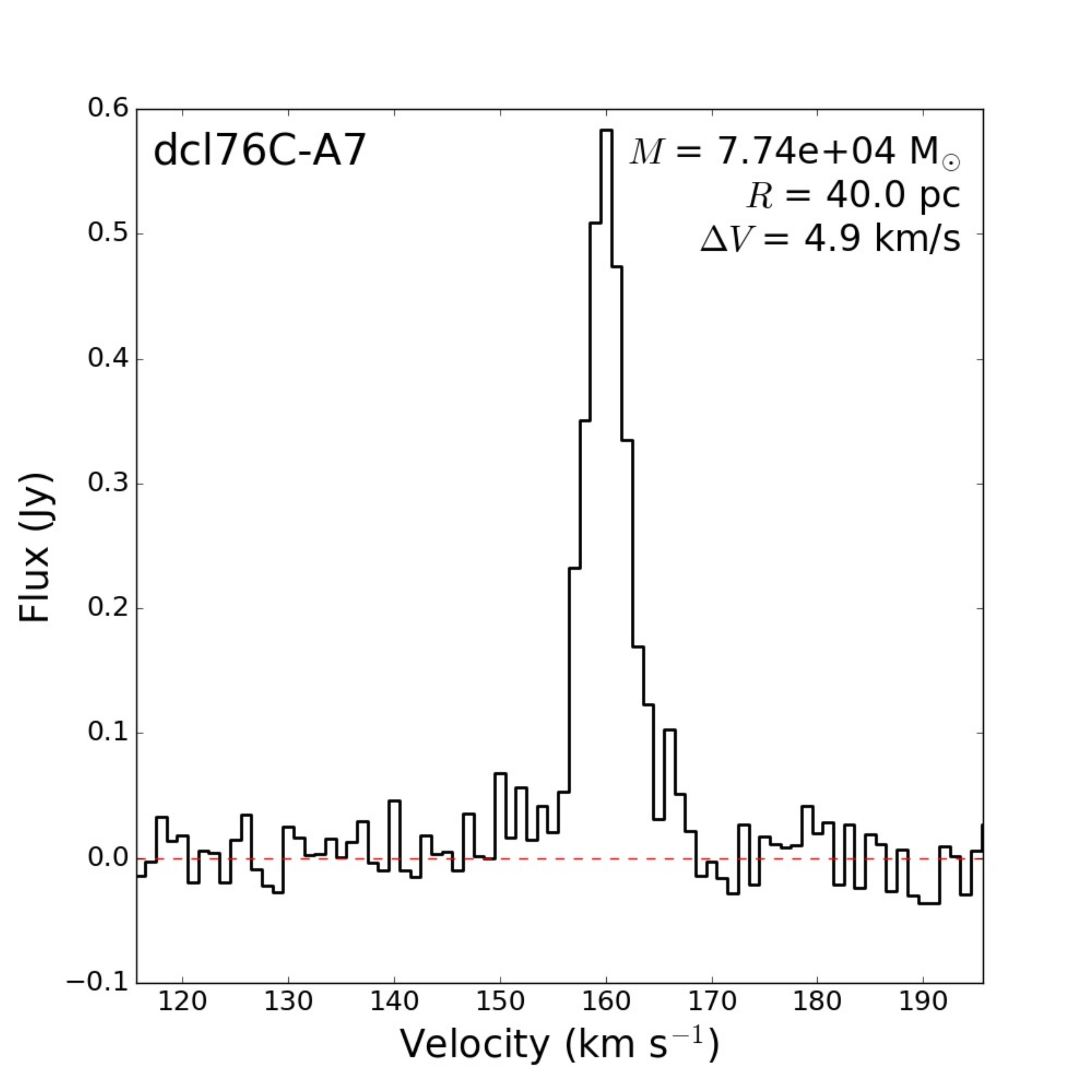} \\
\includegraphics[width=0.5\linewidth]{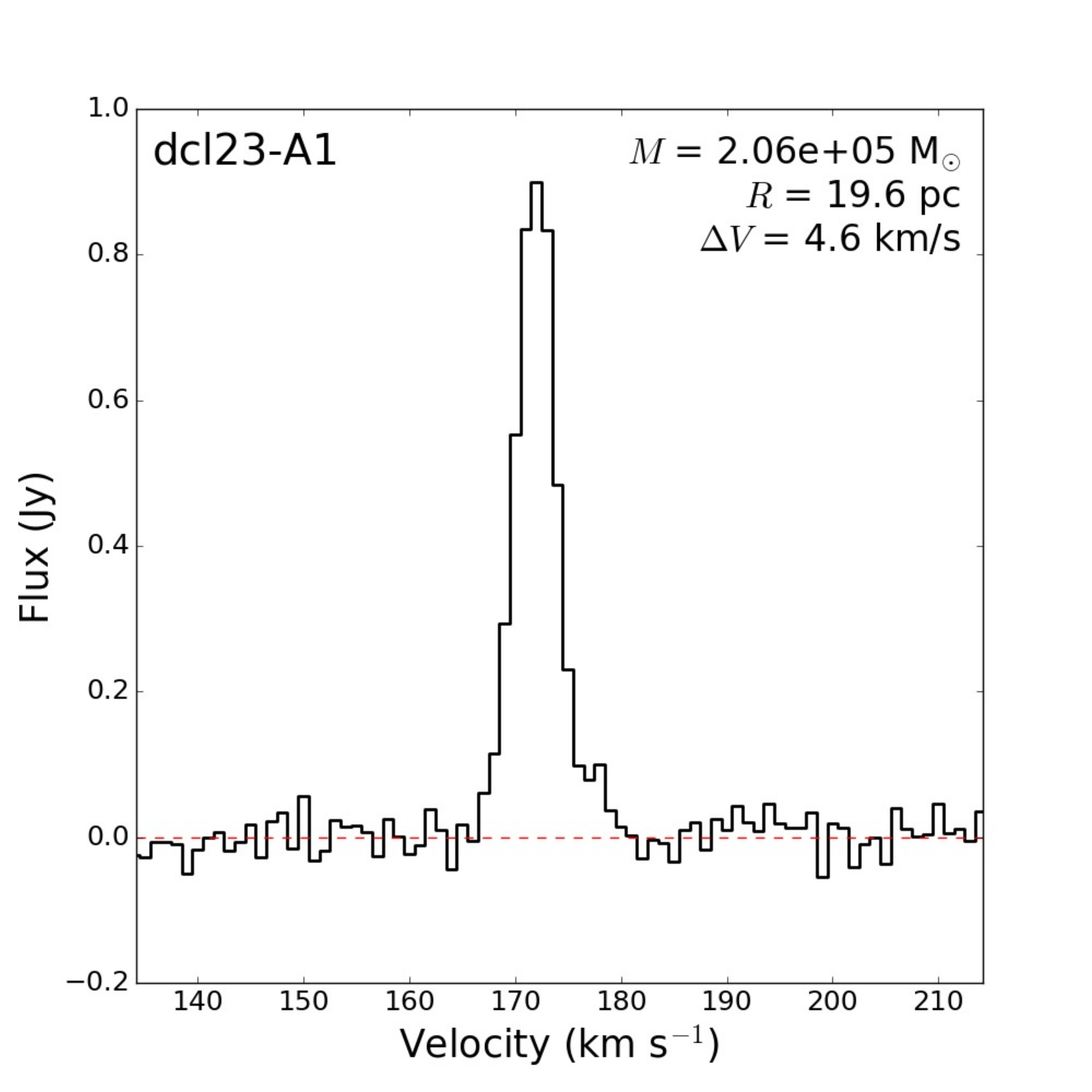} &
\includegraphics[width=0.5\linewidth]{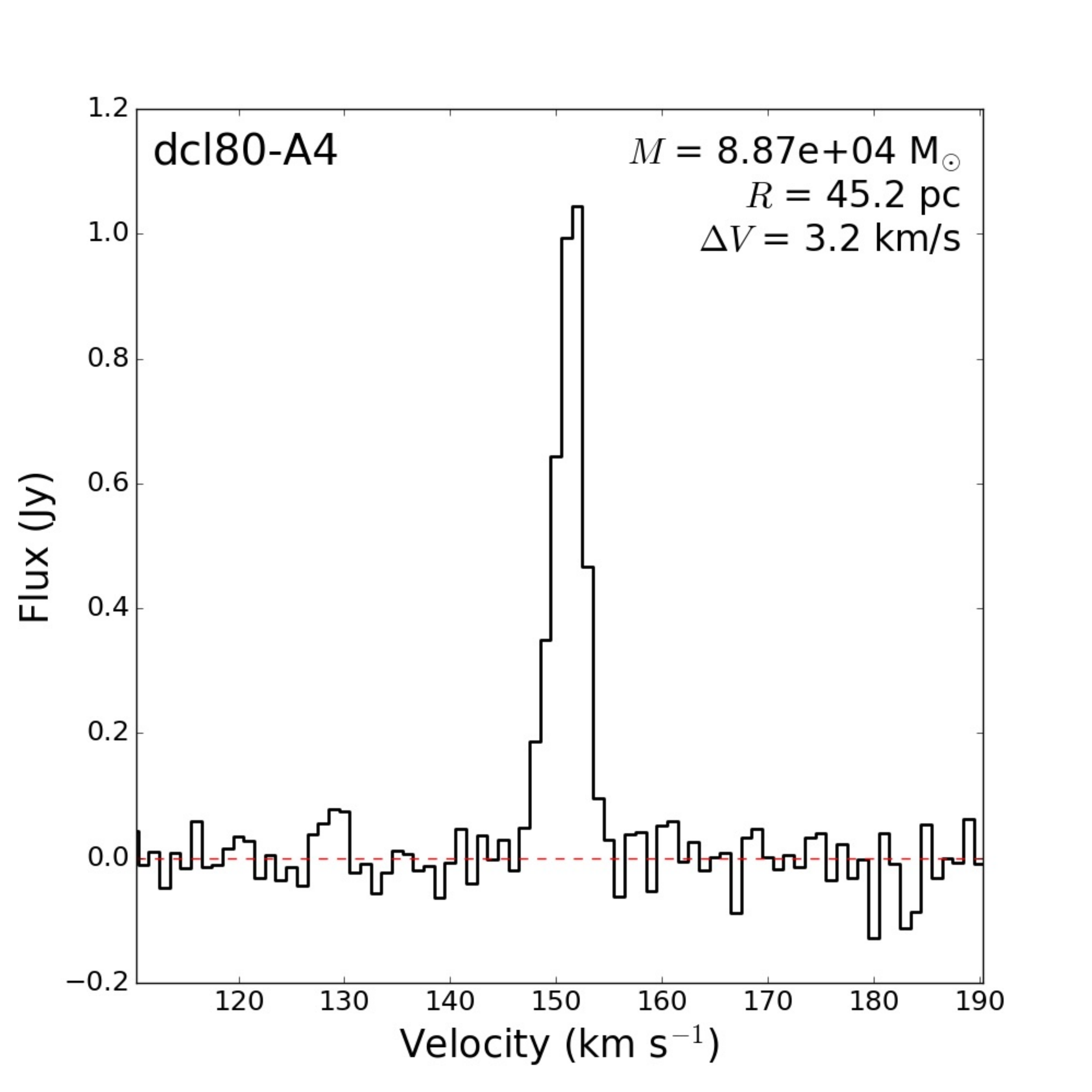} \\
\end{array}$
\caption{One-dimensional spectra for four representative GMCs in our final sample: dcl77-A3 (upper left), dcl76C-A7 (upper right), dcl23-A1 (lower left), and dcl80-A4 (lower right), with each cloud's mass $M$, size $R$, and linewidth $\Delta V$ indicated in the upper right of the panel. The vast majority of GMCs have near-Gaussian, single component spectra similar qualitatively to spectra of GMCs in the Milky Way. See text for a description of how the spectra were constructed.}
\label{fig:specs}
\end{figure}

Our post-smoothing 1~{\kms} velocity resolution is sufficient to spectrally resolve GMCs. We have constructed one-dimensional spectra for each cloud in the sample, and present four representative examples in Figure~\ref{fig:specs}. To make these spectra for a given GMC, we first masked each spectral plane within the data cube for that region to include only pixels belonging to that GMC. We then combined this cubic mask along the velocity axis with a logical `or' (i.e., the final mask has two spatial dimensions in which all pixels with emission in at least one velocity channel in the cloud are included, and all pixels without any such emission are excluded). Finally, we applied this mask to each image plane and averaged the unmasked emission to determine the flux density per channel for the GMC.

\subsection{GMC multiplicity and clustering}

\begin{figure}
\includegraphics[width=\linewidth]{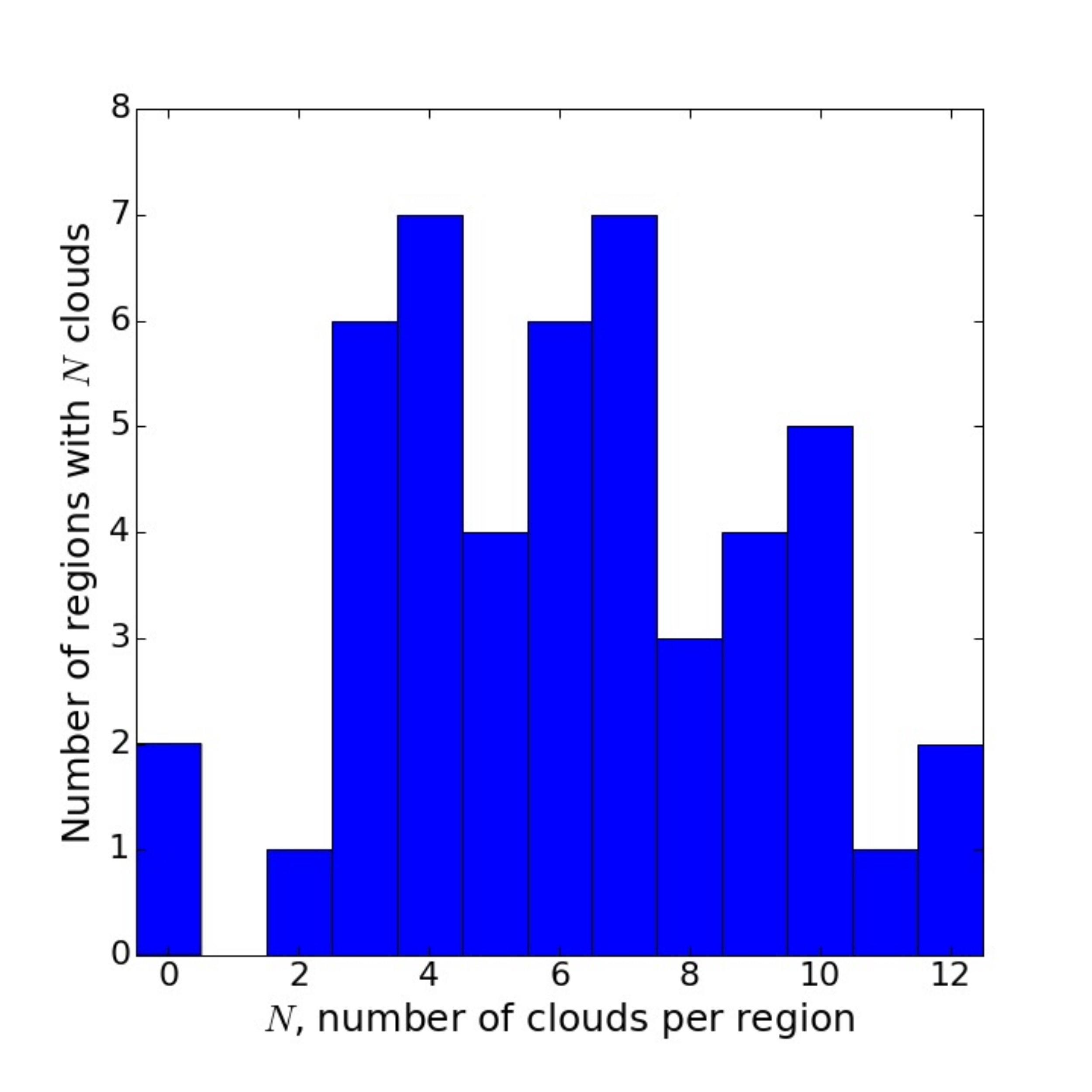}
\caption{Histogram of number of \texttt{CPROPS}-detected clouds in our sample of 48 observed regions.}
\label{fig:nclouds}
\end{figure}

Individual regions have from as few as two up to as many as twelve clouds. We present a histogram of number of clouds per observed region in Figure~\ref{fig:nclouds}. The median number of clouds per region is six, and the majority (76\%) of the regions have between three and eight clouds.  In Figure~\ref{fig:ICOimages} we present {\CO} integrated intensity images of all 48 observed regions with the \texttt{CPROPS} clouds overlaid. The observed distribution of molecular gas shows that (a) the primary beam area filling factor varies, but is generally small ($< 0.5$), at least to the extent that there is substructure on scales less than our short spacing limit of $\sim 100$~pc, and (b) the mostly single-component spectra observed by F14 at $\sim250$~pc resolution smooth out a great degree of substructure. Cloud morphology varies greatly region-to-region, from large complexes of overlapping clouds (e.g. DCL41, DCL 77, DCL114, DCL137C), to chains or clusters of small clouds (e.g. DCL37, DCL49, DCL61, DCL112), to long ($>100$~pc) filaments (e.g. DCL80, DCL88, DCL93).

Additionally, clouds tend to be strongly clustered within a region. The median distance from a cloud to its nearest neighbor is $\sim 5\arcsec$ (47~pc). If six clouds (the median number per region) were distributed randomly within a beam with FWHM $26\arcsec$ (240~pc), their typical separation would be approximately $26/\sqrt{6} \approx 11\arcsec$ (100~pc) -- more than a factor of two larger than the actual median separation. It is conceivable that this clustering may have a physical origin. For example, regions with more, smaller clouds may be examples of gas fragmentation (potentially due to gravitational instabilities) and/or dispersal (potentially due to the energy and momentum input from massive stars or supernovae). Investigating these possibilities further is beyond the scope of this paper.

\subsection{The GMC distribution across NGC~300}
\label{sec:galrad}

Our sample of observed GMCs spans a range of galactocentric radius, from near the center of NGC 300 out to almost 3.5 kpc, as well as a range of environments, from near the galaxy center, to spiral arms, to interarm regions. To explore any potential variations of GMC properties within our sample, we have visually assigned each observed region to one of four different environmental ``zones'' by comparing its position in the context of available H~I, H$\alpha$, \textit{Spitzer}~24~$\mu$m, and \textit{Herschel} 250 and 500 $\mu$m images of NGC~300. The ``spiral arm'' zone consists of the extended spiral features observed in essentially all ISM tracers. The ``interarm'' zone is comprised of areas between the spiral features in which there is little or no gas or dust as seen in the \textit{Herschel} and H~I images. The ``transition zone'' is the border between these two; clouds are assigned to the spiral arm zone if they are mostly within the gas/dust arms, while clouds are assigned to the transition zone if they are not mostly within the arm but overlap it in any way. The ``center'' zone consists of the inner $\sim 0.5$~kpc of the galaxy in which there is no noticeable non-axisymmetric structure in the ISM. Since the above definitions are somewhat subjective, we also examine the variation of cloud properties with galactocentric radius.

\begin{figure*}
\includegraphics[trim=0 4.6in 0 0, width=\linewidth]{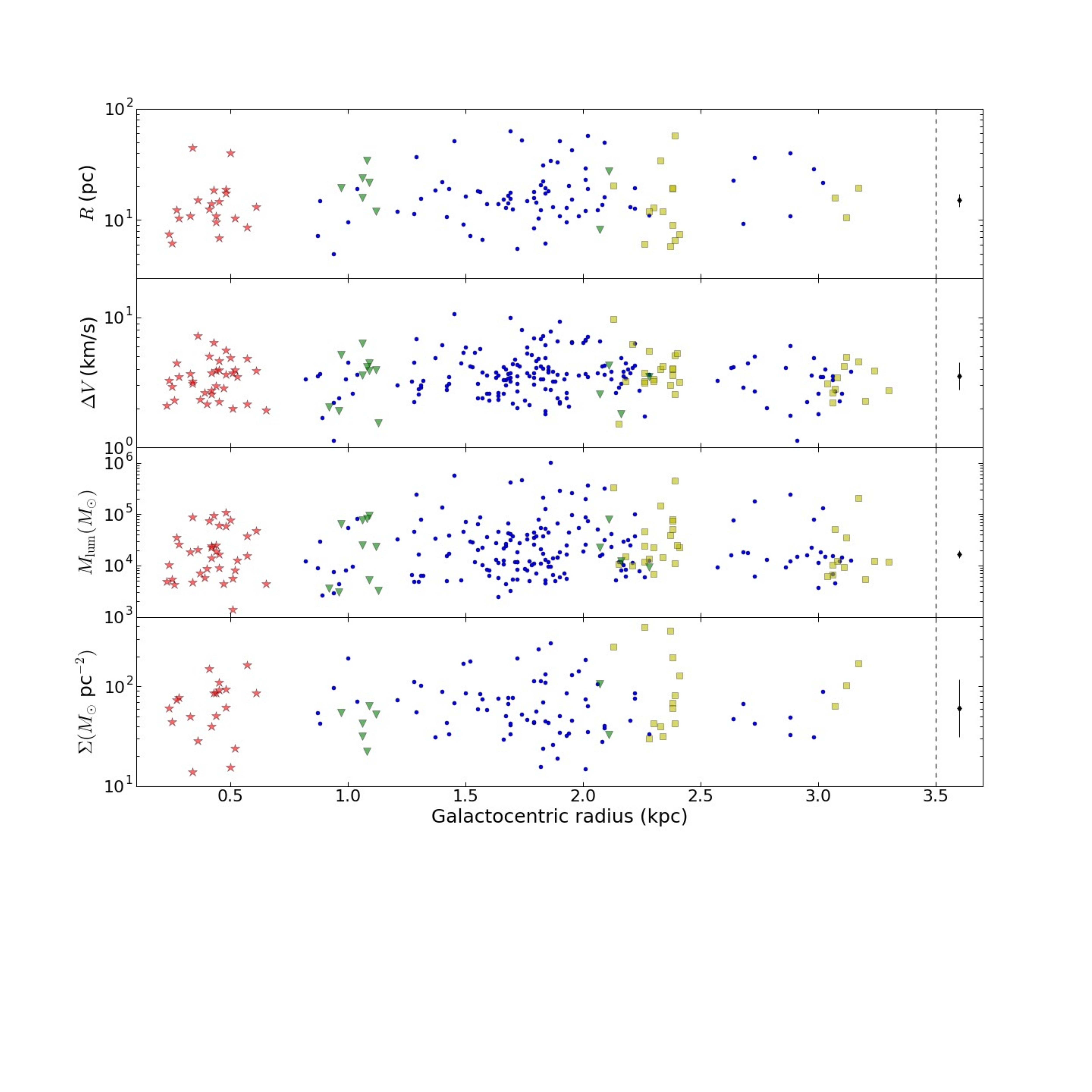}
\caption{From top to bottom panel: size $R$, linewidth $\Delta V$, luminous mass {\Mlum}, and mass surface density $\Sigma$ as a function of galactocentric radius in our sample of GMCs in NGC 300. The different symbols represent clouds in different galactic ``zones'': spiral arms (blue circles), interarm (yellow squares), transition zone (green triangles), or galaxy center (red stars). {\Mlum} and $\Delta V$ are shown for the full sample (with duplicates and clouds beyond beam removed), while $R$ and $\Sigma$ are for the final sample only. The vertical bars at the far right of the plots show the median and median uncertainties in these four parameters. There are large variations but no significant trends in any of these properties with galactocentric radius or with galactic zone.}
\label{fig:rgal}
\end{figure*}

Figure~\ref{fig:rgal} shows $R$, $\Delta V$, $\Mlum$, and $\Sigma$ plotted vs. $R_{\rm gal}$, the deprojected distance\footnote{calculated using the Python code kindly provided at \url{https://gist.github.com/jonathansick/9399842}} from the center of NGC~300, with the four different zones indicated. We adopt the same parameters for NGC 300 for these calculations as in F14, i.e. central coordinates of [$00^{\rm h}54^{\rm m}53^{\rm s}.48$, $-37^{\circ}41\arcmin03\arcsec.8$] (J2000), an inclination of $39.8^{\circ}$, a galactic major axis position angle of $114.3^{\circ}$ \citep{Paturel:2003jm}, and a distance of 1.93~Mpc \citep{2004AJ....128.1167G}.

$R$, $\Delta V$, $\Mlum$, and $\Sigma$ all show large variations but no apparent systematic trends with galactocentric radius, implying that GMC properties do not vary systematically across NGC~300. We do note that the most massive clouds are found between about 1.5 and 2.5 kpc, mostly in spiral arms, which is similar to the case of M51 in which spiral arm GMCs are more massive than interarm clouds \citep{2014ApJ...784....3C}. We also find no significant impact on cloud properties as a function of zone, except to note that zones appear to be clearly delineated by galactocentric radius (note that this is by definition in the case of the ``center'' zone). However, we do only detect clouds out to a maximum deprojected distance of 3.3~kpc, i.e. still within the rising portion of the rotation curve of the galaxy \citep{1990AJ....100.1468P,2011MNRAS.410.2217W}, and so the range of environments sampled in this study is somewhat limited. Note that if we were to use a single value for the CO-to-H$_2$ conversion factor $\alpha_{\rm CO}$ instead of accounting for the decreasing radial metallicity gradient the mass distribution would show an apparent decrease with galactocentric radius.

\section{Empirical GMC diagnostics}
\label{sec:GMCpop}

In this section we present an analysis for our NGC 300 GMC sample of four major empirical diagnostics of cloud populations: the mass spectrum, and the three Larson relations (sizelinewidth, virial, and size-mass). We conclude with a synthesis of the Larson relations in the size-linewidth coefficient -- surface density plane.

\subsection{GMC Mass Spectrum}
\label{sec:massspec}

The GMC mass spectrum describes how clouds are distributed by mass within a population and encodes information about GMC formation and destruction processes \citep[e.g.][]{1996ApJ...471..816E,Kobayashi:2017it}. The mass spectrum is often expressed in differential form as a power law, i.e. $f(M) = dN/dM \propto M^{\gamma}$. Here, like many recent authors, we will discuss the cumulative mass distribution, which has been shown to better recover the descriptive parameters of a GMC population than the differential formulation, particularly in the case of small sample sizes \citep{Rosolowsky:2005gt}. The cumulative mass spectrum describes the number of clouds above a reference mass $M$ as a function of $M$, which is simply the integral of the differential form, i.e.
\begin{equation}
\label{eqn:CMFnotrunc}
N(M^{\prime} > M) = \left ( \frac{M}{M_0} \right )^{\gamma+1}.
\end{equation}
Previous studies have revealed some evidence for mass spectra steepening or truncating at high mass \citep[e.g.,][]{2001PASJ...53L..41F,Gratier:2012km,2014ApJ...784....3C}, in which case it is useful to incorporate the possibility of a truncation at some maximum mass $M_0$~\citep[e.g.,][]{1997ApJ...476..166W,Rosolowsky:2005gt}:
\begin{equation}
\label{eqn:CMFtrunc}
N(M^{\prime} > M) = N_u \left[ \left ( \frac{M}{M_0}\right)^{\gamma+1}-1 \right],
\end{equation}
where $N_u$ represents how many clouds are at upper end of the distribution. Formally, $N_u$ is the number of clouds with mass greater than $M = M_u \equiv 2^{1/{\gamma+1}}M_0$. This algebraic value for $M_u$ can be determined by setting $N = N_u$ and solving for $M$ in Equation~(\ref{eqn:CMFtrunc}).

Most observations in the Milky Way and nearby galaxies reveal mass distributions in which the majority of the clouds by number are at low mass ($\gamma < 0$) but the majority of the mass is in high-mass clouds \citep[$\gamma > -2$; e.g.,][and references therein]{KennicuttJr:2012ey}. The exception may be the outer regions of galaxies, in which $\gamma < -2$, as seen in the Milky Way and M33 \citep{Gratier:2012km,Rice:2016ko}, and other low-density environments such as between spiral arms in M51 \citep{2014ApJ...784....3C}.

\begin{figure}
\includegraphics[width=\linewidth]{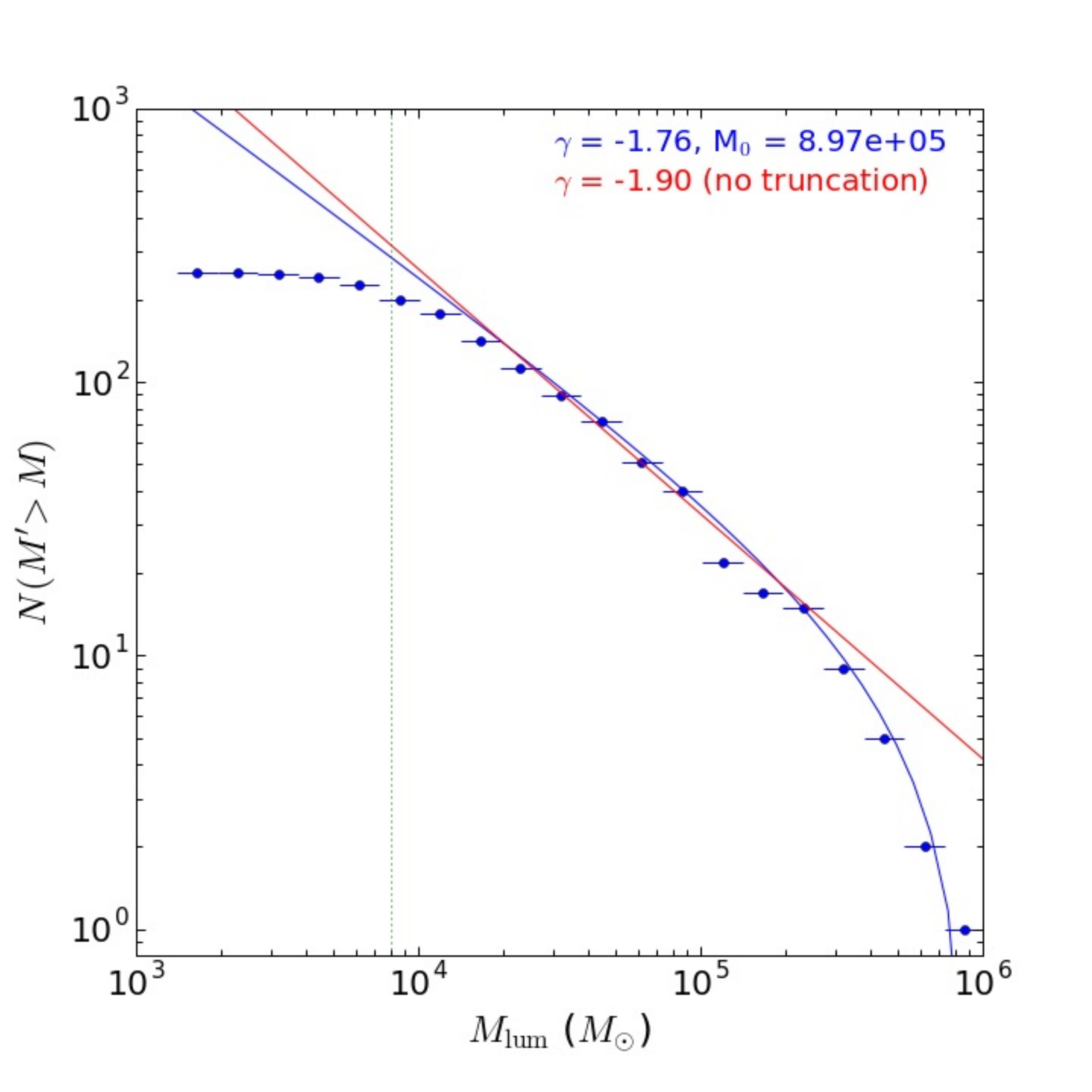}
\caption{Cumulative mass distribution of GMCs in our full sample. The solid line shows the best truncated power law fit to the data above our completeness limit of $8\times10^3~\Msun$. We find a power law slope of $\gamma=-1.76\pm0.07$ with a mass truncation at about $9\times10^5~\Msun$. The red line shows the best fit no-truncation power law, which is not a good fit for these data.}
\label{fig:massspec}
\end{figure}

We fit the cumulative mass distribution of {\Mlum} in our full sample of GMCs using the \texttt{mspecfit.pro} IDL code, which implements a maximum likelihood algorithm to account for uncertainties in both cloud mass and the number distribution \citep{Rosolowsky:2005gt}. We adopt a completeness limit of $8\times10^3~\Msun$ based on our observational sensitivity. To arrive at this number, we take our empirical $3\sigma$ mass surface density sensitivity of $13~\Msun$~pc$^{-2}$ (Section~\ref{sec:imaging}) and assume a spherical cloud with projected area equal to twice the synthesized beam FWHM area. We fit both a truncated (Equation~[\ref{eqn:CMFtrunc}]) and non-truncated (Equation~[\ref{eqn:CMFnotrunc}]) power law to the data above the completeness limit, and present the results in Figure~\ref{fig:massspec}. The truncated power law is a significantly better fit to this cloud population than the standard form, with clear deviation in the data from the best-fit non-truncated power law at high masses. For the truncated power law, we derive a slope of $\gamma = -1.76\pm0.07$ with a truncation mass $M_0 = 9.0\pm2.2\times10^5~\Msun$ and $N_u = 8.3\pm3.0$ clouds. The fact that $N_u$ is significantly greater than unity is additional quantitative evidence of a truncation. The best-fit straight power law has an exponent of $\gamma=-1.90\pm0.05$, but this model clearly does not represent the data as well as the truncated form.

Great caution is required in the comparison of mass spectra between data sets with differing resolution and sensitivity, and for GMCs identified and characterized with different algorithms \citep{2008ApJ...675..330S,Wong:2011ib}. We therefore focus our comparison on the studies that also utilized the cumulative approach, the same algorithm, and even the same fitting code, where possible. The mass spectrum slope of $\gamma = -1.76\pm0.07$ we find in NGC~300 is very similar to slopes inferred in other star-forming environments, both within and beyond the Milky Way. \cite{Rosolowsky:2005gt}, who pioneered the fitting method used here and developed the first version of the code we employed, found $\gamma=-1.53\pm0.06$ and $\gamma=-1.41\pm0.12$ in their study re-analyzing the inner Milky Way catalogs of \cite{Solomon:1987uq} and \cite{1987ApJS...63..821S}, respectively. These latter two studies both note the existence of a high-mass cutoff at 3--$4\times 10^6~\Msun$. Using the same algorithm but more comprehensive and recent CO observations from \cite{2001ApJ...547..792D}, \cite{Rice:2016ko} find $\gamma=-1.59\pm0.11$ and $M_0 \approx 10^7~\Msun$ for their composite inner Milky Way (quadrants I and IV) sample -- within the uncertainties of our slope. Our derived mass spectrum slope is also statistically identical to those found in the disks of several other spiral galaxies. Again using the same algorithm, \cite{2014ApJ...784....3C} found a mass spectrum slope of $\gamma=-1.63$ to $-1.79$ in the molecular ring and spiral arms of M51, while \cite{2007ApJ...654..240R} derived $\gamma=-1.55\pm0.20$ for his sample of clouds in M31. \cite{Gratier:2012km} found a similar slope of $\gamma=1.6\pm0.2$ for the inner 2.2~kpc of M33 using a fit to the cumulative mass spectrum with a different algorithm.

However, the NGC~300 mass spectrum slope we find is significantly shallower than that found in the outer Milky Way by \cite{Rosolowsky:2005gt} ($\gamma=-2.1\pm0.2$) and \cite{Rice:2016ko} ($\gamma=-2.2\pm0.1$), with no evidence for a high mass truncation found in either study. Similarly, \cite{Wong:2011ib} find $\gamma=-2.33\pm0.16$ in the LMC for their ``islands'' decomposition of their CO(1-0) map (and even steeper slopes for more refined decompositions). They also find no evidence for a truncation in the mass spectrum. Note that all three of these studies utilized the same algorithm as us. Additionally, the GMC mass spectra beyond $2.2$~kpc of M33 and in the interarm regions of M51 both demonstrate steeper slopes than NGC~300~\citep[$\gamma=2.3\pm0.2$ and $\gamma \approx -2.5\pm0.2$, respectively;][]{Gratier:2012km,2014ApJ...784....3C}. Furthermore, \cite{2015ApJ...803...16U} used the same algorithm to find a mass spectrum slope of $\gamma=2.39$ in the lenticular galaxy 4526 -- again much steeper than the slope we find in NGC 300's GMC mass spectrum. We note that all of these steeper slopes describe GMCs in environments with a relative paucity of molecular gas (e.g., the outer disks or interarm regions of galaxies), and/or the galaxy has a non-spiral morphology (i.e., the LMC is an irregular galaxy, NGC 4526 is lenticular while the others are spirals). We will discuss this point further in Section~\ref{sec:disc}.

\subsection{The Larson scaling relations in NGC 300}

The three Larson relations \citep{Larson:1981vv} represent now-ubiquitous empirical diagnostics of GMC populations, both within and beyond the Milky Way. In this section we present our analysis of the size-linewidth relation, the virial-luminous mass relation, and the size-mass relation for our NGC 300 final sample of 121 clouds.

\subsubsection{Size-linewidth relation}

The relation between cloud size and linewidth is typically expressed as
\begin{equation}
\sigma_v = C_{R \Delta V} R^b,
\end{equation}
where $\Delta V = 2.355 \times \sigma_v$ is the linewidth, $b$ is the power law exponent, and $C_{R \Delta V}$ is the size-linewidth coefficient (Equation~[\ref{eqn:szlwcoeff}]). As CO linewidths are always observed to be turbulent at pc and larger scales, the existence of a size-linewidth relation is interpreted as evidence for a turbulent cascade of energy within the molecular ISM such that through whatever mechanism(s) that form clouds, they inherit the turbulent kinetic energy appropriate for their size scale. According to this interpretation, the value of the power law slope is related to the nature of the turbulence, with incompressible turbulence theory predicting a slope of $1/3$, while a slope of $0.5$ suggests compressible (Burger's) turbulence \citep[e.g.,][]{McKee:2007bd}.

\begin{figure}
\includegraphics[width=\linewidth]{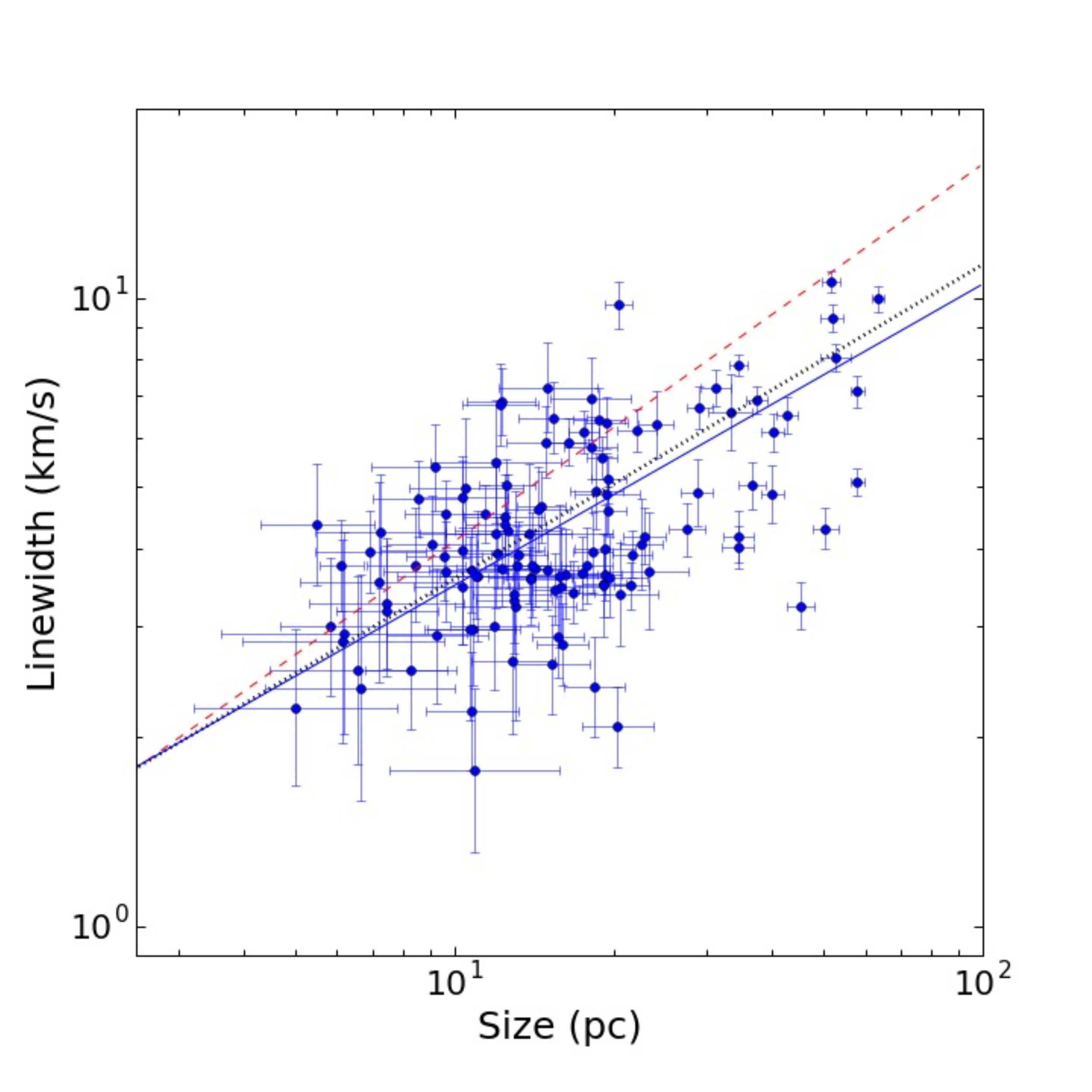}
\caption{The size-linewidth relation in the NGC~300 final sample. The solid blue line is our orthogonal distance regression fit (power law slope $0.48\pm0.05$), the black dotted line is from the Milky Way sample of \cite{2001ApJ...551..852H}, while the red dashed line is the extragalactic sample of \cite{2008ApJ...686..948B}. Our results are consistent with a defined size-linewidth relation with slope $\sim0.5$ and a similar scaling coefficient as has been observed in the Milky Way.}
\label{fig:sizelinewidth}
\end{figure}

Figure~\ref{fig:sizelinewidth} shows the size-linewidth relation for our final GMC sample. We see a clear if noisy trend in our data such that larger clouds have larger linewidths on average. We formally fit a linear function to the logarithms of these two quantities using orthogonal distance regression (ODR), which finds the maximum likelihood power law slope while accounting for errors in both variables simultaneously. From this analysis, we find a slope of $0.48\pm0.05$. To assess the degree to which these quantities are quantitatively correlated, we also compute the Pearson correlation coefficient $r_P$. As a preliminary, we note that the statistical distributions of sizes and linewidths are approximately lognormal, and that a linear function appears to qualitatively describe the logarithm of the data well (i.e., the data are well-described by a power law). Thus the Pearson coefficient should provide a reasonable statistical estimate of correlation. For the size-linewidth relation, we find $r_P=0.55$. The corresponding ``$p$-value'', or probability of an uncorrelated data set producing at least the apparent correlation by chance, is $4\times10^{-11}$, though we caution that this may not be entirely reliable for a data set of this size. Nevertheless, these results provide strong evidence for real correlation between GMC size and linewidth in our sample. We present a summary of the derived slopes and correlation coefficients for this and the other Larson relations in Table~\ref{tab:Larson}.

A size-linewidth relation having exponent $\sim 0.5$ is fully consistent with the scenario of compressible turbulence in the molecular ISM. Furthermore, since there appears to be no change in slope or cutoff at any size scale, the turbulence injection and dissipation scales appear to be beyond the range of spatial scales probed here. Our results are fully consistent with Milky Way GMC studies both old \citep{1986ApJ...305..892D,1987ApJS...63..821S,Solomon:1987uq} and more recent \citep{2001ApJ...551..852H,2014ApJS..212....2G,Rice:2016ko}, and including both the inner and outer galaxy (though note that \cite{2001ApJ...551..852H} find a lack of correlation for clouds with sizes less than 7~pc in the outer Milky Way).

The recent extragalactic literature is highly divided as to the presence or absence of a size-linewidth relation in GMCs. Early studies of M33 and M31 showed that GMCs in these galaxies fell in the same range of parameter space as Milky Way clouds, but the small sample sizes and large measurement uncertainties precluded the ability to derive independent relations \citep[e.g.,][]{2003ApJ...599..258R,2007ApJ...654..240R,2008ApJ...675..330S}.  \cite{2008ApJ...686..948B} jointly analyzed a sample of nearby dwarf and spiral galaxies (including M31 and M33) and found a size-linewidth exponent of $0.60\pm0.10$ for the full multi-galaxy GMC population. However, their inclusion of dwarf galaxies and combination of mixed resolution data sets may complicate the interpretation of these results~\citep[e.g.,][]{2013ApJ...779...46H}. Nevertheless, these studies taken together with the Milky Way literature of the past several decades suggest a consistent size-linewidth relation with exponent near one-half.

However, other extragalactic studies have not found strong correlations between GMC size and linewidth. For example, \cite{Gratier:2012km} analyzed IRAM data on M33 at 50~pc resolution and concluded that there was no correlation between these quantities. Similarly, \cite{2014ApJ...784....3C} do not see a size-line width relation in M51 GMCs at 40~pc resolution. \cite{2013ApJ...779...46H} demonstrated the importance in matching resolution between data sets taken with different facilities and of galaxies at vastly different distances, but note that there does not appear to be a single or individual galaxy-based size-linewidth relation in the LMC, M33, or M51. Importantly, however, none of the aforementioned studies achieve sufficient spatial resolution to sample more than a factor of $\sim$~2 to 3 in spatial dynamic range, limiting their ability to detect logarithmic scaling relations. Our results, which reach 10~pc resolution and thus a factor of $\sim10$ dynamic range, and which agree with Galactic cloud studies, suggest that the Larson size-linewidth relation in NGC~300 is identical to that in the Milky Way. We will further discuss the implications of these results in Section~\ref{sec:disc}.

\begin{deluxetable}{l | c c c | c}
\tabletypesize{\scriptsize}
\tablecolumns{5}
\tablewidth{0pt}
\tablecaption{Larson relation fits \label{tab:Larson}}
\tablehead{
	\colhead{Relation} &
	\colhead{Slope} &
	\colhead{$r_P$} &
	\colhead{$p$-value} &
	\colhead{MW slope}
}
\startdata
Size-linewidth &	$0.48\pm0.05$ &	0.55 &	$4\times10^{-11}$ &	$0.5\tablenotemark{a}$ \\
Virial relation &		$1.00\pm0.05$ &	0.84 &	$1\times 10^{-33}$ &	$0.8\tablenotemark{a}$ \\
Mass-size &		$2.00\pm0.12$ &	0.81 &	$3\times 10^{-29}$ &	$2.0\tablenotemark{a}$ \\
$C_{R\Delta V}$-$\Sigma$ &	$0.43\pm0.04$ &	0.55 &	$4\times10^{-11}$ & 0.5\tablenotemark{b}
\enddata
\tablenotetext{a}{\cite{Solomon:1987uq}}
\tablenotetext{b}{\cite{Heyer:2009ii}}
\end{deluxetable}

\subsubsection{Virial equilibrium}

\begin{figure}
\includegraphics[width=\linewidth]{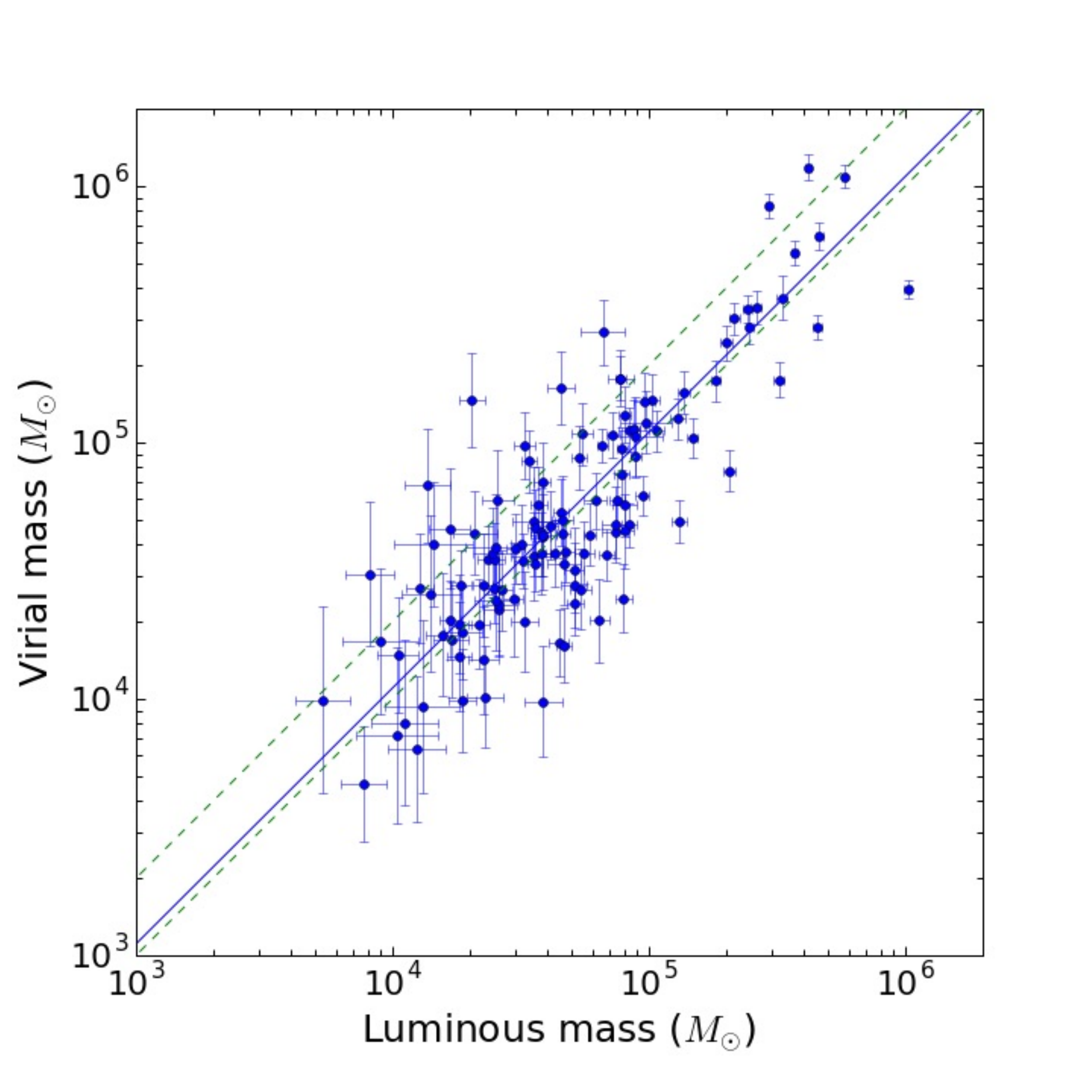}
\caption{Luminous vs. virial masses in our final sample. The solid line, which has a slope of $1.00\pm0.05$, is the fit to these data. The green dashed lines show the one-to-one relation and two-to-one relations. The majority of GMCs in our sample appear to be gravitationally bound.}
\label{fig:masscomp}
\end{figure}

The majority of GMCs in the Milky Way have been observed to be in self-gravitational equilibrium -- i.e., their gravitational potential energy $\mathcal{W}$ and kinetic energy $\mathcal{T}$ are in approximate balance \citep[e.g.,][]{Larson:1981vv,Solomon:1987uq,Heyer:2009ii,2014ApJS..212....2G,Heyer:2015ee}. One consequence of this relationship is a direct correlation between the mass measured through the simple virial theorem (i.e., {\Mvir}) and the mass measured through some other independent method (most often CO luminosity). In Figure~\ref{fig:masscomp} we show {\Mlum} vs. {\Mvir} for our final sample of 121 GMCs in NGC~300. There is again a clear visual correlation between these quantities. Quantitatively, we find (again via ODR fitting) a slope of $1.00\pm0.05$ and an offset fully consistent with zero. The Pearson correlation coefficient is $0.84$ -- the highest among the relations we examine here, with a vanishingly small $p$-value. This suggests that the GMCs in our sample are overwhelmingly in gravitational equilibrium, as is the case for Milky Way clouds. To examine the potential effects of our assumptions in converting CO luminosity to mass, we also compare {\Mvir} to {\LCO}. The correlation is similarly robust as the {\Mvir}-{\Mlum} relation, with a slightly higher scatter and lower $p$-value that likely reflect the fact that we compute {\Mlum} for each region individually based on the galactocentric radius-appropriate conversion factor instead of utilizing a single value for $\alpha_{\rm CO}$.

\begin{figure}
\includegraphics[width=\linewidth]{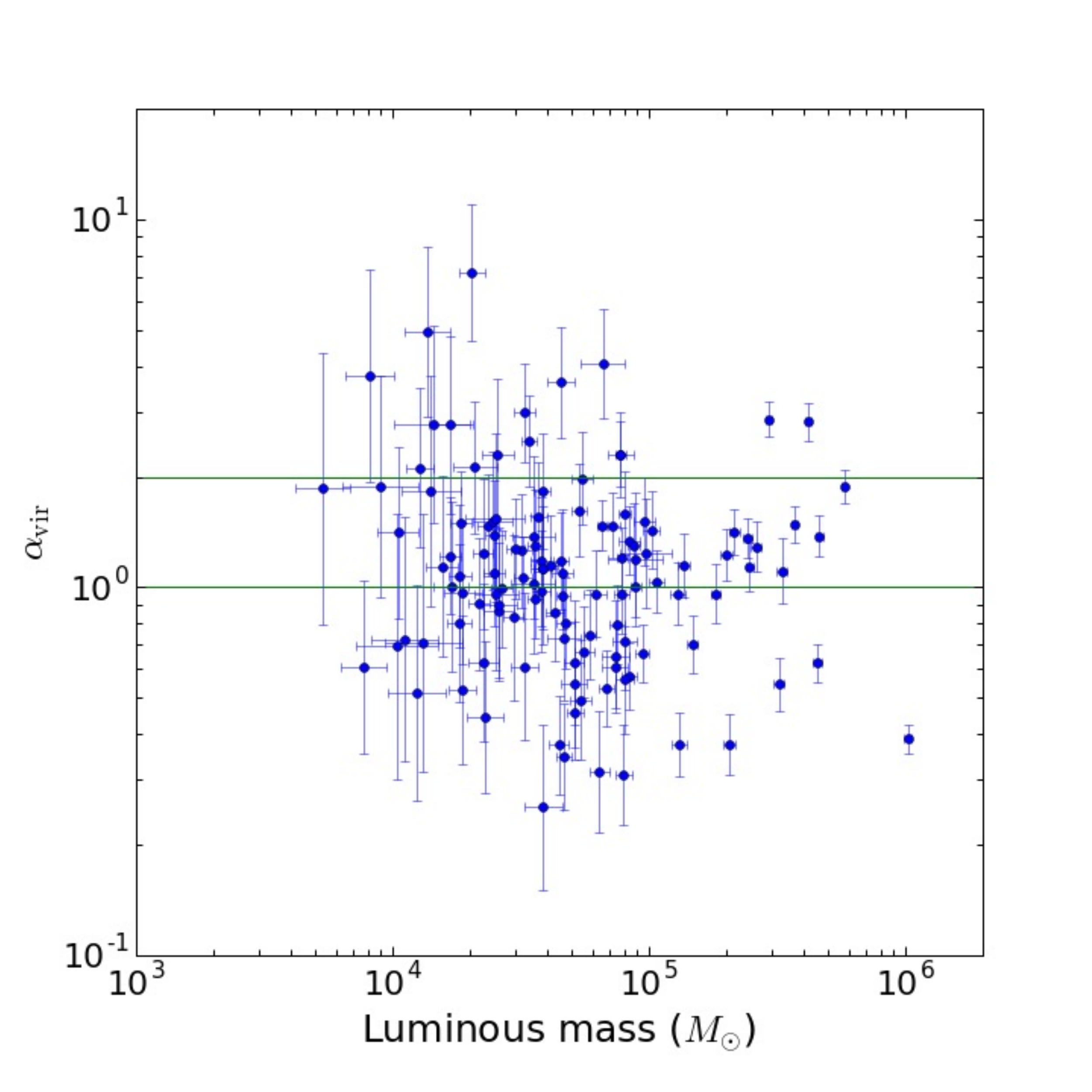}
\caption{Virial parameter $\alpha_{\rm vir}$ vs. CO-derived mass {\Mlum} for our final sample of 121 GMCs in NGC 300. The horizontal  lines delineate the range in which clouds are in approximate virial equilibrium. The vast majority of clouds are within $2\sigma$ of this zone, suggesting that they are gravitationally bound.}
\label{fig:alphavir}
\end{figure}

The relationship between gravitational and kinetic energy is often parameterized by the virial parameter $\alpha_{\rm vir}$ (Equation~[\ref{eqn:alphavir}]), with values near or just above unity denoting virial equilibrium~\citep[Section~\ref{sec:derivedprops}; see also][]{1992ApJ...395..140B}. Clouds with high virial parameters may by confined by external pressure~\citep[e.g.,][]{2011MNRAS.416..710F}, while those with sub-unity $\alpha_{\rm vir}$ may have additional magnetic support that inhibits collapse \citep{2003ApJ...585..850M} or potentially be in a state of dynamical collapse, though the short dynamical times in GMCs suggest the latter to be an unlikely scenario. We show in Figure~\ref{fig:alphavir} that the virial parameter in our NGC~300 final sample of GMCs is generally between $1$ and $2$, and is approximately constant with {\Mlum} across our sample. The strong correlation of {\Mvir} with {\Mlum} and low scatter in $\alpha_{\rm vir}$ near unity suggests that these GMCs are generally gravitationally bound and that additional effects such as magnetic or external pressure are subdominant.

\subsubsection{The mass-size relation}
\label{sec:masssize}

In his compilation of Milky Way cloud data, \cite{Larson:1981vv} noted an inverse correlation between Milky Way GMC volume density $\rho$ and size $R$, i.e. $\rho \propto R^{-1}$. This has been interpreted as an indication that GMCs have constant surface density since $\Sigma \propto M/R^2 \propto \rho R$, which combined with the above yields $\Sigma =$ constant. This apparent universality in cloud structure was verified in subsequent Galactic studies~\citep{Solomon:1987uq,Heyer:2009ii,2010A&A...519L...7L}. A number of extragalactic investigations have also found correlations between GMC size and mass (or CO luminosity)~\citep[e.g.,][]{2003ApJ...599..258R,2008ApJ...686..948B,2010MNRAS.406.2065H}, although given the low sensitivity of such observations, the majority of pixels in GMCs are near the sensitivity threshold and thus these correlations may or may not be physically meaningful. In contrast, other studies have presented arguments for GMCs exhibiting a large range in surface densities, particularly between different environments in galaxies \citep[e.g.,][]{2008ApJ...686..948B,2015ApJ...803...16U}.

\begin{figure}
\includegraphics[width=\linewidth]{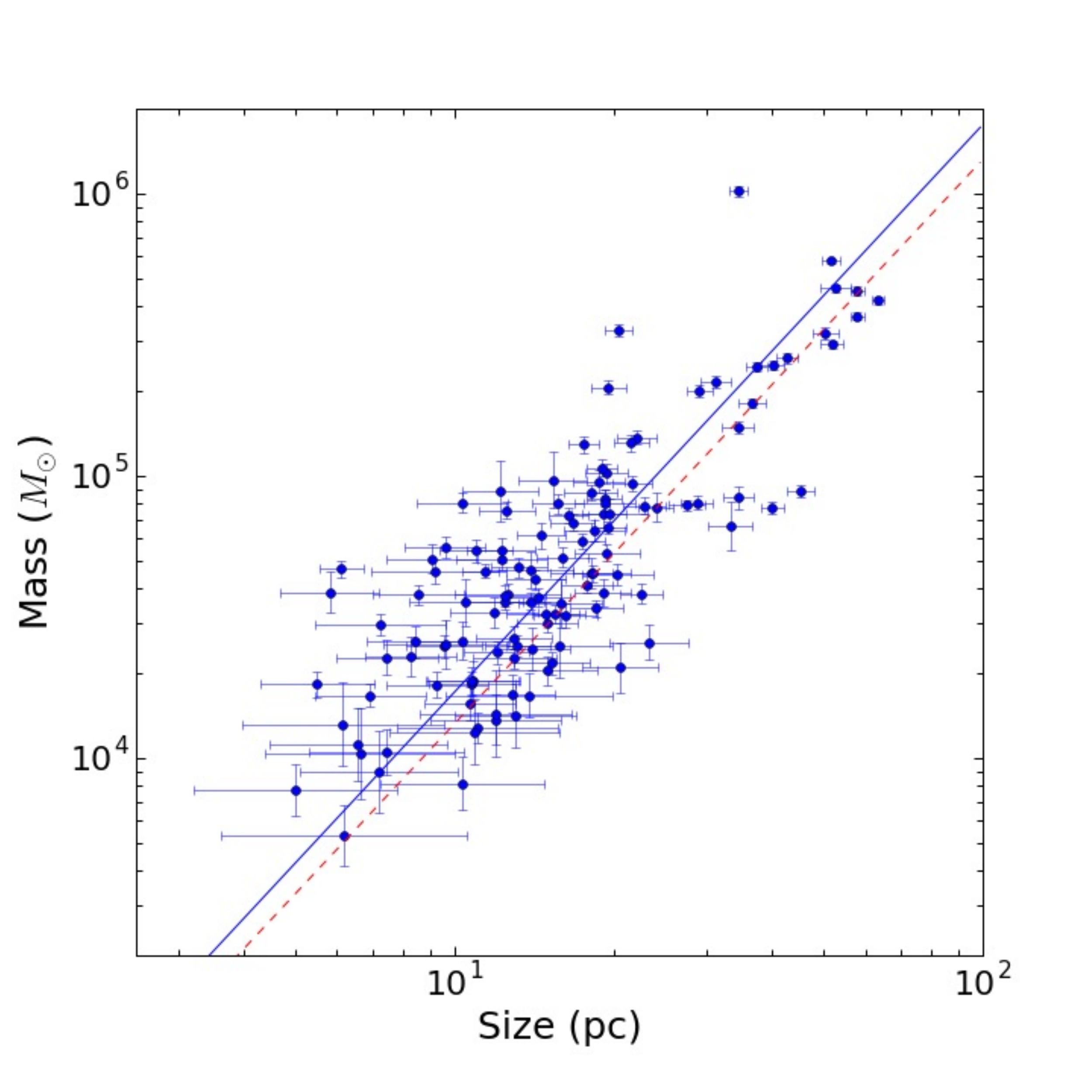}
\caption{The size-mass relation in the NGC~300 final sample. The solid line is our best fit relation, which has a power law slope of $2.00\pm0.12$, implying a median surface density of $61~\Msun$~pc$^{-2}$. The dashed red line shows the relation from the Milky Way cloud sample of \cite{2010A&A...519L...7L}. Our results are consistent to within the formal uncertainties with the GMCs in NGC~300 having a constant surface density.}
\label{fig:sizemass}
\end{figure}

In Figure~\ref{fig:sizemass}, we present the mass-size relation for our final GMC sample. Fitting our data with ODR as described in the previous sections, we find a power law slope of $2.00\pm0.12$. The Pearson correlation coefficient is $r_P = 0.81$, suggesting a high degree of correlation. While the scatter is significant (0.29 dex), the median measurement uncertainty in $\Sigma$ is equivalently large (also 0.29 dex). Therefore our results are statistically consistent with the population of NGC 300 GMCs having a constant surface density, as has been measured in Milky Way clouds.

The median surface density within our final sample is 60.6~$\Msun$~pc$^{-2}$, which is shown as the solid line in the figure. This is somewhat higher than the median surface density of $\sim40~\Msun$~pc$^{-2}$ in local Milky Way clouds~\citep{Heyer:2009ii,2010A&A...519L...7L}. However, \cite{2010A&A...519L...7L}, who utilized high-fidelity extinction measurements to derive mass surface density, noted that the precise value of the zero-point of the mass size relation (i.e., the value of the characteristic constant surface density) depends sensitively on where a cloud's boundaries are defined. Our median $\Sigma$ lies between the that derived by \cite{2010A&A...519L...7L} for extinction thresholds of $A_K=0.1$ and $0.2$ mag (these thresholds are equivalent to about $15$ -- $30~\Msun$~pc$^{-2}$). Considering the differences in techniques, and the level of scatter in our measurements, our results appear to be consistent with a similar characteristic surface density of GMCs in NGC~300 as in the Milky Way. However, GMCs in certain different environments -- both Galactic and extragalactic -- do appear to show real, physical differences in surface density~\citep{2015ApJ...803...16U,Heyer:2015ee}. For example, NGC 300 and Milky Way disk GMCs appear to have much lower surface densities than those in M51~\citep{2014ApJ...784....3C}. We will return to this question in the context of galactic environment in Section~\ref{sec:disc}.

The amount and origin of the scatter in GMC surface densities have been the topic of much debate. Larson's original results and subsequent Milky Way disk GMC studies have suggested that the range in surface density is quite narrow, and that the minimal scatter reflects a universality in cloud structure. In contrast, predictions from turbulent simulations have suggested that GMC surface densities should vary by up to two orders of magnitude, and that observational techniques are biased to a particular narrow range in surface density~\citep[e.g.,][]{1997ApJ...474..292V,2006MNRAS.372..443B}. Since the effective emissivity of a single CO line transition changes as a function of volume density, single-transition observational studies may only be sensitive to the range of density over which that line emits efficiently~\citep[e.g.,][]{Leroy:2017cv}. However, Galactic observations utilizing extinction techniques, which do not suffer from the above biases, find an exceptionally tight correlation between GMC mass and size, and a power law exponent of two, suggesting that at least in the nearby Milky Way clouds in that sample, the surface density is indeed constant~\citep{2010A&A...519L...7L}. In the following section, we further explore the scatter in $\Sigma$ in our data.

\subsection{The size-linewidth coefficient and virial equilibrium}
\label{sec:szlwsigma}

\begin{figure}
\includegraphics[width=\linewidth]{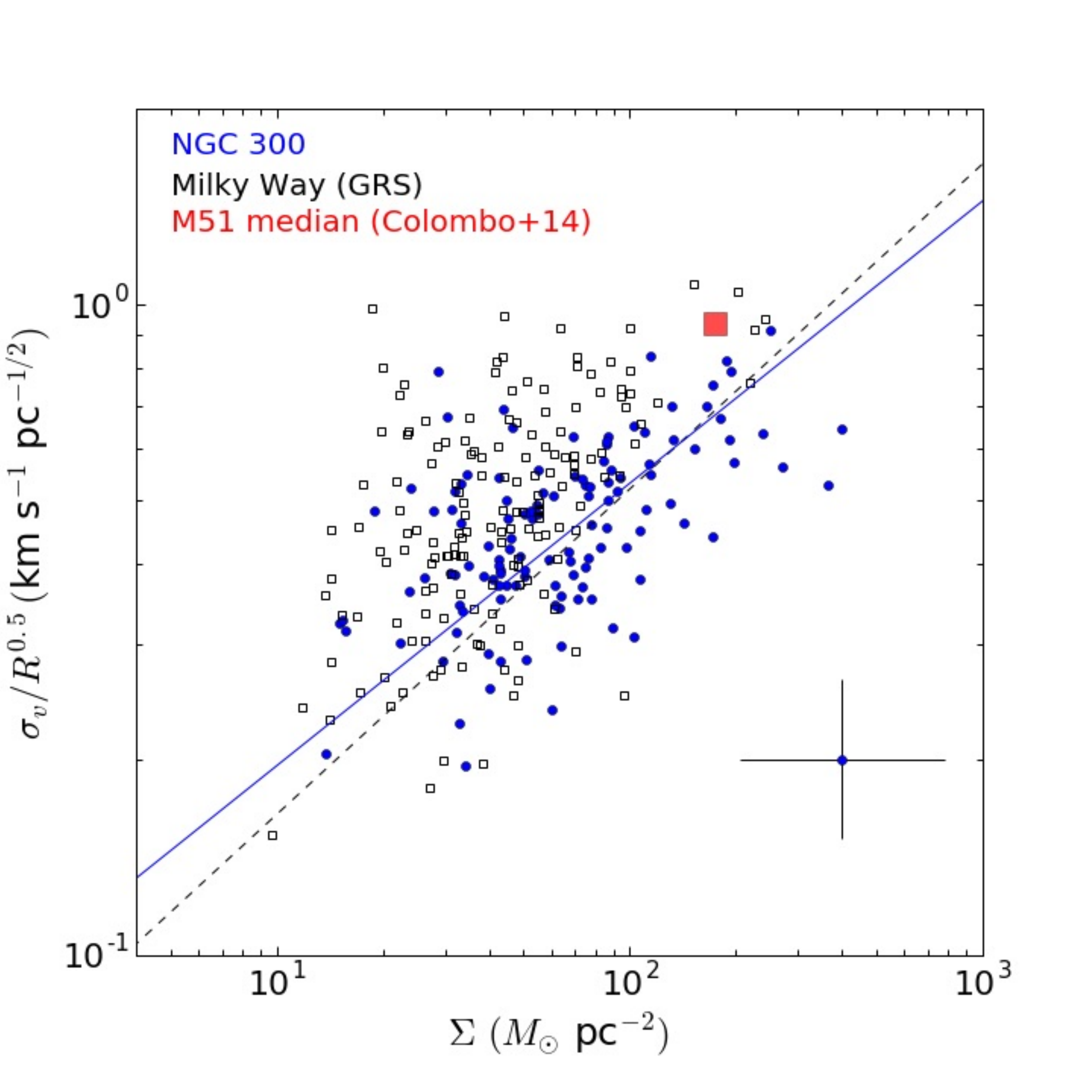}
\caption{Size-linewidth coefficient vs. GMC surface density for our final sample (blue circles) and clouds from the Milky Way Galactic Ring Survey GRS \citep[black open squares;][]{2006ApJS..163..145J}. The median values of these quantities for M51 clouds from \cite{2014ApJ...784....3C} is shown as a red square. The dashed line shows the locus defining virial equilibrium in a turbulent medium ($\sigma_v/R^{0.5} \propto \Sigma^{0.5}$ from \citealp{Heyer:2009ii}). The solid line shows the ODR fit to our data (slope $0.43\pm0.04$). The error bars in the lower right reflect the average uncertainty within the sample. Our data suggest that the observed correlation is most likely a result of the Larson relations holding in the presence of observational uncertainties (see the text).}
\label{fig:szlwsigma}
\end{figure}

\cite{Heyer:2009ii}, in re-examining the GMCs catalogued by \cite{Solomon:1987uq}, noted that the scatter in GMC surface density appeared to be systematic in that surface density scaled with the coefficient of the velocity structure function (i.e., the size-linewidth relation) as $C_{R\Delta V} \propto \Sigma^{1/2}$. Figure~\ref{fig:szlwsigma} shows that we find a similar correlation in our NGC~300 cloud sample. This scaling closely matches the prediction for clouds in virial equilibrium in a compressively-turbulent medium~\citep{Larson:1981vv,Solomon:1987uq,Heyer:2009ii}, or for clouds in pressure equilibrium for low to modest pressures~\citep{2011MNRAS.416..710F}. Put more formally, if one assumes that $\Mlum = \Mvir$ (or equivalently that $\alpha_{\rm vir} \approx 1$ and that $\Delta V \propto R^{0.5}$ (i.e., if the first two Larson relations hold), the relation $C_{R \Delta V} = \sqrt{\pi G/5}\, \Sigma^{0.5}$ naturally emerges. This exact relation is shown as a dashed line in the Figure, and it appears to be consistent with the NGC 300 data as well as those of \cite{Heyer:2009ii}. The same trend is also seen in the extragalactic sample of \cite{2008ApJ...686..948B}. Using orthogonal distance regression to fit our data as described for the other three relations described above, we find a best-fit slope of $0.43\pm0.04$ -- within 2$\sigma$ of the expected result of 0.5 for clouds in gravitational equilibrium within a turbulent medium. The Pearson correlation coefficient is $r_P = 0.55$. 

If all GMCs in NGC 300 had the same surface density and followed the size-linewidth relation with the same scaling coefficient, and if there were no observational uncertainties, there would be a single point in Figure~\ref{fig:szlwsigma}. In the presence of uncertainties, any variation in the measured value of $\Sigma$ would lead to a proportional change in $C_{R \Delta V}$ given by Equation~(\ref{eqn:szlwcoeff}), on average, and thus a cloud of points scattered about the trend line given by Equation~(\ref{eqn:szlwcoeff}). Thus the observed correlation could simply be a natural consequence of the Larson relations in the presence of observational uncertainties, and such a correlation would be expected to be visible in any sample of GMCs for which these relations hold. Given that our results are consistent (to within the uncertainties) with GMCs in NGC 300 having a constant surface density, and with the size-linewidth relation, this would be the simplest explanation for the trend observed in our data. Our results thus suggest a level of caution in interpreting the $C_{R \Delta V}$ -- $\Sigma$ trend, particularly in extragalactic studies where measurement uncertainties are typically large and can readily explain the observed correlation without resort to any other physical explanation.

However, examining GMC properties in the $\Sigma$ -- $C_{R \Delta V}$ plane may still be informative in comparing to other studies where measurable differences in these quantities are apparent and may reflect meaningful differences in the physical environments of GMCs. For example, the GMCs in M51 have on average much higher surface densities and are consistent with a larger size-linewidth coefficient \citep{2014ApJ...784....3C}. The median values of $\Sigma$ and $C_{R \Delta V}$ from \cite{2014ApJ...784....3C} are shown in the Figure, and this point clearly lies above the trend defined by the Milky Way and NGC 300 data. We will explore this further in Section~\ref{sec:disc}.

Since $\Sigma \equiv \Mlum/(\pi R^2)$, there are powers of $R$ in the denominators of both axes, and so there is some intrinsic covariance between the quantities plotted in Figure~\ref{fig:szlwsigma}. To test if this is driving the observed correlation, we constructed a sample of $10^4$ model clouds with size, linewidth, and mass each independently drawn from a random uniform distribution defined by the range of these parameters in our sample. We fit this model population as described above, and now find a slope of 0.27 with a correlation coefficient of 0.37. Thus while the intrinsic covariance by definition contributes somewhat to the observed correlation in the model population, we take these results as evidence that it is not the major driver of the trend seen in the Figure, as the slope is significantly shallower and the level of correlation poorer. The correlation seen in Figure~\ref{fig:szlwsigma} is thus most likely driven by the presence of measurement uncertainties in the observed parameters of a population of GMCs governed by the Larson relations.

\section{Discussion}
\label{sec:disc}

Our results imply that the GMC populations of NGC 300 and (at least the inner disk of) the Milky Way are similar in both distribution by mass and in the Larson scaling relations, despite significant differences (e.g., mass, morphology, average metallicity) between these two galaxies. This suggests that these macroscopic properties may be subdominant in setting the properties of a spiral galaxy disk's GMC population, and may be indicative of certain universal characteristics of GMCs in spiral galaxies. However, previous studies of other spiral galaxies have underlined the importance of galaxy environment in controlling GMC properties, and revealed measurable differences in these properties between different galactic environments \citep[e.g.,][]{2013ApJ...779...46H,2014ApJ...784....3C,2015ApJ...803...16U}. In this section we make comparisons between our study and previous Galactic and extragalactic work in order to investigate which environmental parameters play dominant roles in setting the key empirical diagnostics of GMC populations.

\subsection{GMCs in NGC 300 and the Milky Way}

Our study of of molecular gas in NGC 300 at 10 pc and 1 {\kms} resolution allows the most direct comparison to-date between the well-studied GMC population of our own Milky Way and those in an external spiral galaxy. This comparison is particularly salient because NGC 300 is significantly different from the Milky Way in many of its global properties. While both galaxies are spirals, NGC 300 (class SAs(d)) has weak, loosely wound spiral arm features, no bar, and almost no bulge, in contrast to the Milky Way (class SB(rs)bc), which has a significant bulge component, a bar, and more prominent, tightly wound spiral arms. Additionally, NGC 300's disk stellar mass ($M_*$) of $2.1\times10^9$~\citep{2010PASP..122.1397S} and star formation rate (SFR) of $\sim0.11~\Msun$~yr$^{-1}$~\citep{2004ApJS..154..253H} are each more than an order of magnitude lower than those of the Milky Way, which has a stellar mass of $5.17\times10^{10}~\Msun$ and SFR of $1.65~\Msun$~yr$^{-1}$ (Licquia \& Newman 2015). Furthermore, NGC 300 has an average metallicity of approximately 60\% solar over the range of galactocentric radius spanned by our sample \citep{Deharveng:1988wh}. Note that since the differences in SFR and $M_*$ between the two galaxies are approximately the same factor, both these galaxies have similar specific star formation rates (SSFR~$=$~SFR~$/$~$M_*$) -- i.e. they are both on the galaxy star-forming main sequence.

Despite these global differences, our results have revealed that the GMC populations of (at least the inner disks of) these two galaxies are almost indistinguishable. This suggests that global properties such as galaxy mass, morphology, and average metallicity do not directly control the properties of GMCs within spiral galaxies, at least for galaxies on the star-forming main sequence. This hypothesis is supported by lower-resolution studies of the inner disk of the Local Group spiral M33, which find to the limits of relatively high measurement uncertainties, correspondence with the Milky Way GMC population in the size-linewidth relation \citep{2003ApJ...599..258R} and GMC mass spectrum for the inner 2 kpc of M33 \citep{Gratier:2012km}. Similar conclusions have been reached in comparing GMCs between the Milky Way, M31, and M33 \citep{2007ApJ...654..240R,2008ApJ...675..330S}. Our data, which are the first to achieve 10~pc scales in an external spiral galaxy, provide crucial additional evidence for this scenario.

\subsection{The role of environment: NGC 300 vs. M51}

If global galaxy characteristics do not control GMC properties, what mechanism(s) can explain the apparent measurable differences in empirical diagnostics of GMC populations in different galaxies, or in different regions within galaxies? For example, the GMC mass spectrum slope in NGC 300 is significantly shallower than that of the LMC \citep{Wong:2011ib}, or the outer Milky Way \citep[][see also Section~\ref{sec:massspec}]{Rice:2016ko}, while the median surface density and size-linewidth coefficient in NGC 300 are both significantly smaller than in M51 \citep{2014ApJ...784....3C}. Since GMCs are localized to specific regions within galaxies, could ``local'' properties more effectively influence empirical GMC diagnostics than global ones? This has been suggested as an explanation for the differing Larson and mass spectrum power law slopes among the spiral arm, interarm, and nuclear regions of M51 \citep{2014ApJ...784....3C}. Since this study of M51 is one of the most comprehensive GMC investigations in an external galaxy, and since its population shows differences in several diagnostics as compared with NGC 300 (and the Milky Way), we focus our efforts here in comparing our work with theirs. We have downloaded the catalog kindly made publicly available by \cite{2014ApJ...784....3C} and discuss the comparison below.

\begin{figure}
\includegraphics[width=\linewidth]{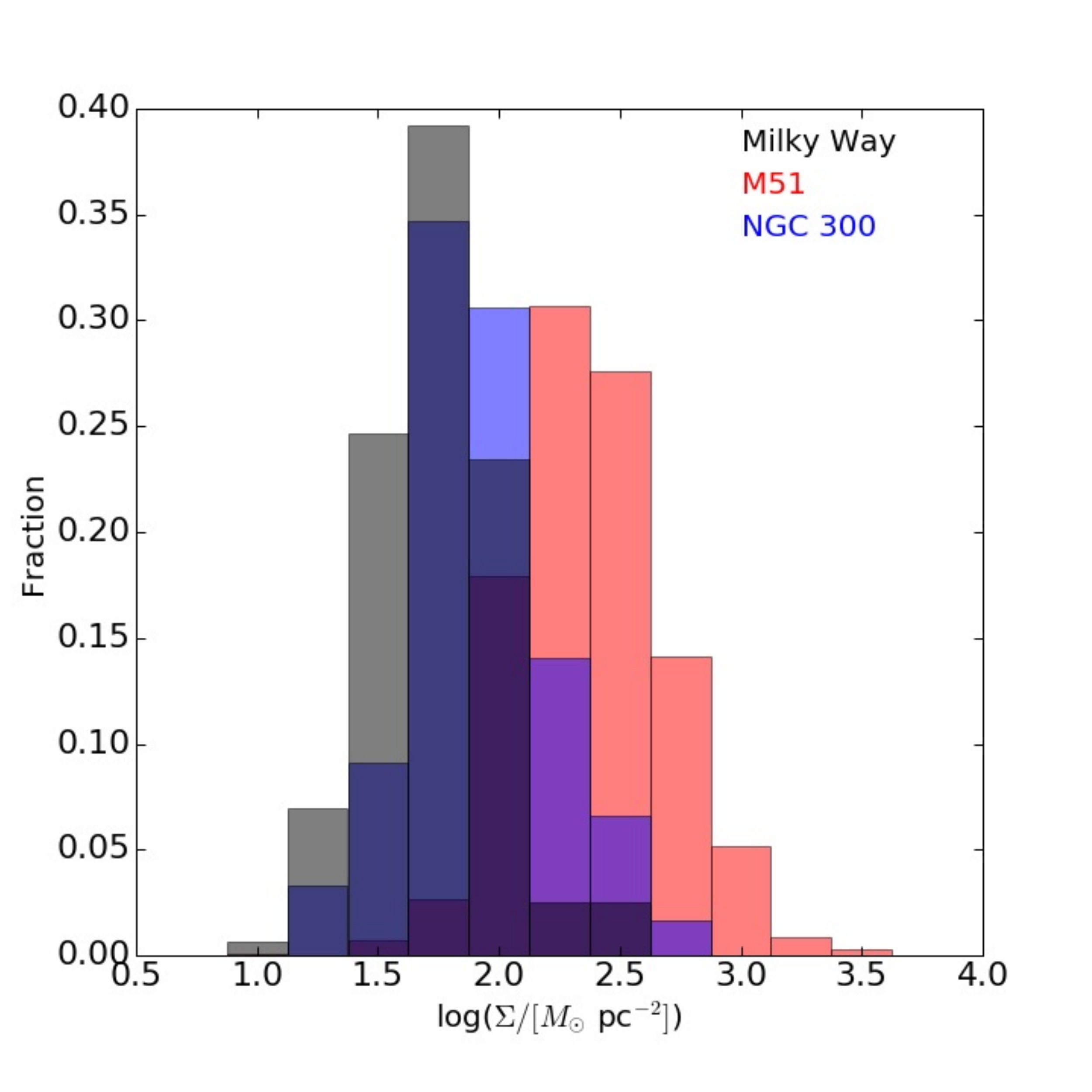}
\caption{Histogram of GMC mass surface densities in NGC 300 (blue), the Milky Way (gray), and M51 (red). The median GMC surface density in M51 is a factor of two higher than that in NGC 300 and the Milky Way. This discrepancy is likely due to physical differences between the environments in which the GMCs reside.}
\label{fig:comparesurfacedensity}
\end{figure}

As previously alluded, there are a number of key differences in the GMC populations between M51 and NGC 300. For one, there is a significant offset in the size-linewidth relation between these galaxies: the median $C_{R \Delta V}$ in NGC 300 is 0.46~{\kms}~pc$^{-1/2}$ with a scatter of 0.14, while in M51 it is 0.94~{\kms}~pc$^{-1/2}$. Furthermore, the median mass surface density is also more than a factor of two higher in M51. Figure~\ref{fig:comparesurfacedensity} presents histograms of the GMC mass surface density in NGC 300, M51, and the Milky Way (from the GRS), and demonstrates that M51's GMCs appear to be on average significantly more dense than those in NGC 300 or the Milky Way.

\subsection{The effects of resolution}

The M51 observations have significantly lower resolution (40 pc, 2~{\kms}) than the present study (10 pc, 1~{\kms}). To test whether the differences in GMC population discussed above are effects of resolution or truly physical, we smoothed our data and re-ran our entire analysis pipeline, as follows. First, we used the \texttt{CASA} tasks \texttt{IMSMOOTH}, \texttt{SPECSMOOTH}, and \texttt{IMREBIN} to convolve our fully cleaned data cubes to a spatial and spectral resolution matching the PdBI (Plateau de Bure Interferometer) Arcsecond Whirlpool Survey (PAWS) of M51 (i.e., 40 pc and 2~{\kms}), while also rebinning to $0.6\arcsec$ pixels. Basic tests demonstrated that the flux in the original data was fully preserved in the convolution. We then ran \texttt{CPROPS} on the smoothed data cubes and constructed a new NGC 300 GMC catalog at 40 pc / 2~{\kms} resolution. Our smoothed sample ended up having 110 clouds in total, of which 41 met the criteria established in Section~\ref{sec:results} for inclusion in the ``final'' sample (i.e., deconvolved size larger than half the physical resolution, no more than 10\% beyond the primary beam FWHM, no duplicate clouds between regions).

\begin{figure}
\includegraphics[width=\linewidth]{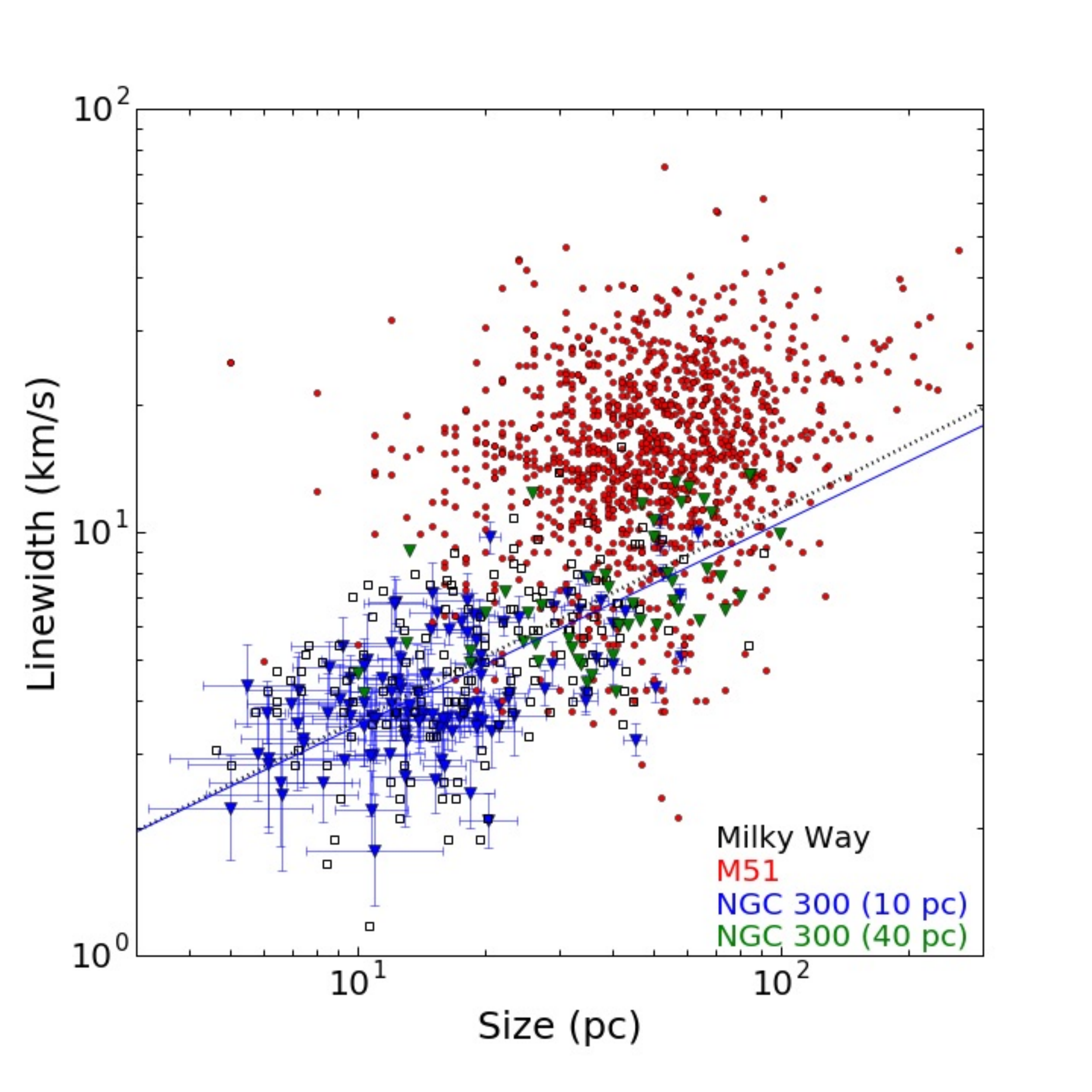}
\caption{The size-linewidth relation in NGC 300 (blue triangles: 10 pc resolution, green triangles: 40 pc resolution), the Milky Way (black open squares), and M51 (red circles). The GMCs in M51 have a signficantly higher size-linewidth coefficient than those in NGC 300 or the Milky Way. The fact that this offset persists when we re-analyze our data at the same physical resolution as the M51 data demonstrates that the difference is not simply due to resolution, and is likely a real physical difference between the environments in these galaxies. The blue solid line shows the fit to our full resolution final sample, and the black dotted line is the Milky Way relation (see Figure~\ref{fig:sizelinewidth}).}
\label{fig:comparesizelinewidth}
\end{figure}

Figure~\ref{fig:comparesizelinewidth} shows the size-linewidth relation for our smoothed sample alongside that of the M51 GMCs from \cite{2014ApJ...784....3C}, as well as our full resolution sample and Milky Way clouds from the GRS. The smoothing process apparently shifts clouds along the relation derived at full resolution toward higher size and linewidth, but preserves the size-linewidth coefficient (the median $C_{R \Delta V} = 0.43$~{\kms}~pc$^{-1/2}$ in the smoothed sample, with a $1\sigma$ scatter of 0.15). Our smoothed sample has a median surface density \textit{lower} than our full resolution sample: $45~\Msun$~pc$^{-2}$. This further exacerbates the discrepancy in surface density between these galaxies' GMCs. It is thus clear that spatial and spectral resolution are not responsible for the differences seen between these GMC populations.

One alternative possibility is that in the PAWS analysis, GMCs are more often superposed along the line-of-sight, leading to blending and potentially artificially increasing measured linewidths (while not increasing measured sizes). While this is not out of the question, \cite{2014ApJ...784....3C} appear to be successful in recovering GMCs that overlap spatially (see their Figure 3), and so it seems unlikely that any super-resolution blending is occurring at the level necessary to explain the differences seen in Figures~\ref{fig:comparesurfacedensity} and \ref{fig:comparesizelinewidth}.

Finally, it is worth exploring the potential effects of differing sensitivity between PAWS \citep{2014ApJ...784....3C} and the present work. They achieved a final sensitivity of 0.4 K per 5~{\kms} channel, while we achieve $\sim0.13$~K per 1~{\kms} channel. They utilized an edge threshold of $1.5$ times the noise in their data cubes, while we used a threshold of $2$ times the noise (the \texttt{EDGE} parameter in \texttt{CPROPS}, meaning that in terms of channel noise in K, we were a factor of $\sim2.3$ more sensitive (though their wider spectral resolution means this discrepancy is greater for lines narrower than 5~{\kms}). However, we note that (a) NGC 300 is at lower average metallicity, meaning a higher median CO-to-H$_2$ conversion factor is required than for M51, reducing our relative effective sensitivity to H$_2$; and, (b) we observed the CO(2-1) transition, while \cite{2014ApJ...784....3C} observed CO(1-0), meaning our effective sensitivity is further reduced by a factor of the (2-1) to (1-0) line ratio (approximately a factor of 1.4 for the line ratio of 0.7 we used). Taken together these effects add up to produce almost identical effective molecular gas sensitivities between our work and PAWS, assuming CO lines are not significantly narrower than 5~{\kms}. In addition, the extrapolation procedure in \texttt{CPROPS} (theoretically) accounts for all emission out to the 0~K contour, and so given sufficient signal-to-noise, the algorithm should still recover a cloud's full extent. We thus cannot fully rule out observational effects producing the results we discussed above, but based on these calculations expect the effects of superposition and sensitivity to be minor.

\subsection{Midplane disk pressure}

One local environmental factor that may affect GMC properties in spiral galaxies is pressure due to the surrounding ISM~\citep[e.g.,][]{2011MNRAS.416..710F,2013ApJ...779...46H,2015ApJ...803...16U}. Pressure may act to confine GMCs, increasing the level of turbulent support necessary to stabilize a cloud, thereby increasing their observed linewidths. Additionally, pressure may compress GMCs, increasing their surface densities. For example, \cite{2013ApJ...779...46H} hypothesized that the higher pressure in the ISM of M51 as compared to M33 and the LMC explains the relatively higher surface densities of M51's GMCs. \cite{2015ApJ...803...16U} argue that pressure-bound clouds should have higher linewidths than clouds bound by gravity alone. If this is the case, then the ISM pressure in NGC 300 should be compatible with that of the Milky Way to explain the similar GMC surface densities and size-linewidth coefficient between these two galaxies.

To test this scenario, we have estimated the ISM pressure in NGC 300 and the Milky Way. Following \cite{1989ApJ...338..178E}, we take the midplane gas pressure $P_{\rm ISM}$ to be
\begin{equation}
P_{\rm ISM} = \frac{\pi G}{2} \Sigma_g \left(\Sigma_g + \frac{\sigma_g}{\sigma_*} \Sigma_* \right),
\label{eqn:pressure}
\end{equation}
where $\Sigma_g$ and $\Sigma_*$ are total (atomic plus molecular) gas and the stellar mass surface densities, respectively, and $\sigma_g$ and $\sigma_*$ are the corresponding velocity dispersions. Since there have been no published direct measurements of the stellar velocity dispersion in NGC~300, we make a series of assumptions and substitutions to arrive at an equation in which the right hand side consists entirely of measured quantities. We first assume a self-gravitating stellar disk with scale height $h_* = (\sigma_*^2 / 4\pi G \rho_*)^{1/2}$, where $\rho_*$ is the stellar volume density. In such a disk, $\Sigma_* = 2\sqrt{2} \rho_* h_*$, and we can then substitute these two definitions into Equation~(\ref{eqn:pressure}) to obtain
\begin{equation}
P_{\rm ISM} = \frac{\pi G}{2} \Sigma_g \left[\Sigma_g + \sigma_g \left(\frac{\Sigma_*}{\sqrt{2}\pi G \, h_*} \right)^{1/2} \ \right].
\label{eqn:pressure2}
\end{equation}
Note that this equation reduces to within a small correction factor of Equation (2) from \cite{2004ApJ...612L..29B} if one assumes $\Sigma_* >> \Sigma_g$. Such an assumption is reasonable in the center of NGC 300, but we choose to use the full Equation~(\ref{eqn:pressure2}) in our analysis as this assumption breaks down beyond the inner galaxy.

To estimate $\Sigma_g$, we assume that the atomic gas dominates the molecular gas over large scales, as is the case in the Milky Way (see below). \cite{2011MNRAS.410.2217W} mapped NGC 300 in the \ion{H}{1} 21~cm line with the Australian Telescope Compact Array, and modeled the gas surface density as a function of galactocentric radius. They found a nearly constant value of $M_g \approx 7~\Msun$~pc$^{-2}$ (including the contribution from He) out to 6~kpc, well beyond the positions we observed in CO. They also used \textit{Spitzer} Infrared Array Camera (IRAC) data to compute the radial $\Sigma_*$ profile of NGC~300. We use their results (from their Table 2) for these two quantities. \cite{2011MNRAS.410.2217W} also measured $\sigma_g$ to be 10--15~{\kms} across the inner disk, with no trend with radius; we use the average of 12.5~{\kms} at all radii at which we compute the pressure (0 to 4~kpc). There are no direct measurements of the stellar scale height in NGC~300, however we use the empirical relation between $h_*$ and maximum rotational velocity $v_{\rm rot,max}$ for spiral galaxies from \citep[][Equation (3)]{vanderKruit:2011cg} to estimate it. For the rotational velocity we again draw upon \cite{2011MNRAS.410.2217W}, who find $v_{\rm rot, max} = 98.8~\kms$. This yields a stellar scale height of $h_* = 310$~pc, which we treat as constant with radius, motivated by the observed general constancy of stellar scale heights in spiral disks~\citep[][and references therein]{vanderKruit:2011cg}.

\begin{figure}
\includegraphics[width=\linewidth]{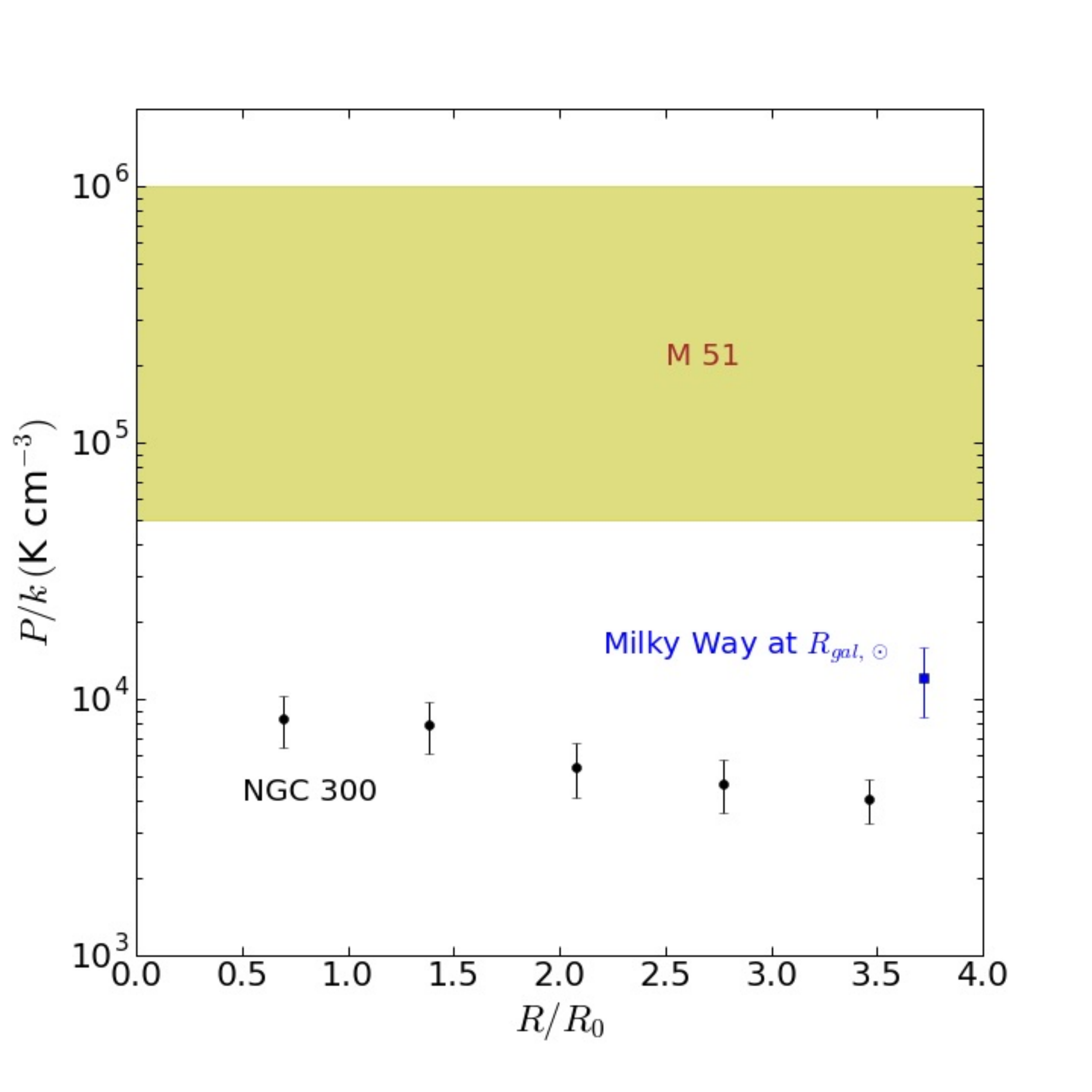}
\caption{Midplane disk pressure as a function of scaled galactocentric radius in NGC 300, calculated as described in the text. The pressure decreases slightly with radius but is in the range $P/k \sim 4$--$8\times10^3$~K~cm$^{-3}$ over the portion of the galaxy containing the GMCs we study here. The pressure in the Milky Way at the solar radius is shown as a blue square, and is similar to the pressures we find in the inner NGC~300 disk. Radii for both galaxies are scaled by the individual galaxy's exponential scale length $R_0$. The shaded region indicates the range of midplane pressures in M51 calculated by \cite{2014ApJ...784....3C}. The order-of-magnitude pressure difference between M51 and NGC 300 (as well as the Milky Way) may potentially explain the differences in median GMC surface density and size-linewidth scaling coefficient observed in these galaxies.}
\label{fig:pressure}
\end{figure}

With all necessary variables accounted for, we present the radial profile of ISM pressure in Figure~\ref{fig:pressure}. Pressure decreases monotonically but only marginally with radius from a value of $P/k \approx 8.4 \times 10^3$~K~cm$^{-3}$ at 0.9~kpc to $4.1 \times 10^3$~K~cm$^{-3}$ at 3.7~kpc, where $k = 1.38 \times 10^{-16}$~erg~K$^{-1}$ is Boltzmann's constant. Propagating uncertainties gives a median error of $1.3 \times 10^3$. Since of the input variables to Equation~(\ref{eqn:pressure2}) only $\Sigma_*$ changes appreciably with radius, it is the decrease in stellar surface density that drives our derived midplane pressure trend.

In order to perform a salient comparison, we compute the ISM pressure for the solar neighborhood of the Milky Way in the same manner as for NGC 300. While it is difficult to measure macroscopic properties of our Galaxy from within the disk, several studies have made efforts to do so. \cite{1990MNRAS.244...25G} found $\Sigma_* \approx 54~\Msun$~pc$^{-2}$, while more recently, \cite{2012ApJ...751..131B} derived a significantly lower value of $30~\Msun$~pc$^{-2}$ for the solar radius. We take the average of these two results as our best estimate for $\Sigma_*$ in the Milky Way (noting the significant uncertainty). For $\Sigma_g$, we use the result from \cite{Kalberla:2009gm}, who find $\Sigma_g \approx 10~\Msun$~pc$^{-2}$ at the solar circle (and also find that it is relatively constant with radius, as it is in NGC~300). To maintain consistency with the calculation for NGC~300, we do not incorporate the molecular gas contribution, which is subdominant (approximately $1~\Msun$~pc$^{-2}$ \citep{1987ApJ...322..706D,Heyer:2015ee}). \cite{1995ApJ...448..138M} measured the Milky Way gas velocity dispersion to be 9.2~{\kms}, independent of radius, and we use this as our estimate for $\sigma_g$. Finally, we assume a stellar scale height of 300~pc, the value for the thin disk from \cite{2006A&A...451..515M} (see also \cite{2014ApJ...794...90K}. Using these numbers, we find the ISM midplane pressure at the solar radius is $1.2\times10^4$~K~cm$^{-3}$. Using the lower or upper values for $\Sigma_*$ discussed above, and incorporating the molecular gas gives us an approximate plausible range of 8.5--19.8$\times 10^3$~K~cm$^{-3}$. Our result is consistent with the modeled estimate of the kinetic pressure at the solar radius by \cite{1990ApJ...365..544B}, who found $P/k \approx 10^4$~K~cm$^{-3}$.

The ISM pressure in the Milky Way at the solar radius appears to be consistent with that in NGC~300 within 3~kpc. If pressure plays a role in setting GMC properties, this is precisely what one would expect given the similarities between the GMC properties in NGC~300 and nearby Milky Way clouds. Since the latter galaxy is significantly smaller (radial scale length of 1.35~kpc \citep{2005ApJ...629..239B}, as compared with 2.15~kpc for the Milky Way \citep{2013ApJ...779..115B}), in terms of scale lengths our observations actually probe the range of radii not far inside NGC~300's equivalent of the solar radius. It is thus perhaps not surprising that the stellar and gas surface densities and velocity dispersions are similar in the disks of these galaxies.

In contrast, M51 has an ISM pressure of 0.5--$10\times10^5$~K~cm$^{-3}$ \citep{2013ApJ...779...46H} -- an order of magnitude or more higher than in the Milky Way and NGC~300. This provides a plausible explanation for the notable differences in GMC populations -- in particular, the higher size-linewidth coefficient and higher average surface density. We present the difference in pressure between these systems visually in Figure~\ref{fig:pressure}; see also Figure~\ref{fig:szlwsigma}, which shows the size-linewidth coefficient -- surface density plane. Taken together, these results suggest that pressure potentially plays a key role in setting two important empirical parameters of a GMC population: the size-linewidth coefficient, and the characteristic mass surface density. A similar argument was presented to explain the differences between GMCs in the LMC, M33, M51, and NGC 4526 \citep{2013ApJ...779...46H,2015ApJ...803...16U}. This scenario also explains the size-linewidth coefficient in the GMCs within the Milky Way's Central Molecular Zone (CMZ), which is significantly higher than that in the disk \cite{2001ApJ...562..348O}, and even higher than in M51. The pressure in the CMZ has been estimated to between $5 \times 10^6$ and $1\times 10^8$~K~cm$^{-3}$ \citep{1992Natur.357..665S,2000ApJ...536..357M}, again quite a bit higher than any of the environments discussed here.

Given the arguments above, it is also possible that local differences in disk midplane pressure may contribute to the distribution of points in the $\Sigma$ - $C_{R \Delta V}$ plane within our sample as well as the Milky Way cloud sample of \cite{Heyer:2009ii} (Figure~\ref{fig:szlwsigma}), though we again note that our data favor the simpler explanation that the trend reflects the combined effects of the Larson relations in the presence of observational scatter. Furthermore, given the decreasing pressure with galactocentric radius we find in NGC 300, one might expect $\Sigma$ to show a similar radial trend. We do not see such a trend (Figure~\ref{fig:rgal}), though the large scatter may be masking a weak underlying dependence. Nevertheless, there could be real, if minor, region-to-region differences in $P_{\rm ISM}$ that are driven by non-axisymmetric processes in spiral galaxies such as spiral arm density waves, bars, and streaming motions, or highly localized events such as supernovae. Given that our calculation of $P_{\rm ISM}$ is based on large-scale, azimuthally binned measurements, further investigation into this intriguing possibility is beyond the scope of this paper.

\section{Summary and Conclusions}
\label{sec:summary}

We have utilized ALMA CO(2-1) interferometric observations of forty-eight 250-pc size (FWHM) pointings in the general vicinity of {\HII} regions to study the properties of GMCs in the nearby spiral galaxy NGC~300 at the highest resolution (10~pc, 1~{\kms}) achieved to-date in an external galaxy beyond the Local Group. Using the \texttt{CPROPS} algorithm tailored to be sensitive to GMCs while recovering the full emission from all spatial scales sampled ($\sim10-100$~pc), we have identified and characterized 250 clouds at high ($>10$) signal-to-noise ratio. We focus on a final sample of 121 well-resolved objects for the majority of our analysis. In particular:

\begin{enumerate}
\item{The GMCs in our sample span a similar range in physical properties to those in the Milky Way, with sizes of 5--60~pc, linewidths of 1--10~{\kms}, and masses of $\sim10^4$ -- $10^6~\Msun$.}
\item{Clouds show multiplicity and preferential clustering, with the vast majority of 250~pc sized regions observed having 4--8 clouds. The median axis ratio is 2.4, and 75\% of the final sample have axis ratios smaller than 3.5, demonstrating that clouds tend to be more elliptical than circular in cross section.}
\item{Cloud properties show wide variation but no significant trends with galactocentric radius or as a function of whether or not they are located in a spiral arm feature. We note that all the most massive clouds are located between 1.5-2.5 kpc, and that our observations only sample the inner (rising) portion of NGC~300's rotation curve.}
\item{The GMC mass spectrum above our sensitivity limit of $8\times10^3~\Msun$ is consistent with a truncated power law with slope $-1.76\pm0.07$ and a truncation mass of $9\times10^5~\Msun$. This is similar to mass spectra observed in the inner Milky Way and other galaxy disks, but shallower than those in the outer regions of galaxies.}
\item{GMC size and linewidth are well-correlated in NGC 300, with a power law slope of $0.48\pm0.05$. This is similar to the relation observed in the Milky Way in both slope and offset, suggesting a level of universality in cloud structure between these two different galaxies. These results are consistent with the hypothesis of compressible hierarchical turbulence within the molecular ISM.}
\item{Virial and luminous masses are linearly correlated, while the virial parameter ranges from about one to two and does not appear to vary with any parameter we examined. These results suggest that the vast majority of GMCs in our sample are gravitationally bound.}
\item{Cloud mass scales as size with a power law exponent of $2.00\pm0.12$, with the zero-point of this relation denoting  a median surface density of 61~{\Msun}~pc$^{-2}$, slightly larger than that found in Milky Way disk GMCs (40~{\Msun}~pc$^{-2}$). The scatter in the mass-size relation is large but similar in magnitude to the measurement uncertainties, and thus our results are statistically consistent with the NGC 300 GMC population having a constant surface density. This apparent constancy may reflect a similarity in the general physical conditions between the disks of the Milky Way and NGC 300.}
\item{We find a correlation between the GMC size-linewidth (i.e., velocity structure function) coefficient and surface density, similar to earlier studies. We demonstrate that this observed trend can be readily explained by the Larson relations holding in the presence of observational scatter, in contrast to previous interpretations that have suggested a modification to these relations to explain this correlation.}
\item{The general similarity between the NGC 300 GMCs and clouds in the solar neighborhood suggests that global galaxy properties such as stellar mass, morphology, or average metallicity are not responsible for setting the properties of the GMC population in main sequence spiral galaxy disks.}
\item{We smooth our data to 40 pc and 2~{\kms} and re-analyze it to show that the observed difference in size-linewidth coefficient and median surface density between NGC 300 and M51 is not an artifact of resolution, but likely a result of the different environments present in these galaxy disks.}
\item{We calculate the ISM pressure in NGC 300 and the solar neighborhood and find that the range of values is similar in these two galaxies, but  lower in both cases than in the spiral M51 by more than an order of magnitude. This variation in pressure may help explain the clear observed difference between GMC size-linewidth coefficient and surface density in NGC 300 (as well as the Milky Way disk) and M51. We confirm the results of previous studies that suggest that ISM pressure is a key environmental variable in setting GMC properties among galaxies.}
\end{enumerate}

To further test the role of global and local properties of galaxies in setting the characteristics GMC populations, it is necessary to conduct additional high-resolution studies such as the one presented here in additional galaxies, both on and beyond the star-forming main sequence. We hope that ALMA continues its transformative role in connecting Galactic and extragalactic studies of the ISM and star formation.

C.M.F. would like to deeply thank Bruce Elmegreen for his thoughtful and generous help in thinking through some of the key points in the discussion section regarding pressure and its potential effects on GMCs. The authors also owe a debt of gratitude to Erik Rosolowsky for assistance and insights regarding \texttt{CPROPS} in particular and extragalactic GMC observations in general. We also acknowledge Alyssa Goodman, Charlie Conroy, and Qizhou Zhang for helpful discussions. C.M.F. spent a week at NRAO-Charlottesville and wishes to thank the staff there, particularly support scientist Drew Brisbin, for assistance with re-reducing and re-imaging ALMA data. Finally, we very much appreciate the thoughtful and helpful comments from the anonymous referee, which greatly improved the quality and rigor of this manuscript.

Support for this work was provided by the NSF through award SOSPA2-019 from the NRAO. C.M.F. also acknowledges support from a National Science Foundation Graduate Research Fellowship under Grant No. DGE-1144152.

This paper makes use of the following ALMA data: ADS/JAO.ALMA\#2013.1.00586.S (PI. C. Lada). ALMA is a partnership of ESO (representing its member states), NSF (USA) and NINS (Japan), together with NRC (Canada), MOST and ASIAA (Taiwan), and KASI (Republic of Korea), in cooperation with the Republic of Chile. The Joint ALMA Observatory is operated by ESO, AUI/NRAO and NAOJ. The National Radio Astronomy Observatory is a facility of the National Science Foundation operated under cooperative agreement by Associated Universities, Inc. This research has made use of the NASA/IPAC Extragalactic Database (NED) which is operated by the Jet Propulsion Laboratory, California Institute of Technology, under contract with the National Aeronautics and Space Administration. \textit{Herschel} is an ESA space observatory with science instruments provided by European-led Principal Investigator consortia and with important participation from NASA.

\appendix
\section{CPROPS parameter experiments}
\label{sec:cpropsex}

The importance of the choice of cloud decomposition algorithm has been clearly demonstrated in the literature \citep[e.g.,][]{2008ApJ...675..330S,Wong:2011ib,2013ApJ...779...46H}. The decomposition of data cubes into clouds clearly represents a source of significant uncertainty beyond that captured in the bootstrapped uncertainties in cloud parameters. However, this additional uncertainty is extremely difficult to quantify, as structures in one run of the decomposition algorithm do not always map one-to-one to those in other runs. It is thus more instructive to look at statistical trends in the population of clouds and in the scaling relations derived from that population. In order to test the effects of varying key \texttt{CPROPS} parameters on derived cloud properties, we created multiple realizations of our cloud catalog. In particular, we investigated (1) the use of the modified \texttt{CLUMPFIND} in assigning emission to clouds and (2) the effects of varying \texttt{SIGDISCONT} (the parameter controlling the minimum allowable change in moments across a merge level). In this Appendix we provide context for the parameter choices used in our analysis and explore the potential pitfalls and uncertainties associated with the cloud identification algorithm \texttt{CPROPS}.

In assigning emission to the surviving clouds (the last step in cloud identification), \texttt{CPROPS} by default will only assign emission to a given local maximum that is uniquely associated with it. This implies that for islands decomposed into more than one cloud, some of the low-level emission will not be assigned to any clouds. While this is potentially useful for determining the locations and properties of clumps and substructure within clouds, or for detecting clouds at low resolution and/or signal-to-noise, this approach is not ideal for our goal of detecting and characterizing the full extent of the star-forming molecular ISM. Furthermore, the flux loss can be severe, up to 50\% or more in some cases, based on comparisons we conducted between runs using this default method and runs using the modified \texttt{CLUMPFIND} described below. To generate our cloud catalog, we instead apply a modified \texttt{CLUMPFIND} algorithm to assign emission to the final set of local maxima that survive step (3) of the \texttt{CPROPS} workflow (see Section~\ref{sec:clouds}). In addition to assigning emission uniquely associated with individual local maxima to those maxima, this approach also divides the watershed emission between clouds in the island according to three-dimensional boundaries extending from the merge contours, assigning all significant emission to one cloud in the island.

\texttt{CPROPS} also allows flexibility in defining how islands of emission are decomposed into clouds via the \texttt{SIGDISCONT} parameter, and the choice can affect the derived characteristics of a cloud population significantly \citep[e.g.,][]{Wong:2011ib}. As expected, increasing \texttt{SIGDISCONT} leads to an increasingly conservative cloud decomposition, resulting in fewer clouds in the final catalog. We systematically tested \texttt{SIGDISCONT} values of 0, 1, 2, 5, and 999. The first of these is equivalent to keeping every candidate cloud that passes the size and peak-to-merge threshold, while the last amounts to keeping only the islands (i.e., no decomposition). We generate full cloud catalogs for each of these five \texttt{CPROPS} runs and run our full analysis on each catalog. Table~\ref{tab:cpropstest} summarizes the results of this experiment. We find that median cloud size and mass increase very slightly with increasing \texttt{SIGDISCONT}, while median linewidth remains constant. The decreased size and mass toward small values of the parameter reflect that complexes of emission of a given size are broken up into more clouds, and thus each is smaller and slightly less massive.

The derived mass spectrum slope $\gamma$ and cutoff mass $M_0$ are both nearly monotonic functions of \texttt{SIGDISCONT}: $\gamma$ generally decreases with increasing \texttt{SIGDISCONT}, while $M_0$ increases. The decrease in $\gamma$ is only significant below \texttt{SIGDISCONT} $=2$; the slope is statistically identical above this value. The general trend likely again reflects the fact that more aggressive decomposition effectively removes clouds from high-mass bins and adds clouds to low-mass bins, leading to a steeper effective slope. Note that the formal uncertainty on $\gamma$ is relatively similar across these different \texttt{CPROPS} runs, in the 0.07 -- 0.10 range in all cases. This means that the slope at \texttt{SIGDISCONT}$ = 0$ is only marginally different from that at \texttt{SIGDISCONT}$ = 999$, given the mutual uncertainty.

The Larson size-linewidth exponent is essentially the same for each run in the experiment, to within the uncertainties, which are about 0.05-0.06. However the Pearson correlation coefficient gets gradually smaller (i.e., a worse correlation) the more aggressive the decomposition scheme is (i.e., the lower \texttt{SIGDISCONT}). This likely reflects a combination of decreased dynamic range (as large clouds are replaced with multiple smaller clouds) and over-decomposing clouds (i.e., mistakenly dividing a single cloud into sub-objects). The Larson mass-size exponent is significantly steeper for \texttt{SIGDISCONT}$ = 0$ and 1, and quite different from that for the higher \texttt{SIGDISCONT} values. The major effect here is likely again over-decomposing clouds. Again the Pearson correlation coefficient decreases as the decomposition becomes increasingly aggressive.

Ultimately, since we are interested in the global properties of entire GMCs and not the substructure within them, we select the most conservative value for \texttt{SIGDISCONT} of 5 which still separates clearly different clouds that happen to overlap in space or velocity. This choice was further motivated by inspecting the integrated intensity images and channel maps for the cubes produced by each run of \texttt{CPROPS}, in which we note that more aggressive decomposition schemes (i.e. lower \texttt{SIGDISCONT}) lead to often arbitrary decomposition, separating into multiple clouds objects that appear to be continuous by eye. We choose \texttt{SIGDISCONT}$=5$ instead of 999 because there were some cases where overlap occurs at very low contour levels, and so a conservative but reasonable value of \texttt{SIGDISCONT} was warranted.

As noted by \cite{Wong:2011ib}, the choice of a conservative value of \texttt{SIGDISCONT} preserves CO emission that is spatially correlated over large scales as single structures and thus spans a range of parameter space similar to that seen in Milky Way GMC studies \citep[e.g.,][]{Rice:2016ko}. Note that our basic conclusions remain unchanged provided we use \texttt{SIGDISCONT} $ \geq 2$ (2 is the default \texttt{CPROPS} value).

\begin{deluxetable}{c | c c | c c c | c c | c c}
\centering
\tabletypesize{\scriptsize}
\tablecolumns{10}
\tablewidth{0pt}
\tablecaption{\texttt{CPROPS} Parameter Experiments \label{tab:cpropstest}}
\tablehead{
	\colhead{\texttt{SIGDISCONT}} &
	\colhead{\# Clouds\tablenotemark{a}} &
	\colhead{Resolved cloud} &
	\colhead{Median($R$)} &
	\colhead{Median($\Delta V$)} &
	\colhead{Median($\Mlum$)} &
	\multicolumn{2}{c}{Mass spectrum} &
	\multicolumn{2}{c}{Larson exponents} \\
	\colhead{} &
	\colhead{} &
	\colhead{fraction} &
	\colhead{(pc)} &
	\colhead{({\kms})} &
	\colhead{($10^4~\Msun$)} &
	\colhead{Slope} &
	\colhead{$M_0 (\Msun)$} &
	\colhead{$R-\Delta V$} &
	\colhead{$R-\Mlum$} 
}
\startdata
0 &	367 &	0.54 &	12.4 &	3.5 &		$1.60$ &	$-1.91$ &	$3.37 \times 10^5$ &	$0.61$ &	$2.73 $ \\
1 &	352 &	0.54 &	12.6 &	3.5 &		$1.61$ &	$-1.86$ &	$3.46 \times 10^5$ &	$0.52 $ &	$2.52$ \\
2 &	286 &	0.51 &	14.3 &	3.5 &		$1.68$ &	$-1.74$ &	$6.01 \times 10^5$ &	$0.46 $ &	$2.09$ \\
5 &	253 &	0.51 &	14.4 &	3.5 &		$1.67$ &	$-1.76$ &	$8.97 \times 10^5$ &	$0.48 $ &	$2.01$ \\
999 &	233 &	0.49 &	15.0 &	3.5 &	$1.67$ &	$-1.72$ &	$10.38 \times 10^5$ &	$0.49 $ &	$1.97$
\enddata
\tablenotetext{a}{not including duplicates or clouds with central pixel beyond the primary beam FWHM}
\end{deluxetable}

\begin{figure}
$\begin{array}{cc}
\includegraphics[width=0.53\linewidth]{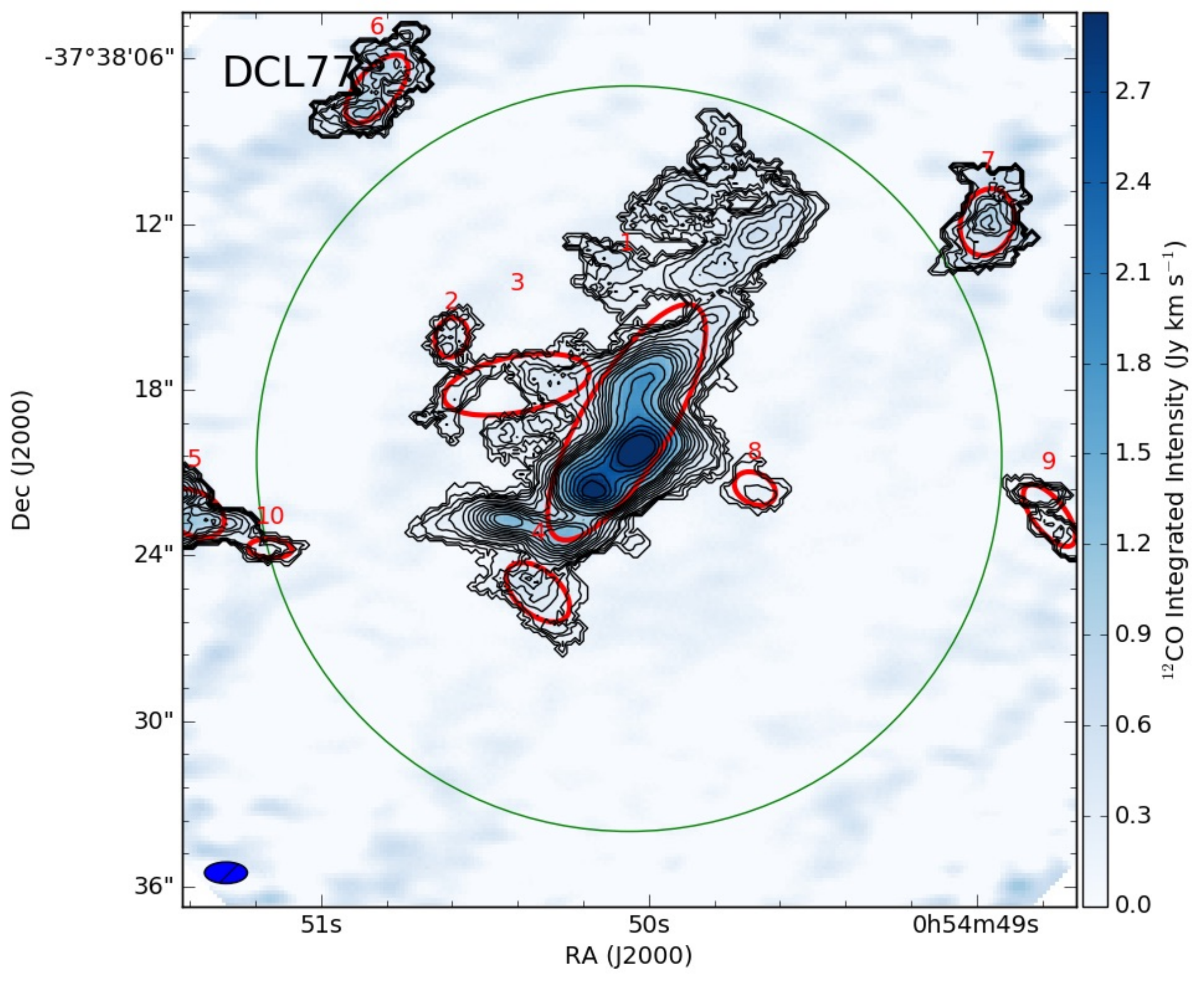} &
\includegraphics[width=0.47\linewidth]{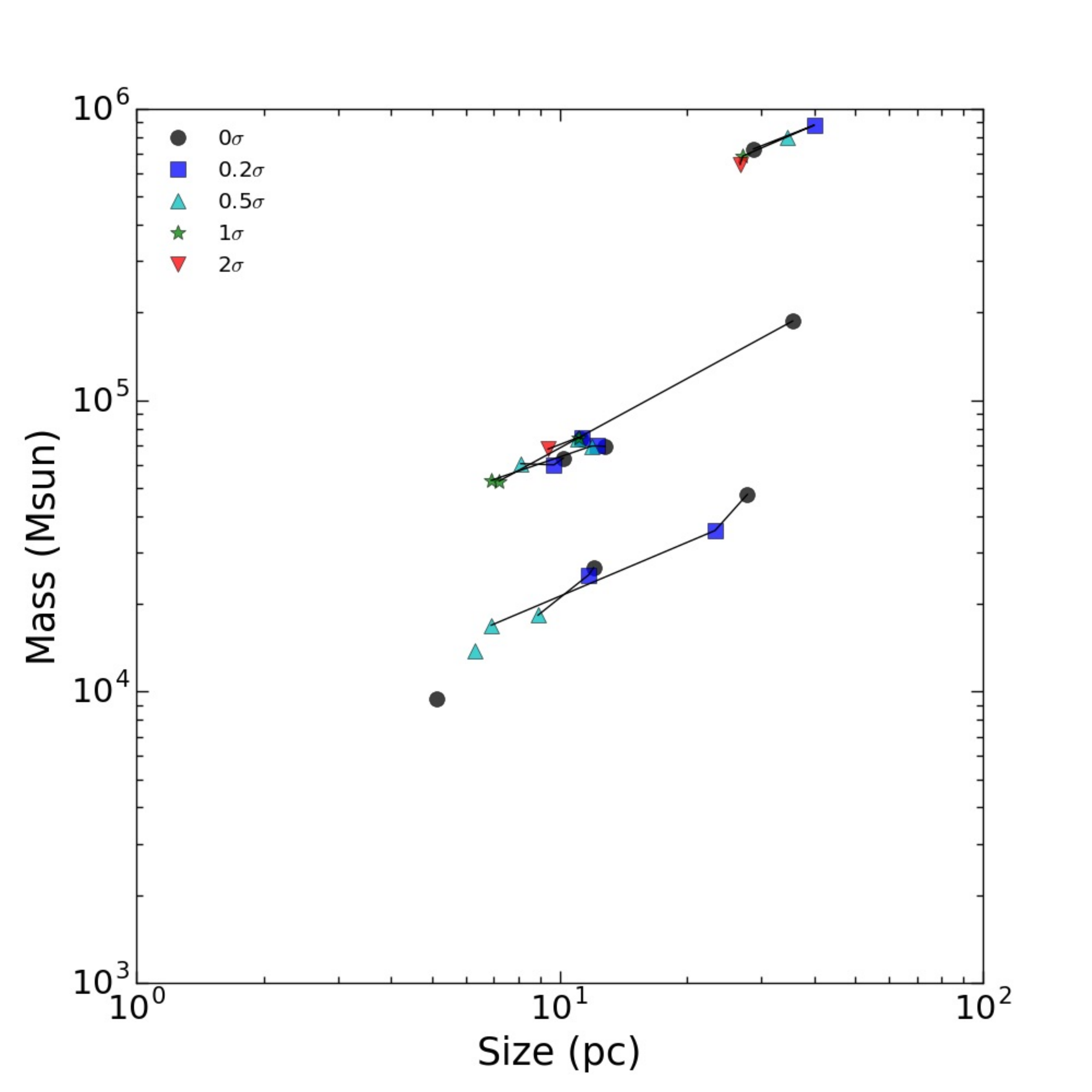} \\
\end{array}$
\caption{Left: CO(2-1) integrated intensity image of the 0.2$\sigma$ added noise realization of DCL77 in colorscale, with contours showing the outlines of the clouds detected by \texttt{CPROPS}. Red ellipses show the FWHM sizes and orientations of the clouds. The cloud labeled ``1'' in this figure contains additional emission previously assigned to a separate object. Right: mass vs. size for the added noise realizations. Each symbol-color pair refers to a realization with some amount of added noise, from 0$\sigma$ (the original cube) to $2\sigma$. Lines connect spatially matched clouds between realizations. The mass and size of matched clouds fluctuates as noise is added and cloud identification changes, but the overall statistics of the population remains similar, at least until 1$\sigma$ added noise.}
\label{fig:addnoise}
\end{figure}

Finally, to explore the effects of varying noise levels on derived cloud properties, we have conducted an additional experiment using our highest signal-to-noise data cube, that of DCL88-77. We created multiple realizations of this cube by adding pixel-level random Gaussian noise to each, corresponding to additional noise of 0.2, 0.5, 1, 2, and 5 times the signal-free RMS noise level of the original cube. We then ran \texttt{CPROPS} on each realization and compared the cloud identification and characterization across the set of runs. Table \ref{fig:addnoise} presents a summary of the results of this experiment. As noise is added, progressively fewer clouds are identified, the total mass in clouds decreases, and the mass and size of the largest GMC fluctuate slightly as pixels are differently assigned to different clouds in each realization. Changes do not become severe until 1$\sigma$ added noise (i.e., double the initial noise), at which point only half the original clouds are identified, and about 70\% of the original mass in clouds included. We argue that at noise levels below this, the cloud properties are robust in a statistical sense.

To demonstrate why the properties of the most massive cloud change, we show in the left panel of Figure~\ref{fig:addnoise} the CO integrated intensity image of the realization with 0.2$\sigma$ added noise in grayscale and the \texttt{CPROPS} cloud boundaries in contour. The red ellipses show the FWHM sizes and orientations of the identified clouds. In this realization, the cloud labeled ``1'' in the original image has now been subsumed into what was previously labeled cloud ``3'' and is in this image labeled cloud ``1'', while former cloud ``2'' has been split into clouds labeled ``2'' and ``3'' in this realization (compare with Appendix~\ref{sec:ICOimages}). The remaining cloud identification appears to hold between the two realizations. While this may appear to be a drastic change, the overall effects on the GMC masses and sizes are modest. The right panel of the Figure shows the mass and size of the GMCs identified in each realization, where each symbol-color pair represents one GMC, and lines connect matched GMCs between realizations. The masses and sizes of some clouds matched by position between realizations can either increase or decrease based on the results of the identification, and occasionally a cloud present in one realization is not present in another. However, the portion of mass-size space spanned by the set of clouds in a given realization is relatively constant. We thus conclude that while the cloud identification and decomposition procedure can lead to some uncertainty in individual cloud properties, the statistics of the population are robust to added noise, \textit{provided the noise is not more than about 50\% more than the original noise level in the data cube}. This again underscores the importance of high signal-to-noise in extragalactic observations of molecular gas.

\begin{deluxetable}{c c c c c | c c c c}
\centering
\tabletypesize{\scriptsize}
\tablecolumns{9}
\tablewidth{0pt}
\tablecaption{Effects of added noise \label{tab:addnoise}}
\tablehead{
	\colhead{$n\sigma$} &
	\colhead{RMS noise} &
	\colhead{N$_{\rm islands}$} &
	\colhead{N$_{\rm clouds}$} &
	\colhead{Total {\Mlum}\tablenotemark{a}} &
	\multicolumn{4}{c}{Most massive cloud properties} \\
	\colhead{} &
	\colhead{(K)} &
	\colhead{} &
	\colhead{} &
	\colhead{($10^5 \Msun$)} &
	\colhead{peak SNR} &
	\colhead{$R$ (pc)} &
	\colhead{$\Delta V~(\kms)$} &
	\colhead{$\Mlum~(\Msun)$}
}
\startdata
0 &	0.13 &	7 &	10 & 12.2 &	48 &	28.7 &	7.8 &		$7.3\times10^5$ \\
0.2 &	0.16 &	7 &	10 &	11.8 &	40 &	40.0 &	7.5 &		$8.8\times10^5$ \\
0.5 &	0.20 &	7 &	8 &	10.6 &	30 &	34.5 &	7.5 &		$8.0\times10^5$ \\
1 &	0.26 &	5 &	5 &	8.8 &		20 &	27.1 &	7.5 &		$6.9\times10^5$ \\
2 &	0.39 &	2 &	2 &	7.1 &		13 &	26.7 &	7.4 &		$6.5\times10^5$ \\
5 &	0.78 &	0 &	0 &	0.0 &		0 &	\nodata &	\nodata &	\nodata
\enddata
\tablenotetext{a}{in GMCs across the region}
\end{deluxetable}

\section{Deconvolved sizes at the limit of spatial resolution}
\label{sec:sizes}

As noted in Section~\ref{sec:results}, the deconvolution of the spatial beam from the measured size can result in corrected sizes that are formally much smaller than the beam size. The spatial deconvolution is defined as
\begin{equation}
\sigma_{r, \rm{dc}} = [(\sigma_{\rm maj,ex}^2 - \sigma_{\rm maj,beam}^2)^{1/2} (\sigma_{\rm min,ex}^2 - \sigma_{\rm min,beam}^2)^{1/2})]^{1/2},
\end{equation}
where $\sigma_{\rm ex}$ refers to the extrapolated spatial moments, $\sigma_{\rm beam}$ is the beam size, and ``maj'' and ``min'' refer to the major and minor axes of the cloud. Note that these generally are not the same as the beam major and minor axes, and must be determined by solving for the beam dimensions along an arbitrary angle. This procedure is performed in the version of \texttt{CPROPS} implemented here.

\begin{figure}
\plottwo{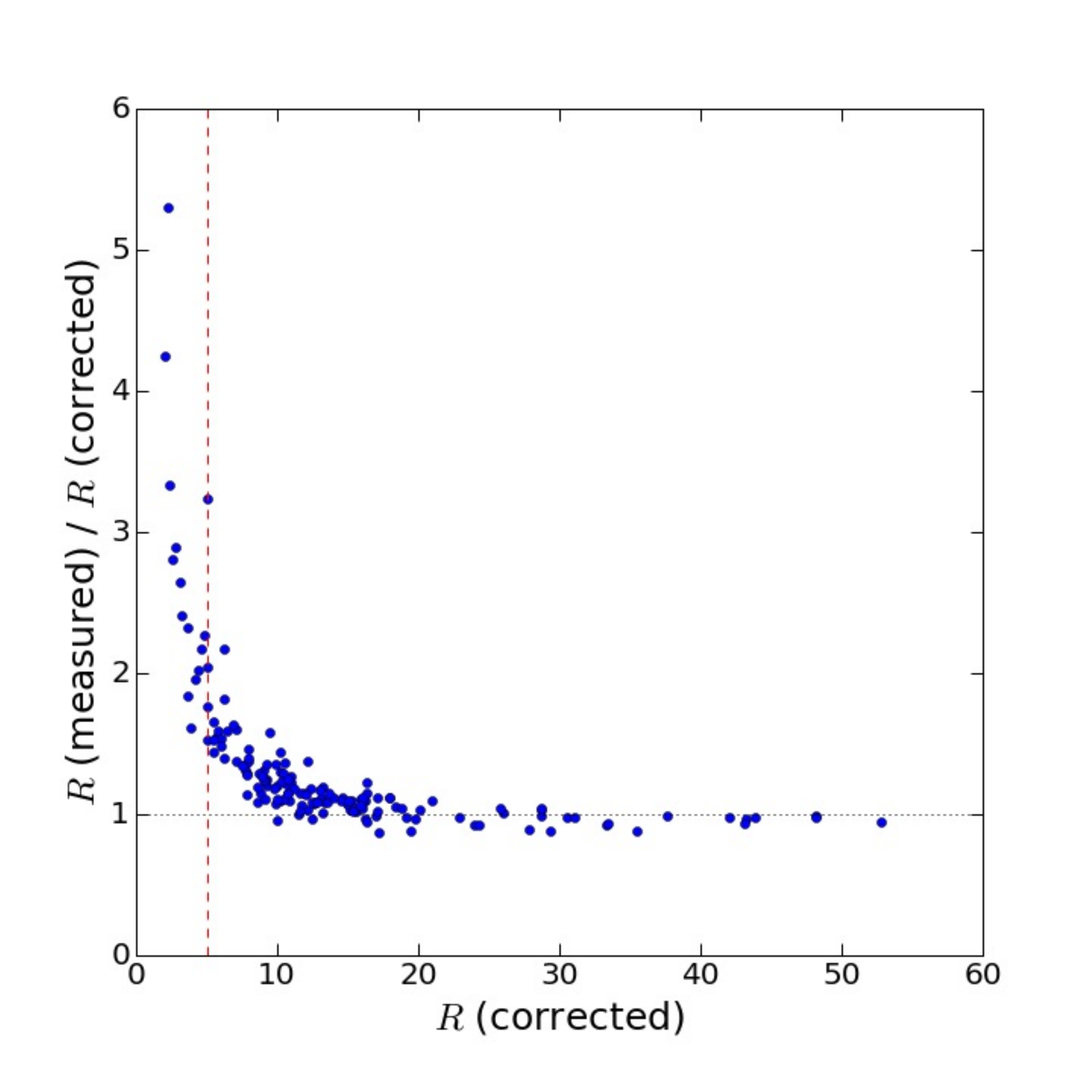}{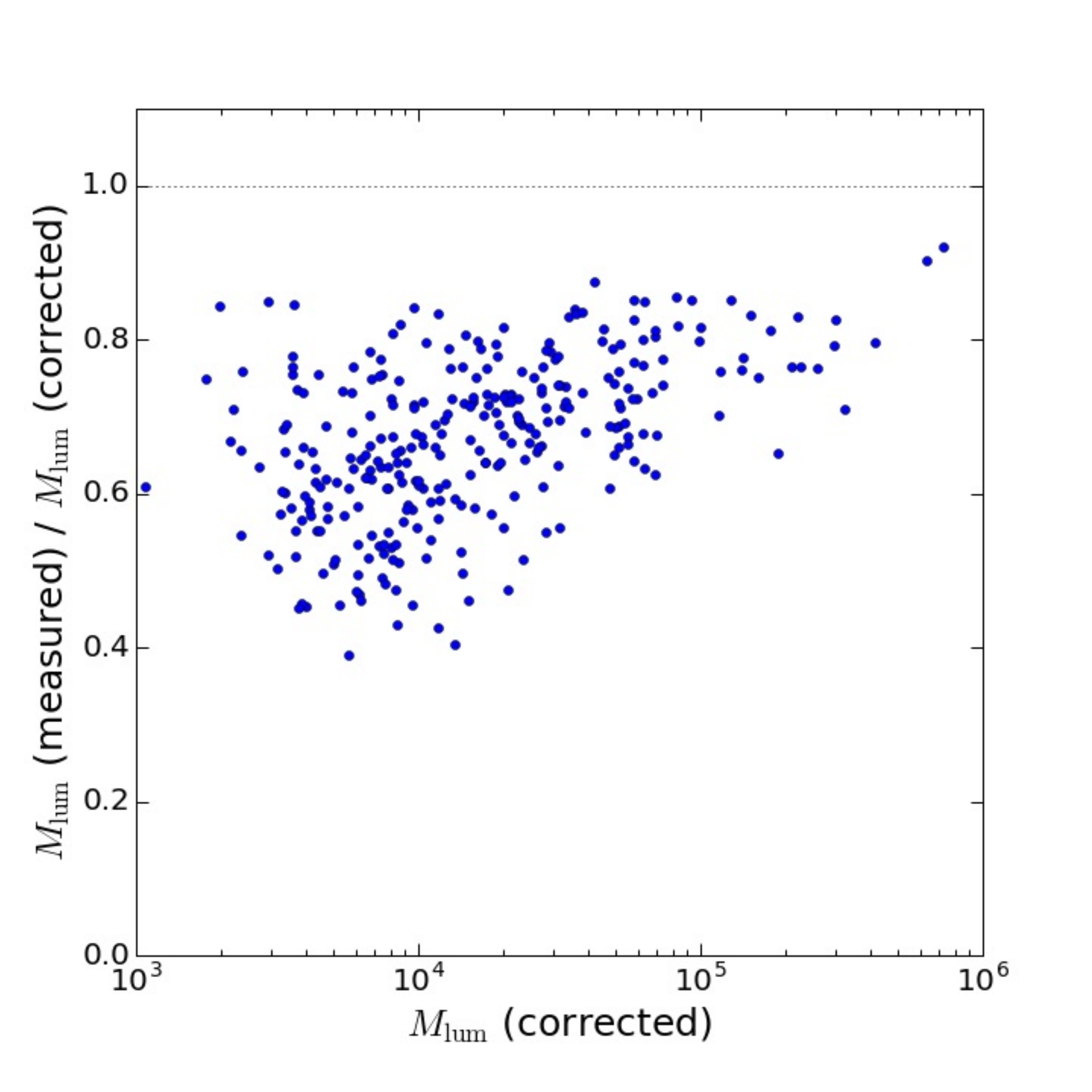}
\caption{Left: ratio of measured to corrected (i.e., deconvolved, extrapolated) size for all resolved clouds in our catalog. Due to the nonlinear nature of the spatial deconvolution of the beam, this ratio becomes significantly greater than unity below the beam resolution. The vertical red line indicates 5~pc, below which the ratio typically exceeds 2:1, and which is our chosen cutoff for inclusion in the final sample. Right: ratio of measured to extrapolated luminous mass. The median correction is 32\% and can be slightly more significant for low mass clouds, where a sizable fraction of the mass resides in the outer regions of the cloud. This correction is essential for ensuring recovery of the total cloud mass.}
\label{fig:sizecompare}
\end{figure}

For spatial moments similar to the beam size, the deconvolved spatial moment (and thus the size) can become vanishingly small. Furthermore, uncertainties in the size are not directly accounted for in the deconvolution, rendering the deconvolved size unreliable and potentially unphysical (if the extrapolated size minus uncertainty is smaller than the beam, as is often the case). In practice, these considerations simply reflect the difficulty in determining the true size of a marginally resolved object (i.e., one where the measured size is very close to the beam size). The effect is most pronounced for clouds with very small (corrected) size. We show in Figure~\ref{fig:sizecompare} the ratio of measured to corrected size plotted against the corrected size, which demonstrates that below the beam resolution, the ratio becomes progressively larger (note that the beam size is $\sim 10$~pc for our observations at the distance of NGC~300). In particular, below about 5~pc the ratio is often larger than a factor of two, or even higher. Since it is in practice not possible to fully correct for this effect, we instead discard clouds with measured sizes below 5~pc from our final sample. We still do include these clouds in the GMC Mass Spectrum (since their masses should be accurately determined) as well as the full catalog. Note that a deconvolved size smaller than the beam size is not in itself unphysical: a point source would be observed as a source with size exactly equal to the beam size, while a resolved source with observed size only slightly larger than the beam could have a deconvolved size smaller than the beam.

The effect just described is only a major issue for the size, and not the linewidth or luminous mass. In the former case, the spectral resolution (1~{\kms}) is significantly smaller than even the smallest linewidth, and so the correction factor is typically quite small (and the corrected linewidth is never smaller than the resolution). For the mass, there is no deconvolution involved, as resolution does not directly limit the ability to measure CO intensity (and thus estimate mass). The extrapolation procedure is quadratic for the mass (as opposed to linear for the size and linewidth), and can be significant: the median correction is 32\% with respect to the extrapolated mass (see the right panel of Figure~\ref{fig:sizecompare}). It is particularly important for low-mass clouds, in which a large fraction of the flux can reside near cloud boundaries. RL06 showed that with signal-to-noise greater than about four, this correction allows the recovery of essentially all flux in model clouds.

\section{GMC moment 0 images}
\label{sec:ICOimages}

In this appendix, we present CO(2-1) moment-0 (integrated intensity) images of all 48 regions we observed with ALMA. Each panel shows a different region, with the name indicated in the upper left. The colorscale shows the full CO cube integrated over the full range of velocities in which GMCs have at least one voxel, or over 20~{\kms} centered on the \ion{H}{1} velocity in the case of regions with no detected clouds (DCL6, DCL115). Contours show the \texttt{CPROPS} cloud boundaries and the integrated intensity within the GMCs. The red ellipses and numbers indicate the individual GMCs in each region. The ellipse sizes and orientations reflect the observed major and minor axis lengths and position angles for each cloud. The green circle on each panel represents the $\sim26\arcsec$ (FWHM) primary beam of ALMA (we do not include clouds more than 10\% beyond the primary beam FWHM in any of our analysis).

\clearpage
\begin{figure*}

\begin{minipage}{0.50\linewidth}
\includegraphics[width=\linewidth]{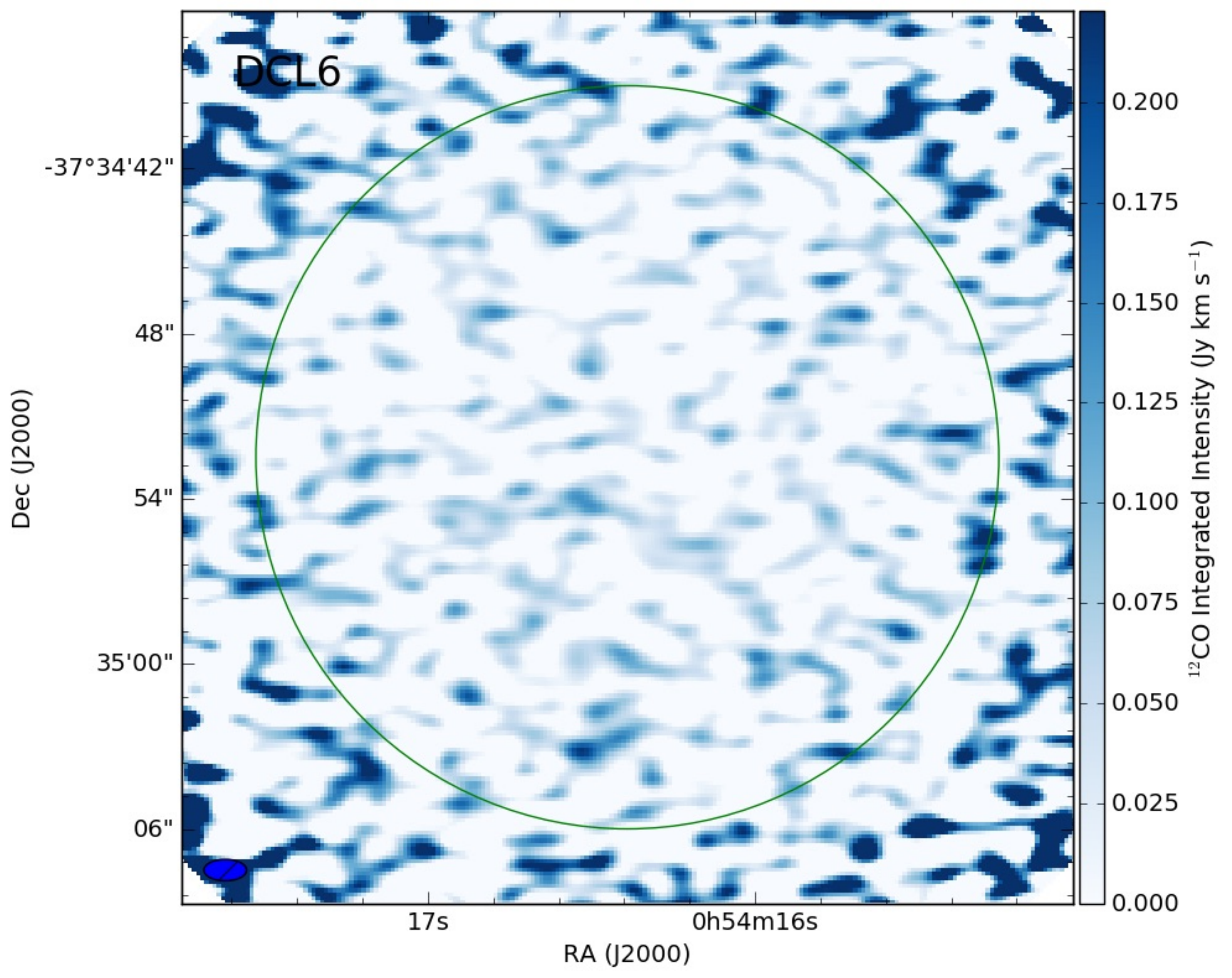}
\end{minipage}
\begin{minipage}{0.50\linewidth}
\includegraphics[width=\linewidth]{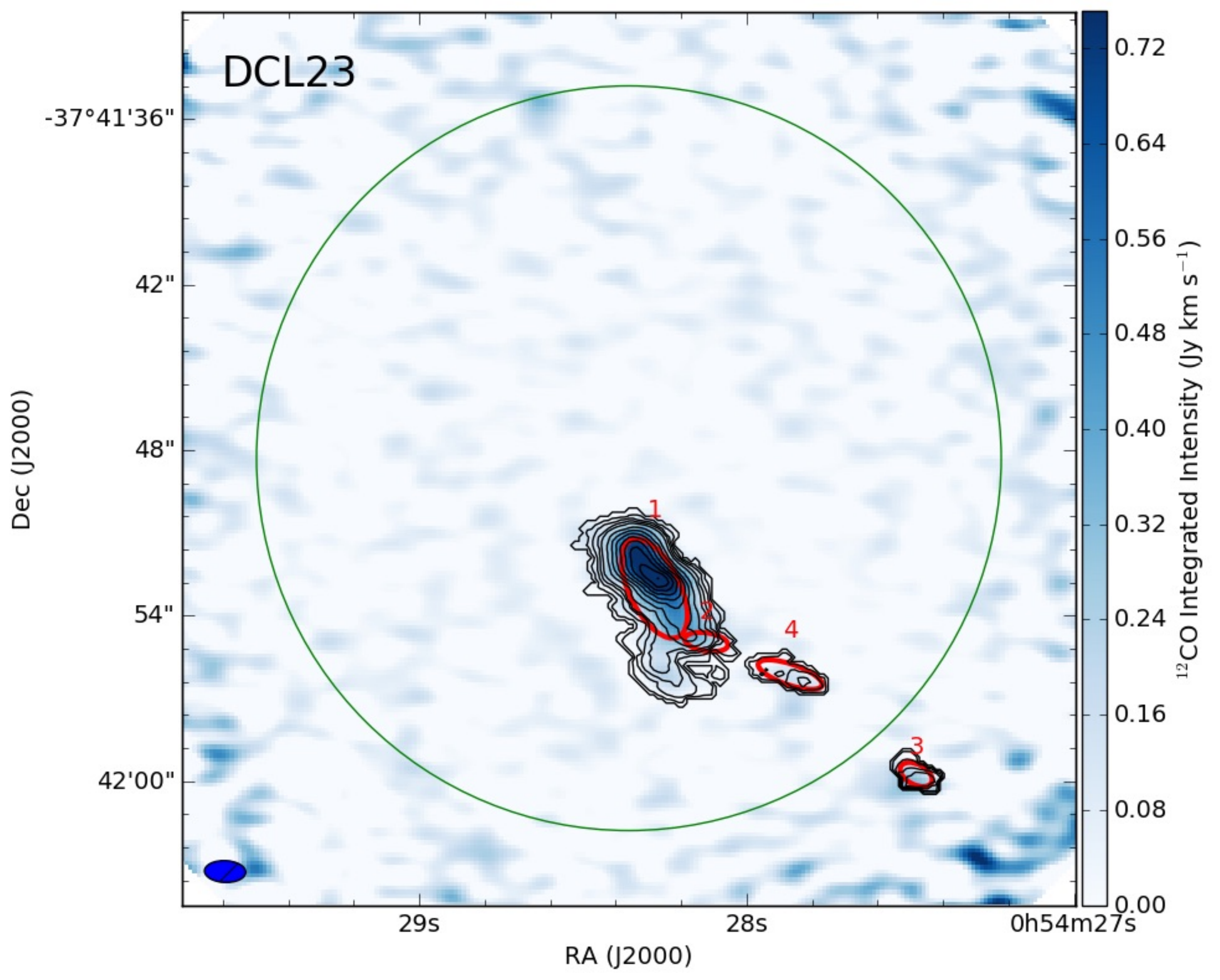}
\end{minipage}

\begin{minipage}{0.50\linewidth}
\includegraphics[width=\linewidth]{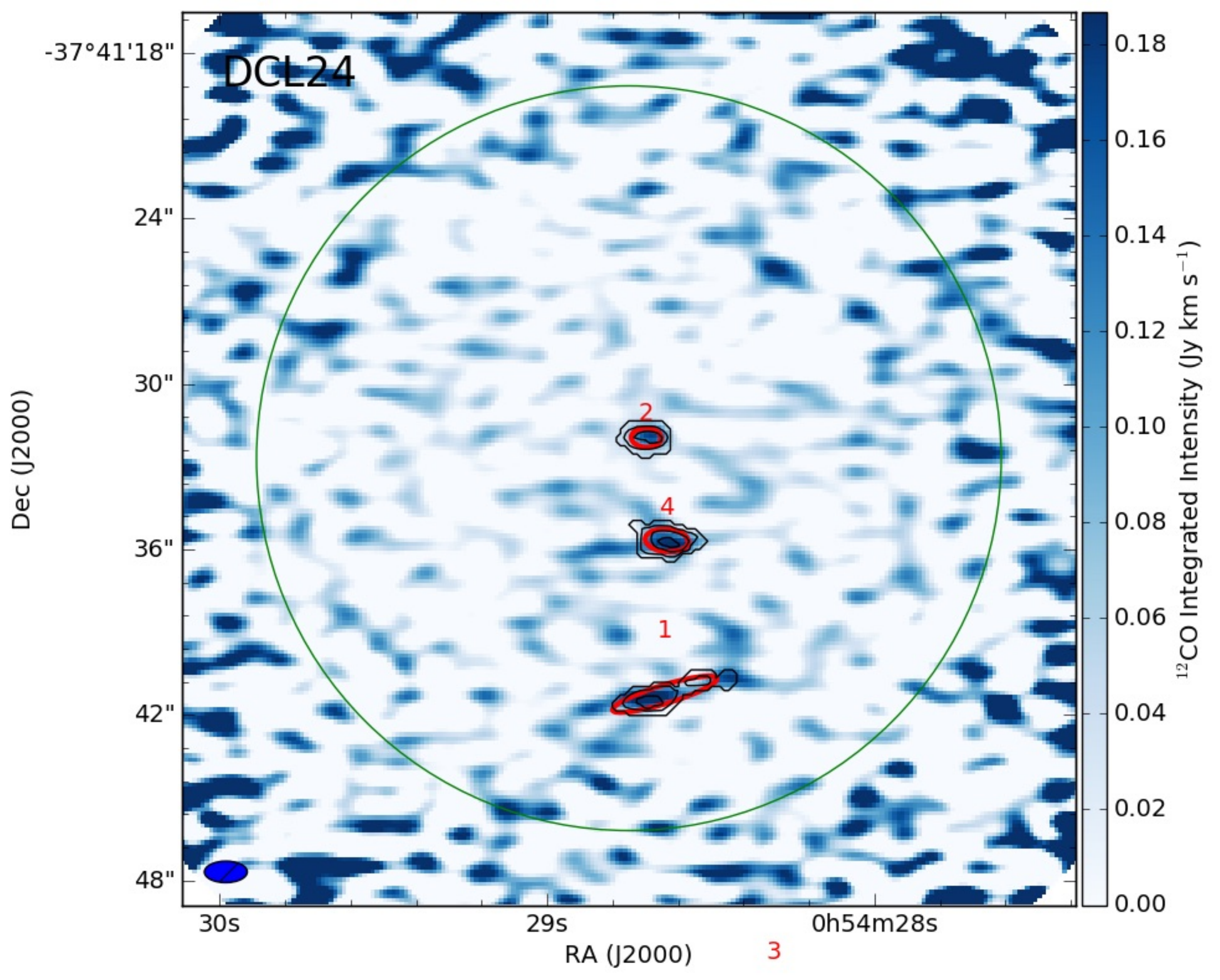}
\end{minipage}
\begin{minipage}{0.50\linewidth}
\includegraphics[width=\linewidth]{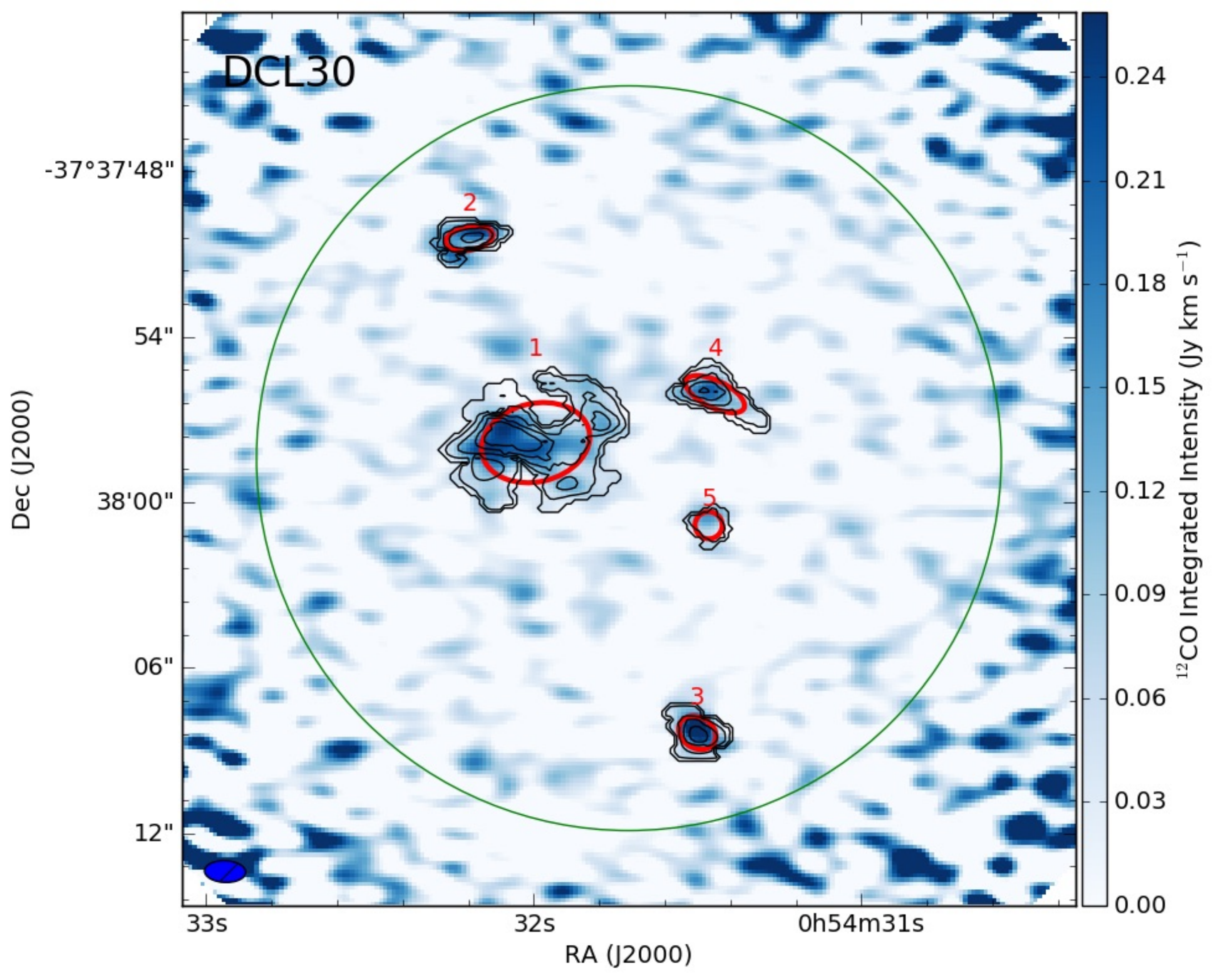}
\end{minipage}

\begin{minipage}{0.50\linewidth}
\includegraphics[width=\linewidth]{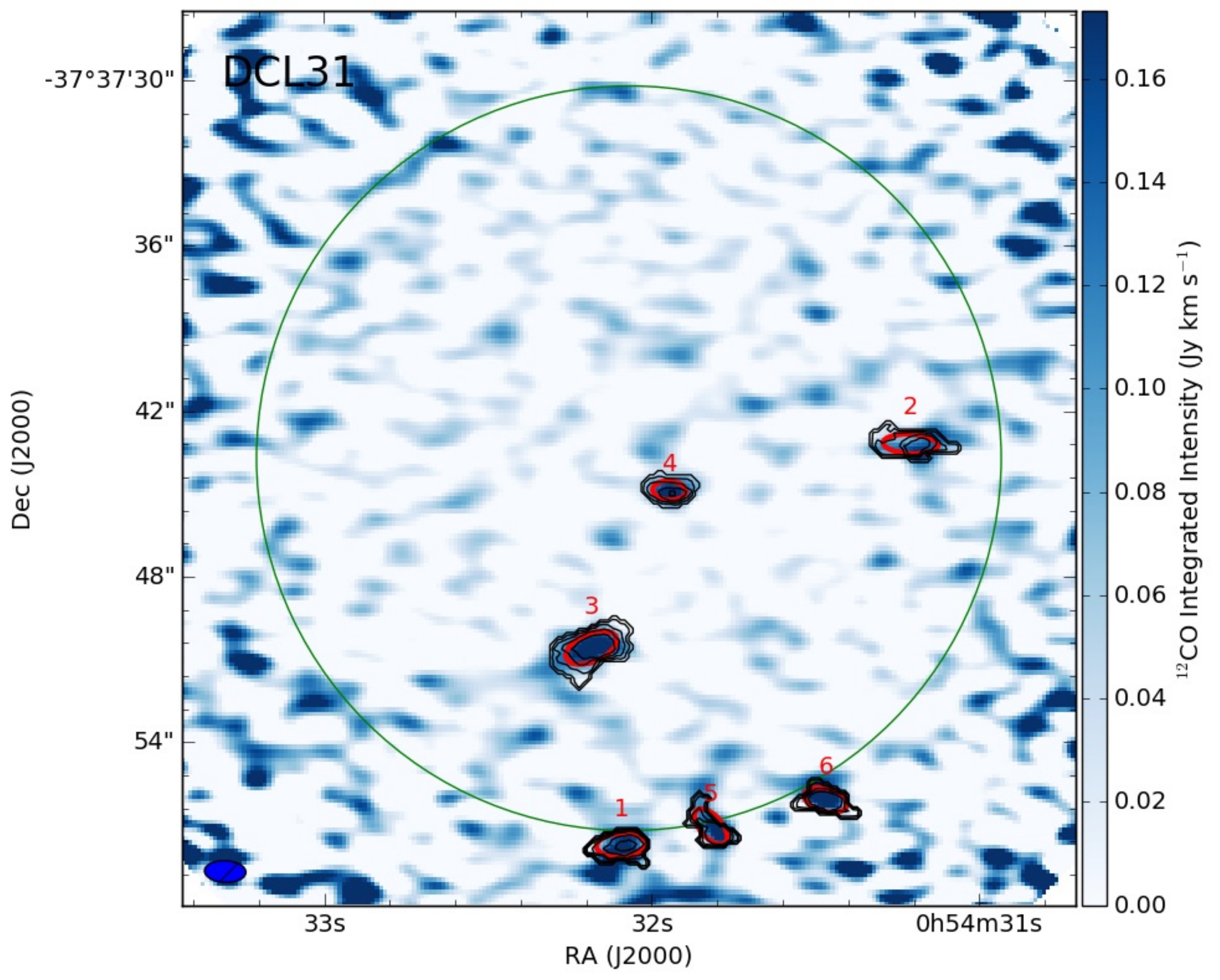}
\end{minipage}
\begin{minipage}{0.50\linewidth}
\includegraphics[width=\linewidth]{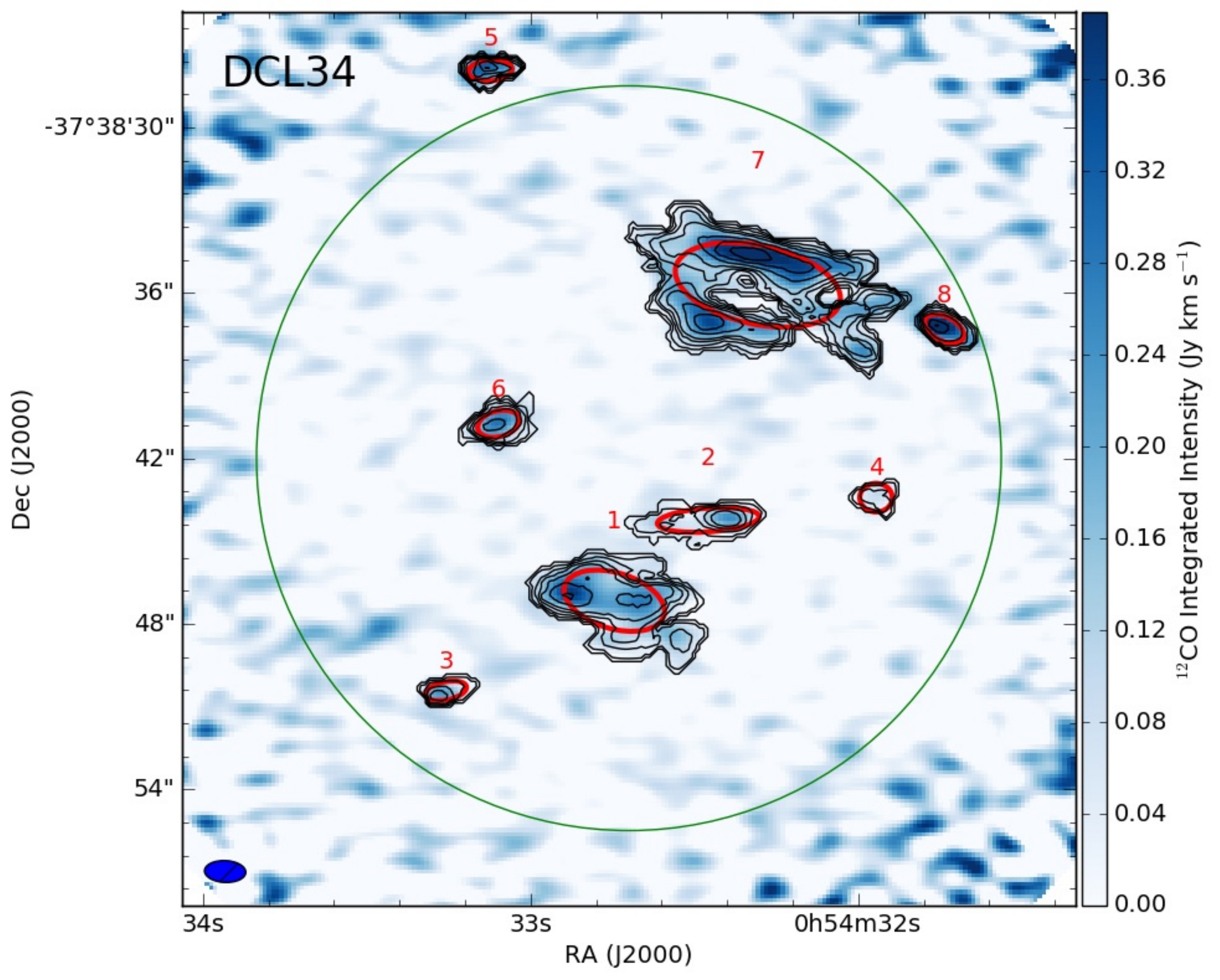}
\end{minipage}

\caption{ALMA {\CO} integrated intensity maps of the 48 regions observed with ALMA in grayscale, with \texttt{CPROPS}-identified GMCs overplotted as black contours. Contour levels are in integer multiples of the integrated intensity RMS noise beginning at $1\sigma$. The colorscale shows the integrated intensity of the full primary beam-corrected data cube over the  velocity range containing GMCs, in linear stretch from 0 Jy~{\kms}~to 80\% the image maximum. Red ellipses show the FWHM sizes and orientations of all \texttt{CPROPS} clouds, and numbers refer to the cloud designations listed in Table~\ref{tab:gmcprops}. The synthesized beam in each image is indicated by the ellipse in the lower left. The large green circle in each panel indicates the APEX $27\arcsec$ ($\sim250$~pc) FWHM pointing from \cite{Faesi:2014ib}.}
\label{fig:ICOimages}
\end{figure*}

\clearpage
\begin{figure*}

\begin{minipage}{0.50\linewidth}
\includegraphics[width=\linewidth]{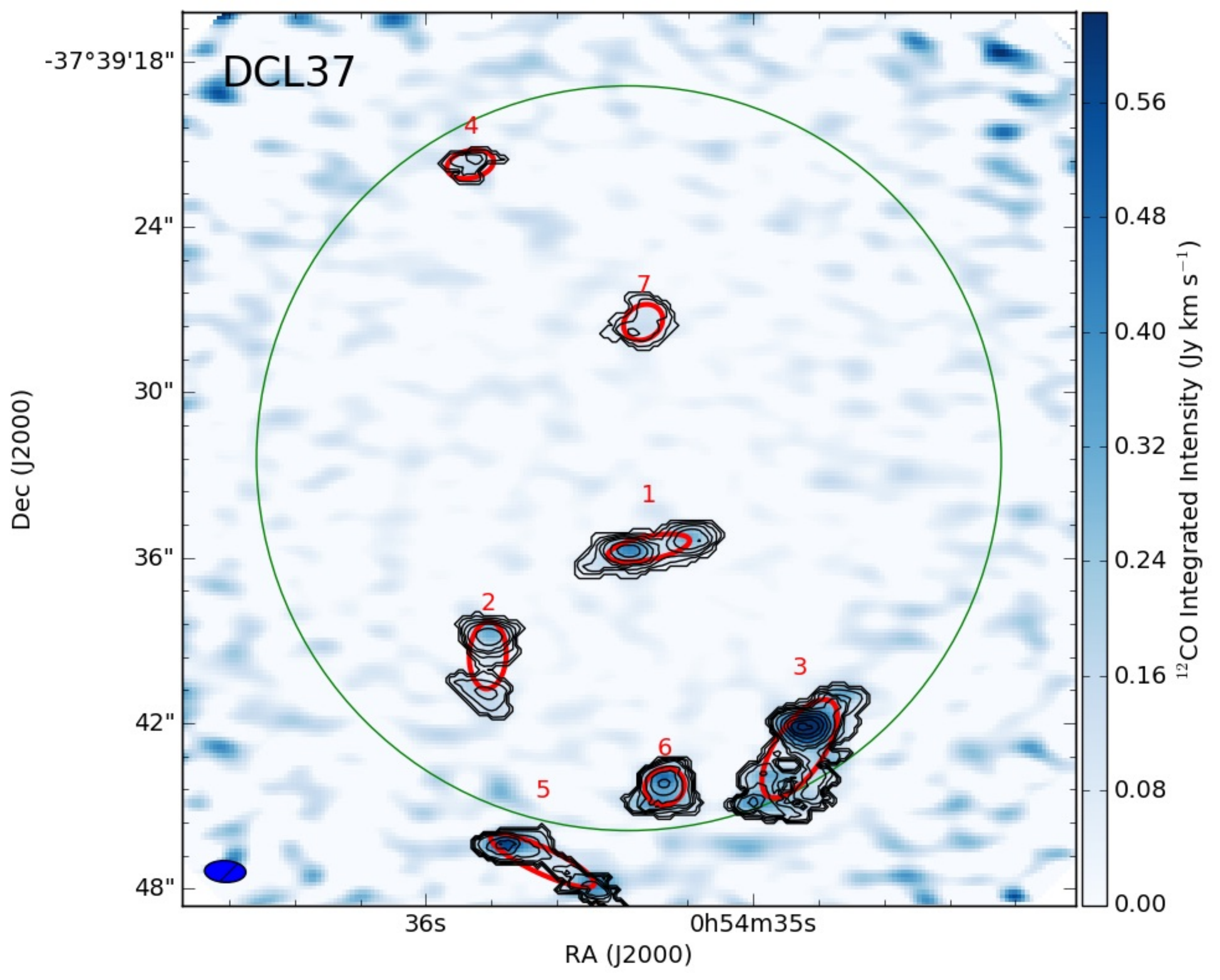}
\end{minipage}
\begin{minipage}{0.50\linewidth}
\includegraphics[width=\linewidth]{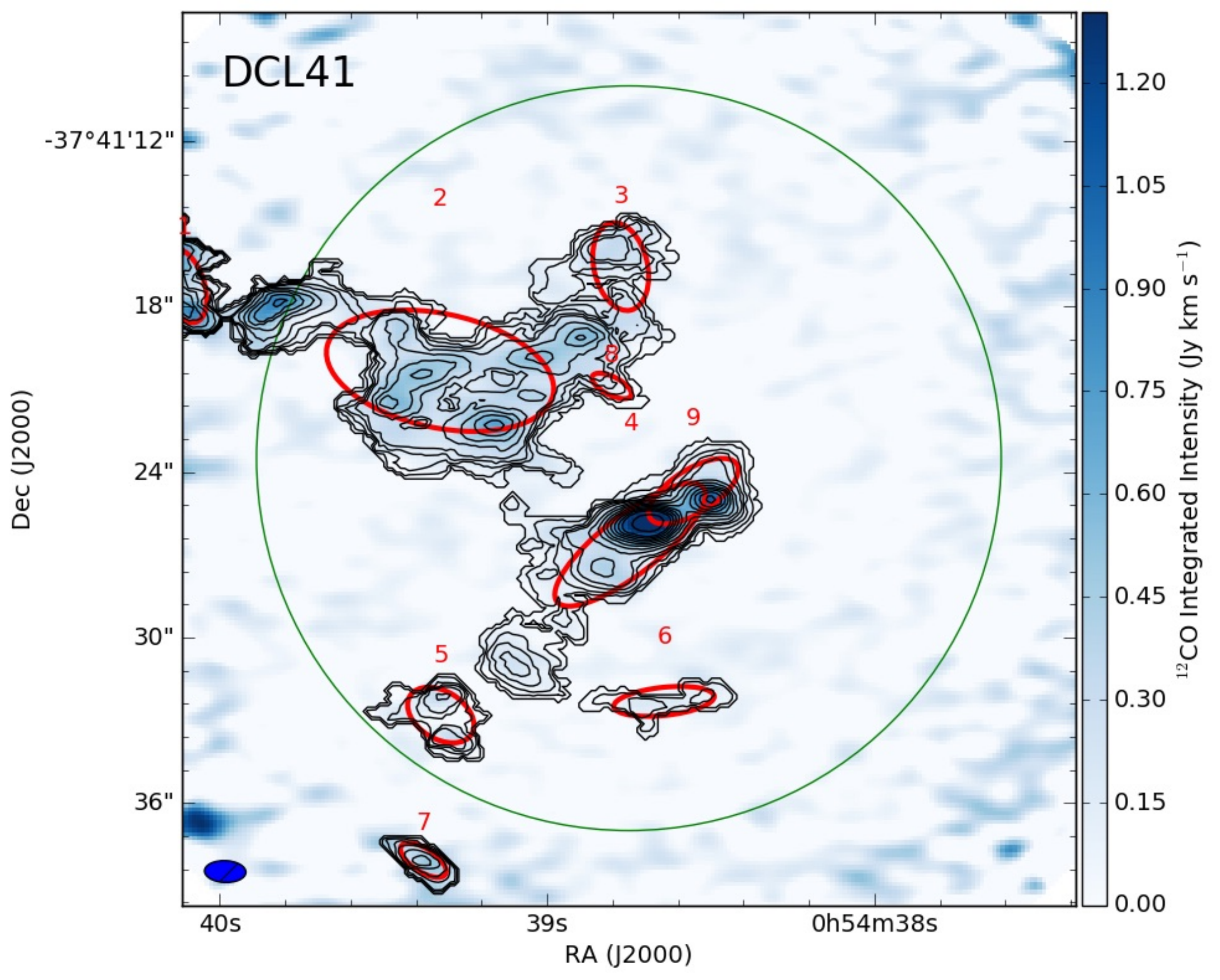}
\end{minipage}

\begin{minipage}{0.50\linewidth}
\includegraphics[width=\linewidth]{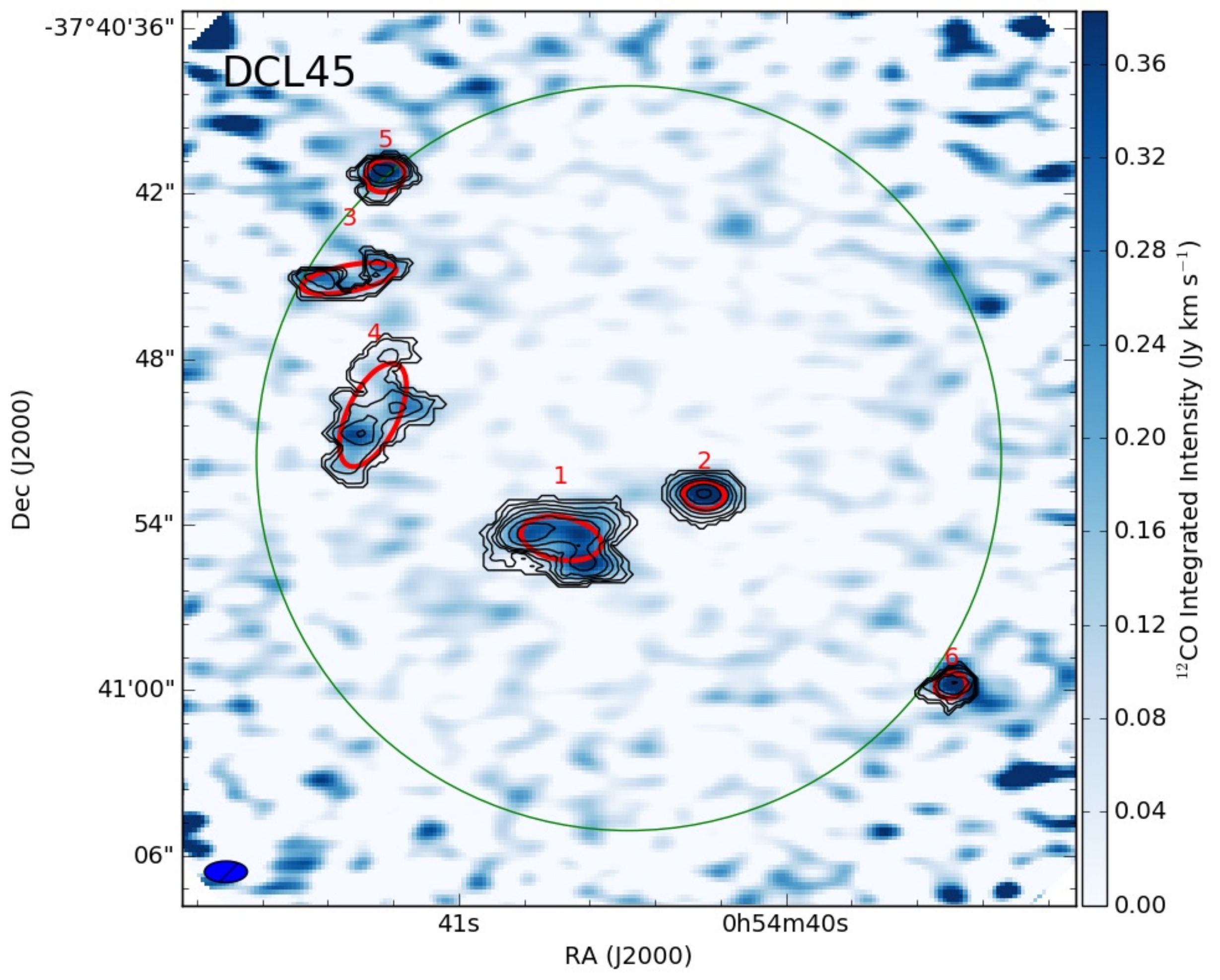}
\end{minipage}
\begin{minipage}{0.50\linewidth}
\includegraphics[width=\linewidth]{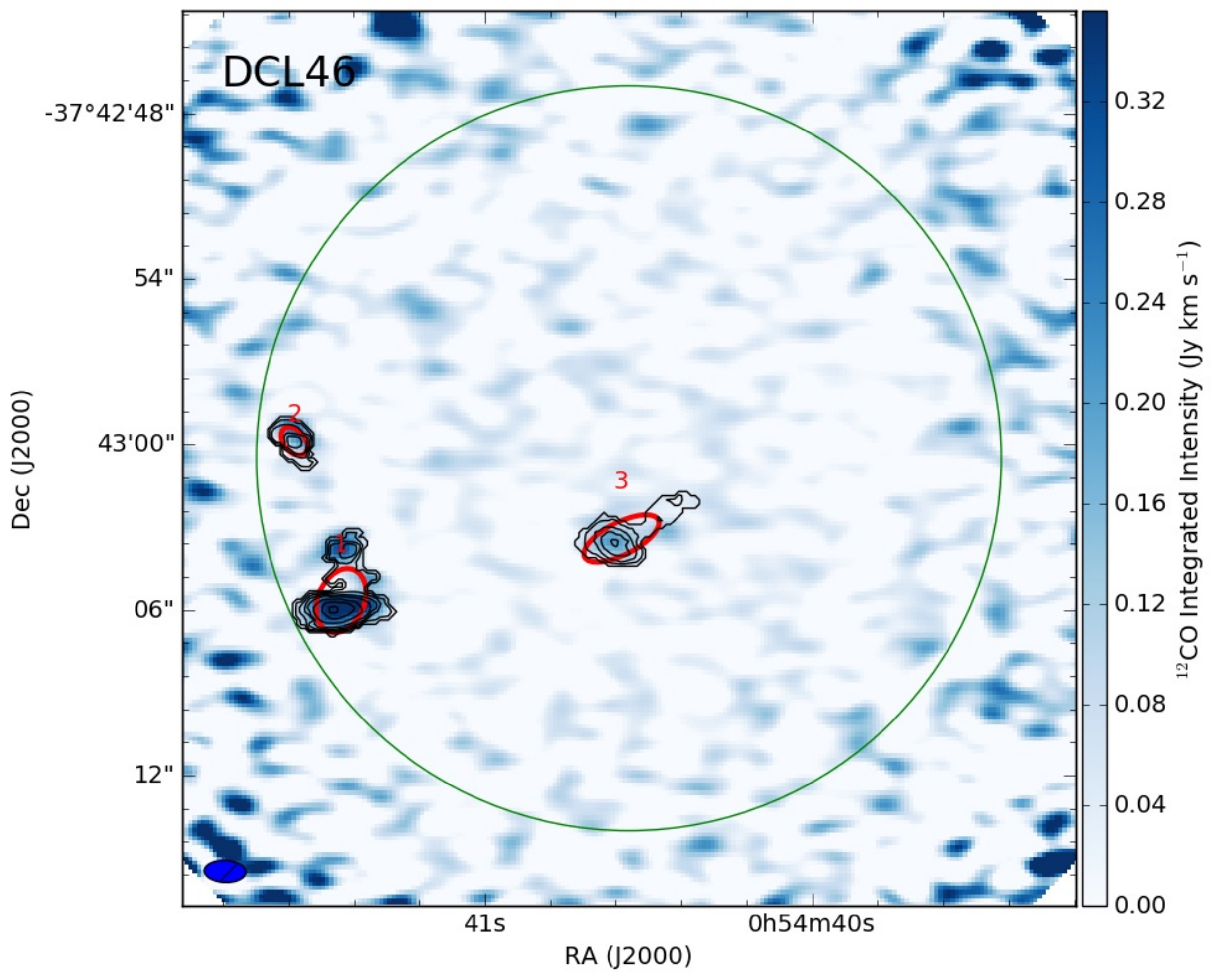}
\end{minipage}

\begin{minipage}{0.50\linewidth}
\includegraphics[width=\linewidth]{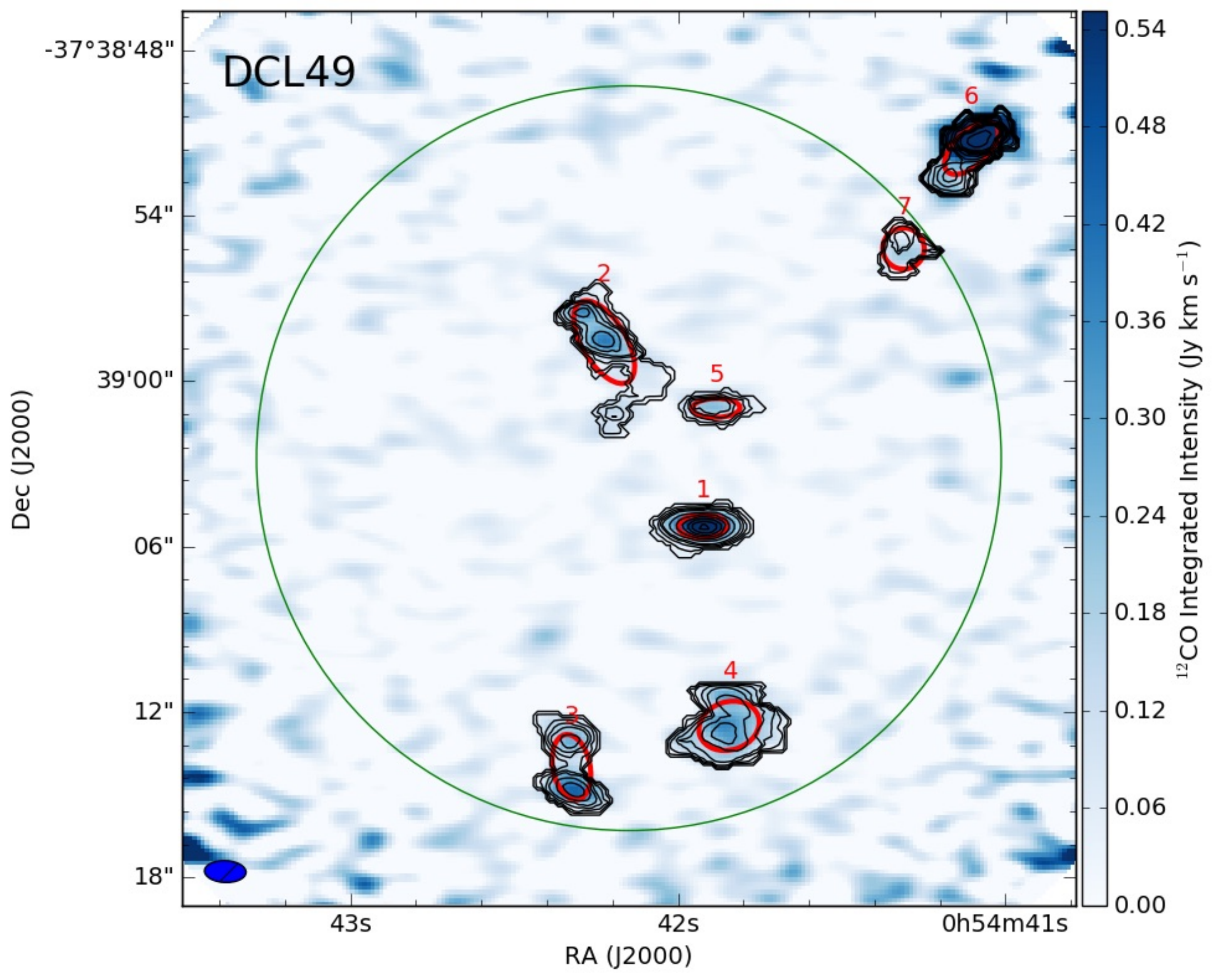}
\end{minipage}
\begin{minipage}{0.50\linewidth}
\includegraphics[width=\linewidth]{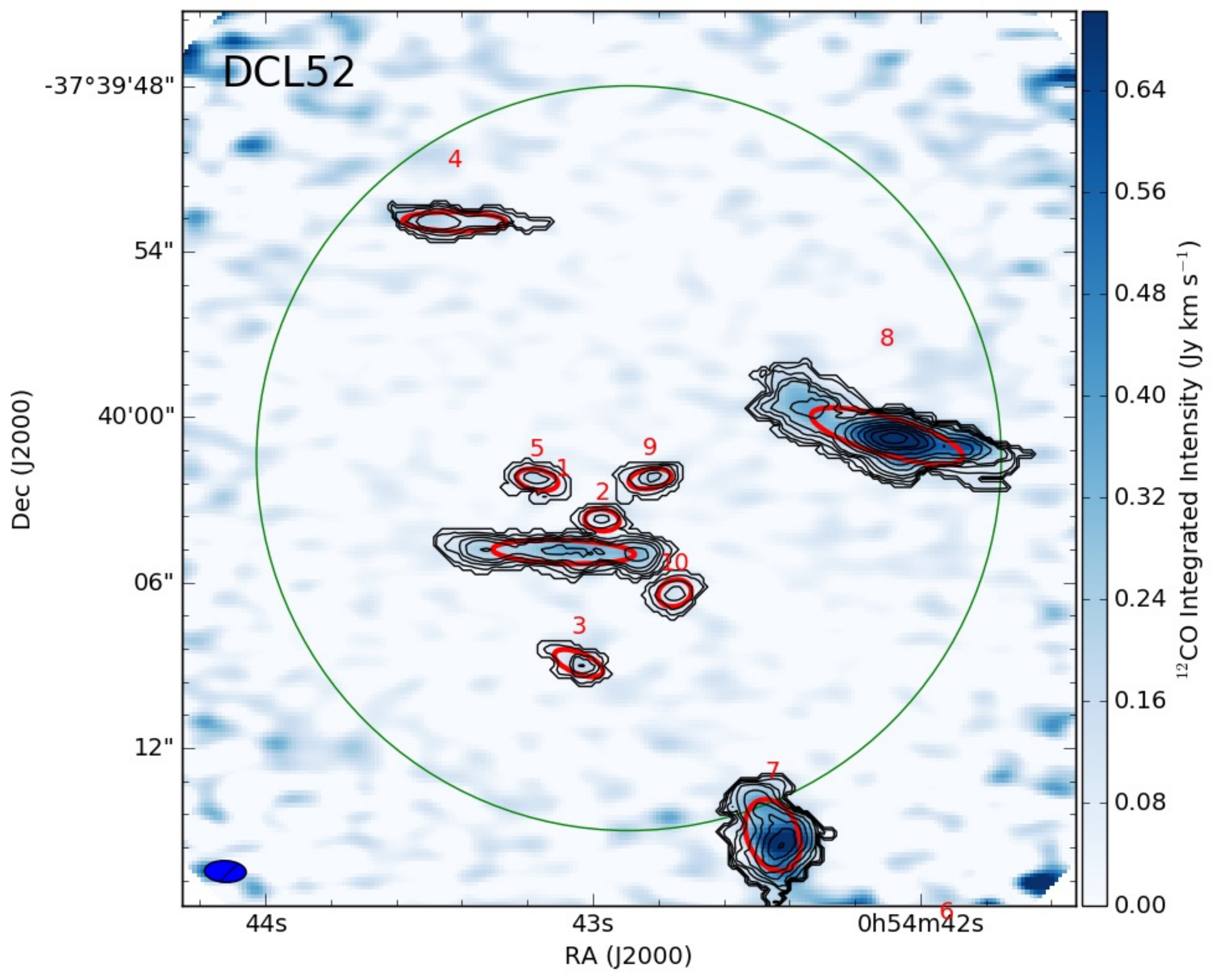}
\end{minipage}

\caption{CO integrated intensity images, continued.}
\end{figure*}

\clearpage
\begin{figure*}

\begin{minipage}{0.50\linewidth}
\includegraphics[width=\linewidth]{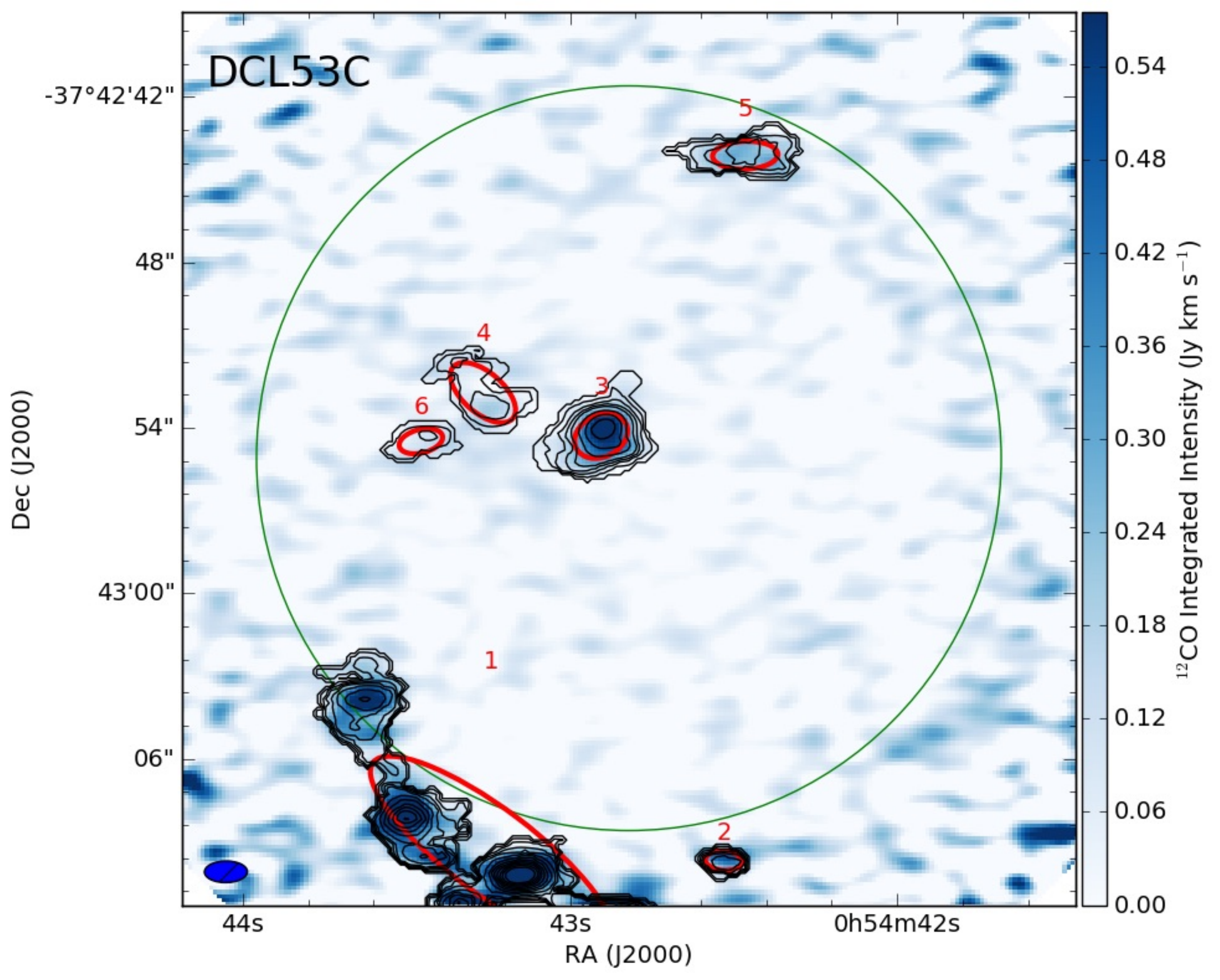}
\end{minipage}
\begin{minipage}{0.50\linewidth}
\includegraphics[width=\linewidth]{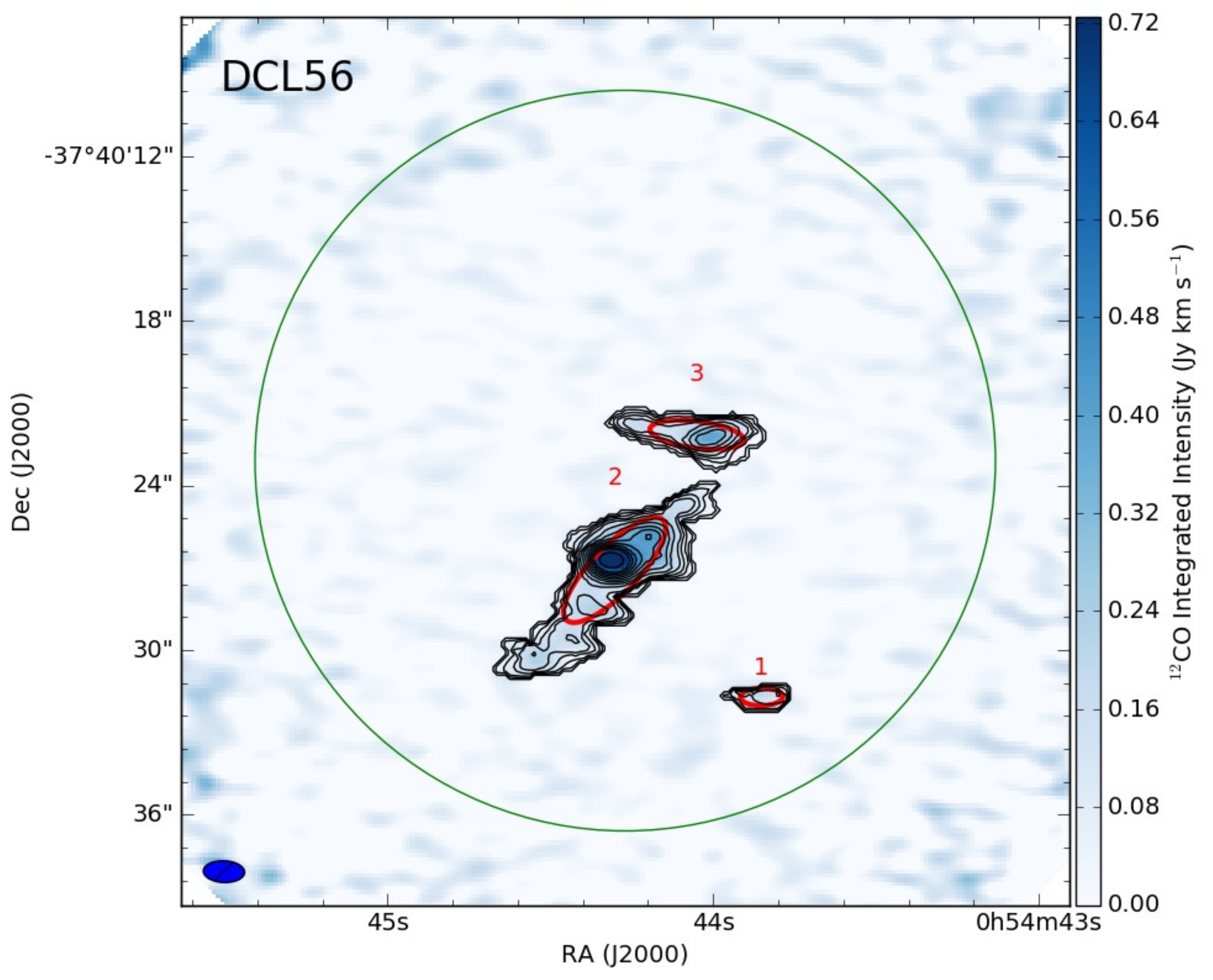}
\end{minipage}

\begin{minipage}{0.50\linewidth}
\includegraphics[width=\linewidth]{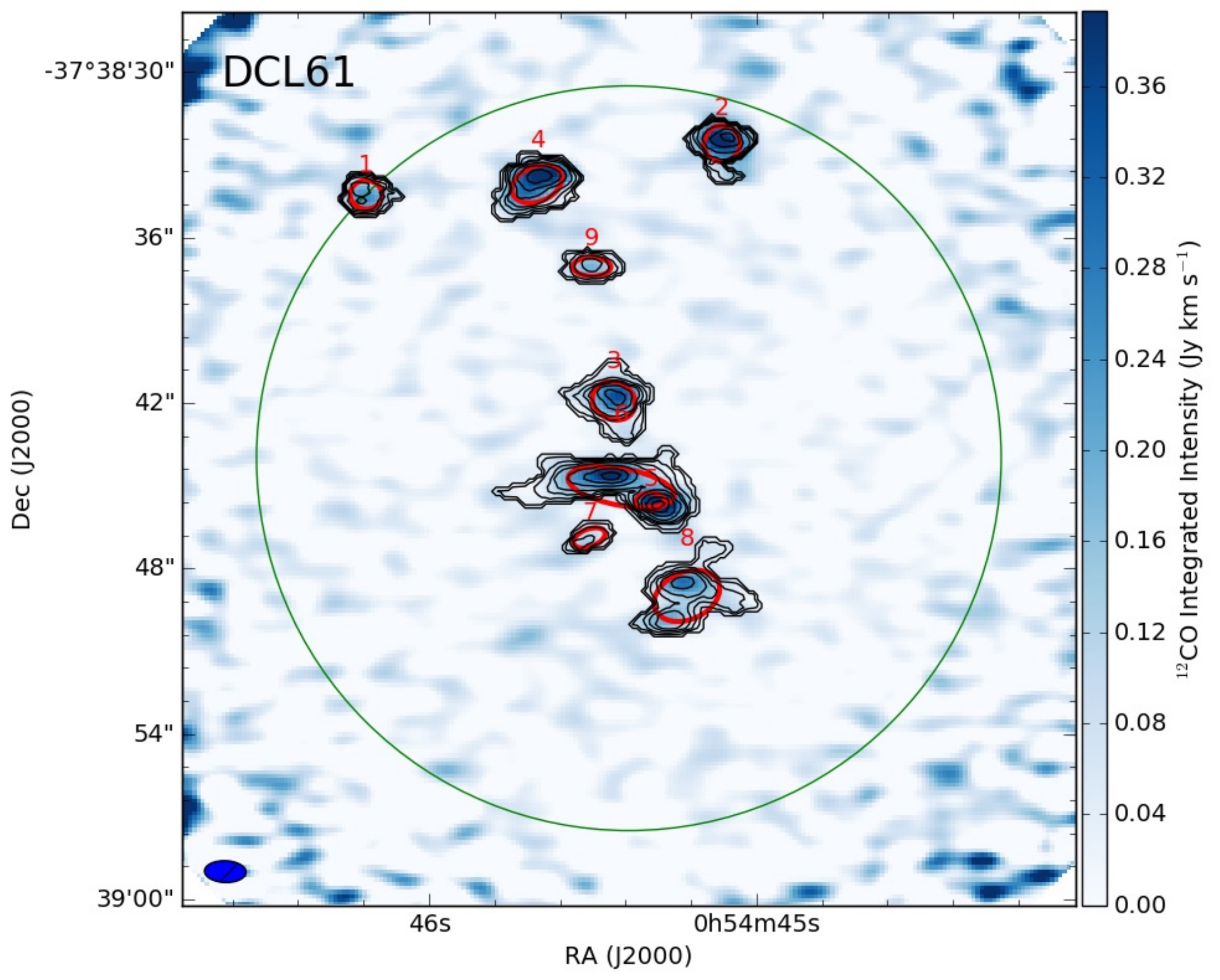}
\end{minipage}
\begin{minipage}{0.50\linewidth}
\includegraphics[width=\linewidth]{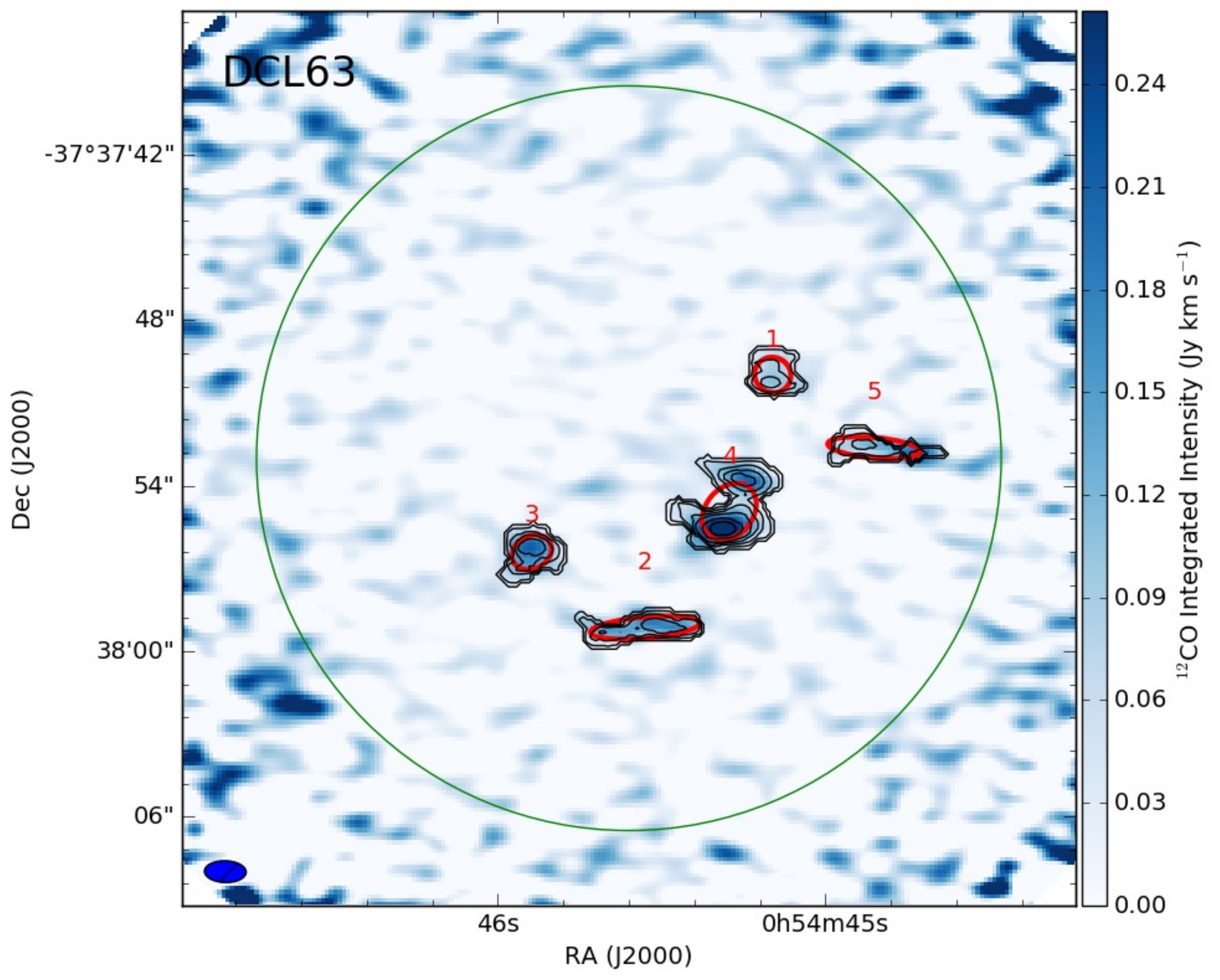}
\end{minipage}

\begin{minipage}{0.50\linewidth}
\includegraphics[width=\linewidth]{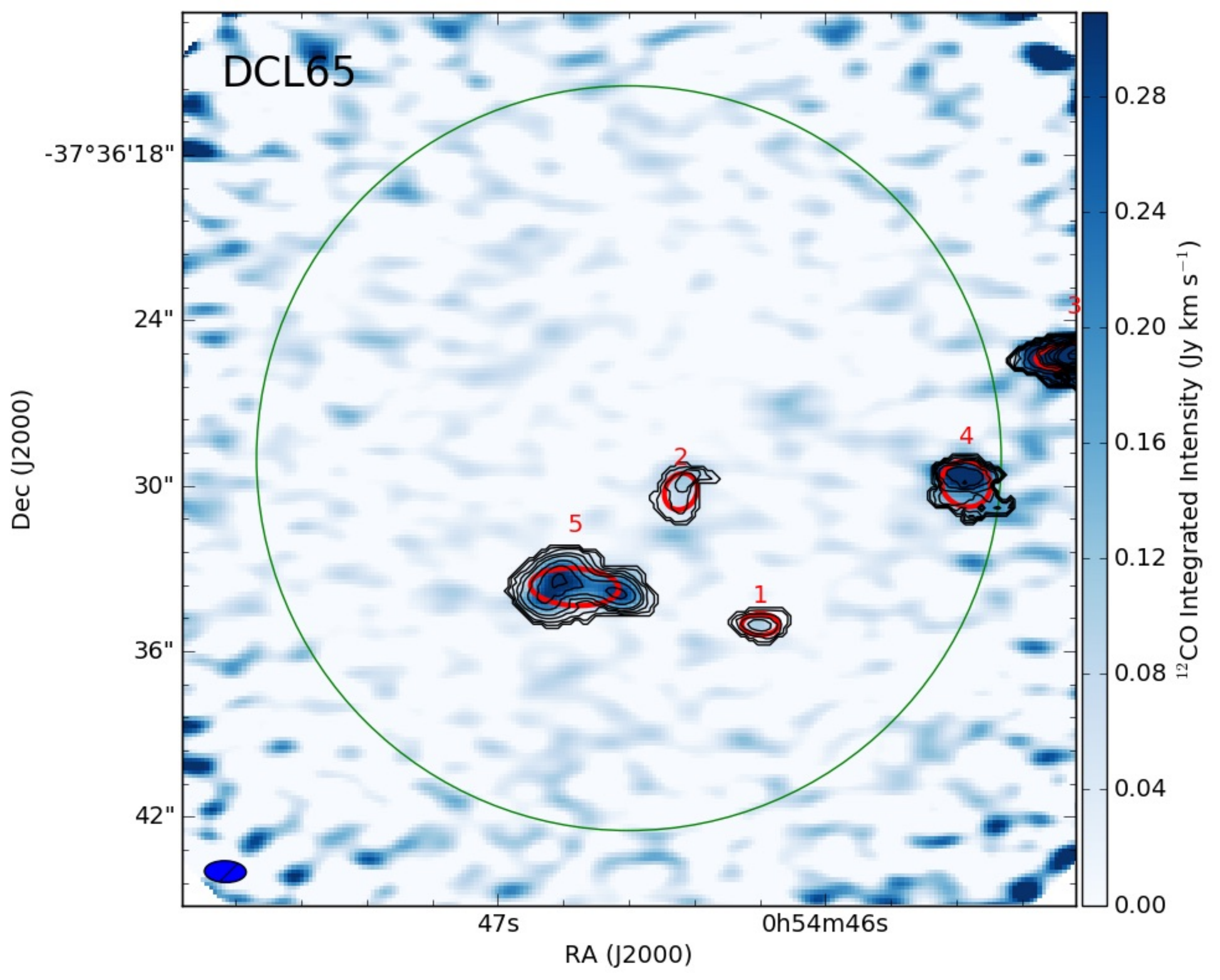}
\end{minipage}
\begin{minipage}{0.50\linewidth}
\includegraphics[width=\linewidth]{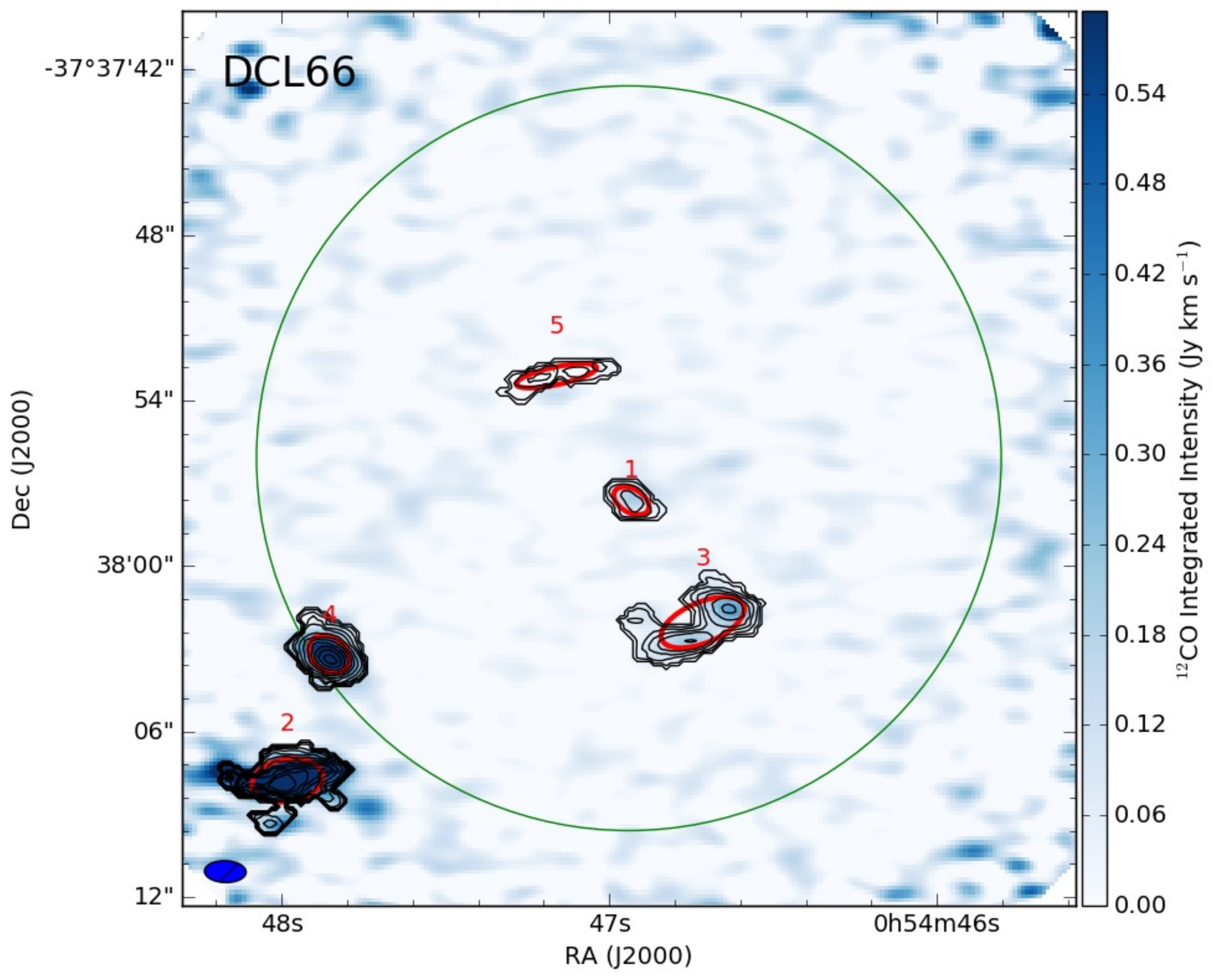}
\end{minipage}

\caption{CO integrated intensity images, continued.}
\end{figure*}

\clearpage
\begin{figure*}

\begin{minipage}{0.50\linewidth}
\includegraphics[width=\linewidth]{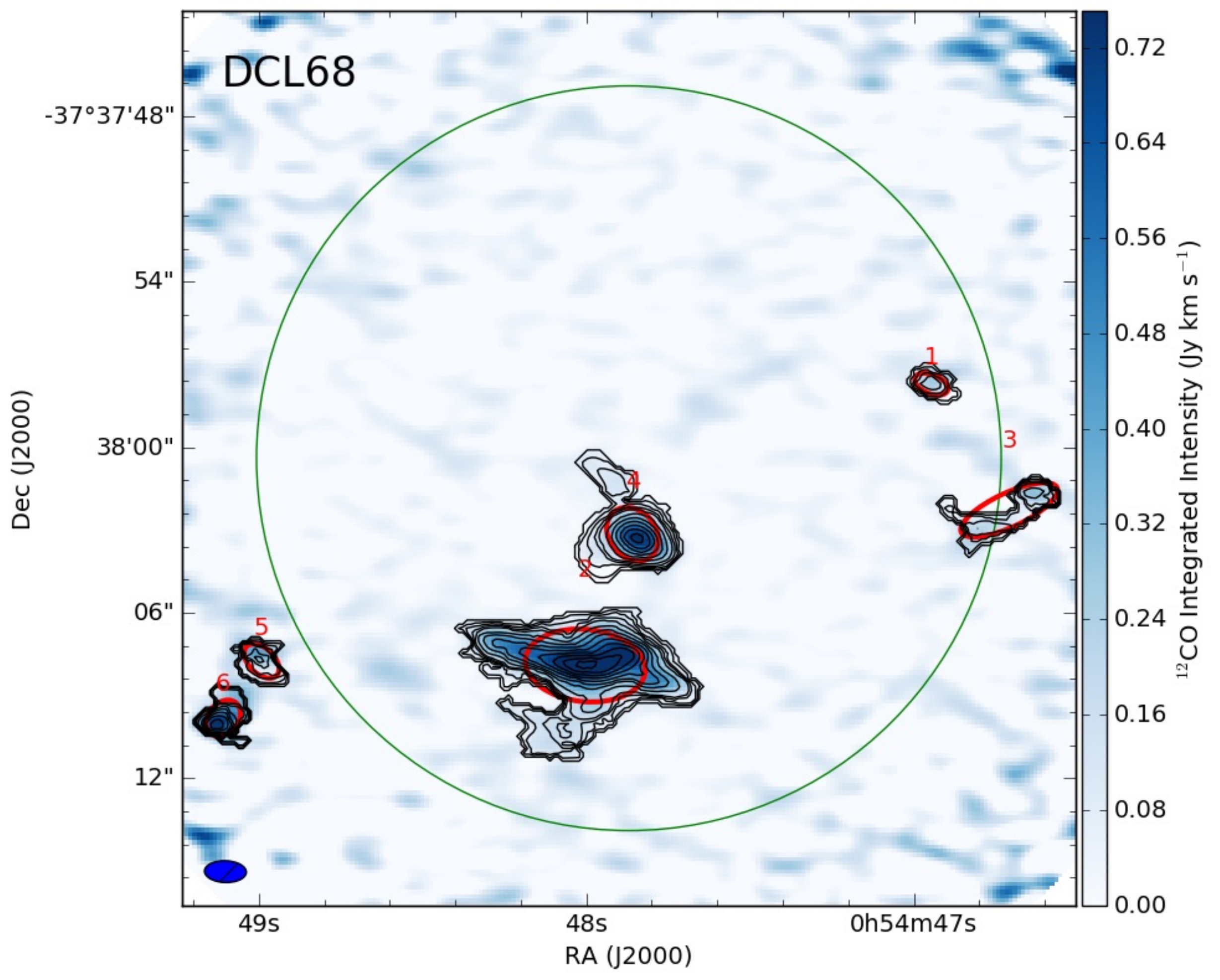}
\end{minipage}
\begin{minipage}{0.50\linewidth}
\includegraphics[width=\linewidth]{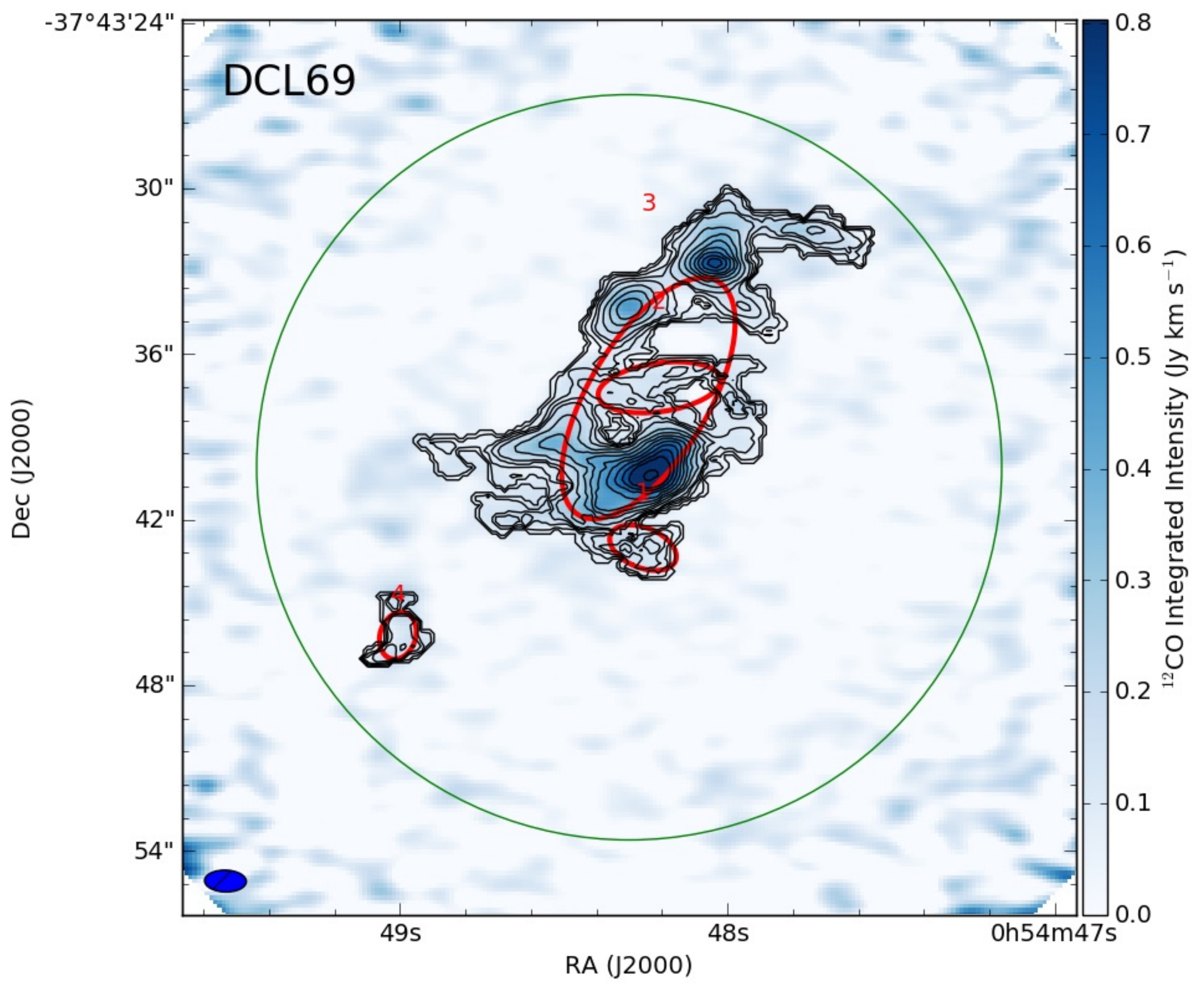}
\end{minipage}

\begin{minipage}{0.50\linewidth}
\includegraphics[width=\linewidth]{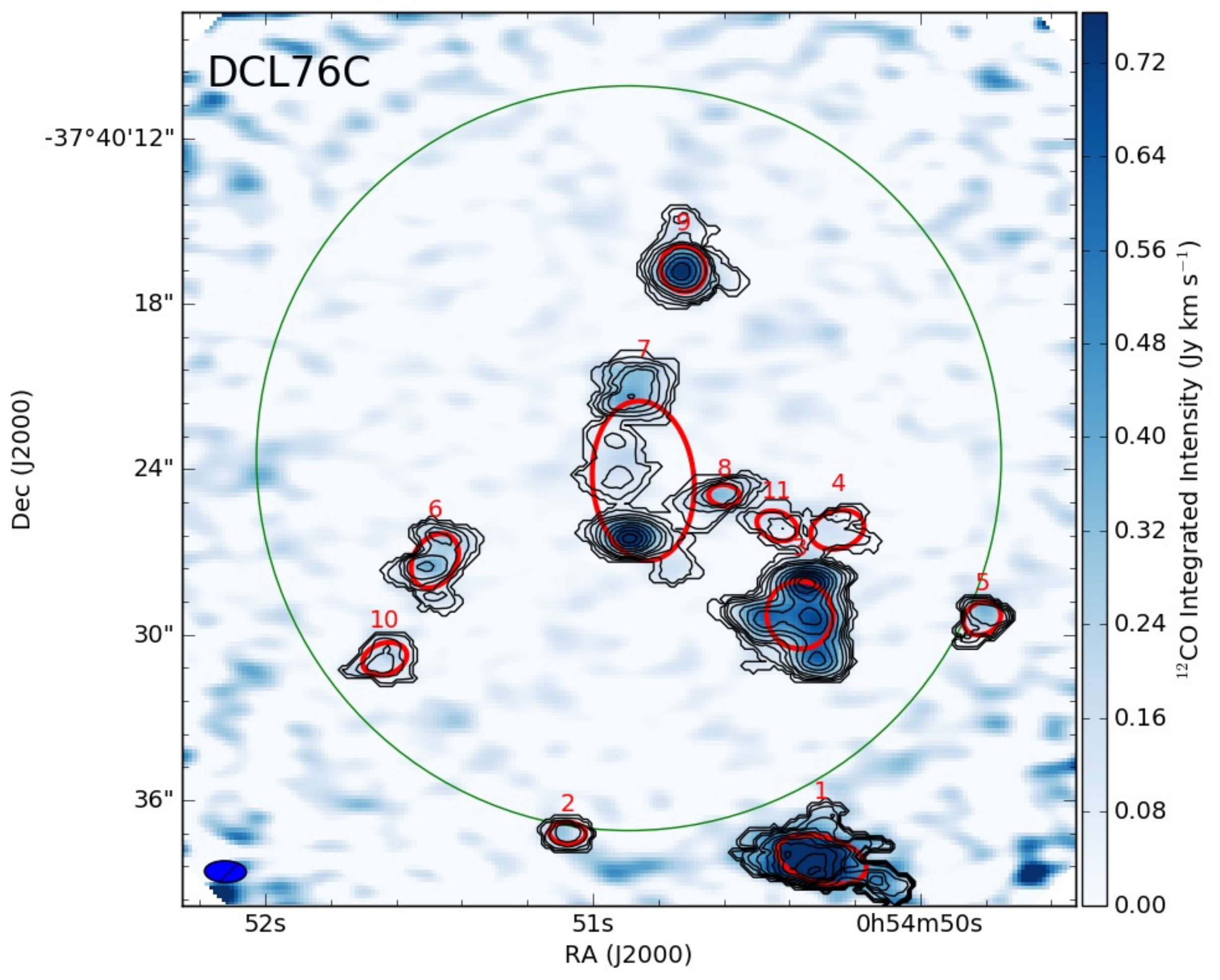}
\end{minipage}
\begin{minipage}{0.50\linewidth}
\includegraphics[width=\linewidth]{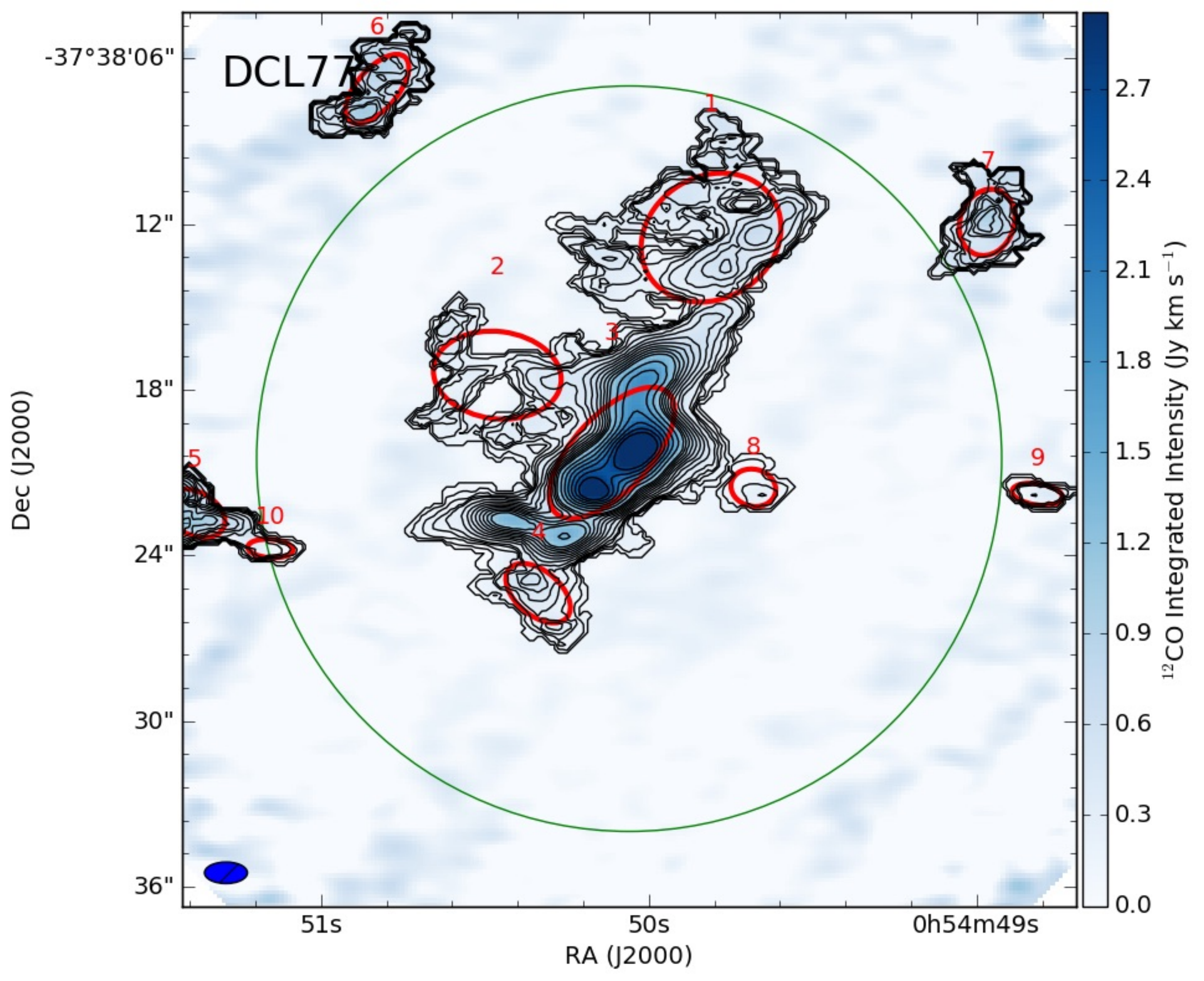}
\end{minipage}

\begin{minipage}{0.50\linewidth}
\includegraphics[width=\linewidth]{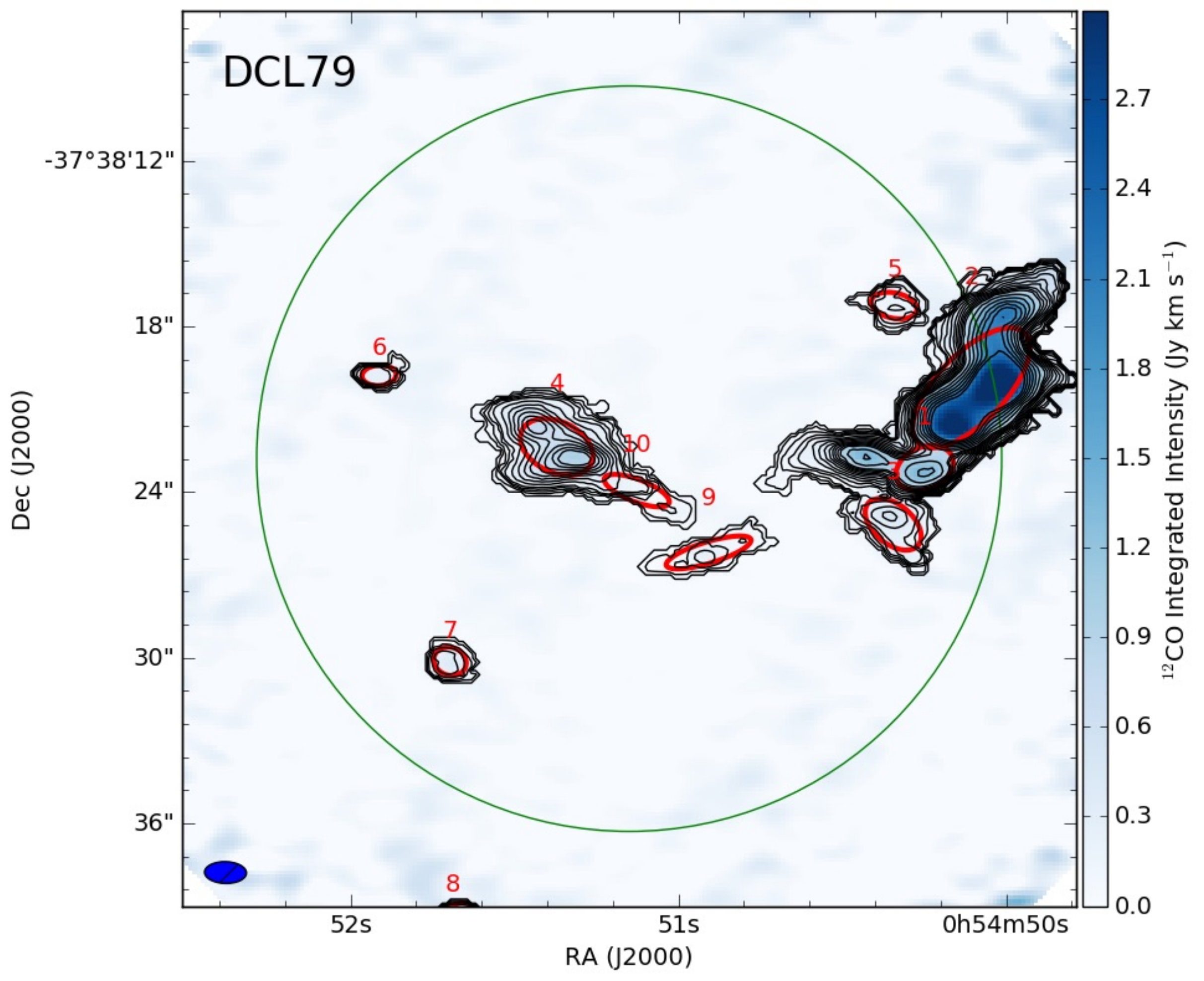}
\end{minipage}
\begin{minipage}{0.50\linewidth}
\includegraphics[width=\linewidth]{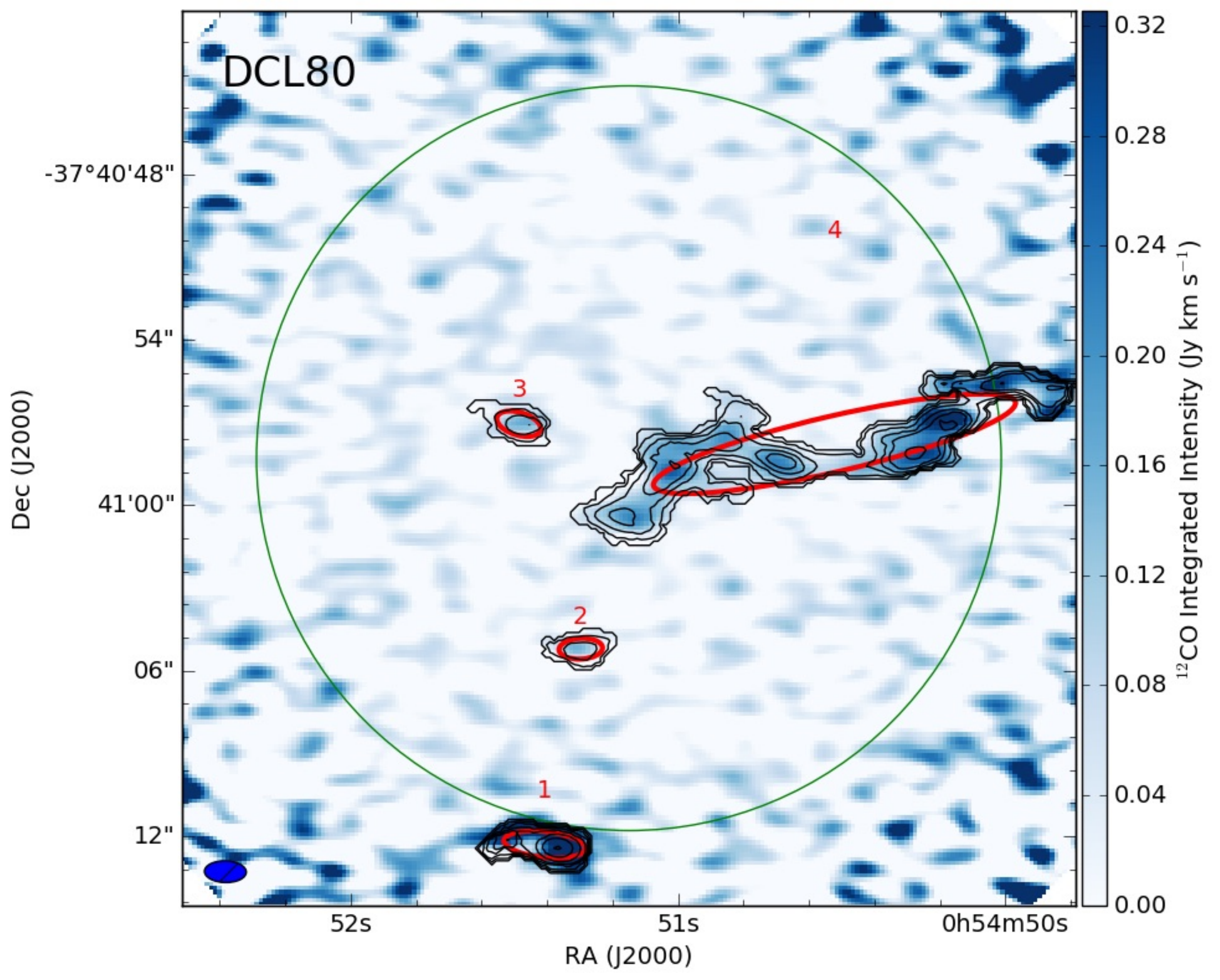}
\end{minipage}

\caption{CO integrated intensity images, continued.}
\end{figure*}

\clearpage
\begin{figure*}

\begin{minipage}{0.50\linewidth}
\includegraphics[width=\linewidth]{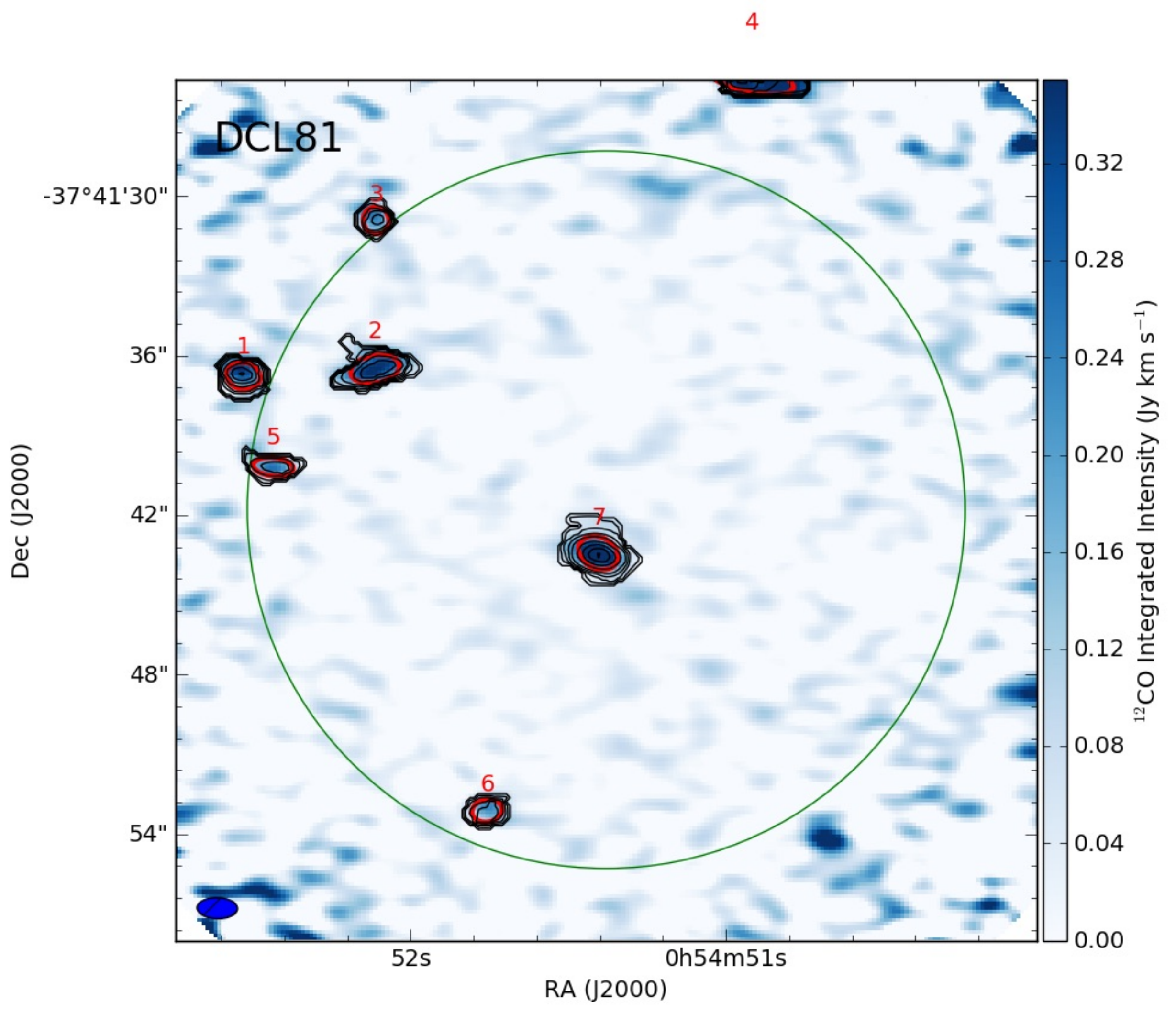}
\end{minipage}
\begin{minipage}{0.50\linewidth}
\includegraphics[width=\linewidth]{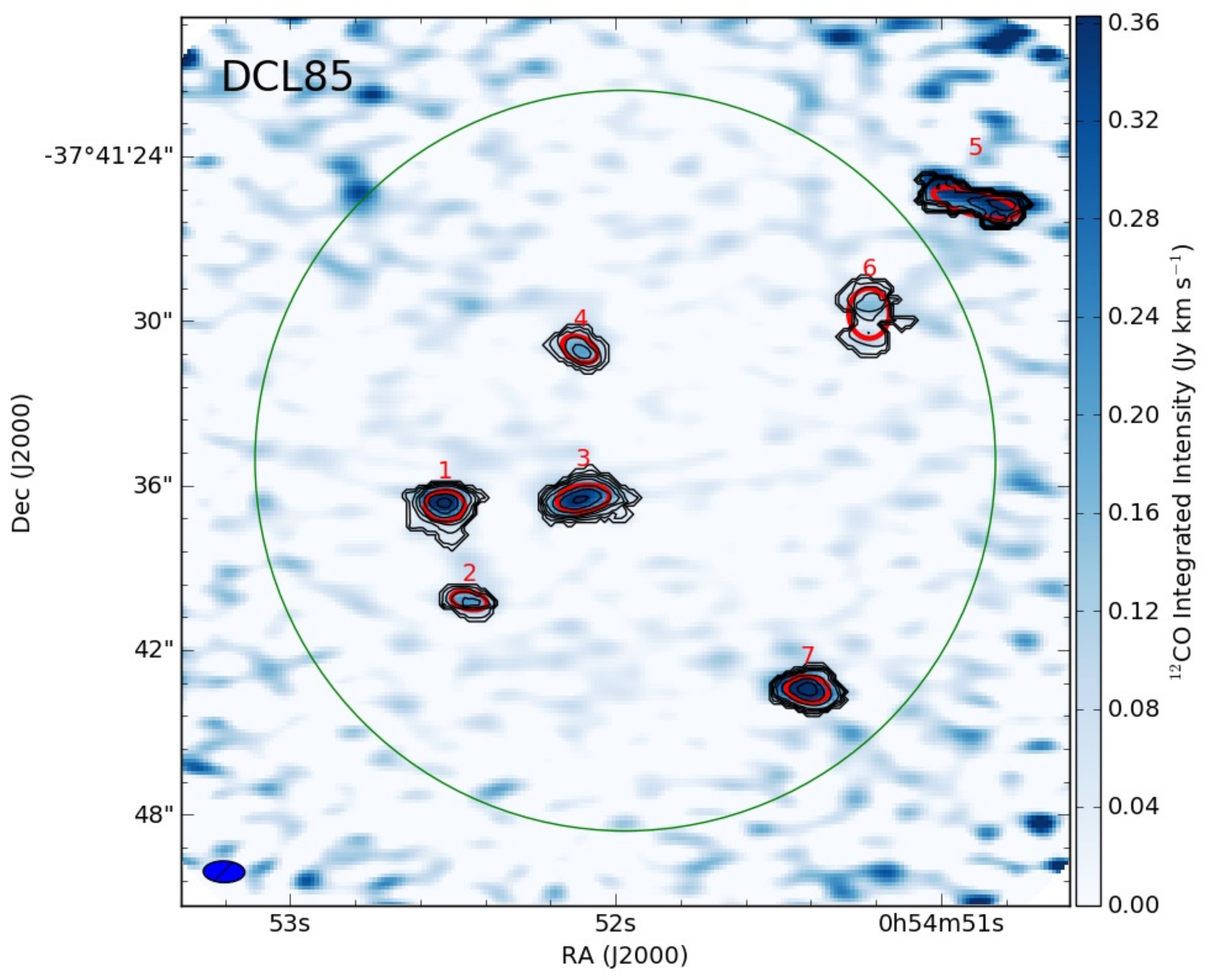}
\end{minipage}

\begin{minipage}{0.50\linewidth}
\includegraphics[width=\linewidth]{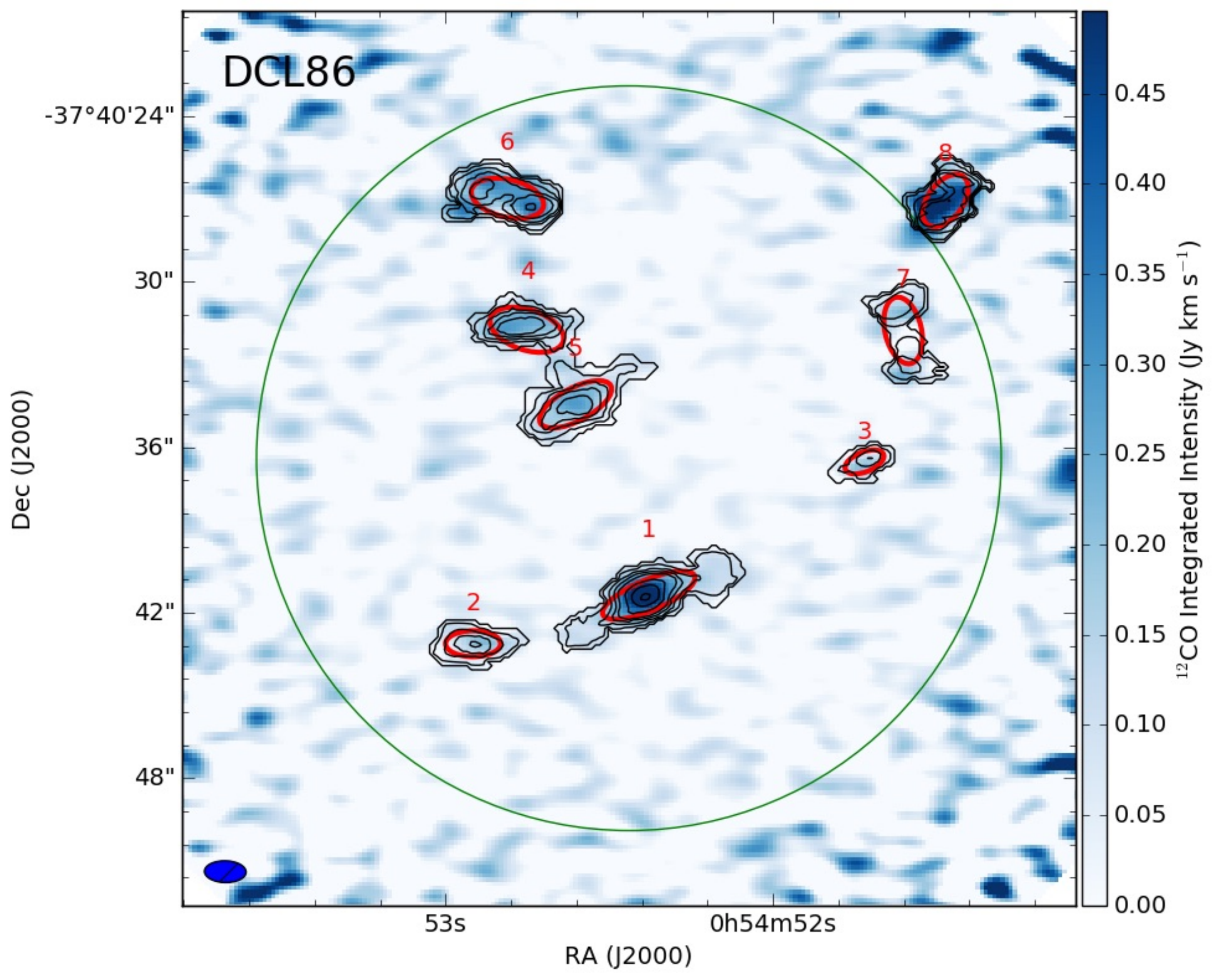}
\end{minipage}
\begin{minipage}{0.50\linewidth}
\includegraphics[width=\linewidth]{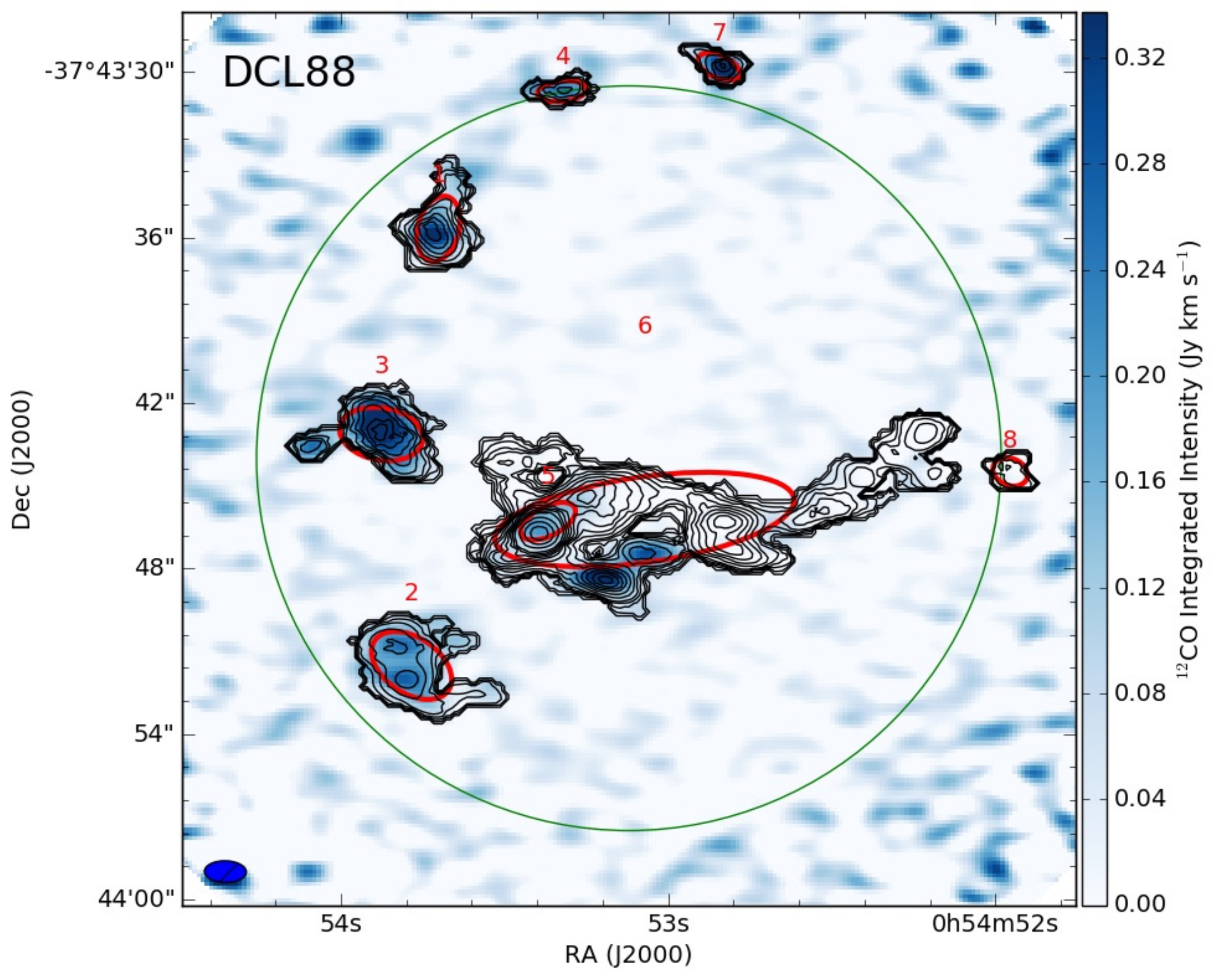}
\end{minipage}

\begin{minipage}{0.50\linewidth}
\includegraphics[width=\linewidth]{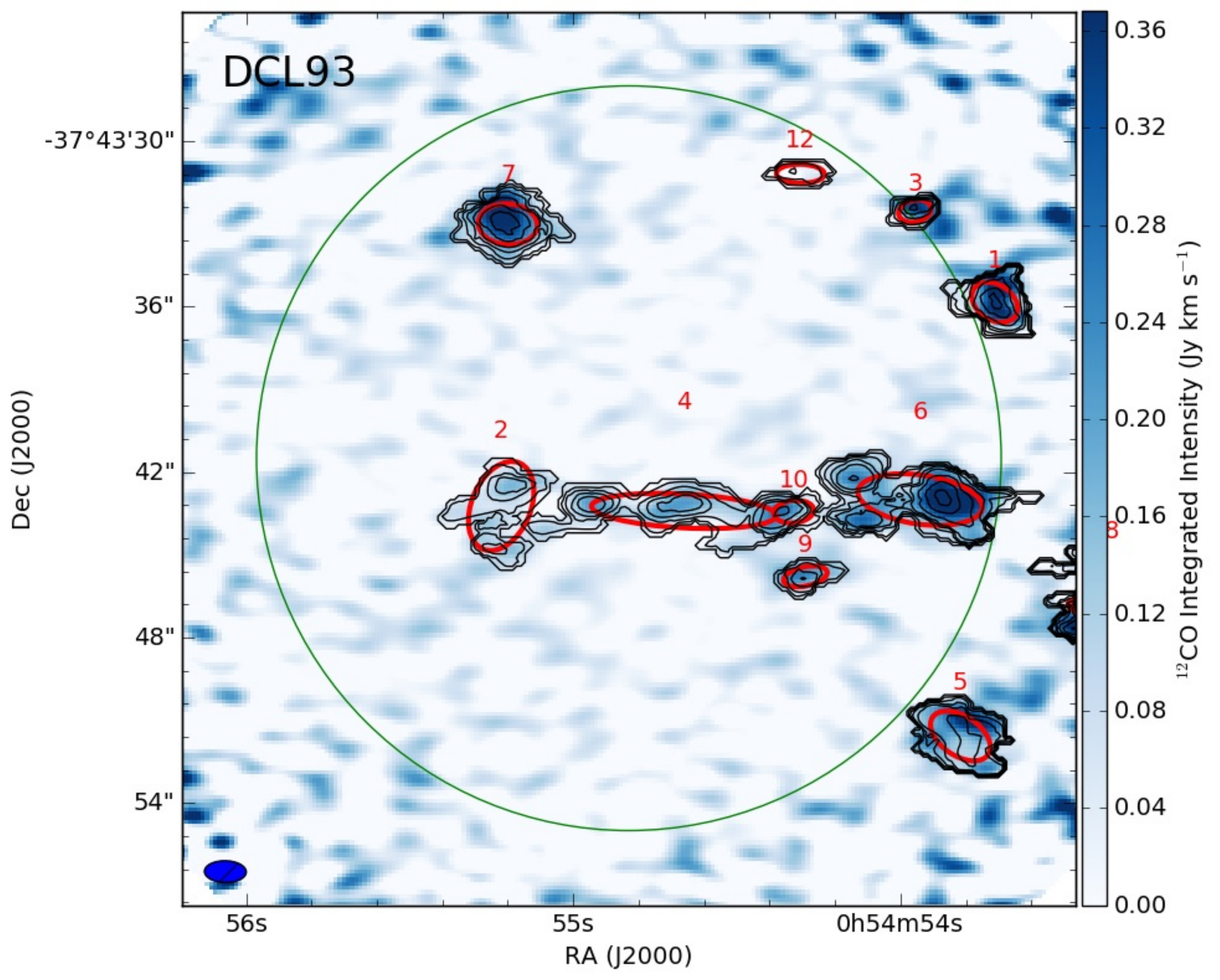}
\end{minipage}
\begin{minipage}{0.50\linewidth}
\includegraphics[width=\linewidth]{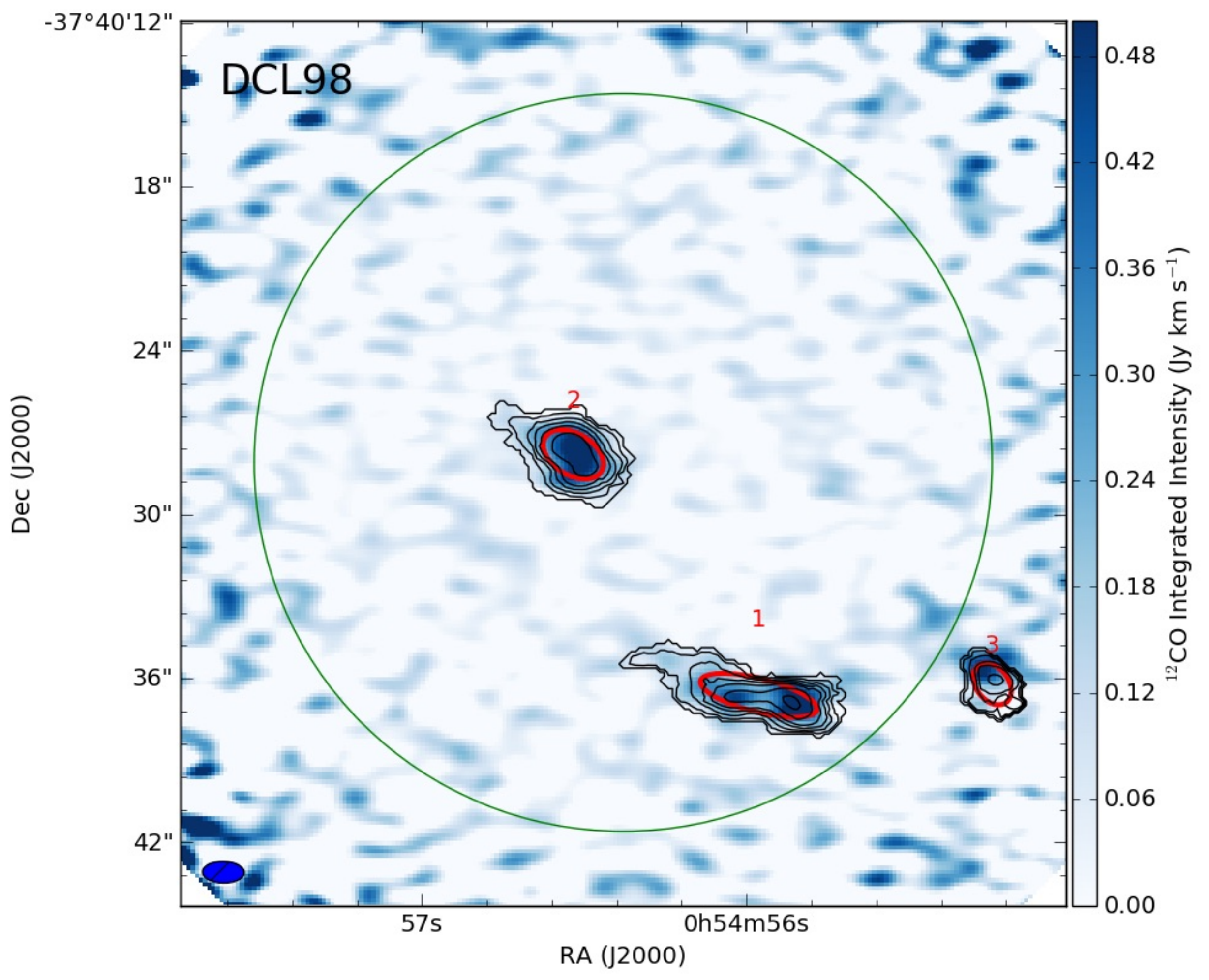}
\end{minipage}

\caption{CO integrated intensity images, continued.}
\end{figure*}

\clearpage
\begin{figure*}

\begin{minipage}{0.50\linewidth}
\includegraphics[width=\linewidth]{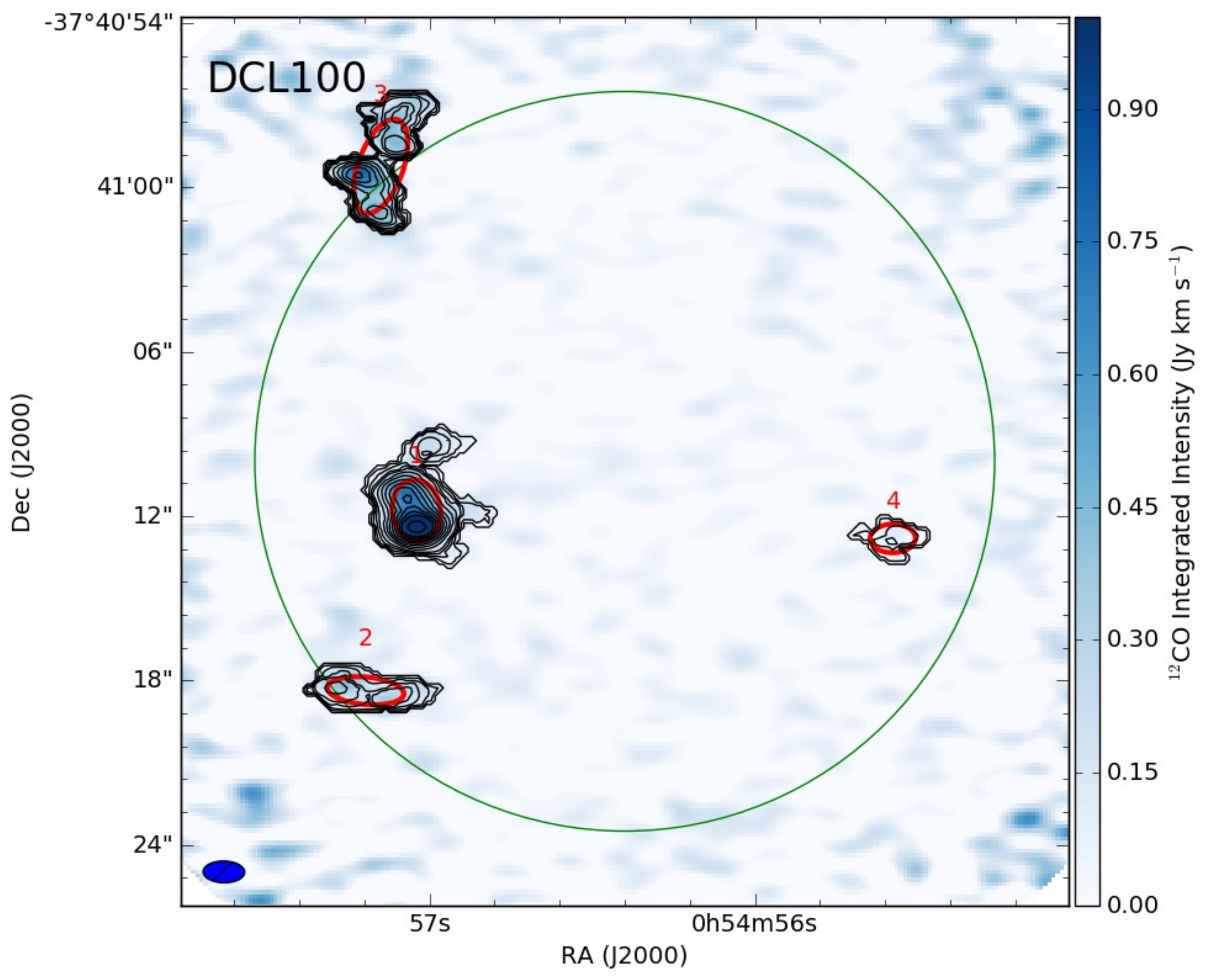}
\end{minipage}
\begin{minipage}{0.50\linewidth}
\includegraphics[width=\linewidth]{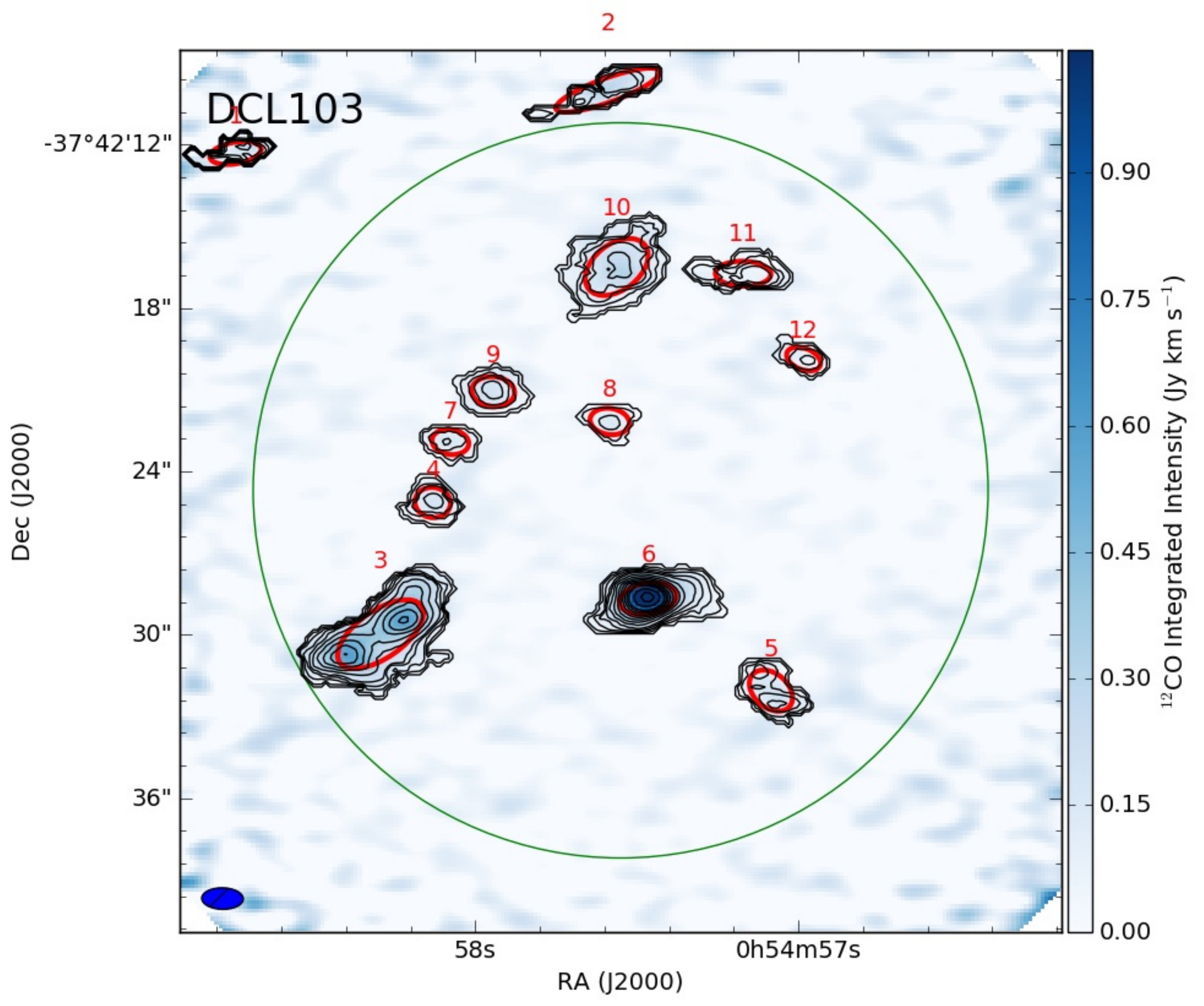}
\end{minipage}

\begin{minipage}{0.50\linewidth}
\includegraphics[width=\linewidth]{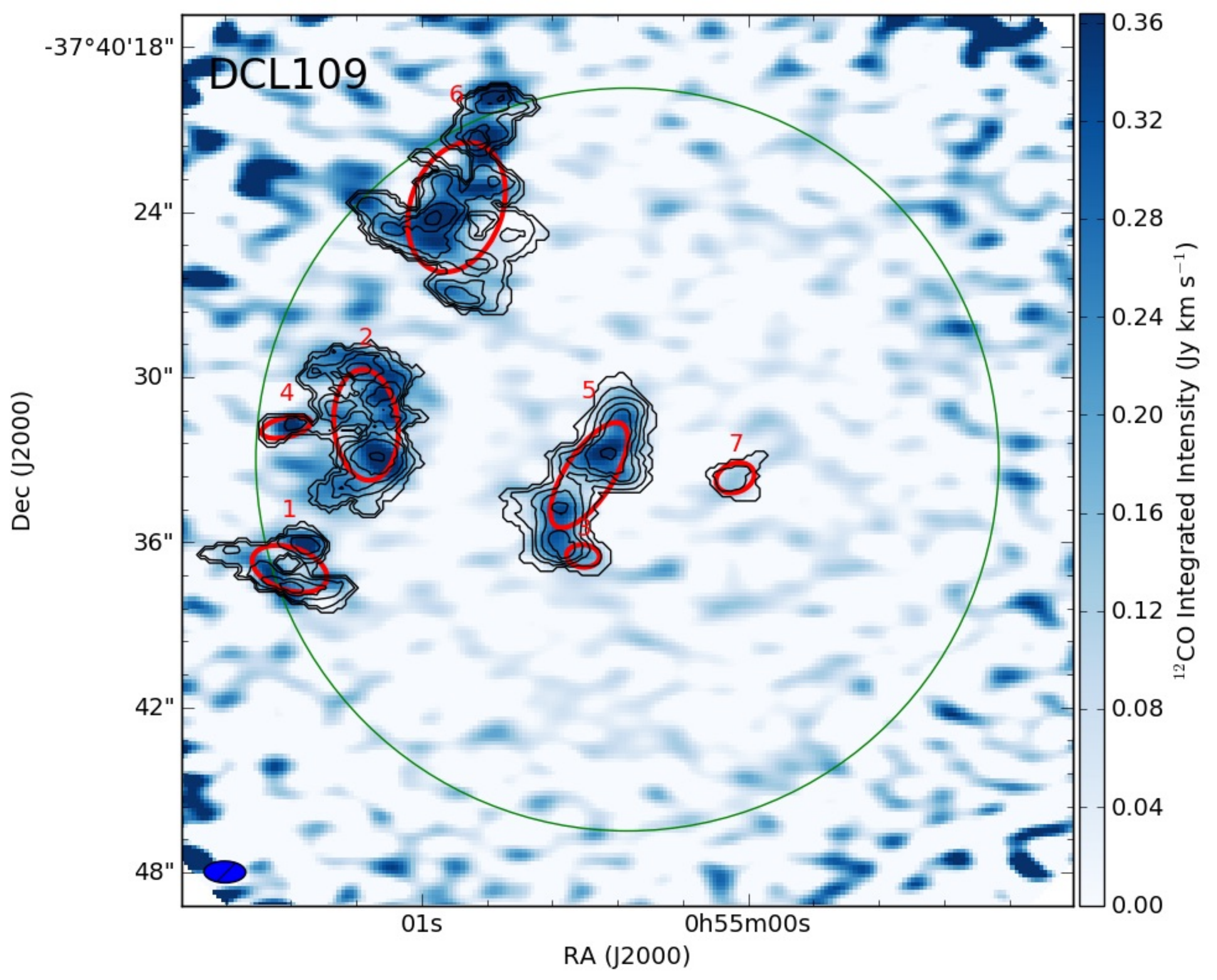}
\end{minipage}
\begin{minipage}{0.50\linewidth}
\includegraphics[width=\linewidth]{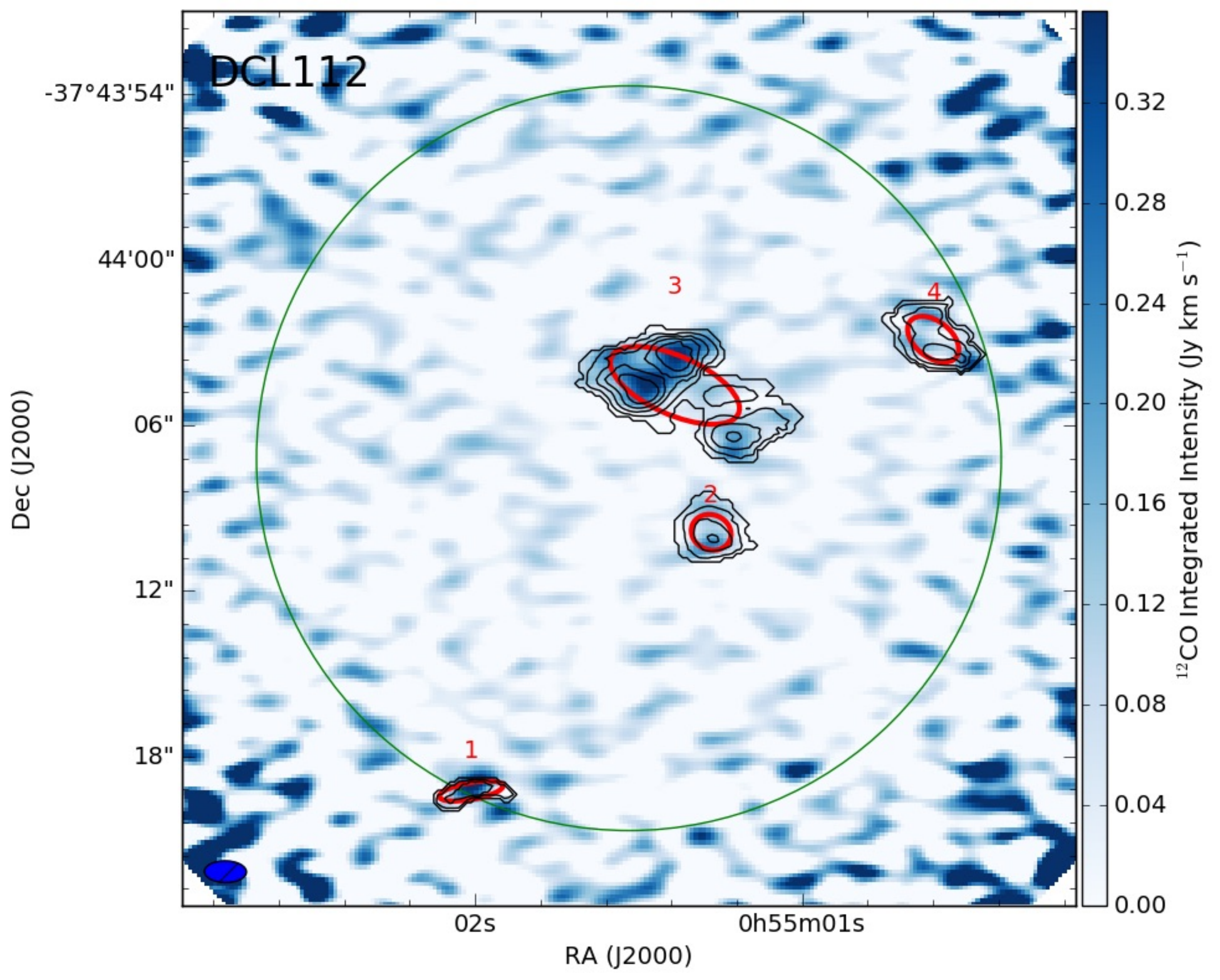}
\end{minipage}

\begin{minipage}{0.50\linewidth}
\includegraphics[width=\linewidth]{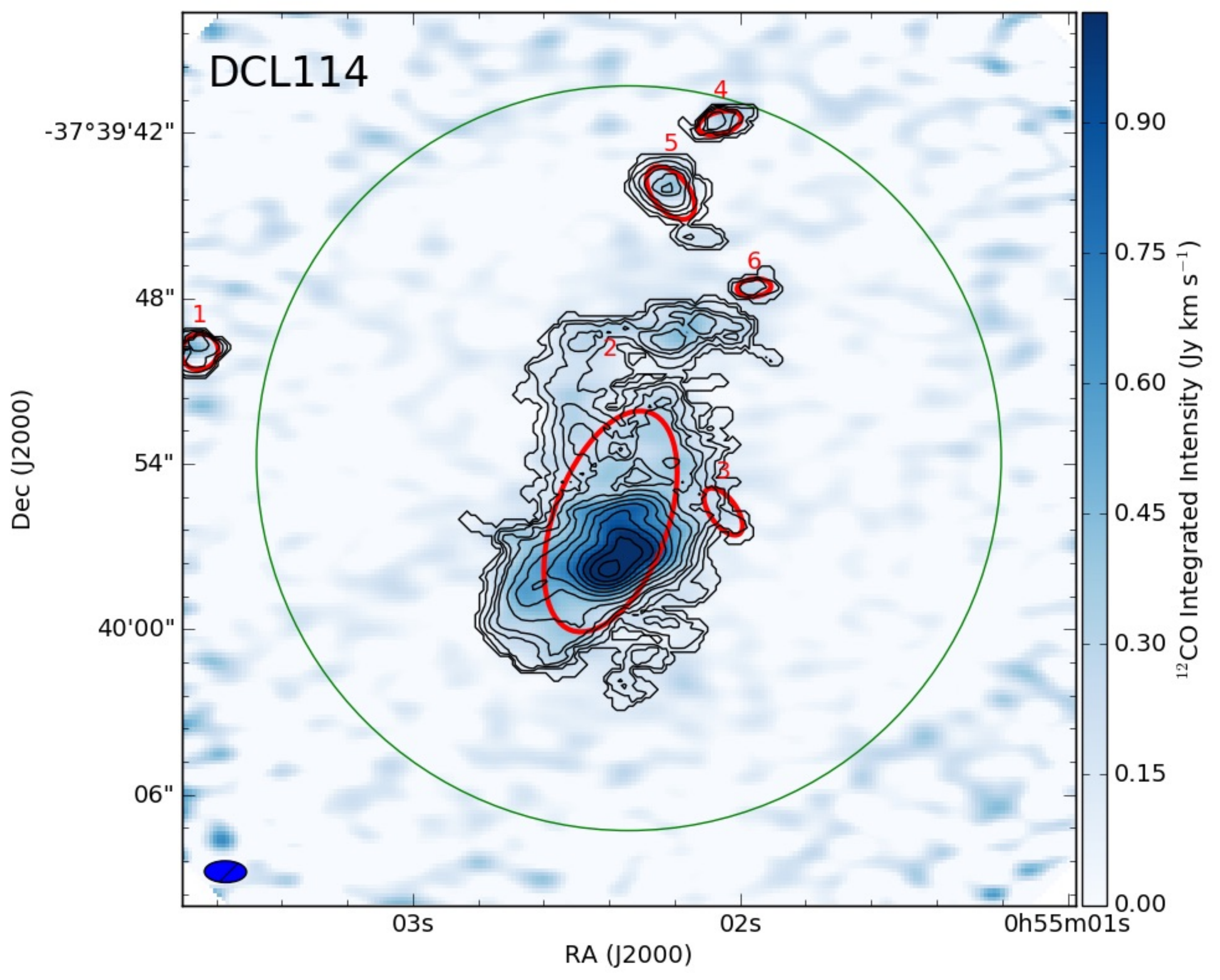}
\end{minipage}
\begin{minipage}{0.50\linewidth}
\includegraphics[width=\linewidth]{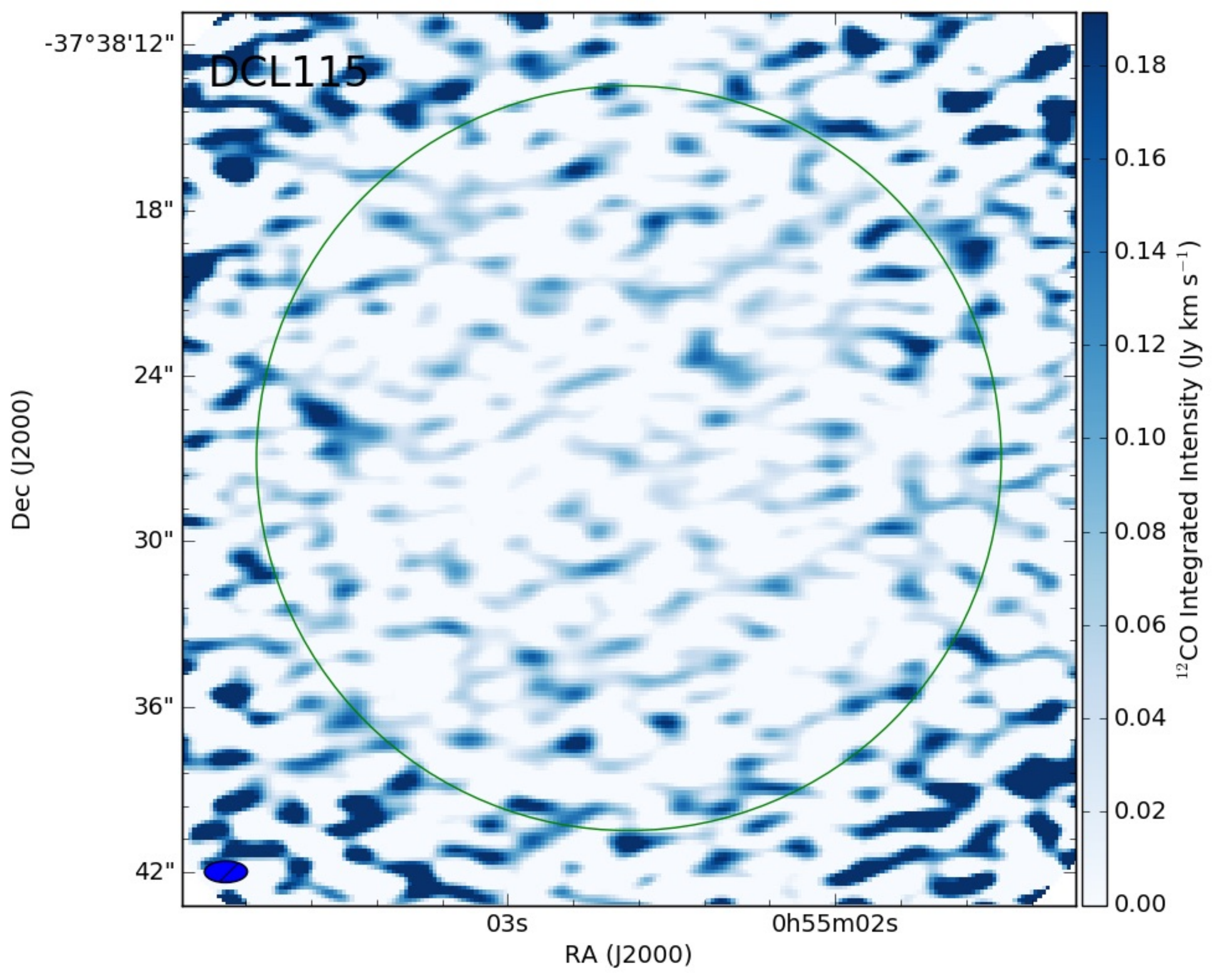}
\end{minipage}

\caption{CO integrated intensity images, continued.}
\end{figure*}

\clearpage
\begin{figure*}

\begin{minipage}{0.50\linewidth}
\includegraphics[width=\linewidth]{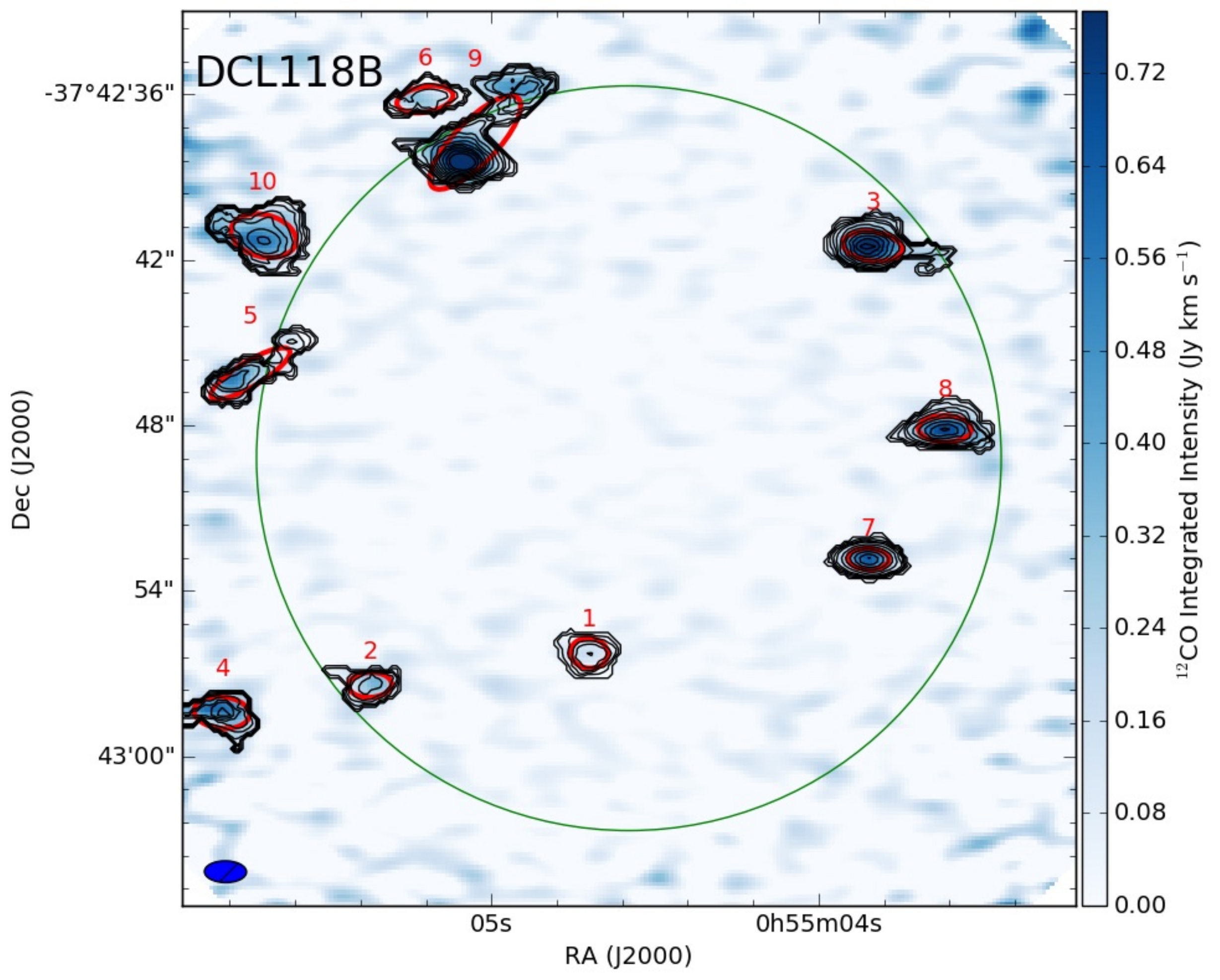}
\end{minipage}
\begin{minipage}{0.50\linewidth}
\includegraphics[width=\linewidth]{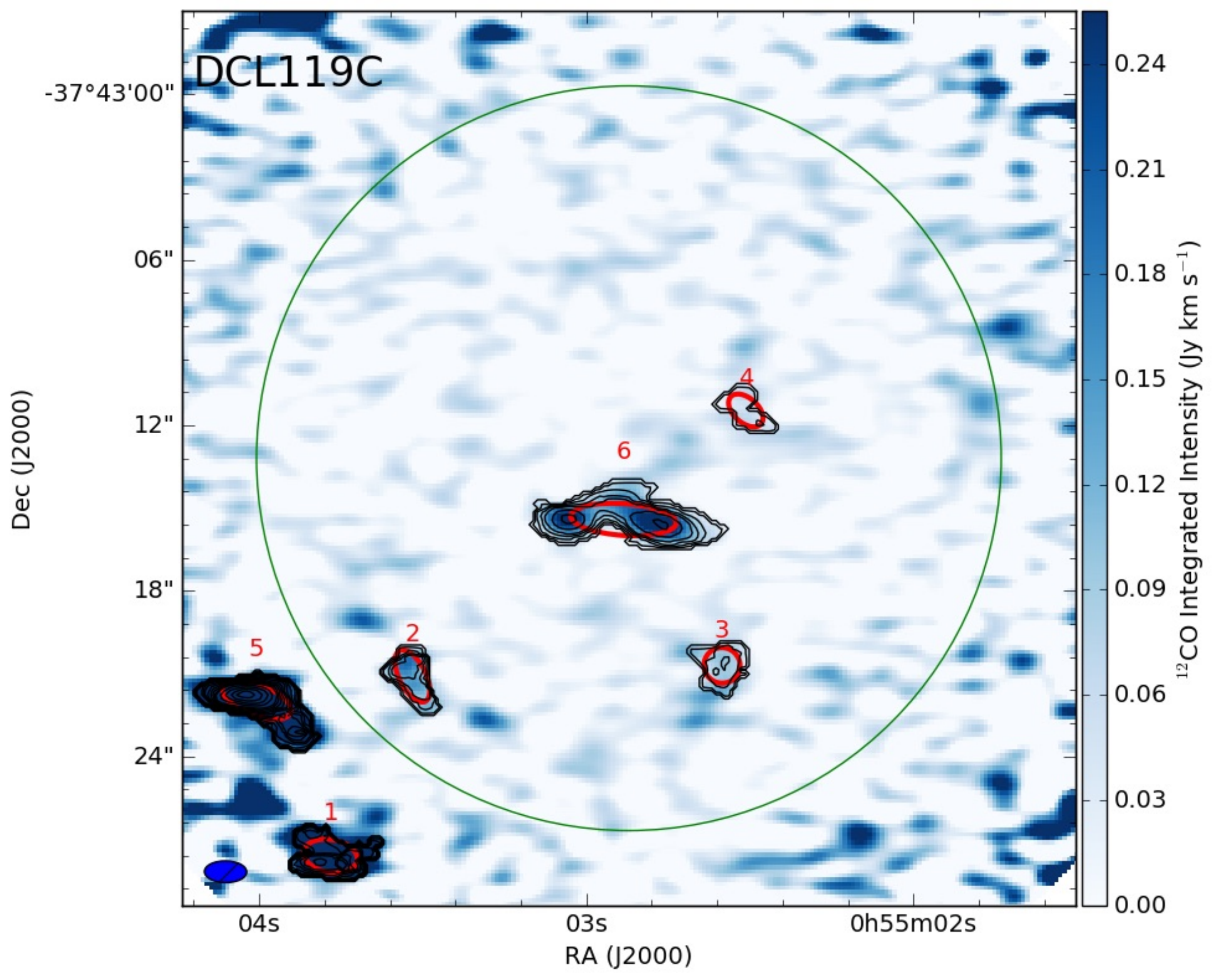}
\end{minipage}

\begin{minipage}{0.50\linewidth}
\includegraphics[width=\linewidth]{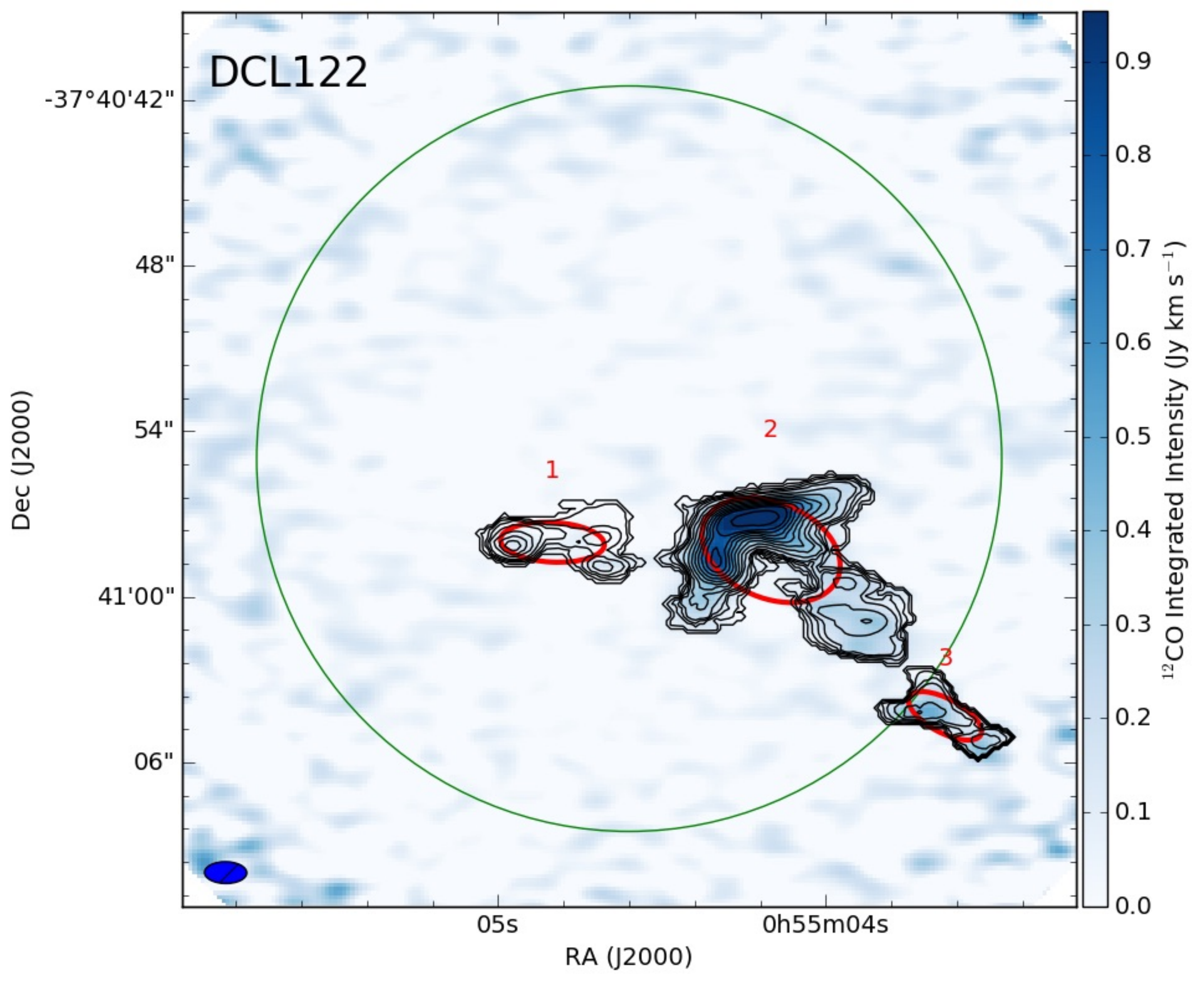}
\end{minipage}
\begin{minipage}{0.50\linewidth}
\includegraphics[width=\linewidth]{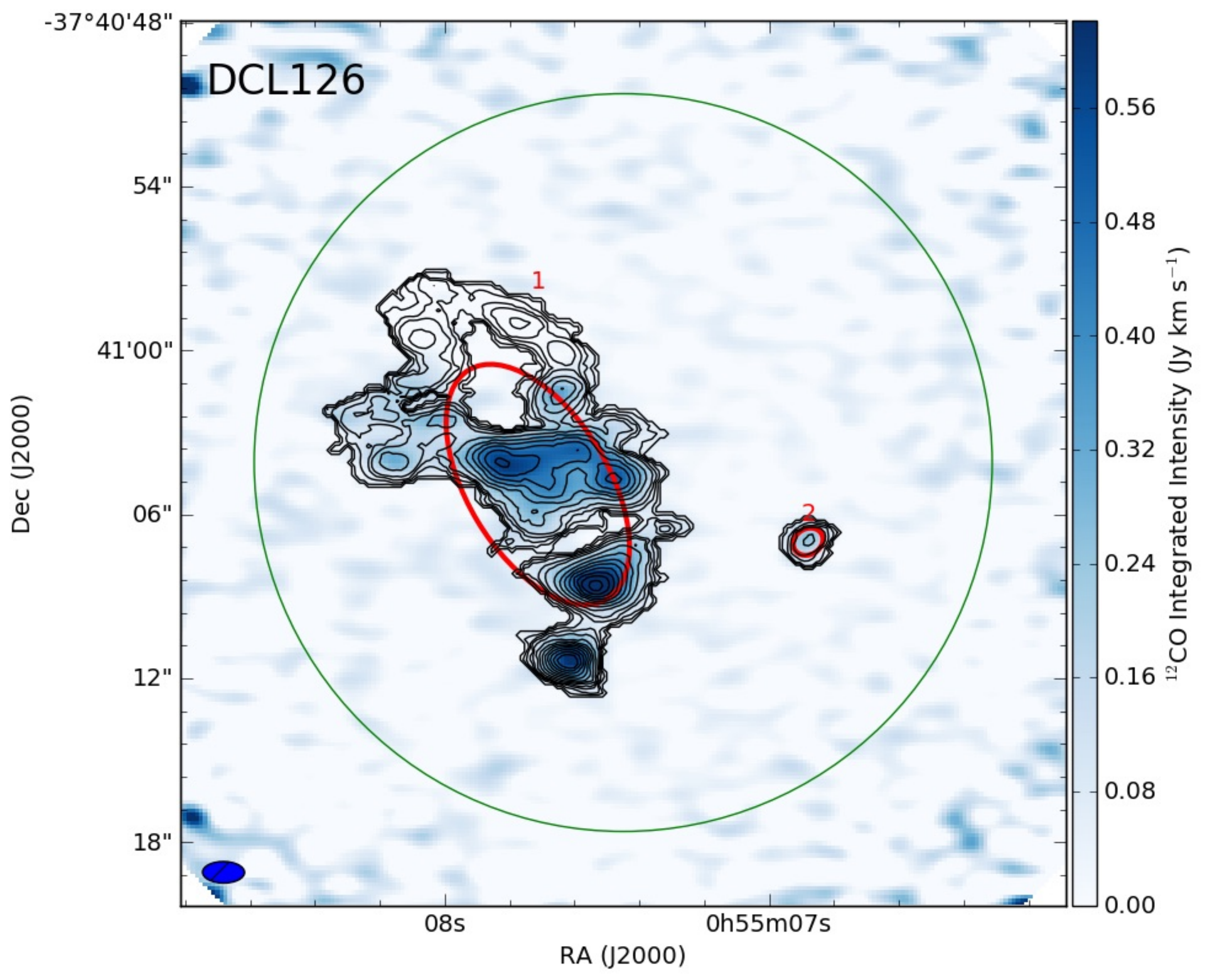}
\end{minipage}

\begin{minipage}{0.50\linewidth}
\includegraphics[width=\linewidth]{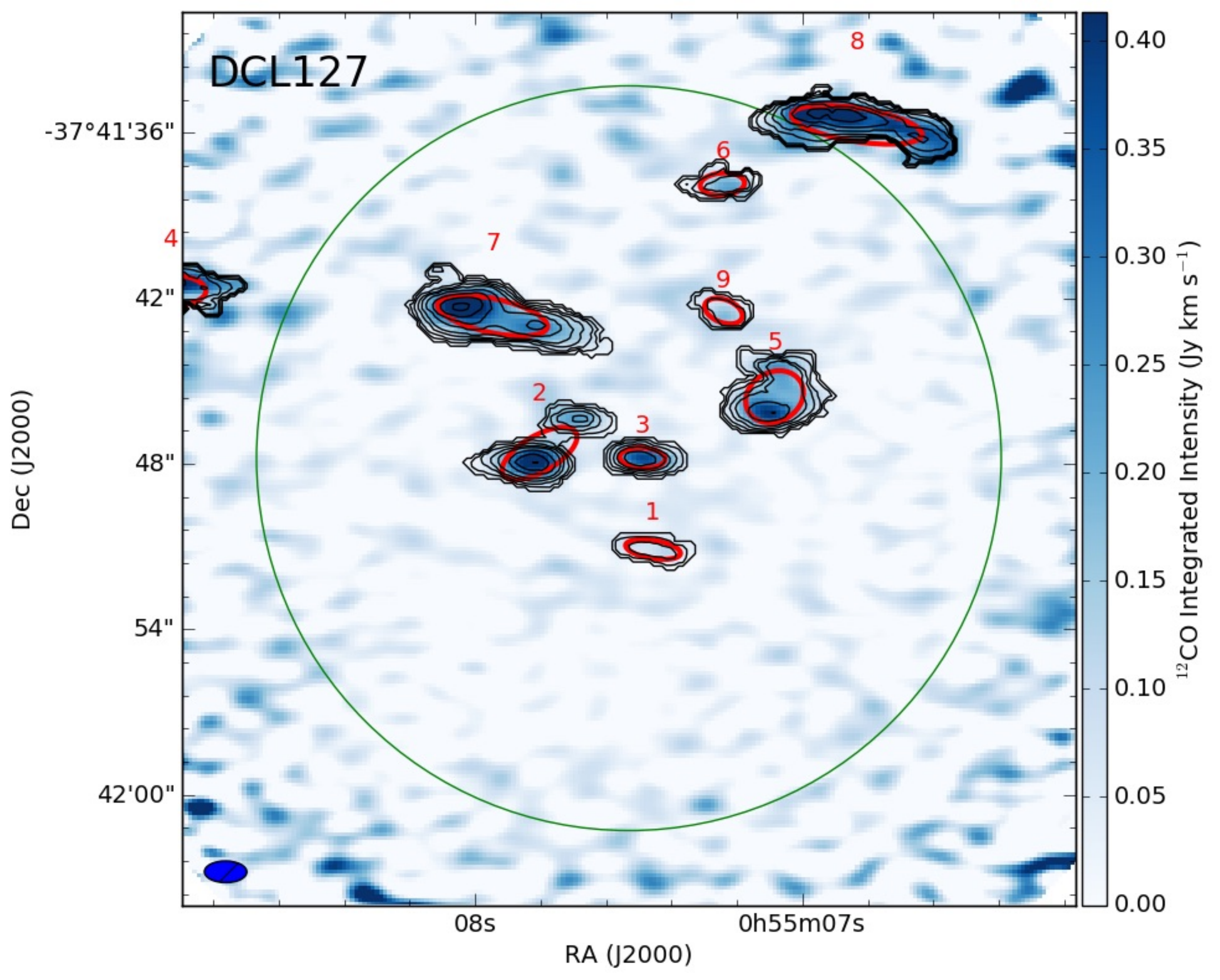}
\end{minipage}
\begin{minipage}{0.50\linewidth}
\includegraphics[width=\linewidth]{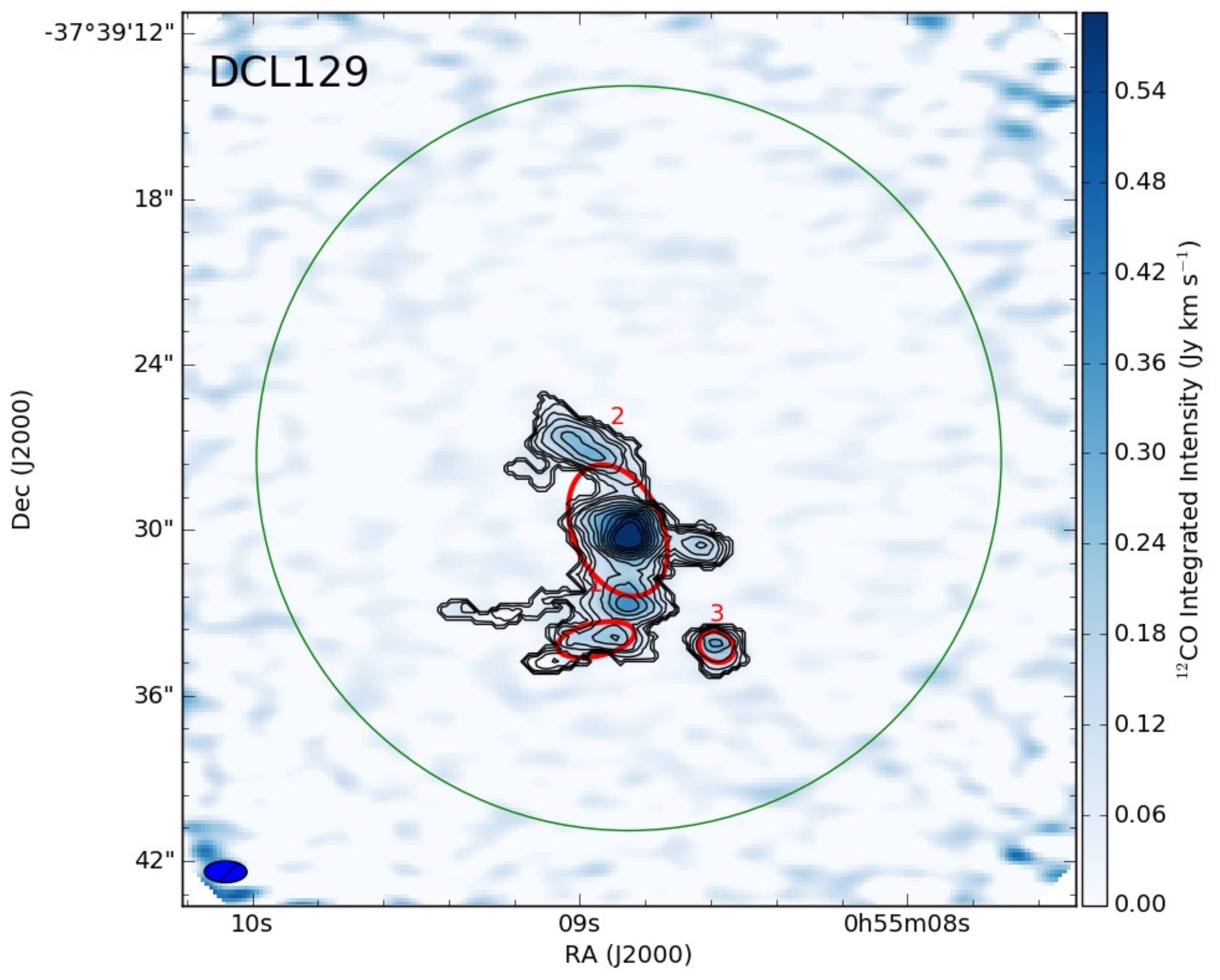}
\end{minipage}

\caption{CO integrated intensity images, continued.}
\end{figure*}

\clearpage
\begin{figure*}

\begin{minipage}{0.50\linewidth}
\includegraphics[width=\linewidth]{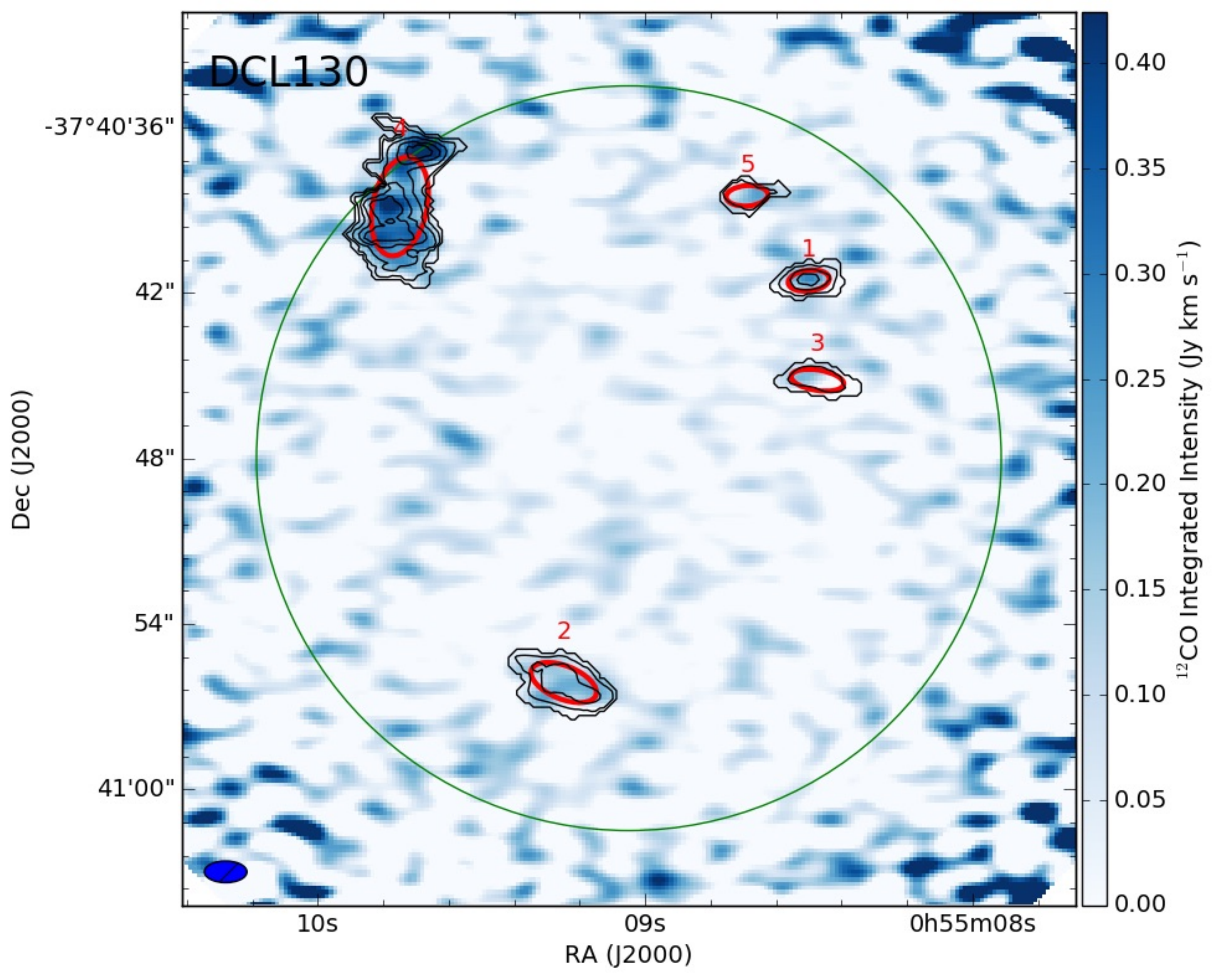}
\end{minipage}
\begin{minipage}{0.50\linewidth}
\includegraphics[width=\linewidth]{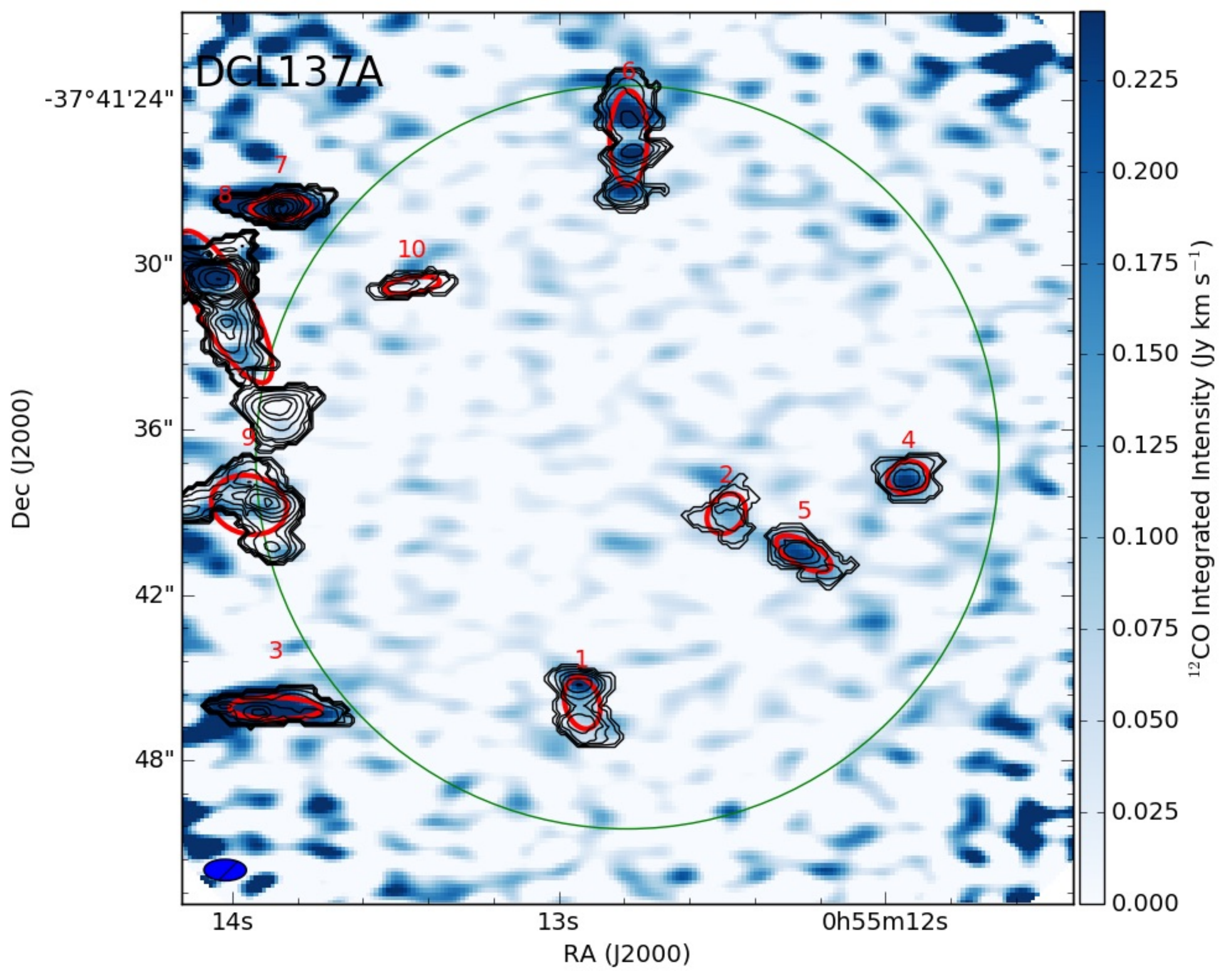}
\end{minipage}

\begin{minipage}{0.50\linewidth}
\includegraphics[width=\linewidth]{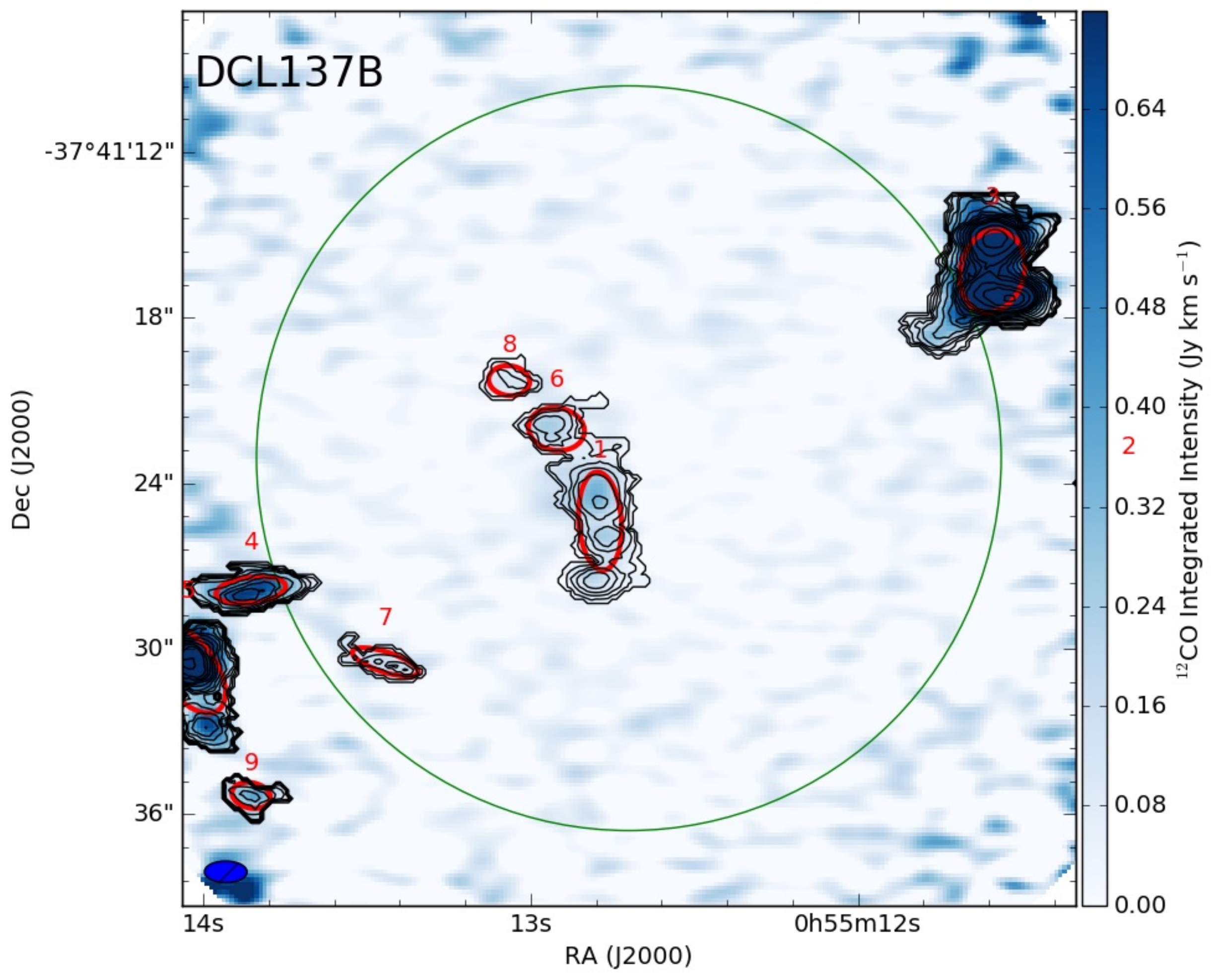}
\end{minipage}
\begin{minipage}{0.50\linewidth}
\includegraphics[width=\linewidth]{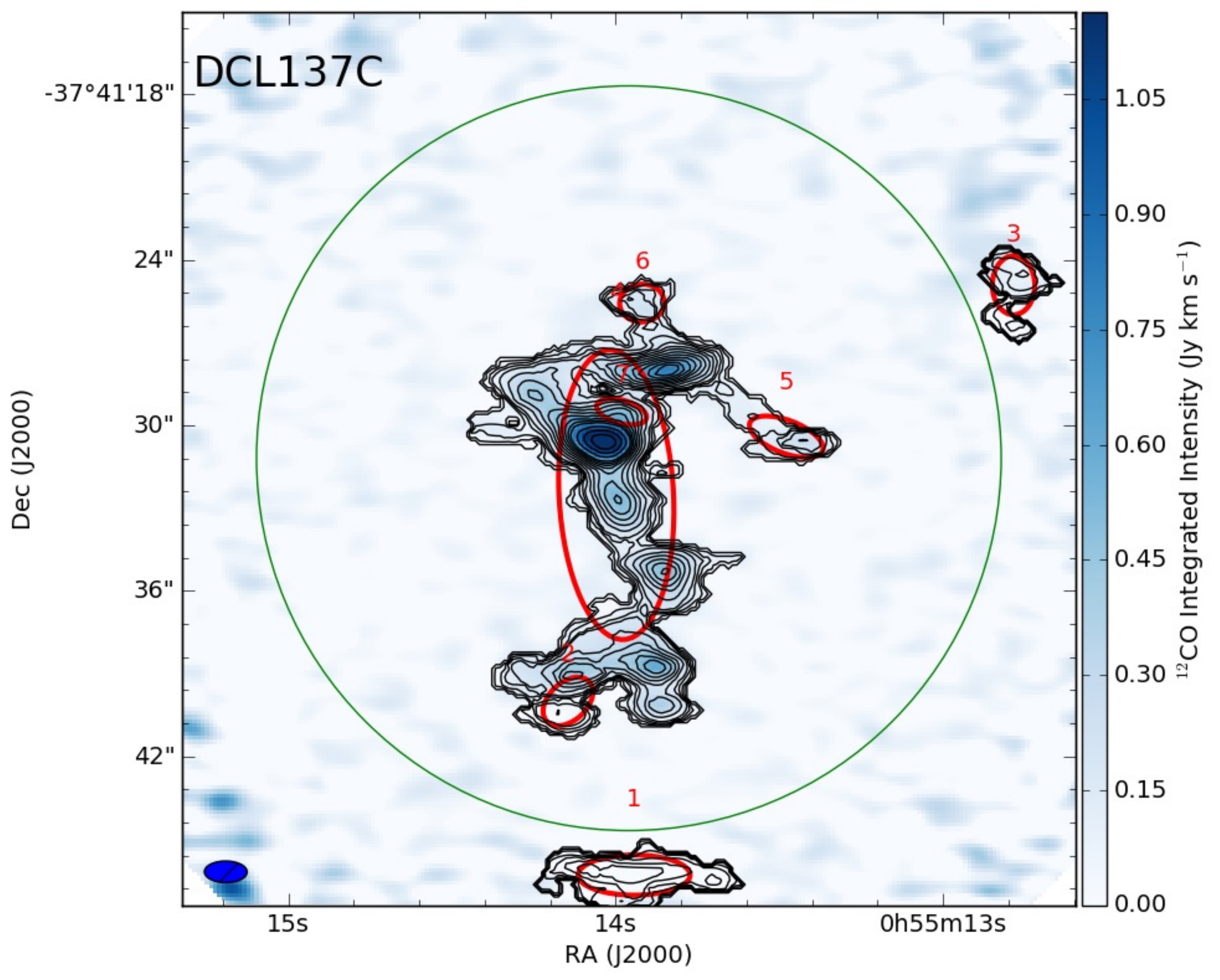}
\end{minipage}

\begin{minipage}{0.50\linewidth}
\includegraphics[width=\linewidth]{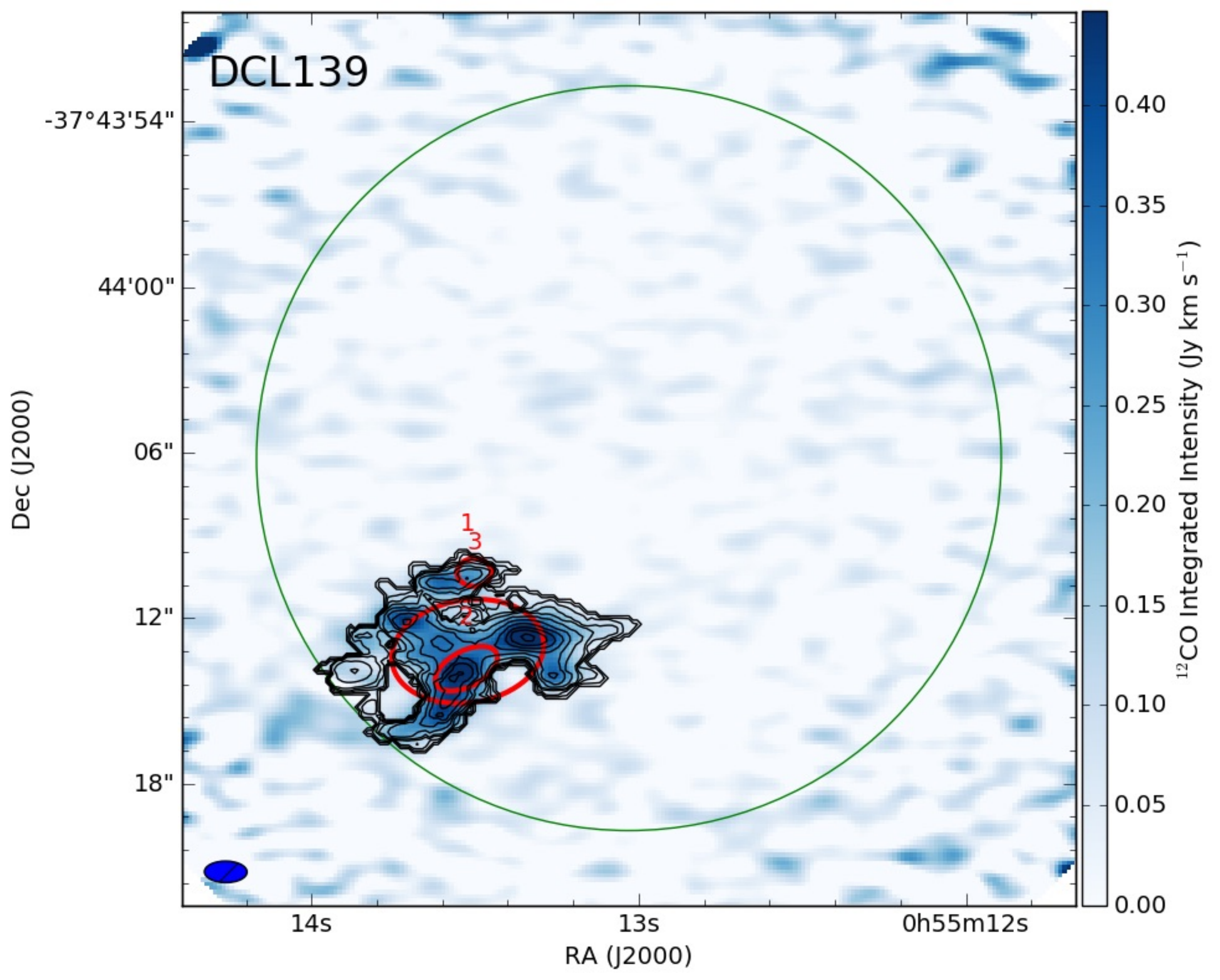}
\end{minipage}
\begin{minipage}{0.50\linewidth}
\includegraphics[width=\linewidth]{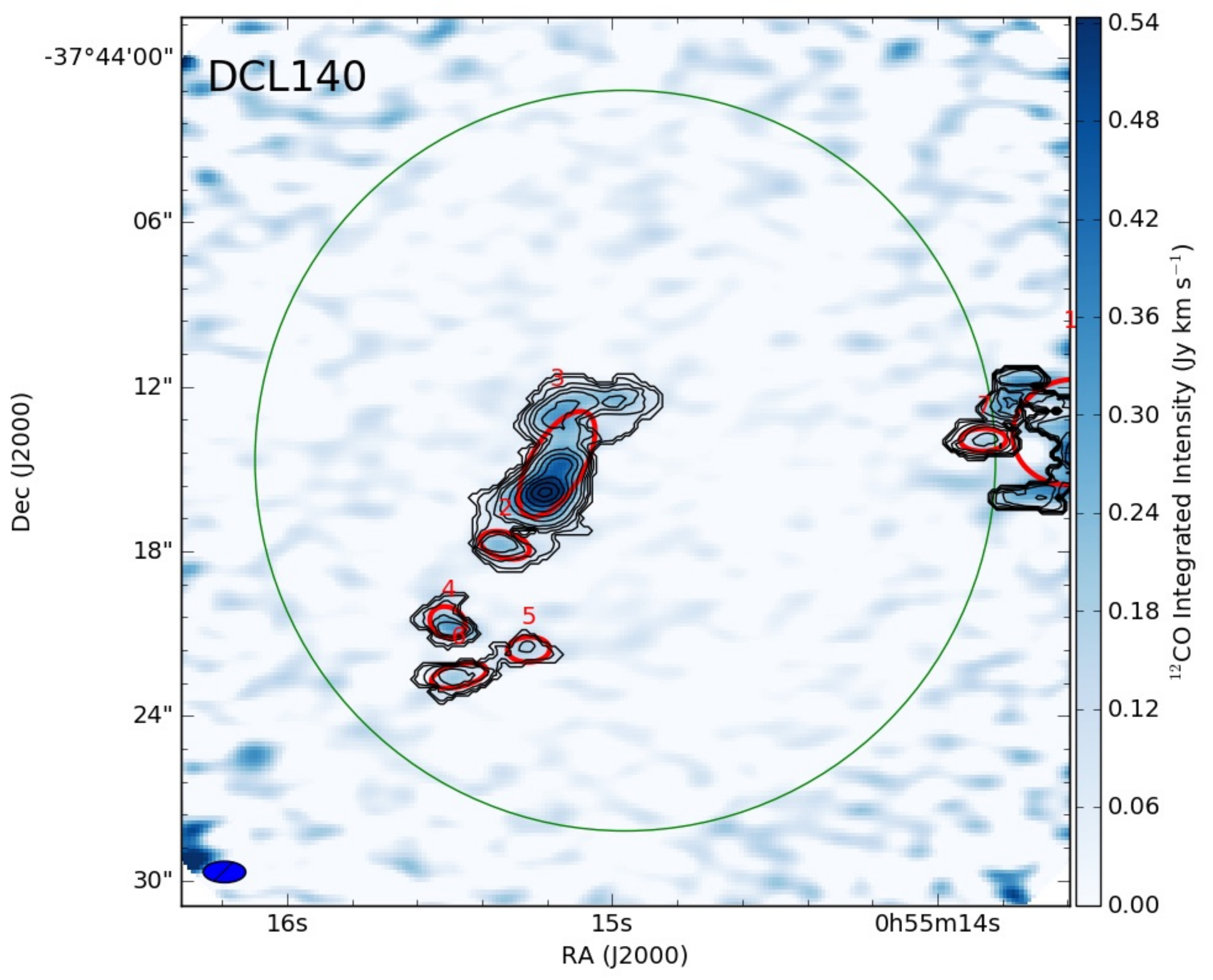}
\end{minipage}

\caption{CO integrated intensity images, continued.}
\end{figure*}

\clearpage
\bibliography{CMF_Papers}

\clearpage
\LongTables
\centering
\begin{landscape}
\begin{deluxetable*}{cccccccccccccc}
\tablecaption{GMC Catalog\tablenotemark{a}\label{tab:gmcprops}}
\tabletypesize{\scriptsize}
\tablehead{\colhead{Cloud ID} & \colhead{R.A} & \colhead{decl.} & \colhead{v$_0$} & \colhead{$(a/b)_{\rm dc}$} & \colhead{PA$_{\rm dc}$} & \colhead{$R$} & \colhead{$\Delta V$} & \colhead{$M_{\rm lum}$} & \colhead{$M_{\rm vir}$} & \colhead{$L_{\rm CO}$} & \colhead{Tmax} & \colhead{S2N} & \colhead{Note\tablenotemark{b}} \\
\colhead{} & \colhead{(J2000)} & \colhead{(J2000)} & \colhead{({\kms})} & \colhead{} & \colhead{($^{\circ}$)} & \colhead{(pc)} & \colhead{({\kms})} & \colhead{($10^3~\Msun$)} & \colhead{($10^3~\Msun$)} & \colhead{(K~{\kms}~pc$^2$)} & \colhead{(K)} & \colhead{} & \colhead{}
}
\startdata
DCL23-A1 & 00h54m28.28s & -37d41m53.0s & 171.8 & 3.10 & 23.1 & 19.6$\pm$1.6 & 4.6$\pm$0.4 & 206.0$\pm$12.4 & 77.4$\pm$15.5 & 116.0$\pm$7.0 & 4.296 & 28.2 & F \\
DCL23-A2 & 00h54m28.12s & -37d41m55.0s & 178.1 & \nodata & \nodata & \nodata & 2.3$\pm$1.3 & 5.4$\pm$2.9 & \nodata & 3.0$\pm$1.6 & 1.001 & 6.3 &  \\
DCL23-A3 & 00h54m27.48s & -37d41m59.8s & 180.9 & \nodata & \nodata & \nodata & 2.7$\pm$1.7 & 11.9$\pm$6.4 & \nodata & 6.7$\pm$3.6 & 2.005 & 5.7 &  \\
DCL23-A4 & 00h54m27.87s & -37d41m56.1s & 183.6 & \nodata & \nodata & \nodata & 3.9$\pm$1.7 & 12.2$\pm$4.3 & \nodata & 6.9$\pm$2.4 & 1.288 & 6.9 &  \\
DCL24-A1 & 00h54m28.64s & -37d41m41.3s & 167.4 & \nodata & \nodata & \nodata & 3.5$\pm$1.9 & 12.2$\pm$3.7 & \nodata & 7.0$\pm$2.1 & 1.323 & 7.7 &  \\
DCL24-A2 & 00h54m28.70s & -37d41m32.0s & 168.2 & \nodata & \nodata & \nodata & 3.1$\pm$1.6 & 6.3$\pm$1.4 & \nodata & 3.6$\pm$0.8 & 1.181 & 8.9 &  \\
DCL24-A3 & 00h54m28.31s & -37d41m52.2s & 171.1 & \nodata & \nodata & \nodata & 3.3$\pm$1.2 & 81.4$\pm$17.9 & \nodata & 47.2$\pm$10.4 & 6.467 & 7.6 & O; DCL23-A1 \\
DCL24-A4 & 00h54m28.63s & -37d41m35.7s & 176.8 & \nodata & \nodata & \nodata & 2.6$\pm$0.9 & 10.3$\pm$2.1 & \nodata & 6.0$\pm$1.2 & 1.074 & 7.9 &  \\
DCL30-A2 & 00h54m32.20s & -37d37m50.4s & 195.6 & \nodata & \nodata & \nodata & 2.3$\pm$0.7 & 11.5$\pm$3.4 & \nodata & 7.0$\pm$2.1 & 1.432 & 7.4 & DCL31-A3 \\
DCL30-A3 & 00h54m31.50s & -37d38m08.4s & 197.4 & \nodata & \nodata & \nodata & 2.2$\pm$0.8 & 16.0$\pm$2.6 & \nodata & 9.7$\pm$1.6 & 2.375 & 11.5 &  \\
DCL30-A4 & 00h54m31.45s & -37d37m56.1s & 199.6 & \nodata & \nodata & \nodata & 4.0$\pm$1.1 & 15.0$\pm$2.1 & \nodata & 9.2$\pm$1.3 & 1.375 & 9.9 &  \\
DCL30-A5 & 00h54m31.47s & -37d38m00.8s & 200.3 & \nodata & \nodata & \nodata & 1.8$\pm$1.6 & 3.7$\pm$1.8 & \nodata & 2.2$\pm$1.1 & 0.805 & 6.2 &  \\
DCL31-A1 & 00h54m32.09s & -37d37m57.7s & 191.9 & 4.23 & -74.8 & 4.5$\pm$3.1 & 3.6$\pm$1.3 & 22.6$\pm$7.2 & 10.8$\pm$12.0 & 13.7$\pm$4.4 & 2.082 & 7.0 & R \\
DCL31-A2 & 00h54m31.21s & -37d37m43.2s & 192.4 & \nodata & \nodata & \nodata & 3.9$\pm$2.0 & 12.6$\pm$3.9 & \nodata & 7.7$\pm$2.4 & 1.441 & 7.2 &  \\
DCL31-A3 & 00h54m32.18s & -37d37m50.6s & 196.0 & \nodata & \nodata & \nodata & 3.5$\pm$1.0 & 18.6$\pm$2.6 & \nodata & 11.4$\pm$1.6 & 1.653 & 10.0 &  \\
DCL31-A4 & 00h54m31.95s & -37d37m44.9s & 195.2 & \nodata & \nodata & \nodata & 3.3$\pm$1.4 & 6.8$\pm$2.0 & \nodata & 4.2$\pm$1.2 & 1.12 & 8.6 &  \\
DCL31-A5 & 00h54m31.82s & -37d37m57.1s & 196.4 & \nodata & \nodata & \nodata & 2.6$\pm$1.7 & 11.4$\pm$4.3 & \nodata & 7.0$\pm$2.6 & 1.674 & 5.9 &  \\
DCL31-A6 & 00h54m31.47s & -37d37m56.1s & 199.8 & \nodata & \nodata & \nodata & 2.7$\pm$1.3 & 10.3$\pm$7.5 & \nodata & 6.3$\pm$4.6 & 1.56 & 4.9 & DCL30-A4 \\
DCL34-A1 & 00h54m32.74s & -37d38m47.2s & 186.9 & 1.91 & 69.9 & 22.9$\pm$2.3 & 4.2$\pm$0.5 & 78.3$\pm$5.5 & 75.0$\pm$20.2 & 51.2$\pm$3.6 & 2.067 & 14.0 & F \\
DCL34-A2 & 00h54m32.46s & -37d38m44.2s & 185.5 & 10.90 & -85.1 & 9.3$\pm$2.3 & 2.9$\pm$0.8 & 18.2$\pm$2.0 & 14.6$\pm$9.2 & 11.9$\pm$1.3 & 1.545 & 11.2 & F \\
DCL34-A3 & 00h54m33.26s & -37d38m50.4s & 186.8 & \nodata & \nodata & \nodata & 3.3$\pm$1.5 & 9.4$\pm$3.6 & \nodata & 6.2$\pm$2.3 & 1.499 & 7.3 &  \\
DCL34-A4 & 00h54m31.95s & -37d38m43.4s & 186.4 & \nodata & \nodata & \nodata & 2.7$\pm$1.7 & 6.1$\pm$4.3 & \nodata & 4.0$\pm$2.8 & 0.903 & 5.0 &  \\
DCL34-A5 & 00h54m33.12s & -37d38m27.9s & 188.0 & \nodata & \nodata & \nodata & 4.5$\pm$1.5 & 18.1$\pm$6.5 & \nodata & 11.8$\pm$4.2 & 2.374 & 6.7 &  \\
DCL34-A6 & 00h54m33.10s & -37d38m40.8s & 189.8 & \nodata & \nodata & \nodata & 4.1$\pm$0.8 & 15.9$\pm$2.7 & \nodata & 10.4$\pm$1.8 & 1.395 & 9.2 &  \\
DCL34-A7 & 00h54m32.31s & -37d38m35.7s & 192.2 & 2.16 & 77.1 & 36.6$\pm$2.2 & 5.0$\pm$0.5 & 181.0$\pm$7.2 & 174.0$\pm$34.8 & 118.0$\pm$4.7 & 2.575 & 14.7 & F \\
DCL34-A8 & 00h54m31.74s & -37d38m37.3s & 194.2 & \nodata & \nodata & \nodata & 2.0$\pm$1.1 & 13.0$\pm$4.0 & \nodata & 8.5$\pm$2.6 & 1.739 & 6.6 &  \\
DCL37-A1 & 00h54m35.32s & -37d39m35.7s & 181.0 & \nodata & \nodata & \nodata & 4.0$\pm$0.8 & 31.3$\pm$2.8 & \nodata & 23.3$\pm$2.1 & 1.936 & 13.5 &  \\
DCL37-A2 & 00h54m35.81s & -37d39m39.5s & 180.3 & \nodata & \nodata & \nodata & 3.4$\pm$0.7 & 23.1$\pm$3.5 & \nodata & 17.1$\pm$2.6 & 2.604 & 13.9 &  \\
DCL37-A3 & 00h54m34.86s & -37d39m42.9s & 186.0 & 3.53 & -31.1 & 19.4$\pm$1.9 & 6.3$\pm$0.6 & 102.0$\pm$8.2 & 146.0$\pm$38.0 & 75.9$\pm$6.1 & 3.357 & 13.3 & F \\
DCL37-A4 & 00h54m35.86s & -37d39m21.7s & 181.3 & 3.76 & -50.4 & 4.7$\pm$4.7 & 4.5$\pm$2.4 & 8.3$\pm$4.7 & 17.5$\pm$25.9 & 6.2$\pm$3.5 & 1.2 & 4.9 & R \\
DCL37-A5 & 00h54m35.64s & -37d39m47.0s & 184.6 & \nodata & \nodata & \nodata & 3.8$\pm$1.3 & 41.6$\pm$10.4 & \nodata & 30.9$\pm$7.7 & 2.682 & 7.2 &  \\
DCL37-A6 & 00h54m35.27s & -37d39m44.3s & 186.0 & \nodata & \nodata & \nodata & 3.5$\pm$0.7 & 30.1$\pm$3.6 & \nodata & 22.4$\pm$2.7 & 2.429 & 9.8 &  \\
DCL37-A7 & 00h54m35.34s & -37d39m27.5s & 185.4 & \nodata & \nodata & \nodata & 4.1$\pm$1.4 & 11.5$\pm$2.9 & \nodata & 8.6$\pm$2.1 & 1.226 & 8.2 &  \\
DCL41-A1 & 00h54m40.11s & -37d41m17.2s & 158.7 & 10.29 & 13.6 & 7.5$\pm$1.3 & 4.3$\pm$0.7 & 90.9$\pm$10.9 & 26.4$\pm$9.0 & 73.4$\pm$8.8 & 4.592 & 9.3 & O \\
DCL41-A2 & 00h54m39.33s & -37d41m20.3s & 163.5 & 2.03 & 78.0 & 52.6$\pm$3.7 & 8.0$\pm$0.4 & 462.0$\pm$13.9 & 637.0$\pm$82.8 & 373.0$\pm$11.2 & 3.191 & 18.9 & F \\
DCL41-A3 & 00h54m38.77s & -37d41m16.6s & 173.3 & 2.47 & 9.9 & 18.1$\pm$3.4 & 6.9$\pm$1.1 & 45.2$\pm$5.9 & 163.0$\pm$63.6 & 36.5$\pm$4.7 & 1.293 & 7.8 & F \\
DCL41-A4 & 00h54m38.75s & -37d41m26.6s & 168.4 & 3.70 & -51.3 & 31.2$\pm$2.2 & 7.2$\pm$0.5 & 215.0$\pm$10.8 & 304.0$\pm$45.6 & 173.0$\pm$8.7 & 4.696 & 31.2 & F \\
DCL41-A5 & 00h54m39.32s & -37d41m32.8s & 168.3 & 1.70 & 45.7 & 15.9$\pm$3.2 & 3.5$\pm$0.8 & 35.3$\pm$7.8 & 36.0$\pm$17.6 & 28.5$\pm$6.3 & 1.729 & 6.6 & F \\
DCL41-A6 & 00h54m38.64s & -37d41m32.3s & 170.6 & 5.49 & -82.7 & 13.1$\pm$3.9 & 3.2$\pm$1.7 & 14.1$\pm$4.2 & 25.7$\pm$26.0 & 11.3$\pm$3.4 & 1.307 & 6.7 & F \\
DCL41-A7 & 00h54m39.38s & -37d41m38.1s & 176.5 & \nodata & \nodata & \nodata & 2.6$\pm$0.9 & 21.7$\pm$5.9 & \nodata & 17.6$\pm$4.8 & 2.868 & 6.9 & O \\
DCL41-A8 & 00h54m38.80s & -37d41m20.9s & 177.0 & \nodata & \nodata & \nodata & 4.1$\pm$1.4 & 7.1$\pm$2.0 & \nodata & 5.8$\pm$1.6 & 1.01 & 7.2 &  \\
DCL41-A9 & 00h54m38.55s & -37d41m24.7s & 178.6 & 3.00 & -54.3 & 18.2$\pm$1.6 & 3.9$\pm$0.6 & 45.2$\pm$3.2 & 53.2$\pm$18.6 & 36.5$\pm$2.6 & 2.051 & 14.4 & F \\
DCL45-A1 & 00h54m40.69s & -37d40m54.5s & 161.4 & 1.94 & 76.1 & 16.4$\pm$1.5 & 5.9$\pm$0.5 & 72.5$\pm$4.3 & 107.0$\pm$23.5 & 62.0$\pm$3.7 & 2.436 & 17.1 & F \\
DCL45-A2 & 00h54m40.25s & -37d40m53.0s & 161.2 & \nodata & \nodata & \nodata & 3.1$\pm$0.6 & 20.5$\pm$2.7 & \nodata & 17.5$\pm$2.3 & 2.908 & 21.4 &  \\
DCL45-A3 & 00h54m41.33s & -37d40m45.1s & 168.5 & 6.93 & -77.3 & 10.7$\pm$2.4 & 3.0$\pm$1.2 & 15.6$\pm$2.3 & 17.8$\pm$13.2 & 13.3$\pm$2.0 & 2.184 & 9.6 & F \\
DCL45-A4 & 00h54m41.26s & -37d40m50.0s & 169.7 & 3.56 & -22.8 & 19.1$\pm$2.5 & 3.5$\pm$0.7 & 38.6$\pm$4.6 & 43.6$\pm$19.6 & 33.0$\pm$4.0 & 1.74 & 9.1 & F \\
DCL45-A5 & 00h54m41.22s & -37d40m41.4s & 173.5 & \nodata & \nodata & \nodata & 3.1$\pm$1.2 & 17.5$\pm$5.1 & \nodata & 15.0$\pm$4.3 & 2.099 & 7.4 &  \\
DCL45-A6 & 00h54m39.50s & -37d40m59.8s & 179.4 & \nodata & \nodata & \nodata & 3.3$\pm$1.7 & 12.8$\pm$3.3 & \nodata & 10.9$\pm$2.8 & 2.353 & 7.5 &  \\
DCL46-A1 & 00h54m41.44s & -37d43m05.7s & 165.4 & 2.64 & -16.2 & 12.7$\pm$3.3 & 4.3$\pm$1.0 & 38.3$\pm$3.1 & 43.2$\pm$26.4 & 29.2$\pm$2.3 & 2.98 & 13.4 & F \\
DCL46-A2 & 00h54m41.58s & -37d42m59.9s & 165.0 & \nodata & \nodata & \nodata & 3.1$\pm$1.6 & 8.3$\pm$2.4 & \nodata & 6.3$\pm$1.8 & 1.884 & 7.5 &  \\
DCL46-A3 & 00h54m40.58s & -37d43m03.4s & 173.7 & 4.92 & -57.4 & 11.1$\pm$4.6 & 3.6$\pm$1.0 & 12.8$\pm$1.7 & 27.1$\pm$17.1 & 9.7$\pm$1.3 & 1.267 & 9.5 & F \\
DCL49-A1 & 00h54m41.92s & -37d39m05.3s & 169.3 & 4.14 & -83.8 & 4.4$\pm$1.5 & 4.2$\pm$0.8 & 28.1$\pm$3.1 & 14.9$\pm$8.3 & 23.9$\pm$2.6 & 3.217 & 23.7 & R \\
DCL49-A2 & 00h54m42.23s & -37d38m58.6s & 169.4 & 3.85 & 26.6 & 14.9$\pm$2.8 & 5.9$\pm$0.8 & 32.6$\pm$3.3 & 97.3$\pm$34.1 & 27.8$\pm$2.8 & 1.256 & 8.8 & F \\
DCL49-A3 & 00h54m42.33s & -37d39m14.0s & 170.8 & \nodata & \nodata & \nodata & 3.5$\pm$0.8 & 28.2$\pm$4.5 & \nodata & 24.1$\pm$3.9 & 2.051 & 9.0 &  \\
DCL49-A4 & 00h54m41.84s & -37d39m12.5s & 171.5 & 1.29 & -43.6 & 14.3$\pm$2.0 & 3.7$\pm$0.6 & 42.9$\pm$3.4 & 36.8$\pm$12.9 & 36.7$\pm$2.9 & 3.04 & 14.7 & F \\
DCL49-A5 & 00h54m41.88s & -37d39m01.0s & 175.2 & \nodata & \nodata & \nodata & 5.3$\pm$1.8 & 8.3$\pm$1.8 & \nodata & 7.1$\pm$1.6 & 1.011 & 7.5 &  \\
DCL49-A6 & 00h54m41.10s & -37d38m51.5s & 177.1 & \nodata & \nodata & \nodata & 6.9$\pm$1.0 & 77.0$\pm$9.2 & \nodata & 65.9$\pm$7.9 & 3.461 & 7.9 & O \\
DCL49-A7 & 00h54m41.31s & -37d38m55.2s & 179.0 & \nodata & \nodata & \nodata & 3.8$\pm$1.7 & 9.5$\pm$5.4 & \nodata & 8.1$\pm$4.6 & 1.335 & 5.2 &  \\
DCL52-A1 & 00h54m43.09s & -37d40m04.9s & 176.1 & 15.33 & 88.0 & 11.4$\pm$1.1 & 4.5$\pm$0.5 & 46.1$\pm$2.8 & 43.9$\pm$10.1 & 41.5$\pm$2.5 & 2.09 & 14.2 & F \\
DCL52-A2 & 00h54m42.97s & -37d40m03.7s & 175.5 & \nodata & \nodata & \nodata & 2.8$\pm$1.3 & 4.8$\pm$1.4 & \nodata & 4.4$\pm$1.2 & 1.08 & 7.6 &  \\
DCL52-A3 & 00h54m43.04s & -37d40m08.9s & 174.0 & \nodata & \nodata & \nodata & 3.2$\pm$1.4 & 6.5$\pm$2.8 & \nodata & 5.9$\pm$2.5 & 1.007 & 6.0 &  \\
DCL52-A4 & 00h54m43.42s & -37d39m52.9s & 173.7 & \nodata & \nodata & \nodata & 2.6$\pm$0.7 & 16.6$\pm$3.2 & \nodata & 14.9$\pm$2.8 & 1.753 & 8.4 &  \\
DCL52-A5 & 00h54m43.17s & -37d40m02.3s & 177.4 & \nodata & \nodata & \nodata & 2.2$\pm$1.5 & 4.9$\pm$0.8 & \nodata & 4.5$\pm$0.8 & 1.535 & 11.6 &  \\
DCL52-A6 & 00h54m41.92s & -37d40m19.4s & 177.1 & 4.07 & -25.3 & 5.3$\pm$2.5 & 3.4$\pm$1.0 & 75.8$\pm$18.2 & 11.4$\pm$7.6 & 68.4$\pm$16.4 & 9.746 & 8.2 & O \\
DCL52-A7 & 00h54m42.45s & -37d40m15.2s & 178.2 & 2.15 & 15.2 & 15.7$\pm$1.9 & 2.9$\pm$0.5 & 79.7$\pm$6.4 & 24.6$\pm$8.9 & 71.8$\pm$5.7 & 4.553 & 13.9 & F \\
DCL52-A8 & 00h54m42.10s & -37d40m00.7s & 184.6 & 5.15 & 73.3 & 22.1$\pm$2.0 & 6.2$\pm$0.5 & 137.0$\pm$8.2 & 157.0$\pm$33.0 & 123.0$\pm$7.4 & 3.979 & 21.1 & F \\
DCL52-A9 & 00h54m42.83s & -37d40m02.2s & 181.1 & \nodata & \nodata & \nodata & 3.3$\pm$1.0 & 6.5$\pm$1.4 & \nodata & 5.8$\pm$1.3 & 1.165 & 8.6 &  \\
DCL52-A10 & 00h54m42.75s & -37d40m06.4s & 183.8 & \nodata & \nodata & \nodata & 3.0$\pm$1.0 & 6.5$\pm$1.4 & \nodata & 5.8$\pm$1.3 & 1.524 & 10.2 &  \\
DCL53C-A1 & 00h54m43.24s & -37d43m09.4s & 156.9 & 3.73 & 52.4 & 50.4$\pm$3.0 & 4.3$\pm$0.3 & 321.0$\pm$16.1 & 175.0$\pm$29.8 & 260.0$\pm$13.0 & 6.011 & 15.8 & F \\
DCL53C-A2 & 00h54m42.53s & -37d43m09.7s & 157.0 & \nodata & \nodata & \nodata & 4.9$\pm$1.8 & 11.8$\pm$6.8 & \nodata & 9.6$\pm$5.6 & 1.941 & 5.5 &  \\
DCL53C-A3 & 00h54m42.90s & -37d42m54.3s & 163.8 & 1.68 & -30.0 & 11.0$\pm$1.9 & 3.6$\pm$0.6 & 54.2$\pm$5.4 & 26.8$\pm$11.8 & 43.7$\pm$4.4 & 4.124 & 31.1 & F \\
DCL53C-A4 & 00h54m43.27s & -37d42m52.7s & 162.3 & 3.32 & 40.0 & 12.9$\pm$2.6 & 2.6$\pm$0.8 & 16.9$\pm$2.9 & 17.0$\pm$11.6 & 13.6$\pm$2.3 & 1.267 & 8.4 & F \\
DCL53C-A5 & 00h54m42.46s & -37d42m44.1s & 166.4 & 2.83 & -87.5 & 9.6$\pm$3.0 & 3.7$\pm$0.7 & 25.3$\pm$5.6 & 24.2$\pm$13.6 & 20.4$\pm$4.5 & 1.919 & 8.6 & F \\
DCL53C-A6 & 00h54m43.46s & -37d42m54.5s & 166.5 & \nodata & \nodata & \nodata & 2.4$\pm$1.1 & 5.6$\pm$2.4 & \nodata & 4.5$\pm$1.9 & 1.059 & 6.4 &  \\
DCL56-A1 & 00h54m43.85s & -37d40m31.7s & 162.5 & \nodata & \nodata & \nodata & 1.6$\pm$1.3 & 3.2$\pm$1.4 & \nodata & 3.1$\pm$1.4 & 1.352 & 6.9 &  \\
DCL56-A2 & 00h54m44.30s & -37d40m27.1s & 167.8 & 4.32 & -42.1 & 21.6$\pm$1.9 & 3.9$\pm$0.4 & 94.1$\pm$5.6 & 62.2$\pm$11.8 & 89.3$\pm$5.4 & 3.498 & 24.1 & F \\
DCL56-A3 & 00h54m44.05s & -37d40m22.2s & 169.8 & 5.60 & 82.3 & 12.0$\pm$1.8 & 3.9$\pm$0.6 & 23.7$\pm$2.4 & 34.9$\pm$12.6 & 22.5$\pm$2.2 & 1.899 & 14.2 & F \\
DCL61-A1 & 00h54m46.19s & -37d38m34.4s & 156.6 & \nodata & \nodata & \nodata & 3.5$\pm$1.0 & 12.4$\pm$3.2 & \nodata & 10.8$\pm$2.8 & 2.098 & 7.8 &  \\
DCL61-A2 & 00h54m45.11s & -37d38m32.6s & 158.3 & \nodata & \nodata & \nodata & 3.3$\pm$0.9 & 17.3$\pm$2.9 & \nodata & 15.1$\pm$2.6 & 2.016 & 8.1 &  \\
DCL61-A3 & 00h54m45.44s & -37d38m41.9s & 159.9 & 3.76 & 10.7 & 5.5$\pm$1.5 & 4.4$\pm$1.1 & 18.3$\pm$2.0 & 19.6$\pm$10.8 & 15.9$\pm$1.7 & 2.186 & 16.3 & F \\
DCL61-A4 & 00h54m45.67s & -37d38m34.1s & 159.0 & 2.24 & -41.0 & 8.5$\pm$2.0 & 3.8$\pm$0.9 & 25.9$\pm$3.6 & 22.4$\pm$11.4 & 22.6$\pm$3.2 & 2.093 & 10.0 & F \\
DCL61-A5 & 00h54m45.32s & -37d38m45.6s & 157.5 & \nodata & \nodata & \nodata & 3.2$\pm$2.1 & 3.3$\pm$1.2 & \nodata & 2.9$\pm$1.1 & 1.009 & 7.0 &  \\
DCL61-A6 & 00h54m45.41s & -37d38m45.0s & 162.7 & 3.37 & 78.5 & 17.8$\pm$2.1 & 3.8$\pm$0.6 & 41.1$\pm$2.9 & 47.3$\pm$17.0 & 35.6$\pm$2.5 & 2.104 & 15.7 & F \\
DCL61-A7 & 00h54m45.51s & -37d38m46.9s & 161.0 & \nodata & \nodata & \nodata & 4.0$\pm$2.1 & 4.4$\pm$2.7 & \nodata & 3.8$\pm$2.4 & 0.783 & 5.5 &  \\
DCL61-A8 & 00h54m45.21s & -37d38m49.0s & 161.3 & 1.58 & -51.2 & 15.3$\pm$2.8 & 2.6$\pm$0.5 & 21.7$\pm$2.2 & 19.6$\pm$9.8 & 19.0$\pm$1.9 & 1.881 & 12.5 & F \\
DCL61-A9 & 00h54m45.50s & -37d38m37.1s & 161.2 & \nodata & \nodata & \nodata & 2.1$\pm$0.7 & 5.0$\pm$1.3 & \nodata & 4.3$\pm$1.1 & 1.376 & 8.9 &  \\
DCL63-A1 & 00h54m45.16s & -37d37m50.0s & 149.8 & \nodata & \nodata & \nodata & 1.7$\pm$0.9 & 6.1$\pm$1.3 & \nodata & 4.8$\pm$1.0 & 1.4 & 9.5 &  \\
DCL63-A2 & 00h54m45.55s & -37d37m59.1s & 152.5 & \nodata & \nodata & \nodata & 2.9$\pm$0.9 & 11.9$\pm$3.8 & \nodata & 9.4$\pm$3.0 & 0.929 & 6.2 &  \\
DCL63-A3 & 00h54m45.90s & -37d37m56.4s & 152.0 & \nodata & \nodata & \nodata & 3.8$\pm$1.2 & 10.3$\pm$1.2 & \nodata & 8.1$\pm$1.0 & 1.723 & 12.2 &  \\
DCL63-A4 & 00h54m45.29s & -37d37m54.9s & 152.9 & 1.99 & -24.6 & 13.2$\pm$2.0 & 3.8$\pm$0.6 & 25.0$\pm$2.2 & 34.8$\pm$13.6 & 19.8$\pm$1.8 & 1.75 & 12.8 & F \\
DCL63-A5 & 00h54m44.85s & -37d37m52.6s & 153.8 & \nodata & \nodata & \nodata & 2.8$\pm$1.1 & 7.9$\pm$2.4 & \nodata & 6.2$\pm$1.9 & 1.378 & 8.7 &  \\
DCL65-A1 & 00h54m46.20s & -37d36m35.0s & 154.0 & \nodata & \nodata & \nodata & 2.2$\pm$0.8 & 6.5$\pm$2.2 & \nodata & 4.4$\pm$1.5 & 1.427 & 8.4 &  \\
DCL65-A2 & 00h54m46.44s & -37d36m30.2s & 157.6 & \nodata & \nodata & \nodata & 4.2$\pm$1.4 & 9.3$\pm$3.0 & \nodata & 6.2$\pm$2.0 & 0.802 & 5.9 &  \\
DCL65-A3 & 00h54m45.24s & -37d36m25.3s & 159.2 & 4.82 & -83.6 & 9.4$\pm$2.4 & 5.2$\pm$1.0 & 75.8$\pm$16.7 & 47.9$\pm$20.1 & 50.6$\pm$11.1 & 3.469 & 8.0 & O \\
DCL65-A4 & 00h54m45.57s & -37d36m29.9s & 159.7 & 1.61 & 9.3 & 10.5$\pm$2.9 & 5.0$\pm$1.4 & 35.7$\pm$7.5 & 49.0$\pm$30.9 & 23.9$\pm$5.0 & 1.649 & 6.6 & F \\
DCL65-A5 & 00h54m46.77s & -37d36m33.7s & 159.3 & 2.60 & 86.4 & 16.0$\pm$1.9 & 2.8$\pm$0.4 & 51.5$\pm$4.1 & 23.6$\pm$8.0 & 34.5$\pm$2.8 & 2.3 & 15.1 & F \\
DCL66-A1 & 00h54m46.93s & -37d37m57.6s & 148.7 & \nodata & \nodata & \nodata & 2.6$\pm$1.0 & 5.3$\pm$2.0 & \nodata & 4.3$\pm$1.6 & 1.185 & 8.8 &  \\
DCL66-A2 & 00h54m47.98s & -37d38m07.8s & 153.5 & 1.68 & -73.2 & 15.1$\pm$2.6 & 5.5$\pm$1.0 & 82.6$\pm$12.4 & 86.4$\pm$35.4 & 66.8$\pm$10.0 & 3.749 & 8.0 & O; DCL68-A2 \\
DCL66-A3 & 00h54m46.71s & -37d38m02.1s & 152.3 & 2.67 & -61.8 & 16.2$\pm$2.4 & 3.6$\pm$0.7 & 32.0$\pm$3.2 & 40.2$\pm$16.9 & 25.9$\pm$2.6 & 1.979 & 12.6 & F \\
DCL66-A4 & 00h54m47.85s & -37d38m03.2s & 154.3 & \nodata & \nodata & \nodata & 3.5$\pm$0.6 & 40.6$\pm$4.1 & \nodata & 32.8$\pm$3.3 & 4.13 & 15.5 & DCL68-A4 \\
DCL66-A5 & 00h54m47.16s & -37d37m53.1s & 164.8 & \nodata & \nodata & \nodata & 3.4$\pm$1.0 & 6.3$\pm$2.2 & \nodata & 5.1$\pm$1.8 & 0.895 & 6.1 &  \\
DCL68-A1 & 00h54m46.95s & -37d37m57.7s & 148.6 & \nodata & \nodata & \nodata & 3.8$\pm$1.6 & 7.2$\pm$2.9 & \nodata & 5.9$\pm$2.4 & 1.386 & 6.4 & DCL66-A1 \\
DCL68-A2 & 00h54m48.00s & -37d38m07.9s & 153.8 & 1.65 & 83.1 & 29.1$\pm$1.7 & 6.7$\pm$0.5 & 200.0$\pm$10.0 & 245.0$\pm$41.7 & 165.0$\pm$8.2 & 3.226 & 18.0 & F \\
DCL68-A3 & 00h54m46.71s & -37d38m02.3s & 151.7 & 5.06 & -61.7 & 14.5$\pm$4.2 & 3.3$\pm$0.9 & 19.2$\pm$5.4 & 29.2$\pm$19.3 & 15.8$\pm$4.4 & 1.959 & 6.8 & DCL66-A3 \\
DCL68-A4 & 00h54m47.86s & -37d38m03.2s & 154.5 & 1.88 & 18.6 & 12.3$\pm$2.7 & 3.7$\pm$0.6 & 50.9$\pm$4.6 & 31.9$\pm$15.0 & 41.8$\pm$3.8 & 3.533 & 24.4 & F \\
DCL68-A5 & 00h54m48.99s & -37d38m07.8s & 156.8 & \nodata & \nodata & \nodata & 6.4$\pm$2.5 & 19.2$\pm$8.4 & \nodata & 15.8$\pm$7.0 & 2.077 & 5.8 &  \\
DCL68-A6 & 00h54m49.11s & -37d38m09.8s & 158.4 & \nodata & \nodata & \nodata & 4.2$\pm$1.7 & 29.5$\pm$9.4 & \nodata & 24.4$\pm$7.8 & 2.825 & 6.0 & O \\
DCL69-A1 & 00h54m48.26s & -37d43m43.0s & 137.5 & 1.93 & 58.4 & 13.8$\pm$6.1 & 4.2$\pm$1.2 & 16.7$\pm$3.3 & 46.2$\pm$32.8 & 13.9$\pm$2.8 & 0.933 & 6.8 & F \\
DCL69-A2 & 00h54m48.21s & -37d43m37.2s & 139.4 & 2.70 & -81.1 & 23.4$\pm$4.4 & 3.7$\pm$0.9 & 25.7$\pm$3.9 & 59.2$\pm$34.3 & 21.6$\pm$3.2 & 1.094 & 7.5 & F \\
DCL69-A3 & 00h54m48.24s & -37d43m37.6s & 141.9 & 2.39 & -30.2 & 57.8$\pm$1.7 & 7.1$\pm$0.4 & 369.0$\pm$11.1 & 547.0$\pm$65.6 & 309.0$\pm$9.3 & 3.936 & 27.2 & F \\
DCL69-A4 & 00h54m49.01s & -37d43m46.2s & 147.0 & \nodata & \nodata & \nodata & 6.5$\pm$2.4 & 15.5$\pm$6.2 & \nodata & 13.0$\pm$5.2 & 1.197 & 5.8 &  \\
DCL76C-A1 & 00h54m50.30s & -37d40m38.2s & 147.8 & 2.04 & 72.6 & 18.7$\pm$2.2 & 6.4$\pm$0.8 & 95.6$\pm$9.6 & 144.0$\pm$43.2 & 103.0$\pm$10.3 & 3.013 & 7.6 & F \\
DCL76C-A2 & 00h54m51.08s & -37d40m37.2s & 143.9 & \nodata & \nodata & \nodata & 2.3$\pm$1.2 & 7.0$\pm$2.9 & \nodata & 7.6$\pm$3.1 & 2.072 & 7.2 &  \\
DCL76C-A3 & 00h54m50.37s & -37d40m29.3s & 150.5 & 1.25 & 2.3 & 19.0$\pm$1.3 & 5.6$\pm$0.4 & 107.0$\pm$7.5 & 111.0$\pm$22.2 & 115.0$\pm$8.1 & 5.348 & 29.2 & F \\
DCL76C-A4 & 00h54m50.25s & -37d40m26.2s & 156.4 & 1.52 & -46.0 & 10.4$\pm$4.5 & 4.0$\pm$1.6 & 8.1$\pm$1.9 & 30.6$\pm$27.8 & 8.7$\pm$2.1 & 1.494 & 8.6 & F \\
DCL76C-A5 & 00h54m49.81s & -37d40m29.4s & 152.1 & \nodata & \nodata & \nodata & 3.5$\pm$1.4 & 12.7$\pm$4.6 & \nodata & 13.6$\pm$4.9 & 2.136 & 6.6 &  \\
DCL76C-A6 & 00h54m51.48s & -37d40m27.3s & 155.5 & 3.22 & -18.2 & 9.6$\pm$1.7 & 3.9$\pm$0.7 & 24.9$\pm$2.5 & 27.2$\pm$10.3 & 26.7$\pm$2.7 & 2.013 & 11.7 & F \\
DCL76C-A7 & 00h54m50.85s & -37d40m24.4s & 160.0 & 1.71 & 4.4 & 40.0$\pm$2.0 & 4.9$\pm$0.5 & 77.4$\pm$3.9 & 178.0$\pm$39.2 & 83.0$\pm$4.2 & 3.982 & 28.9 & F \\
DCL76C-A8 & 00h54m50.60s & -37d40m24.9s & 166.5 & \nodata & \nodata & \nodata & 2.0$\pm$1.8 & 1.4$\pm$1.7 & \nodata & 1.5$\pm$1.8 & 0.575 & 4.5 &  \\
DCL76C-A9 & 00h54m50.72s & -37d40m16.7s & 156.3 & 2.46 & 4.0 & 8.6$\pm$1.9 & 4.8$\pm$0.7 & 37.9$\pm$3.0 & 37.0$\pm$16.7 & 40.6$\pm$3.2 & 3.914 & 24.4 & F \\
DCL76C-A10 & 00h54m51.64s & -37d40m30.9s & 156.5 & 16.00 & -30.6 & 2.4$\pm$1.3 & 2.1$\pm$0.9 & 8.8$\pm$2.6 & 2.1$\pm$1.9 & 9.5$\pm$2.8 & 1.508 & 6.7 & R \\
DCL76C-A11 & 00h54m50.44s & -37d40m26.1s & 162.5 & \nodata & \nodata & \nodata & 3.7$\pm$1.7 & 5.6$\pm$3.2 & \nodata & 6.0$\pm$3.4 & 0.781 & 5.0 &  \\
DCL77-A1 & 00h54m49.81s & -37d38m12.5s & 155.3 & 1.19 & -53.3 & 42.6$\pm$2.1 & 6.5$\pm$0.5 & 261.0$\pm$13.1 & 337.0$\pm$53.9 & 227.0$\pm$11.4 & 2.952 & 11.4 & F \\
DCL77-A2 & 00h54m50.46s & -37d38m17.5s & 156.1 & 1.43 & 81.4 & 33.3$\pm$3.3 & 6.6$\pm$1.0 & 66.2$\pm$14.6 & 269.0$\pm$91.5 & 57.7$\pm$12.7 & 2.035 & 9.0 & F \\
DCL77-A3 & 00h54m50.11s & -37d38m20.3s & 156.3 & 2.37 & -40.9 & 34.5$\pm$1.4 & 7.8$\pm$0.3 & 1020.0$\pm$40.8 & 395.0$\pm$35.5 & 885.0$\pm$35.4 & 8.931 & 48.1 & F \\
DCL77-A4 & 00h54m50.34s & -37d38m25.4s & 161.2 & 2.39 & 39.0 & 14.4$\pm$2.2 & 4.6$\pm$0.8 & 37.1$\pm$3.0 & 57.5$\pm$23.0 & 32.3$\pm$2.6 & 2.268 & 12.4 & F \\
DCL77-A5 & 00h54m51.39s & -37d38m22.5s & 155.2 & 1.76 & 46.4 & 13.4$\pm$2.2 & 5.9$\pm$0.9 & 103.0$\pm$11.3 & 88.8$\pm$33.7 & 89.5$\pm$9.8 & 5.126 & 9.2 & DCL79-A4 \\
DCL77-A6 & 00h54m50.83s & -37d38m07.1s & 156.2 & 4.43 & -33.9 & 12.2$\pm$2.2 & 6.8$\pm$1.1 & 88.3$\pm$24.7 & 105.0$\pm$46.2 & 76.8$\pm$21.5 & 2.885 & 4.9 & F \\
DCL77-A7 & 00h54m48.97s & -37d38m11.9s & 156.0 & 1.84 & -10.5 & 15.4$\pm$2.5 & 6.4$\pm$0.9 & 96.9$\pm$25.2 & 120.0$\pm$38.4 & 84.4$\pm$21.9 & 3.537 & 6.7 & F \\
DCL77-A8 & 00h54m49.68s & -37d38m21.6s & 158.6 & 2.76 & 11.7 & 6.2$\pm$3.4 & 2.8$\pm$1.3 & 13.2$\pm$5.4 & 9.3$\pm$10.9 & 11.5$\pm$4.7 & 1.082 & 5.5 & F \\
DCL77-A9 & 00h54m48.82s & -37d38m21.8s & 164.3 & \nodata & \nodata & \nodata & 4.2$\pm$1.7 & 11.9$\pm$6.8 & \nodata & 10.3$\pm$5.9 & 1.654 & 5.1 &  \\
DCL77-A10 & 00h54m51.15s & -37d38m23.8s & 165.3 & \nodata & \nodata & \nodata & 3.0$\pm$1.5 & 9.0$\pm$4.4 & \nodata & 7.9$\pm$3.9 & 1.732 & 6.6 & DCL79-A10 \\
DCL79-A1 & 00h54m50.25s & -37d38m23.1s & 151.5 & 1.64 & -43.8 & 12.3$\pm$2.0 & 6.9$\pm$0.9 & 54.6$\pm$5.5 & 108.0$\pm$34.6 & 47.4$\pm$4.7 & 2.699 & 12.3 & F \\
DCL79-A2 & 00h54m50.11s & -37d38m20.2s & 156.5 & 2.55 & -42.4 & 31.0$\pm$1.2 & 7.7$\pm$0.4 & 880.0$\pm$35.2 & 342.0$\pm$41.0 & 766.0$\pm$30.6 & 9.699 & 32.5 & DCL77-A3 \\
DCL79-A3 & 00h54m50.34s & -37d38m25.2s & 161.5 & 2.81 & 36.4 & 11.1$\pm$2.2 & 3.6$\pm$0.9 & 29.5$\pm$3.2 & 26.9$\pm$16.4 & 25.8$\pm$2.8 & 2.798 & 13.1 & DCL77-A4 \\
DCL79-A4 & 00h54m51.37s & -37d38m22.4s & 155.2 & 1.59 & 53.9 & 17.6$\pm$1.2 & 6.1$\pm$0.5 & 129.0$\pm$9.0 & 124.0$\pm$24.8 & 112.0$\pm$7.8 & 5.041 & 33.3 & F \\
DCL79-A5 & 00h54m50.34s & -37d38m17.3s & 157.0 & 5.86 & 51.6 & 3.3$\pm$2.8 & 3.6$\pm$1.4 & 14.5$\pm$5.8 & 8.2$\pm$11.8 & 12.6$\pm$5.0 & 1.661 & 6.0 & R \\
DCL79-A6 & 00h54m51.91s & -37d38m19.8s & 158.2 & \nodata & \nodata & \nodata & 3.2$\pm$1.3 & 5.1$\pm$4.2 & \nodata & 4.4$\pm$3.6 & 1.261 & 5.6 &  \\
DCL79-A7 & 00h54m51.70s & -37d38m30.1s & 159.8 & \nodata & \nodata & \nodata & 2.3$\pm$1.0 & 8.7$\pm$2.6 & \nodata & 7.6$\pm$2.3 & 1.479 & 6.7 &  \\
DCL79-A8 & 00h54m51.69s & -37d38m39.3s & 161.7 & \nodata & \nodata & \nodata & 2.2$\pm$1.7 & 9.4$\pm$6.9 & \nodata & 8.2$\pm$6.0 & 2.901 & 5.3 & O \\
DCL79-A9 & 00h54m50.91s & -37d38m26.2s & 162.0 & \nodata & \nodata & \nodata & 3.0$\pm$0.9 & 10.9$\pm$1.6 & \nodata & 9.5$\pm$1.4 & 1.752 & 12.0 &  \\
DCL79-A10 & 00h54m51.13s & -37d38m24.0s & 165.6 & \nodata & \nodata & \nodata & 4.2$\pm$1.3 & 11.1$\pm$1.9 & \nodata & 9.7$\pm$1.6 & 1.379 & 10.7 &  \\
DCL80-A1 & 00h54m51.41s & -37d41m12.3s & 140.1 & 5.06 & 82.6 & 10.3$\pm$2.4 & 3.5$\pm$0.8 & 26.0$\pm$4.4 & 23.4$\pm$13.8 & 28.8$\pm$4.9 & 2.677 & 8.5 & F \\
DCL80-A2 & 00h54m51.30s & -37d41m05.2s & 148.0 & \nodata & \nodata & \nodata & 2.3$\pm$0.9 & 4.3$\pm$1.3 & \nodata & 4.7$\pm$1.5 & 1.123 & 7.3 &  \\
DCL80-A3 & 00h54m51.49s & -37d40m57.1s & 147.7 & \nodata & \nodata & \nodata & 2.1$\pm$0.9 & 4.8$\pm$1.0 & \nodata & 5.4$\pm$1.1 & 1.346 & 9.9 &  \\
DCL80-A4 & 00h54m50.52s & -37d40m57.8s & 151.1 & 7.28 & -76.9 & 45.2$\pm$2.7 & 3.2$\pm$0.3 & 88.7$\pm$4.4 & 88.8$\pm$17.8 & 97.8$\pm$4.9 & 3.176 & 21.2 & F \\
DCL81-A1 & 00h54m52.53s & -37d41m36.8s & 135.1 & \nodata & \nodata & \nodata & 3.1$\pm$1.1 & 12.8$\pm$3.7 & \nodata & 13.5$\pm$3.9 & 2.538 & 7.9 & DCL85-A1 \\
DCL81-A2 & 00h54m52.11s & -37d41m36.5s & 137.3 & 2.42 & -58.9 & 6.8$\pm$2.6 & 4.2$\pm$0.7 & 14.9$\pm$2.7 & 22.6$\pm$11.1 & 15.8$\pm$2.8 & 1.91 & 9.2 & DCL85-A3 \\
DCL81-A3 & 00h54m52.11s & -37d41m30.9s & 136.4 & \nodata & \nodata & \nodata & 3.3$\pm$2.2 & 6.9$\pm$3.8 & \nodata & 7.3$\pm$4.0 & 1.546 & 5.6 & DCL85-A4 \\
DCL81-A4 & 00h54m50.92s & -37d41m25.6s & 139.3 & 4.93 & 76.0 & 11.9$\pm$3.0 & 6.1$\pm$2.1 & 26.2$\pm$7.3 & 83.0$\pm$63.9 & 27.7$\pm$7.8 & 2.944 & 6.2 & O \\
DCL81-A5 & 00h54m52.44s & -37d41m40.2s & 135.5 & \nodata & \nodata & \nodata & 1.1$\pm$0.6 & 3.9$\pm$1.6 & \nodata & 4.1$\pm$1.7 & 1.929 & 7.6 & DCL85-A2 \\
DCL81-A6 & 00h54m51.76s & -37d41m53.1s & 139.4 & \nodata & \nodata & \nodata & 1.9$\pm$1.2 & 4.5$\pm$1.9 & \nodata & 4.7$\pm$2.0 & 1.657 & 6.6 &  \\
DCL81-A7 & 00h54m51.40s & -37d41m43.4s & 140.3 & \nodata & \nodata & \nodata & 2.1$\pm$0.4 & 15.5$\pm$2.0 & \nodata & 16.3$\pm$2.1 & 2.757 & 20.4 &  \\
DCL85-A1 & 00h54m52.52s & -37d41m36.8s & 134.9 & \nodata & \nodata & \nodata & 2.6$\pm$0.6 & 13.8$\pm$1.7 & \nodata & 14.8$\pm$1.8 & 2.754 & 17.2 &  \\
DCL85-A2 & 00h54m52.45s & -37d41m40.2s & 135.2 & \nodata & \nodata & \nodata & 2.9$\pm$1.2 & 4.4$\pm$1.1 & \nodata & 4.7$\pm$1.1 & 1.486 & 8.7 &  \\
DCL85-A3 & 00h54m52.10s & -37d41m36.5s & 137.5 & 2.59 & -71.7 & 6.9$\pm$1.9 & 4.0$\pm$0.6 & 16.7$\pm$1.7 & 20.3$\pm$8.7 & 17.9$\pm$1.8 & 1.812 & 13.0 & F \\
DCL85-A4 & 00h54m52.11s & -37d41m31.0s & 136.6 & \nodata & \nodata & \nodata & 2.6$\pm$1.1 & 5.8$\pm$1.0 & \nodata & 6.2$\pm$1.1 & 1.332 & 9.1 &  \\
DCL85-A5 & 00h54m50.89s & -37d41m25.7s & 138.6 & 11.48 & 73.5 & 7.6$\pm$1.5 & 4.0$\pm$1.2 & 25.0$\pm$4.5 & 22.7$\pm$15.2 & 26.8$\pm$4.8 & 2.842 & 7.4 & DCL81-A4 \\
DCL85-A6 & 00h54m51.22s & -37d41m29.7s & 138.6 & \nodata & \nodata & \nodata & 2.2$\pm$0.8 & 8.9$\pm$3.3 & \nodata & 9.6$\pm$3.5 & 1.347 & 6.4 &  \\
DCL85-A7 & 00h54m51.41s & -37d41m43.4s & 140.5 & 3.06 & 18.5 & 2.9$\pm$1.8 & 3.0$\pm$0.7 & 16.4$\pm$2.0 & 4.8$\pm$4.4 & 17.6$\pm$2.1 & 2.786 & 13.3 & DCL81-A7 \\
DCL86-A1 & 00h54m52.38s & -37d40m41.4s & 141.4 & 6.00 & -63.8 & 12.4$\pm$2.4 & 4.5$\pm$0.6 & 35.8$\pm$2.5 & 46.7$\pm$19.1 & 39.4$\pm$2.8 & 2.819 & 18.5 & F \\
DCL86-A2 & 00h54m52.92s & -37d40m43.1s & 141.9 & 2.56 & -88.3 & 7.4$\pm$3.0 & 3.3$\pm$0.9 & 10.5$\pm$2.1 & 14.8$\pm$10.1 & 11.6$\pm$2.3 & 1.609 & 8.7 & F \\
DCL86-A3 & 00h54m51.72s & -37d40m36.5s & 145.2 & \nodata & \nodata & \nodata & 3.1$\pm$1.6 & 4.8$\pm$1.5 & \nodata & 5.2$\pm$1.7 & 1.184 & 7.0 &  \\
DCL86-A4 & 00h54m52.75s & -37d40m31.7s & 149.9 & 2.05 & 65.4 & 15.0$\pm$3.6 & 7.2$\pm$1.3 & 20.4$\pm$2.4 & 146.0$\pm$75.9 & 22.5$\pm$2.7 & 1.388 & 8.7 & F \\
DCL86-A5 & 00h54m52.60s & -37d40m34.4s & 151.1 & 4.60 & -55.2 & 10.8$\pm$2.3 & 3.7$\pm$0.6 & 18.4$\pm$1.8 & 27.6$\pm$10.2 & 20.4$\pm$2.0 & 1.485 & 11.0 & F \\
DCL86-A6 & 00h54m52.81s & -37d40m27.0s & 150.1 & 2.11 & 71.8 & 14.0$\pm$2.5 & 3.8$\pm$0.8 & 24.5$\pm$4.7 & 37.2$\pm$18.6 & 27.0$\pm$5.1 & 1.89 & 8.7 & F \\
DCL86-A7 & 00h54m51.60s & -37d40m31.8s & 156.1 & \nodata & \nodata & \nodata & 3.0$\pm$1.0 & 13.8$\pm$3.7 & \nodata & 15.2$\pm$4.1 & 1.438 & 6.5 & DCL76C-A1 \\
DCL86-A8 & 00h54m51.47s & -37d40m27.1s & 155.6 & \nodata & \nodata & \nodata & 3.1$\pm$0.8 & 21.9$\pm$5.3 & \nodata & 24.2$\pm$5.8 & 2.682 & 8.1 & DCL76C-A6 \\
DCL88-A1 & 00h54m53.70s & -37d43m35.7s & 122.3 & 53.00 & -10.6 & 2.8$\pm$0.6 & 3.7$\pm$0.6 & 40.0$\pm$4.0 & 7.3$\pm$3.1 & 34.2$\pm$3.4 & 2.86 & 13.4 & R \\
DCL88-A2 & 00h54m53.78s & -37d43m51.5s & 125.6 & 1.75 & 48.2 & 20.4$\pm$3.5 & 2.1$\pm$0.3 & 44.5$\pm$4.0 & 16.6$\pm$5.8 & 38.0$\pm$3.4 & 2.276 & 10.1 & F \\
DCL88-A3 & 00h54m53.88s & -37d43m43.1s & 128.3 & 1.56 & 76.1 & 19.5$\pm$2.9 & 4.9$\pm$0.8 & 53.5$\pm$3.7 & 86.8$\pm$28.6 & 45.9$\pm$3.2 & 2.05 & 10.7 & F \\
DCL88-A4 & 00h54m53.32s & -37d43m30.7s & 125.6 & \nodata & \nodata & \nodata & 4.7$\pm$1.8 & 11.7$\pm$4.0 & \nodata & 10.0$\pm$3.4 & 1.644 & 6.1 &  \\
DCL88-A5 & 00h54m53.37s & -37d43m46.3s & 127.7 & \nodata & \nodata & \nodata & 2.4$\pm$0.7 & 14.6$\pm$1.5 & \nodata & 12.5$\pm$1.2 & 1.544 & 10.7 &  \\
DCL88-A6 & 00h54m53.07s & -37d43m46.2s & 135.8 & 3.64 & -81.9 & 51.8$\pm$2.6 & 9.3$\pm$0.5 & 293.0$\pm$8.8 & 838.0$\pm$92.2 & 251.0$\pm$7.5 & 3.313 & 22.8 & F \\
DCL88-A7 & 00h54m52.85s & -37d43m29.8s & 126.1 & \nodata & \nodata & \nodata & 3.1$\pm$1.4 & 11.9$\pm$4.2 & \nodata & 10.1$\pm$3.5 & 2.524 & 7.8 &  \\
DCL88-A8 & 00h54m51.95s & -37d43m44.5s & 139.5 & \nodata & \nodata & \nodata & 3.4$\pm$2.6 & 7.2$\pm$4.4 & \nodata & 6.1$\pm$3.8 & 1.644 & 5.5 &  \\
DCL93-A1 & 00h54m53.71s & -37d43m35.8s & 122.1 & 3.94 & 27.3 & 6.1$\pm$1.8 & 3.4$\pm$0.7 & 31.3$\pm$4.4 & 13.6$\pm$6.5 & 27.4$\pm$3.8 & 3.024 & 10.0 & DCL88-A1 \\
DCL93-A2 & 00h54m55.22s & -37d43m43.2s & 123.7 & 2.08 & -19.3 & 20.7$\pm$3.7 & 3.4$\pm$0.7 & 20.9$\pm$4.6 & 44.4$\pm$20.4 & 18.2$\pm$4.0 & 0.98 & 6.6 & F \\
DCL93-A3 & 00h54m53.96s & -37d43m32.5s & 123.2 & \nodata & \nodata & \nodata & 3.5$\pm$2.0 & 8.5$\pm$3.4 & \nodata & 7.4$\pm$2.9 & 1.788 & 6.6 &  \\
DCL93-A4 & 00h54m54.66s & -37d43m43.4s & 127.0 & 6.93 & 87.3 & 22.6$\pm$2.3 & 4.1$\pm$0.7 & 38.3$\pm$3.1 & 70.0$\pm$30.1 & 33.2$\pm$2.7 & 1.804 & 12.7 & F \\
DCL93-A5 & 00h54m53.82s & -37d43m51.6s & 125.9 & 2.33 & 42.8 & 12.3$\pm$2.5 & 2.5$\pm$0.6 & 43.4$\pm$5.6 & 14.9$\pm$7.6 & 37.7$\pm$4.9 & 3.486 & 9.4 & DCL88-A2 \\
DCL93-A6 & 00h54m53.94s & -37d43m43.0s & 127.9 & 2.71 & 81.6 & 23.7$\pm$2.4 & 4.0$\pm$0.5 & 72.4$\pm$5.1 & 70.9$\pm$19.9 & 63.0$\pm$4.4 & 2.727 & 13.1 & DCL88-A3 \\
DCL93-A7 & 00h54m55.20s & -37d43m33.0s & 130.7 & 1.22 & 78.9 & 12.5$\pm$2.4 & 4.4$\pm$0.9 & 37.8$\pm$3.4 & 44.6$\pm$21.9 & 32.9$\pm$3.0 & 2.159 & 10.8 & F \\
DCL93-A8 & 00h54m53.36s & -37d43m46.5s & 134.1 & 1.36 & -55.1 & 18.4$\pm$3.1 & 3.1$\pm$0.6 & 65.4$\pm$9.8 & 32.8$\pm$12.8 & 57.0$\pm$8.5 & 4.721 & 9.0 & O; DCL88-A5 \\
DCL93-A9 & 00h54m54.29s & -37d43m45.8s & 133.9 & \nodata & \nodata & \nodata & 3.9$\pm$1.4 & 6.7$\pm$2.9 & \nodata & 5.8$\pm$2.5 & 1.171 & 6.9 &  \\
DCL93-A10 & 00h54m54.33s & -37d43m43.4s & 135.1 & \nodata & \nodata & \nodata & 1.9$\pm$0.9 & 4.6$\pm$1.5 & \nodata & 4.0$\pm$1.3 & 1.209 & 8.1 &  \\
DCL93-A11 & 00h54m53.40s & -37d43m46.9s & 139.9 & \nodata & \nodata & \nodata & 3.1$\pm$2.0 & 16.1$\pm$8.1 & \nodata & 13.9$\pm$7.0 & 3.286 & 6.7 & O; DCL88-A5 \\
DCL93-A12 & 00h54m54.31s & -37d43m31.2s & 139.2 & \nodata & \nodata & \nodata & 2.2$\pm$1.1 & 5.4$\pm$3.3 & \nodata & 4.7$\pm$2.9 & 1.293 & 5.7 &  \\
DCL98-A1 & 00h54m55.96s & -37d40m36.6s & 130.3 & 4.65 & 74.6 & 17.4$\pm$1.9 & 3.6$\pm$0.5 & 58.6$\pm$3.5 & 43.6$\pm$14.0 & 62.1$\pm$3.7 & 2.597 & 13.6 & F \\
DCL98-A2 & 00h54m56.53s & -37d40m27.8s & 133.7 & 1.96 & 44.5 & 13.2$\pm$1.7 & 3.9$\pm$0.5 & 47.3$\pm$3.8 & 37.7$\pm$12.1 & 49.9$\pm$4.0 & 2.95 & 22.4 & F \\
DCL98-A3 & 00h54m55.24s & -37d40m36.2s & 134.8 & \nodata & \nodata & \nodata & 2.8$\pm$0.8 & 22.6$\pm$5.4 & \nodata & 24.0$\pm$5.8 & 3.566 & 9.0 &  \\
DCL100-A1 & 00h54m57.04s & -37d41m11.8s & 122.3 & 2.36 & 4.6 & 12.6$\pm$1.6 & 5.0$\pm$0.5 & 75.4$\pm$5.3 & 59.6$\pm$16.1 & 82.1$\pm$5.7 & 4.038 & 24.0 & F \\
DCL100-A2 & 00h54m57.20s & -37d41m18.4s & 121.4 & 3.96 & 83.2 & 10.9$\pm$2.4 & 3.0$\pm$0.7 & 18.8$\pm$3.4 & 18.1$\pm$9.4 & 20.5$\pm$3.7 & 2.226 & 8.2 & F \\
DCL100-A3 & 00h54m57.15s & -37d40m59.2s & 125.3 & 4.82 & -17.5 & 14.6$\pm$1.6 & 4.7$\pm$0.7 & 61.6$\pm$6.8 & 59.3$\pm$17.2 & 66.9$\pm$7.4 & 3.722 & 12.1 & F \\
DCL100-A4 & 00h54m55.58s & -37d41m12.8s & 128.8 & 1.12 & -41.7 & 6.2$\pm$4.4 & 2.9$\pm$1.3 & 5.3$\pm$1.5 & 9.9$\pm$13.0 & 5.8$\pm$1.6 & 1.466 & 7.6 & F \\
DCL103-A1 & 00h54m58.74s & -37d42m12.3s & 114.5 & \nodata & \nodata & \nodata & 4.7$\pm$2.3 & 18.8$\pm$10.3 & \nodata & 18.7$\pm$10.3 & 2.725 & 4.5 & O \\
DCL103-A2 & 00h54m57.59s & -37d42m10.0s & 116.3 & \nodata & \nodata & \nodata & 3.4$\pm$1.4 & 12.2$\pm$7.4 & \nodata & 12.1$\pm$7.4 & 1.554 & 5.0 &  \\
DCL103-A3 & 00h54m58.29s & -37d42m30.0s & 118.6 & 2.59 & -50.7 & 19.3$\pm$1.5 & 3.6$\pm$0.4 & 83.2$\pm$5.8 & 47.8$\pm$10.5 & 82.7$\pm$5.8 & 3.716 & 18.2 & F \\
DCL103-A4 & 00h54m58.13s & -37d42m25.1s & 118.8 & \nodata & \nodata & \nodata & 3.4$\pm$1.1 & 8.2$\pm$2.1 & \nodata & 8.1$\pm$2.1 & 1.175 & 7.2 &  \\
DCL103-A5 & 00h54m57.08s & -37d42m32.1s & 120.4 & \nodata & \nodata & \nodata & 2.6$\pm$1.0 & 9.8$\pm$2.6 & \nodata & 9.8$\pm$2.6 & 1.357 & 7.3 &  \\
DCL103-A6 & 00h54m57.47s & -37d42m28.7s & 124.1 & 1.99 & -70.1 & 9.6$\pm$1.8 & 4.5$\pm$0.6 & 55.8$\pm$5.0 & 37.2$\pm$11.9 & 55.5$\pm$5.0 & 4.41 & 30.1 & F \\
DCL103-A7 & 00h54m58.08s & -37d42m22.9s & 122.9 & \nodata & \nodata & \nodata & 2.4$\pm$1.2 & 4.3$\pm$1.5 & \nodata & 4.3$\pm$1.5 & 1.035 & 6.7 &  \\
DCL103-A8 & 00h54m57.58s & -37d42m22.2s & 122.5 & \nodata & \nodata & \nodata & 1.1$\pm$0.8 & 2.9$\pm$1.3 & \nodata & 2.9$\pm$1.3 & 0.882 & 6.3 &  \\
DCL103-A9 & 00h54m57.94s & -37d42m21.1s & 125.7 & 2.03 & 9.8 & 5.0$\pm$2.8 & 2.2$\pm$0.7 & 7.7$\pm$1.8 & 4.7$\pm$3.2 & 7.7$\pm$1.8 & 1.394 & 8.9 & F \\
DCL103-A10 & 00h54m57.56s & -37d42m16.5s & 127.8 & 1.99 & -39.5 & 15.0$\pm$2.1 & 3.7$\pm$0.6 & 30.1$\pm$2.4 & 38.5$\pm$13.9 & 29.8$\pm$2.4 & 1.652 & 9.6 & F \\
DCL103-A11 & 00h54m57.17s & -37d42m16.7s & 130.3 & 2.93 & -90.0 & 7.2$\pm$3.0 & 3.5$\pm$1.6 & 8.9$\pm$3.6 & 16.8$\pm$15.5 & 8.9$\pm$3.6 & 1.236 & 6.6 & F \\
DCL103-A12 & 00h54m56.98s & -37d42m19.9s & 130.0 & \nodata & \nodata & \nodata & 1.7$\pm$1.0 & 2.6$\pm$1.6 & \nodata & 2.6$\pm$1.6 & 0.935 & 5.5 &  \\
DCL109-A1 & 00h55m01.40s & -37d40m37.0s & 115.1 & 1.91 & 66.0 & 15.8$\pm$4.1 & 3.6$\pm$1.1 & 25.1$\pm$7.8 & 38.8$\pm$24.1 & 24.5$\pm$7.6 & 1.521 & 5.9 & F \\
DCL109-A2 & 00h55m01.17s & -37d40m31.7s & 116.6 & 2.22 & 4.4 & 24.1$\pm$1.9 & 6.3$\pm$0.8 & 77.5$\pm$9.3 & 178.0$\pm$49.8 & 76.0$\pm$9.1 & 1.625 & 8.4 & F \\
DCL109-A3 & 00h55m00.51s & -37d40m36.5s & 114.4 & \nodata & \nodata & \nodata & 1.9$\pm$1.2 & 3.0$\pm$1.1 & \nodata & 2.9$\pm$1.1 & 1.137 & 8.3 &  \\
DCL109-A4 & 00h55m01.41s & -37d40m31.8s & 114.0 & \nodata & \nodata & \nodata & 4.5$\pm$3.3 & 5.2$\pm$3.3 & \nodata & 5.1$\pm$3.3 & 1.264 & 5.4 &  \\
DCL109-A5 & 00h55m00.49s & -37d40m33.6s & 127.6 & 3.98 & -30.8 & 19.6$\pm$1.6 & 5.1$\pm$0.5 & 65.8$\pm$3.3 & 97.0$\pm$15.5 & 64.6$\pm$3.2 & 1.831 & 13.1 & F \\
DCL109-A6 & 00h55m00.89s & -37d40m23.8s & 126.9 & 1.59 & -17.5 & 34.5$\pm$2.4 & 4.2$\pm$0.4 & 83.6$\pm$8.4 & 112.0$\pm$24.6 & 81.9$\pm$8.2 & 2.087 & 9.1 & F \\
DCL109-A7 & 00h55m00.04s & -37d40m33.7s & 128.5 & \nodata & \nodata & \nodata & 2.0$\pm$1.2 & 3.6$\pm$1.8 & \nodata & 3.6$\pm$1.8 & 0.901 & 6.4 &  \\
DCL112-A1 & 00h55m02.01s & -37d44m19.3s & 96.3 & \nodata & \nodata & \nodata & 3.5$\pm$1.8 & 9.4$\pm$6.0 & \nodata & 7.6$\pm$4.8 & 1.424 & 5.3 &  \\
DCL112-A2 & 00h55m01.28s & -37d44m09.9s & 109.1 & \nodata & \nodata & \nodata & 1.8$\pm$0.4 & 12.4$\pm$1.6 & \nodata & 10.0$\pm$1.3 & 2.0 & 14.4 &  \\
DCL112-A3 & 00h55m01.39s & -37d44m04.6s & 112.8 & 2.55 & 64.6 & 27.5$\pm$2.2 & 4.3$\pm$0.4 & 78.6$\pm$3.9 & 94.4$\pm$19.8 & 63.5$\pm$3.2 & 2.266 & 16.9 & F \\
DCL112-A4 & 00h55m00.60s & -37d44m02.9s & 120.6 & 3.73 & 34.6 & 8.3$\pm$1.8 & 2.6$\pm$0.6 & 22.9$\pm$4.1 & 10.2$\pm$5.9 & 18.5$\pm$3.3 & 2.005 & 8.9 & F \\
DCL114-A1 & 00h55m03.65s & -37d39m49.9s & 106.5 & \nodata & \nodata & \nodata & 3.8$\pm$1.9 & 13.2$\pm$11.9 & \nodata & 11.8$\pm$10.6 & 1.748 & 5.1 &  \\
DCL114-A2 & 00h55m02.40s & -37d39m56.1s & 120.2 & 2.16 & -19.9 & 51.7$\pm$2.1 & 10.6$\pm$0.4 & 578.0$\pm$11.6 & 1090.0$\pm$119.9 & 513.0$\pm$10.3 & 3.683 & 24.6 & F \\
DCL114-A3 & 00h55m02.05s & -37d39m55.7s & 126.6 & \nodata & \nodata & \nodata & 2.8$\pm$1.1 & 5.1$\pm$1.8 & \nodata & 4.5$\pm$1.6 & 0.873 & 6.6 &  \\
DCL114-A4 & 00h55m02.06s & -37d39m41.6s & 130.1 & \nodata & \nodata & \nodata & 5.4$\pm$2.3 & 11.9$\pm$6.2 & \nodata & 10.5$\pm$5.5 & 1.237 & 5.1 &  \\
DCL114-A5 & 00h55m02.21s & -37d39m44.2s & 133.9 & \nodata & \nodata & \nodata & 4.2$\pm$1.1 & 29.0$\pm$2.3 & \nodata & 25.7$\pm$2.1 & 2.873 & 14.5 &  \\
DCL114-A6 & 00h55m01.96s & -37d39m47.6s & 132.0 & \nodata & \nodata & \nodata & 4.5$\pm$2.8 & 5.2$\pm$2.3 & \nodata & 4.6$\pm$2.1 & 0.972 & 6.1 &  \\
DCL118B-A1 & 00h55m04.70s & -37d42m56.3s & 92.4 & \nodata & \nodata & \nodata & 4.2$\pm$1.0 & 10.8$\pm$2.1 & \nodata & 9.4$\pm$1.8 & 1.469 & 10.1 &  \\
DCL118B-A2 & 00h55m05.37s & -37d42m57.4s & 96.1 & 2.50 & -37.2 & 3.1$\pm$3.0 & 2.7$\pm$1.0 & 12.0$\pm$3.1 & 4.3$\pm$6.1 & 10.4$\pm$2.7 & 1.896 & 7.5 & R \\
DCL118B-A3 & 00h55m03.84s & -37d42m41.5s & 98.7 & 2.39 & 77.2 & 9.2$\pm$3.0 & 5.4$\pm$0.9 & 45.8$\pm$4.6 & 50.0$\pm$24.5 & 39.8$\pm$4.0 & 4.345 & 18.6 & F \\
DCL118B-A4 & 00h55m05.82s & -37d42m58.4s & 97.1 & 1.64 & 76.4 & 9.5$\pm$4.7 & 4.0$\pm$1.2 & 27.0$\pm$13.2 & 27.8$\pm$20.6 & 23.6$\pm$11.6 & 2.944 & 6.0 & O \\
DCL118B-A5 & 00h55m05.74s & -37d42m46.1s & 100.0 & \nodata & \nodata & \nodata & 3.0$\pm$1.1 & 24.0$\pm$5.5 & \nodata & 20.9$\pm$4.8 & 2.361 & 7.9 &  \\
DCL118B-A6 & 00h55m05.20s & -37d42m36.2s & 99.2 & 3.80 & -71.0 & 6.6$\pm$3.4 & 2.4$\pm$1.2 & 10.4$\pm$4.7 & 7.2$\pm$8.7 & 9.1$\pm$4.1 & 2.066 & 6.0 & F \\
DCL118B-A7 & 00h55m03.85s & -37d42m52.8s & 102.5 & \nodata & \nodata & \nodata & 3.8$\pm$0.8 & 22.7$\pm$2.7 & \nodata & 19.8$\pm$2.4 & 3.116 & 16.2 &  \\
DCL118B-A8 & 00h55m03.61s & -37d42m48.1s & 103.1 & 2.14 & 88.6 & 7.2$\pm$2.4 & 4.2$\pm$1.0 & 29.6$\pm$2.7 & 24.5$\pm$16.4 & 25.7$\pm$2.3 & 3.312 & 14.7 & F \\
DCL118B-A9 & 00h55m05.05s & -37d42m37.8s & 105.0 & 4.69 & -41.4 & 18.1$\pm$2.2 & 5.8$\pm$0.7 & 87.1$\pm$5.2 & 113.0$\pm$33.9 & 75.7$\pm$4.5 & 4.335 & 16.8 & F \\
DCL118B-A1 & 00h55m05.70s & -37d42m41.1s & 103.0 & 1.55 & 64.7 & 13.9$\pm$2.8 & 3.6$\pm$0.7 & 46.6$\pm$7.5 & 33.8$\pm$15.2 & 40.6$\pm$6.5 & 3.644 & 9.8 & F \\
DCL119C-A1 & 00h55m03.78s & -37d43m27.6s & 106.6 & 1.59 & 63.2 & 9.3$\pm$3.8 & 2.9$\pm$1.1 & 31.7$\pm$8.2 & 14.6$\pm$13.3 & 27.5$\pm$7.2 & 2.867 & 5.1 & O \\
DCL119C-A2 & 00h55m03.53s & -37d43m21.1s & 108.1 & \nodata & \nodata & \nodata & 5.2$\pm$3.0 & 10.5$\pm$3.7 & \nodata & 9.1$\pm$3.2 & 1.442 & 6.8 &  \\
DCL119C-A3 & 00h55m02.59s & -37d43m20.7s & 105.9 & \nodata & \nodata & \nodata & 2.0$\pm$1.1 & 5.5$\pm$3.3 & \nodata & 4.8$\pm$2.9 & 0.901 & 5.4 &  \\
DCL119C-A4 & 00h55m02.51s & -37d43m11.5s & 106.0 & \nodata & \nodata & \nodata & 2.3$\pm$1.0 & 2.5$\pm$1.9 & \nodata & 2.1$\pm$1.6 & 0.803 & 5.8 &  \\
DCL119C-A5 & 00h55m04.01s & -37d43m22.0s & 109.5 & 3.66 & 61.8 & 10.4$\pm$2.3 & 4.8$\pm$0.7 & 80.4$\pm$6.4 & 45.1$\pm$17.6 & 69.9$\pm$5.6 & 5.139 & 13.1 & F \\
DCL119C-A6 & 00h55m02.89s & -37d43m15.4s & 109.0 & 4.22 & 85.1 & 15.5$\pm$1.7 & 3.4$\pm$0.5 & 32.3$\pm$2.6 & 34.4$\pm$9.6 & 28.2$\pm$2.3 & 2.331 & 16.5 & F \\
DCL122-A1 & 00h55m04.83s & -37d40m58.0s & 95.0 & 2.85 & 86.7 & 18.6$\pm$2.2 & 4.9$\pm$0.7 & 33.9$\pm$2.4 & 84.5$\pm$27.0 & 30.6$\pm$2.1 & 1.717 & 12.3 & F \\
DCL122-A2 & 00h55m04.17s & -37d40m58.3s & 107.0 & 1.50 & 64.1 & 37.4$\pm$1.9 & 6.9$\pm$0.3 & 243.0$\pm$9.7 & 331.0$\pm$43.0 & 219.0$\pm$8.8 & 4.965 & 31.9 & F \\
DCL122-A3 & 00h55m03.64s & -37d41m04.3s & 111.3 & 3.89 & 55.3 & 11.9$\pm$2.5 & 3.0$\pm$0.8 & 32.8$\pm$4.3 & 20.0$\pm$11.4 & 29.6$\pm$3.8 & 2.775 & 9.5 & F \\
DCL126-A1 & 00h55m07.71s & -37d41m04.9s & 99.9 & 1.96 & 30.1 & 63.4$\pm$1.9 & 10.0$\pm$0.5 & 419.0$\pm$12.6 & 1180.0$\pm$141.6 & 352.0$\pm$10.6 & 4.031 & 27.3 & F \\
DCL126-A2 & 00h55m06.88s & -37d41m07.0s & 98.3 & \nodata & \nodata & \nodata & 2.6$\pm$0.9 & 8.3$\pm$1.7 & \nodata & 7.0$\pm$1.5 & 1.739 & 9.8 &  \\
DCL127-A1 & 00h55m07.46s & -37d41m51.1s & 92.5 & \nodata & \nodata & \nodata & 2.5$\pm$1.2 & 5.8$\pm$1.7 & \nodata & 4.8$\pm$1.5 & 0.886 & 6.9 &  \\
DCL127-A2 & 00h55m07.80s & -37d41m47.6s & 94.1 & 3.71 & -53.2 & 12.9$\pm$2.3 & 3.3$\pm$0.6 & 26.7$\pm$2.7 & 26.5$\pm$11.4 & 22.4$\pm$2.2 & 2.646 & 19.9 & F \\
DCL127-A3 & 00h55m07.49s & -37d41m47.8s & 94.0 & \nodata & \nodata & \nodata & 2.6$\pm$0.7 & 11.5$\pm$1.4 & \nodata & 9.7$\pm$1.2 & 1.977 & 15.4 &  \\
DCL127-A4 & 00h55m08.93s & -37d41m41.7s & 93.1 & 2.59 & 87.7 & 11.8$\pm$3.3 & 2.4$\pm$0.7 & 39.9$\pm$6.8 & 13.1$\pm$8.4 & 33.3$\pm$5.7 & 4.595 & 8.8 & O \\
DCL127-A5 & 00h55m07.09s & -37d41m45.6s & 96.4 & 1.40 & -27.1 & 14.0$\pm$1.8 & 3.6$\pm$0.4 & 36.0$\pm$2.5 & 33.5$\pm$9.0 & 30.2$\pm$2.1 & 2.117 & 13.6 & F \\
DCL127-A6 & 00h55m07.24s & -37d41m37.9s & 94.9 & \nodata & \nodata & \nodata & 2.6$\pm$1.2 & 8.1$\pm$3.7 & \nodata & 6.8$\pm$3.1 & 1.22 & 6.1 &  \\
DCL127-A7 & 00h55m07.95s & -37d41m42.6s & 99.4 & 4.20 & 76.4 & 16.7$\pm$1.7 & 3.4$\pm$0.4 & 68.4$\pm$4.8 & 36.4$\pm$9.5 & 57.5$\pm$4.0 & 2.875 & 17.3 & F \\
DCL127-A8 & 00h55m06.83s & -37d41m35.7s & 99.6 & 5.06 & 80.6 & 18.4$\pm$2.6 & 2.4$\pm$0.5 & 64.1$\pm$5.8 & 20.2$\pm$9.1 & 53.8$\pm$4.8 & 3.263 & 9.8 & F \\
DCL127-A9 & 00h55m07.24s & -37d41m42.5s & 100.0 & \nodata & \nodata & \nodata & 2.3$\pm$0.7 & 6.3$\pm$1.6 & \nodata & 5.3$\pm$1.4 & 1.348 & 8.1 &  \\
DCL129-A1 & 00h55m08.95s & -37d39m34.0s & 112.0 & 2.86 & -74.5 & 13.0$\pm$2.6 & 3.4$\pm$0.8 & 22.5$\pm$2.0 & 27.8$\pm$16.4 & 16.4$\pm$1.5 & 2.354 & 14.8 & F \\
DCL129-A2 & 00h55m08.88s & -37d39m30.0s & 113.4 & 1.70 & 21.5 & 34.4$\pm$2.4 & 4.0$\pm$0.3 & 148.0$\pm$7.4 & 104.0$\pm$19.8 & 108.0$\pm$5.4 & 3.42 & 24.5 & F \\
DCL129-A3 & 00h55m08.58s & -37d39m34.2s & 114.9 & \nodata & \nodata & \nodata & 3.2$\pm$1.3 & 11.9$\pm$1.8 & \nodata & 8.7$\pm$1.3 & 1.551 & 9.3 &  \\
DCL130-A1 & 00h55m08.50s & -37d40m41.6s & 96.4 & \nodata & \nodata & \nodata & 2.8$\pm$1.3 & 9.7$\pm$1.7 & \nodata & 7.7$\pm$1.3 & 1.868 & 10.1 &  \\
DCL130-A2 & 00h55m09.25s & -37d40m56.1s & 94.9 & 3.02 & 58.4 & 10.8$\pm$2.4 & 2.2$\pm$0.5 & 18.7$\pm$2.4 & 9.8$\pm$5.7 & 14.8$\pm$1.9 & 1.885 & 11.1 & F \\
DCL130-A3 & 00h55m08.47s & -37d40m45.2s & 101.1 & \nodata & \nodata & \nodata & 1.8$\pm$0.7 & 5.2$\pm$2.3 & \nodata & 4.1$\pm$1.8 & 1.048 & 6.6 &  \\
DCL130-A4 & 00h55m09.75s & -37d40m38.9s & 104.6 & 2.81 & -10.3 & 19.2$\pm$1.9 & 3.5$\pm$0.5 & 73.6$\pm$8.8 & 44.8$\pm$12.1 & 58.4$\pm$7.0 & 2.261 & 9.0 & F \\
DCL130-A5 & 00h55m08.69s & -37d40m38.5s & 102.3 & \nodata & \nodata & \nodata & 2.3$\pm$1.2 & 5.9$\pm$2.5 & \nodata & 4.7$\pm$2.0 & 1.209 & 6.2 &  \\
DCL137A-A1 & 00h55m12.93s & -37d41m45.9s & 79.2 & \nodata & \nodata & \nodata & 3.1$\pm$0.7 & 24.6$\pm$3.0 & \nodata & 17.5$\pm$2.1 & 2.011 & 12.0 &  \\
DCL137A-A2 & 00h55m12.49s & -37d41m39.0s & 81.7 & \nodata & \nodata & \nodata & 6.3$\pm$2.1 & 10.1$\pm$4.4 & \nodata & 7.2$\pm$3.2 & 0.67 & 5.0 &  \\
DCL137A-A3 & 00h55m13.87s & -37d41m46.1s & 84.8 & 21.08 & 88.4 & 5.8$\pm$1.4 & 3.0$\pm$0.9 & 38.7$\pm$7.0 & 9.8$\pm$6.4 & 27.6$\pm$5.0 & 2.92 & 7.7 & F \\
DCL137A-A4 & 00h55m11.93s & -37d41m37.7s & 85.5 & \nodata & \nodata & \nodata & 1.5$\pm$0.8 & 10.8$\pm$1.9 & \nodata & 7.7$\pm$1.4 & 1.653 & 8.1 &  \\
DCL137A-A5 & 00h55m12.25s & -37d41m40.5s & 87.0 & \nodata & \nodata & \nodata & 3.2$\pm$1.0 & 14.8$\pm$2.7 & \nodata & 10.6$\pm$1.9 & 1.621 & 9.6 &  \\
DCL137A-A6 & 00h55m12.79s & -37d41m25.4s & 88.7 & \nodata & \nodata & \nodata & 2.9$\pm$0.6 & 40.7$\pm$6.1 & \nodata & 29.2$\pm$4.4 & 2.355 & 10.5 & DCL137B-A \\
DCL137A-A7 & 00h55m13.85s & -37d41m27.9s & 94.8 & 4.71 & -84.8 & 6.6$\pm$2.0 & 4.3$\pm$1.3 & 46.6$\pm$5.6 & 23.3$\pm$13.5 & 33.3$\pm$4.0 & 4.142 & 12.5 & DCL137B-A \\
DCL137A-A8 & 00h55m14.02s & -37d41m31.5s & 97.1 & 4.91 & 28.2 & 25.2$\pm$2.5 & 4.8$\pm$0.6 & 207.0$\pm$14.5 & 108.0$\pm$29.2 & 147.0$\pm$10.3 & 5.904 & 14.9 & DCL137C-A \\
DCL137A-A9 & 00h55m13.95s & -37d41m38.7s & 98.3 & 1.22 & 65.2 & 19.3$\pm$2.7 & 4.0$\pm$0.7 & 80.2$\pm$9.6 & 57.1$\pm$18.8 & 57.3$\pm$6.9 & 2.763 & 9.2 & F \\
DCL137A-A1 & 00h55m13.45s & -37d41m30.8s & 97.3 & \nodata & \nodata & \nodata & 4.1$\pm$1.4 & 8.6$\pm$3.4 & \nodata & 6.1$\pm$2.4 & 1.225 & 6.5 & DCL137C-A \\
DCL137B-A1 & 00h55m12.79s & -37d41m25.4s & 88.7 & 27.08 & 3.3 & 6.1$\pm$0.6 & 3.8$\pm$0.7 & 46.7$\pm$3.3 & 16.2$\pm$6.3 & 33.4$\pm$2.3 & 2.111 & 15.8 & F \\
DCL137B-A2 & 00h55m11.17s & -37d41m24.2s & 91.6 & \nodata & \nodata & \nodata & 4.8$\pm$1.9 & 29.6$\pm$13.9 & \nodata & 21.2$\pm$10.0 & 2.668 & 5.1 & O \\
DCL137B-A3 & 00h55m11.59s & -37d41m16.3s & 95.2 & 1.69 & -9.4 & 20.5$\pm$1.2 & 9.8$\pm$0.9 & 329.0$\pm$16.4 & 365.0$\pm$80.3 & 235.0$\pm$11.8 & 6.476 & 19.3 & F \\
DCL137B-A4 & 00h55m13.85s & -37d41m27.9s & 94.6 & 4.22 & -80.1 & 9.1$\pm$2.0 & 4.0$\pm$0.8 & 50.9$\pm$6.1 & 27.8$\pm$13.1 & 36.4$\pm$4.4 & 3.832 & 11.7 & F \\
DCL137B-A5 & 00h55m14.05s & -37d41m30.7s & 96.9 & 2.46 & 31.7 & 20.5$\pm$2.7 & 5.8$\pm$0.9 & 173.0$\pm$20.8 & 130.0$\pm$42.9 & 123.0$\pm$14.8 & 5.999 & 11.3 & O; DCL137A-A \\
DCL137B-A6 & 00h55m12.92s & -37d41m22.0s & 96.9 & 1.15 & 40.6 & 12.0$\pm$4.0 & 5.5$\pm$1.4 & 13.7$\pm$3.2 & 67.8$\pm$44.7 & 9.8$\pm$2.2 & 0.985 & 7.4 & F \\
DCL137B-A7 & 00h55m13.44s & -37d41m30.5s & 98.4 & \nodata & \nodata & \nodata & 5.1$\pm$2.0 & 12.6$\pm$3.4 & \nodata & 9.0$\pm$2.4 & 1.531 & 7.3 & DCL137C-A \\
DCL137B-A8 & 00h55m13.07s & -37d41m20.3s & 96.8 & \nodata & \nodata & \nodata & 3.2$\pm$1.7 & 6.8$\pm$3.1 & \nodata & 4.8$\pm$2.2 & 0.841 & 6.0 &  \\
DCL137B-A9 & 00h55m13.85s & -37d41m35.3s & 98.0 & \nodata & \nodata & \nodata & 2.4$\pm$1.6 & 21.0$\pm$5.9 & \nodata & 15.0$\pm$4.2 & 4.353 & 7.6 & O \\
DCL137C-A1 & 00h55m13.94s & -37d41m46.3s & 84.2 & 3.05 & 89.6 & 19.6$\pm$3.5 & 3.6$\pm$0.6 & 73.8$\pm$9.6 & 47.8$\pm$19.1 & 51.7$\pm$6.7 & 3.078 & 8.9 & F \\
DCL137C-A2 & 00h55m14.15s & -37d41m40.0s & 89.3 & 4.77 & -28.9 & 7.5$\pm$1.9 & 3.2$\pm$0.9 & 22.7$\pm$3.4 & 14.2$\pm$8.8 & 16.0$\pm$2.4 & 1.724 & 9.2 & F \\
DCL137C-A3 & 00h55m12.78s & -37d41m24.9s & 89.1 & \nodata & \nodata & \nodata & 3.0$\pm$1.1 & 32.7$\pm$6.9 & \nodata & 22.8$\pm$4.8 & 2.709 & 7.8 & DCL137B-A \\
DCL137C-A4 & 00h55m14.00s & -37d41m32.5s & 96.7 & 2.71 & 3.1 & 57.9$\pm$1.7 & 5.1$\pm$0.3 & 451.0$\pm$18.0 & 281.0$\pm$33.7 & 316.0$\pm$12.6 & 5.997 & 38.9 & F \\
DCL137C-A5 & 00h55m13.48s & -37d41m30.4s & 97.9 & 3.39 & 62.4 & 12.0$\pm$4.7 & 4.2$\pm$1.3 & 14.4$\pm$6.0 & 40.0$\pm$30.0 & 10.0$\pm$4.2 & 1.005 & 6.0 & F \\
DCL137C-A6 & 00h55m13.92s & -37d41m25.6s & 96.5 & 2.47 & -3.1 & 6.6$\pm$3.1 & 2.6$\pm$1.0 & 11.2$\pm$3.9 & 8.1$\pm$8.8 & 7.8$\pm$2.7 & 1.02 & 6.1 & F \\
DCL137C-A7 & 00h55m13.98s & -37d41m29.5s & 99.8 & \nodata & \nodata & \nodata & 5.3$\pm$1.1 & 25.0$\pm$2.8 & \nodata & 17.5$\pm$1.9 & 2.04 & 14.6 &  \\
DCL139-A1 & 00h55m13.52s & -37d44m13.2s & 86.2 & 1.47 & -80.1 & 40.1$\pm$2.0 & 6.1$\pm$0.4 & 247.0$\pm$9.9 & 281.0$\pm$45.0 & 161.0$\pm$6.4 & 2.451 & 13.2 & F \\
DCL139-A2 & 00h55m13.53s & -37d44m13.8s & 88.6 & 2.97 & -50.1 & 10.9$\pm$4.9 & 1.8$\pm$0.6 & 12.4$\pm$3.7 & 6.4$\pm$5.9 & 8.1$\pm$2.4 & 1.501 & 7.7 & F \\
DCL139-A3 & 00h55m13.50s & -37d44m10.3s & 91.8 & \nodata & \nodata & \nodata & 4.1$\pm$1.5 & 9.2$\pm$5.1 & \nodata & 6.0$\pm$3.3 & 0.76 & 5.2 &  \\
DCL140-A1 & 00h55m13.59s & -37d44m13.7s & 85.7 & 1.08 & -68.0 & 35.3$\pm$5.3 & 6.3$\pm$1.0 & 106.0$\pm$14.8 & 265.0$\pm$100.7 & 64.4$\pm$9.0 & 3.3 & 8.3 & O; DCL139-A2 \\
DCL140-A2 & 00h55m15.33s & -37d44m17.8s & 83.8 & 7.43 & 55.4 & 3.7$\pm$1.9 & 3.6$\pm$1.2 & 15.7$\pm$2.4 & 8.8$\pm$6.7 & 9.6$\pm$1.4 & 1.468 & 10.2 & R \\
DCL140-A3 & 00h55m15.17s & -37d44m14.8s & 86.0 & 3.15 & -28.0 & 21.6$\pm$1.7 & 3.5$\pm$0.3 & 131.0$\pm$9.2 & 49.2$\pm$10.3 & 79.7$\pm$5.6 & 4.021 & 30.1 & F \\
DCL140-A4 & 00h55m15.51s & -37d44m20.6s & 84.2 & \nodata & \nodata & \nodata & 2.3$\pm$1.0 & 12.2$\pm$2.2 & \nodata & 7.4$\pm$1.3 & 1.942 & 10.4 &  \\
DCL140-A5 & 00h55m15.26s & -37d44m21.6s & 88.3 & \nodata & \nodata & \nodata & 2.7$\pm$1.9 & 4.6$\pm$3.5 & \nodata & 2.8$\pm$2.1 & 0.775 & 5.1 &  \\
DCL140-A6 & 00h55m15.47s & -37d44m22.5s & 91.0 & \nodata & \nodata & \nodata & 2.6$\pm$0.8 & 14.7$\pm$2.1 & \nodata & 8.9$\pm$1.2 & 1.987 & 10.8 &  \\
DCL140-A7 & 00h55m13.86s & -37d44m13.9s & 92.4 & 2.44 & -81.6 & 3.8$\pm$2.7 & 1.1$\pm$0.5 & 15.1$\pm$2.6 & 0.9$\pm$1.1 & 9.2$\pm$1.6 & 2.745 & 10.3 & R
\enddata
\tablenotetext{a}{see the text of Section~\ref{sec:results} for further details on how each column is computed}
\tablenotetext{b}{F: full sample; R: resolved sample; O: outside primary beam by more than 10\% beam FWHM; DCLXX-Y: this cloud is a duplicate of cloud DCLXX-Y, and is not considered independently in the analysis}
\end{deluxetable*}
\clearpage
\end{landscape}

\end{document}